\documentclass[12pt]{report}
 \pdfoutput=1 
\usepackage{xcolor}
\usepackage{framed}
\usepackage[brazil]{babel}
\usepackage{listings}
\usepackage[utf8]{inputenc}
\usepackage[T1]{fontenc}
\usepackage{epsfig}
\usepackage{graphicx}
\usepackage{wrapfig}
\usepackage{appendix}
\usepackage{fancyhdr}
\usepackage{color}
\usepackage{url}
\usepackage{amsmath,amsthm,amssymb,natbib}
 \usepackage{multirow}
\usepackage{adjustbox}
\graphicspath{{./Figuras/}} % Nome da pasta contendo as figuras.

\colorlet{shadecolor}{blue!11}
\renewcommand{\thefigure}{\arabic{chapter}.\arabic{figure}}

\let\oldref\ref
\renewcommand{\ref}[1]{(\oldref{#1})}

\renewcommand{\thefigure}{\arabic{chapter}.\arabic{figure}}

\usepackage{listings}

\begin{document}
\pagenumbering {roman}
\newcounter{step}
\newcounter{list}
%%%%%%%%%%%%%%%%%%%%%%%%%%%%%%%
%%%Capa%%%
\title{\LARGE\bf spsurv: An R package for semi-parametric survival analysis}
\author{\\ [10pt] \large \bf Renato Valladares Panaro \\ [80pt] Departamento de Estatística - ICEx - UFMG}
\date{February 2020}
\maketitle

%%%%%%%%%%%%%%%%%%%%%%%%%%%%%%%%
%%%Folha de rosto%%%

\begin{center}
\setcounter{page}{2} {\Large\bf  spsurv: An R package for semi-parametric survival analysis} \\
\vspace{2cm}
{\large\bf Renato Valladares Panaro}
\end{center}
\vspace{5pt}
{\flushleft \hspace{2cm} Advisor: Vinícius Diniz Mayrink} \\ \vspace{-1cm}
{\flushleft \hspace{2cm} Co-advisor: Fábio Nogueira Demarqui}

\vspace{1cm}

%{\flushleft Dissertação submetida ao Programa de Pós-Graduação em Estatística da Universidade Federal de Minas Gerais, como parte dos requisitos necessários à obtenção do grau de Mestre em Estatística.}

\vspace{3cm}

\begin{center}
Departamento de Estatística\\
Instituto de Ciências Exatas \\
Universidade Federal de Minas Gerais
\end{center}

\vspace{2cm}

\begin{center}
    Belo Horizonte, MG - Brasil \\ February 2020
\end{center}

\newpage

%%%%%%%%%%%%%%%%%%%%%%%%%%%%%%%%
%%% Dedicatória %%%

%\newpage
%%\thispagestyle{empty}
%\noindent \vspace{19cm}
%\begin{flushright}
%Aos meus pais.\\
%\end{flushright}

%%%%%%%%%%%%%%%%%%%%%%%%%%%%%%%%
%%% Frase %%%  Opcional...

%\newpage
%\noindent
%\vspace{18cm}
%%\begin{flushright}
%{\small ``Our world, our life, our destiny, are dominated by
%Uncertainty; this is perhaps the only statement we may assert
%without uncertainty''. \\ 
%\vspace{10pt}
%de Finetti}
%\end{flushright}

%%%%%%%%%%%%%%%%%%%%%%%%%%%%%%%%
%%% Agradecimentos %%%  
% Seja sucinto. Escreva apenas 1 página de agradecimentos. 
% Se recebeu bolsa, agradeça a agência financiadora da bolsa. 

%\newpage
%%\thispagestyle{empty}
%\noindent \vspace{50pt}
%\begin{center}
%{\LARGE\bf Agradecimentos}
%\end{center}
%\vspace{30pt} %\thispagestyle{empty}

%bla bla bla bla bla bla bla bla bla bla bla bla bla bla bla bla
%bla bla bla bla bla bla bla bla bla bla.

%\vspace{10pt}

%bla bla bla bla bla bla bla bla bla bla bla bla bla bla bla bla
%bla bla bla bla bla bla bla bla bla bla.

%\vspace{10pt}

%bla bla bla bla bla bla bla bla bla bla bla bla bla bla bla bla
%bla bla bla bla bla bla.

%\vspace{10pt}

%bla bla bla bla bla bla bla bla bla bla bla bla bla bla bla bla
%bla bla bla bla bla bla bla bla bla bla.

%%%%%%%%%%%%%%%%%%%%%%%%%%%%%%%%
%%% Resumo %%%

%\newpage
%\noindent \vspace{50pt}
%\begin{center}
%{\LARGE\bf Resumo}
%\end{center}
%\vspace{30pt}\\

%{\flushleft Palavras-chave: Palavra 1, Palavra 2, Palavra 3, Palavra 4.}

%%%%%%%%%%%%%%%%%%%%%%%%%%%%%%%%
%%% Abstract %%%

\newpage
\thispagestyle{empty}
 \noindent \vspace{50pt}
\begin{center}
{\LARGE\bf Resumo}
\end{center}
\vspace{30pt}

Avanços na computação e no desenvolvimento de \textit{software} permitiram cálculos mais complexos e menos custosos no que diz respeito a pesquisas médicas (análise de sobrevivência), a estudos de engenharia (confiabilidade) e a observação de eventos sociais (análise de eventos históricos). Assim sendo, muitos esforços de modelagem semi-paramétrica para dados de tempo até o evento surgiram nos últimos anos. Neste contexto, este trabalho apresenta uma estrutura flexível baseada no polinômio de Bernstein para modelagem de dados de sobrevivência. Essa abordagem inovadora é aplicada na estimação de funções de base desconhecidas inerentes de famílias de modelos existentes na literatura, como modelos de riscos proporcionais, chances proporcionais e tempo de falha acelerado. Além da contribuição literária, este trabalho também contribui com rotinas automatizadas inéditas para a comunidade de usuários da linguagem \texttt{R}, com o suporte de algoritmos implementados no  \textit{software} \texttt{Stan}. Ao final do estudo, a implementação das rotinas propostas foi discutida e avaliada através de estudos de simulação. A criação de um pacote \texttt{R} surge como alternativa para agrupar todas essas importantes contribuições. Além disso, os modelos baseados no polinômio de Bernstein de riscos proporcionais,  de chances proporcionais e de tempo de falha acelerado foram ajustados a dados reais de pacientes portadores de câncer, usando tanto o método de estimação por máxima verossimilhança quanto algoritmos Bayesianos. \\
\newpage
\begin{center}
{\LARGE\bf Abstract}
\end{center}
\vspace{30pt}

Software development innovations and advances in computing have enabled more complex and less costly computations in medical research (survival analysis), engineering studies (reliability analysis), and social sciences event analysis (historical analysis). As a result,  many semi-parametric modeling efforts emerged when it comes to time-to-event data analysis. In this context, this work presents a flexible Bernstein polynomial  (BP) based framework for survival data modeling. This innovative approach is applied to existing families of models such as proportional hazards (PH), proportional odds (PO), and accelerated failure time (AFT) models to estimate unknown baseline functions. Along with this contribution, this work also presents new automated routines in \texttt{R}, taking advantage of algorithms available in \texttt{Stan}. The proposed computation routines are tested and explored through simulation studies based on artificial datasets.   The tools implemented to fit the proposed statistical models are combined and organized in an \texttt{R} package.  Also, the BP based proportional hazards (BPPH), proportional odds (BPPO), and accelerated failure time (BPAFT) models are illustrated in real applications related to cancer trial data using maximum likelihood (ML) estimation and  Markov chain Monte Carlo (MCMC) methods. 

%%%%%%%%%%%%%%%%%%%%%%%%%%%%%%%%
%%% Sumário %%% Será inserido automaticamente...

\tableofcontents
\listoffigures
\listoftables
\newpage \thispagestyle{empty}

%%%%%%%%%%%%%%%%%%%%%%%%%%%%%%%%%%%%
%%% Chapters %%%

\pagenumbering {arabic}

% Chapter 1 = Introduction.
\chapter{Introduction}
$~~~~$Lifetime research is one of the earliest fields in statistics. According to \cite{Hacking:2006}, one of its first applications dates back to the 1700s when John Graunt published the first set of analyses upon the London Bills of Mortality. After that, as from Graunt's book review, tabulations also began in France and other western European countries later on. At the beginning of the eighteen century, increasing annuity incomes also demanded expertise in life expectancy calculations. So, De Moivre, Daniel Bernoulli, and other pioneers contributed to the discussion of available techniques to accurately evaluate life tables at that time, giving rise to the field of study called survival analysis \citep{Rviews:2017}.

Nowadays, software development innovations and recent advances in computing have enabled tools to handle more sophisticated statistical techniques to time-to-event data. Indeed, \texttt{R} software libraries \citep{R:2018}  containing specific routines such as \texttt{survival} \citep{survival:2000}, \texttt{survminer} \citep{survminer:2018}, \texttt{timereg} \citep{timereg} and \texttt{flexsurv} \citep{flexsurv:2016} have become essential tools for practitioners, professionals and researchers. Likewise, the \texttt{spsurv} package was designed to contribute with a flexible set of semi-parametric survival regression modelings, including proportional hazards (PH), proportional odds (PO), and accelerated failure time (AFT) models for right-censored data. The proposed package provides extensions based on a fully likelihood-based approach for either Bayesian or maximum likelihood (ML) estimation procedures,  along with smooth estimates for unknown baseline functions based on Bernstein polynomial (BP).

Over the past years, the BP have been widely related to regression modeling \citep{Tenbusch:1997, Chang:2007} and probability density function estimation \citep{Vitale:1975,Petrone:1999, Babu:2002, Choudhuri:2004}. In contrast, few contributions relating to BP with survival analysis regression can be found in the literature. Some applications in this sense have emerged as alternatives to the partial likelihood estimation proposed by \cite{Cox:1972}. \cite{Chang:2005}, for example, estimated the failure rate considering a Beta Process \textit{a priori} to homogeneous populations. The polynomial degree was treated as a random quantity in this reference. Especially, \cite{Osman:2012} addressed the baseline hazard function using BP and provided, among other results, proof on the likelihood log-concavity property for the proposed modeling. This characteristic leads to less costly computational procedures to find Bayesian estimators and is also necessary to guarantee the uniqueness of the ML estimator. The authors also focused on the failure rate but allowed crossing survival curves. In this case, the developed model includes covariates (non-homogeneous population) and a fixed polynomial degree. 

More recently, \cite{Mclain:2013} proposed time transformation models assuming linearity between the survival times and the covariates. \cite{Chen2:2014} used a transformed BP, centered at standard parametric families, in the accelerated hazards model framework. This application assumes a random degree polynomial by applying a Dirichlet process. \cite{Zhou:2017} used BP to approximate the cumulative baseline hazard function. Besides, \cite{Zhou:2018} proposed modeling the baseline hazard function through a prior distribution called \textit{transformed Bernstein polynomial}. In this proposal, the authors assume a parametric distribution as central (e.g., Weibull) and use the Bernstein polynomial structure (linked to the baseline survival function) to allow variations around the central choice. That allows greater flexibility in the format of the baseline hazard function. \cite{Wu:2018} introduced a flexible Bayesian nonparametric procedure to estimate the odds under the case of interval censoring.

The \texttt{spsurv} interfaces with \texttt{Stan} for more flexibility in terms of user-defined modeling. \texttt{Stan} is an open-source platform that has a specific language and many built-in log-probability functions that can be used to define custom likelihood functions and prior specifications \citep{Carpenter:2017}. The program has extensive supporting literature available online such as reference manuals, forums, articles, and books for users and developers. The \texttt{Stan} currently defaults to NUTS sampling, which consists of a Hamiltonian Monte Carlo \citep{Duane:1987} extension that explores the posterior distribution more efficiently  \citep{Hoffman:2014}. In addition, access to \texttt{Stan} can be established through several modules integration: such as \texttt{rstan} \citep{rstan:2018}, PyStan (integrated with Python), MatlabStan (integrated to MATLAB), Stan.jl (integrated with Julia) and StataStan (integrated to Stata). 

$~~~~$The general goal of this work is to present the \texttt{spsurv} package along with technical details and practical aspects of its usage. The specific contributions of the present work are:
\begin{itemize}
    \item Explore the BP approach to semi-parametric modeling in survival analysis. Here, we consider three contexts: PH, PO, and AFT models.
    \item Present a comprehensive simulation study to show that each model performs well under the Bayesian or Frequentist inference approach.
    \item Build a \texttt{R} package called \texttt{spsurv} that shall be used to fit the semi-parametric models discussed in this dissertation.
\end{itemize}

This study is organized as follows: Chapter 2 presents the necessary background on survival analysis, which is essential to comprehend the theoretical basis of this work. Afterward, Chapter 3 consists of a summary of how to address BP in the context of approximation or estimation (survival analysis). The next chapter (Chapter 4) explores implementation issues and statistical inference concerns. Achievements from simulation studies were discussed in Chapter 5. Chapter 6 discusses two real data applications and the main results reached, along with comments on the interpretation of the distinct approaches and frameworks. Chapter 7 summarizes the whole content of this dissertation, highlights the first results, and shows the most significant proposals of future work. Finally, the main \texttt{spsurv} package routines are introduced in Appendix. Here we indicate, how to fit a model, discuss summary elements, and present specific graphs for the survival data analysis in \texttt{R}.

% Chapter 2 = Chapter Title.
\chapter{Survival analysis fundamentals}
$~~~~$Survival analysis is a field of study in statistical science dedicated to solving problems in which the time to an event of interest is of a reasonable importance \citep{Cox:1972}. For this, a random variable $T$ describes the continuously distributed time to a particular event of interest, namely occurrence or failure time. The time-to-event observations are considered incomplete in a manner that each response $Y_i$ is subject to censoring, where $i \in \{1,\dots,n\}$ represents an individual (e.g.: patient or equipment). 

\begin{figure}[!htb]
\centering
     \includegraphics[width=.65\textwidth]{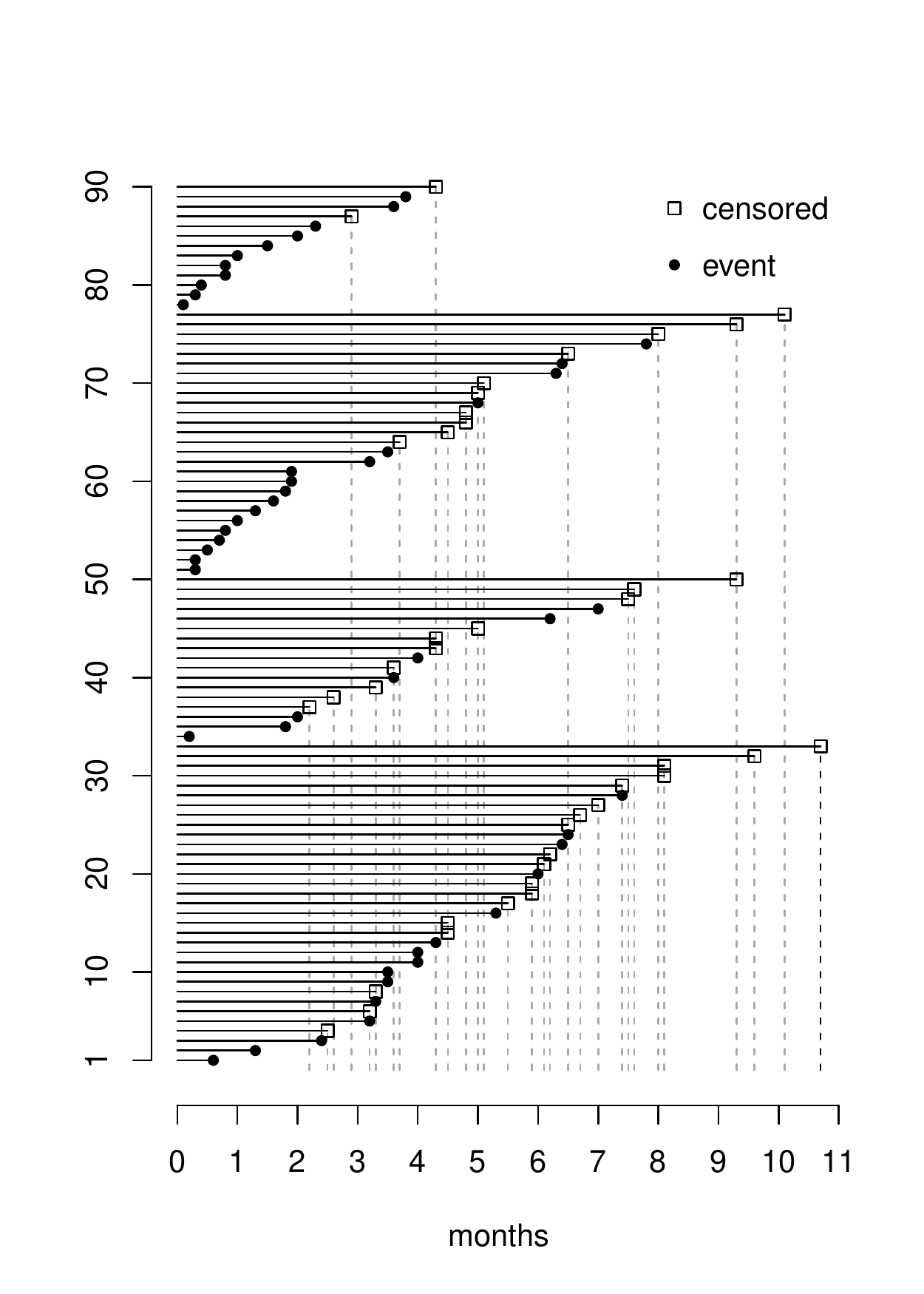}
\caption{Data set about reports on male larynx-cancer patients diagnosed during the 1970s \citep{Kardaun:1983}. The continuously observed time is represented by horizontal lines in which rounded points give failure times, and squared points indicate censored observations. The lifetime is measured in months for 90 patients in the study.}
 \label{Fig:censoringplot}
\end{figure}

An observation is classified as right-censored or left-censored when the event of interest did not occur within the survey period. Particularly, an observation is said to be right-censored if the failure time $T_i$ is greater than the censoring time $C_i$, but it is unknown by how much. The right censoring assumption adopted for this work is properly denoted as \\
\begin{equation}
Y_i =  \begin{cases}
T_i, \text{ if } T_i \le C_i; \\ 
C_i,  \text{ otherwise}.
\end{cases}\label{rightcensor}
\end{equation}\\
In contrast, if the failure time is lower than the censoring time, but it is uncertain by how much, the data is said to be left-censored. Not least, the data is classified as interval-censored when both right and left censoring are likely to happen. In this situation, the event occurs between two censoring times within the period of the survey. In other words, that is to say, that the failure time is lower than an upper censoring time and greater than a lower censoring time. To identify censoring and failure times, let $\delta_i$ be the failure binary indicator, assume that $\delta_i$ = 1 indicates that the $i^{th}$ observed time point is a failure time. Figure \oldref{Fig:censoringplot} shows a real right censoring mechanism example, this data is further explored in Chapter 5. \\  \\
Some ordinary situations might have caused the right-censored records in monthly lifetimes reports of laryngeal cancer patients (see Figure \oldref{Fig:censoringplot}). For instance, a right-censored time might have measured the time that a patient dropped the study, or it might have described the lifetime of a certain patient due to uncontrolled external causes of death. Remarkably, the total research time was reported for patient number 33 in this study because the patient has not experienced the event of interest (death) before the end of the survey.

Henceforth, consider the cumulative distribution function (c.d.f.) \textit{i.e.} $F(t) = P(T \le t)$ so that $S(t)=1-F(t)$ is the survival function. The survival function is of great interest in medical research since it describes the probability of a patient experiencing the event of interest beyond a specified time. According to \cite{Colosismo:2001}, similar survival functions may have
completely different failure rate functions. This occurs because the failure (or hazard) rate function is defined as: \\
$$h(t)=\lim _{\Delta t \rightarrow{0}}\frac{P(t<T\le t+ \Delta t \mid T > t)}{\Delta t} = \frac{f(t)}{S(t)}.$$ \\
Other important functions, such as the cumulative hazard function, $H(t)$, and the odds function, $R(t)$, play a key role in modeling lifetime data. These functions are related to the survival function and they can be obtained from it. For instance, some very useful established relationships are: $H(t) = - \log S(t)$, $H(t)=\int^t_0 h(u)du$, $f(t)=\frac{d [1-S(t)]}{dt}$ and $S(t) = [1+R(t)]^{-1}$. 

\cite{Klein:1997} carefully detail the construction of the likelihood function for survival experiments. In short, consider a survey where $r = \sum \limits _{i=1}^{n}\delta_i$ failures and $n-r$ right censoring  times \ref{rightcensor} have been reported, so that:\\
\begin{equation}
    \begin{aligned}
    L(\boldsymbol{\beta} \mid \boldsymbol{y}, \boldsymbol{\delta}, \boldsymbol{x}) &= \prod_{i = 1}^{n} P(Y_i=t, \delta_i = 1)  ~P(Y_i=t, \delta_i = 0)\\
    &= \prod_{i = 1}^{n} P(Y_i=t, T_i \le C_i, Y_i=T_i) ~P(Y_i=t, T_i > C_i, Y_i=C_i)\\
    &= \prod_{i = 1}^{n} P(T_i=t, C_i \ge t) ~P(C_i=t, T_i > t).\\
     \end{aligned}\label{fulllike}
\end{equation}\\
%\begin{equation}
%    \begin{aligned}
%    L(\boldsymbol{\beta} \mid \boldsymbol{y}, %\boldsymbol{\delta}, \boldsymbol{x}) &= %\prod_{i = 1}^{r} P(Y_i=t, \delta_i = 1) %\prod_{i=r+1}^{n} P(Y_i=t, \delta_i = 0)\\
%    &= \prod_{i = 1}^{r} P(Y_i=t, T_i \le C_i, %Y_i=T_i) \prod_{i=r+1}^{n} P(Y_i=t, T_i > C_i, %Y_i=C_i)\\
%    &= \prod_{i = 1}^{r} P(T_i=t, C_i \ge t) %\prod_{i=r+1}^{n} P(C_i=t, T_i > t).\\
%    \end{aligned}\label{fulllike}
%\end{equation}\\
where $\boldsymbol\beta$ represents the effects of each patient's individual characteristics that might influence on their respective remaining lifetime (failure time), such as age, gender, or disease status. In general, the goal of survival analysis regression models is to draw inferences about the effect of those explanatory variables to the time-to-event response. Also,the present work relies on the assumption of a non-informative censoring mechanism, therefore the failure and censoring times are considered mutually independent in \ref{fulllike}. As importantly, the probability distribution of the censoring times does not bring any information about the failure times distribution \citep{Klein:1997}. As a consequence, the likelihood function for non-informative right-censored data is:\\
\begin{equation}
    \begin{aligned}
    L(\boldsymbol{\beta} \mid \boldsymbol{y}, \boldsymbol{\delta}, \boldsymbol{x}) &= \prod_{i = 1}^{n} P(T_i=t)P(C_i \ge t) ~ P(C_i=t)P(T_i > t)\\
       &\propto \prod_{i = 1}^{n} P(T_i=t)~ P(T_i > t).\\
    \end{aligned}\label{formula:likelihood} 
\end{equation}\\
%\begin{equation}
%    \begin{aligned}
%    L(\boldsymbol{\beta} \mid \boldsymbol{y}, %\boldsymbol{\delta}, \boldsymbol{x}) &= %\prod_{i = 1}^{r} P(T_i=t)P(C_i \ge t) %\prod_{i=r+1}^{n} P(C_i=t)P(T_i > t)\\
%     &= \prod_{i = 1}^{r} f(y_i, \boldsymbol %\beta \mid \boldsymbol x) \prod_{i=r+1}^{n} %S(y_i, \boldsymbol \beta \mid \boldsymbol %x)\prod_{i = 1}^{r} P(C_i \ge t) %\prod_{i=r+1}^{n} P(C_i=t).\\
%             &\propto \prod \limits_{i=1}^n %\left[ f(y_i, \boldsymbol \beta \mid %\boldsymbol x)\right]^{\delta_i} \left[S(y_i, %\boldsymbol \beta \mid \boldsymbol %x)\right]^{1-\delta_i}\\
%             &\propto \prod \limits_{i=1}^n %\left[ h(y_i, \boldsymbol \beta \mid %\boldsymbol x)\right]^{\delta_i} \exp\{-H(y_i, %\boldsymbol \beta \mid \boldsymbol x)\}.
%    \end{aligned}\label{formula:likelihood}
%\end{equation}\\
The likelihood representation in \ref{formula:likelihood} \text is general in the sense that PH, PO and AFT survival models for non-informative right-censored data can be derived from it.

Generally, health studies are focused on finding disease prognosis. Therefore, it is required to include regression coefficients to investigate the impact of the explanatory variables on the time response. In line with this idea, the PH, PO and AFT classes of survival regression will be presented in this chapter for right-censored data; more details on basic aspects of survival analysis can be found in \cite{Klein:1997}, \cite{Ibrahim:2014}, \citep{Collett2015},  \cite{Colosismo:2001},  and \cite{Kalbfleisch:2011}.
 
\section{Proportional hazards model}

$~~~~$The PH model \citep{Cox:1972} incorporates regression-like arguments into the statistical framework by multiplying a non-negative function of the covariates to the hazard rates \citep{Kalbfleisch:2011}. In short, the cross product, between the column vector of features $\boldsymbol{x_i}$ and the column vector of constant coefficients $\boldsymbol\beta$, exponentially increments (or decrements) the reference group hazard function. The referred group (or population) is determined upon common characteristics among patients. For example, the group of patients whose observed covariates are all equal to zero is referred to as the baseline (or reference) group. The baseline hazard function that describes the reference group is $h_0(t)=h(t \mid \boldsymbol{\beta}, \boldsymbol{x_i = 0})$ and the PH model formulation is then given by:\\
\begin{equation}
h(t \mid \eta_i) = h_0(t) \exp\{\eta_i\}. \label{hazardPH}
\end{equation} \\
where $\eta_i = \boldsymbol{\beta^\top x_i}$ ($i^{th}$ patient). The multiplicative form in \ref{hazardPH} is the most popular in the literature. In this case, there is no intercept due to the presence of $h_0$. The intercept can be included using part of the functional structure adopted for the baseline hazard function. For example, in the parametric case, we can parameterize a regression that includes an intercept using the  Weibull hazard function form. Otherwise, if the regression does not include an intercept, the baseline hazard, and its related functions, incorporate this constant term \citep{Colosismo:2001}.

In the proportional hazards context, the ratio between hazard rate functions of individuals belonging to distinct groups is constant over time. The $i^{th}$ and $j^{th}$ patients are compared through hazards ratio (HR):\\
\begin{equation}
\text{HR} = \frac{h_0(t) \exp(\boldsymbol{\beta^\top} \boldsymbol{x_i})}{h_0(t) \exp(\boldsymbol{\beta^\top} \boldsymbol{x_j})} =\exp\{\boldsymbol{\beta^\top}(\boldsymbol{x_i} - \boldsymbol{x_j})\}
\label{formula:hr}
\end{equation}\\
where the vector $\boldsymbol{\beta}$ of regression effects is being replaced by some estimate $\boldsymbol{ \hat \beta}$ in order to obtain the estimated HR. As the short form \ref{formula:hr} contains only the constant regression effects, the ratio between the two patients is proportional over time, leading to the proportional hazards classification. Furthermore, the survival function described before is now written in the presence of regression elements as follows:
 \begin{equation}
    S(t \mid \eta_i) = S_0(t)^{\exp\{\eta_i\}}, \label{survPH}
 \end{equation}
where $S_0(t)$ is the survival function for the reference group, \textit{i.e.} $S(t\mid 0) = S_0(t)$.  \cite{Cox:1972} did not consider any formulation for the survival times distribution. Conversely, the author treated the baseline survival as an unspecified component of the model. The  Cox's PH (CoxPH) model was classified as a semi-parametric model since the baseline time-dependent term $h_0(t)$ do not rely on any functional form. Indeed, it does not assume a parameter based (or any) formulation.
For this reason, it is not feasible to employ fully likelihood-based inference procedures to the CoxPH model. As a consequence, partial ML methods became very popular on the purpose of drawing inferences from the CoxPH model. Afterward, \cite{Breslow:1972} proposed a fully non-parametric step function to estimate the survival baseline functions which preserved its inherent properties and relationships with associated elements. 

The likelihood function for inferences related to the PH model can be derived from \ref{formula:likelihood}. As mentioned, the general likelihood function for right-censored data is adapted to the specific models. 
The likelihood function whose inferences obtained refer to the PH class models is:\\
\begin{equation}
\begin{aligned}
L(\boldsymbol{\beta} \mid \boldsymbol{y}, \boldsymbol{\delta}, \boldsymbol{x})          &\propto \prod_{i = 1}^{n} P(T_i=t)~ P(T_i > t).\\           &= \prod \limits_{i=1}^n \left[ f(y_i, \boldsymbol \beta \mid \boldsymbol x)\right]^{\delta_i} \left[S(y_i, \boldsymbol \beta \mid \boldsymbol x)\right]^{1-\delta_i}\\   &= \prod \limits_{i=1}^n \left[ h(y_i, \boldsymbol \beta \mid \boldsymbol{x})\right]^{\delta_i} S(y_i, \boldsymbol \beta \mid \boldsymbol{x}) \\
             &= \prod \limits_{i=1}^n \left[ h(y_i, \boldsymbol \beta \mid \boldsymbol x)\right]^{\delta_i} \exp\{-H(y_i, \boldsymbol \beta \mid \boldsymbol x)\}\\
             & = \prod \limits_{i=1}^n \left[ {h_0(y_i, \boldsymbol \beta) }\color{black}\exp\{\eta_i\}\right]^{\delta_i} \exp\{-H_0(y_i, \boldsymbol \beta)\exp\{\eta_i\}\}. \\
\end{aligned}
\label{formula:loglikph}
\end{equation}.\\
The PH model have been widely applied due to the fact that the HR \ref{formula:hr} is straightforward to interpret \citep{Colosismo:2001}. Although PH models are very popular in the literature, other classes of frameworks such as the PO and AFT can be considered. As reported by \cite{DeIorio:2009}, the PH model is limited to situations in which the proportional HR assumption is not violated over time. For instance, the PO model is often of great interest for researchers, as it relies on the proportional odds ratio rather than the proportional hazards ratio assumption. Above all, the AFT does not rely on any assumption of proportionality. In this case, the median time ratio is a  consequence from its logarithm relation to the predictors.  

The next section describes the PO case,  built to represent the chance on the occurrence of an event of interest at a given time. The afterward section represents the AFT case; this is one of the most comprehensive classes of survival models  \citep{Collett2015}. At the end of this chapter, we discuss the relation between the studied frameworks.

\section{Proportional odds model}

$~~~$The PO model \citep{Bennett:1983} is intended to situations in which survival curves tend to get closer to each other (but they do not cross) along time, for different groups of subject in the study. The PO model is built on the assumption of constant odds ratio (OR) between groups which consists of the odds of an event occurring in some group divided by the odds on event occurring in another group.  The OR on the occurrence of an event of interest regarding the baseline group (OR$_0$) is given by:\\
$$
\text{OR}_0=\frac{1-S(t \mid \eta_i)}{S(t\mid \eta_i)} \frac{S_0(t)}{1-S_0(t)}=\exp\{\eta_i\}.
$$\\
As the survival function describes the probability of an individual experiencing the event beyond a specific time, the OR of 1 indicates that the event is equally likely to occur in both groups at the specified time. An OR lower than 1 indicates that the event is less likely to occur in the reference group. Conversely, if the OR is higher than 1, the event is more likely to occur in the reference group. Alternatively, we write:\\
\begin{equation}
\begin{aligned}
R(t \mid \eta_i) &= R_0(t) \exp\{\eta_i\}=\frac{F(t \mid \eta_i)}{1-F(t\mid \eta_i)}. \label{propodds2}\\ 
\end{aligned}
\end{equation} \\
Note that, the covariates, together with their effects, assemble the regression argument  $\eta_i$,  that was previously included through the survival function in $\ref{survPH}$. Correspondingly, the regression is also included in \ref{propodds2} as $F(t \mid \eta_i) = 1-S(t \mid \eta_i)$. The odds on occurring an event for the reference group is $R_0(t)=R(t \mid \eta_i = 0)$. Similar to the hazard function in the PH model \ref{hazardPH}, the odds on the occurrence at a time has a similar multiplicative configuration as the one for the baseline hazard. Therefore, two patients can be compared in the absence (decrease) or presence (increase) of certain characteristic through the odds ratio (OR). We write:\\
\begin{equation}
OR = \frac{R_0(t) \exp\{\boldsymbol{\beta^\top} \boldsymbol{x_i}\}}{R_0(t) \exp\{\boldsymbol{\beta^\top} \boldsymbol{x_i}\}} =\exp\{\boldsymbol{\beta^\top}(\boldsymbol{x_i} - \boldsymbol{x_j})\}.\label{formula:or} 
\end{equation}\\
It should be noted that, the short form \ref{formula:or} contains only the fixed estimated effects over time, which means that the OR is proportional over time, leading to the so called proportional odds classification.

Few changes are made to the general likelihood expression \ref{formula:likelihood} in order to obtain the likelihood form related to the PO model. The fully likelihood function that allows inference for the PO model is:\\
\begin{equation}
\begin{aligned}
L(\boldsymbol{\beta} \mid \boldsymbol{y}, \boldsymbol{\delta}, \boldsymbol{x})
&\propto \prod_{i = 1}^{n} P(T_i=t)~ P(T_i > t).\\ 
        &=  \prod \limits_{i=1}^n \left[ f(y_i, \boldsymbol \beta \mid \boldsymbol x)\right]^{\delta_i} \left[S(y_i, \boldsymbol \beta \mid \boldsymbol x)\right]^{1-\delta_i}\\   &= \prod \limits_{i=1}^n \left[ h(y_i, \boldsymbol \beta \mid \boldsymbol{x})\right]^{\delta_i} S(y_i, \boldsymbol \beta \mid \boldsymbol{x}) \\
        &=\prod \limits_{i=1}^n \left[  \frac{r_0(y_i)\exp\{\eta_i\}}{1+R_0(y_i)\exp\{\eta_i\}}\right]^{\delta_i} \left( \frac{1}{1 +  R_0(y_i)\exp\{\eta_i\}}\right),
\end{aligned}
\label{loglikpo}
\end{equation}\\
with $r_0(t)=\frac{\partial R_0(t)}{\partial t}$. It is noteworthy that, the PO model can be formulated under the parametric or semi-parametric frameworks. As an example, we can obtain a PO model with the parametric Log-logistic functional form in the baseline functions. The PO model can also be parameterized to have an intercept in the regression. According to \cite{Bennett:1983}, the partial likelihood approach does not have a simple closed form in this case. As a solution to this issue, the author applies a full likelihood Newton-Raphson method involving the transformation of the failure times to the Log-logistic distribution. Later, \cite{Pettitt:1984} developed an alternative rank based distribution-free inference method to compute the estimates for the PO models. 

The previous two sections were dedicated to present some critical notations and elements related to the PH and PO models. As stated before, the AFT model does not rely on any assumption over hazards or odds proportionality. Instead, the AFT model is interpreted in terms of the median times ratio, commonly referred to as the acceleration factor. The next section
shows the main aspects of the AFT modeling. These three settings (PH, PO, and AFT) are the main statistical approaches to be investigated
in this dissertation. 

\section{Accelerated failure time model}

$~~~~$In the AFT model, the explanatory variables act multiplicatively on the time scale and thus affect the rate at which the survival curve decays along the time \citep{Collett2015}. The AFT modeling can be represented by the following log-linear relationship with the observed times:\\
\begin{equation}
\log T_i = \alpha + \eta_i + \sigma \epsilon_i.
    \label{Formula:AFT}
\end{equation}\\
where $T_i$ is the random variable that describes the time-to-event for the $i^{th}$ patient, $\alpha$ and $\sigma$ are, respectively, the location and scale parameters of the distribution for the random variable $\log T_i$. According to \cite{hosmer2008applied}, the AFT models are predominantly applied under a parametric perspective. In this case, a distribution function is assumed either for the random variable $T_i$ or $\epsilon_i= [\log T_i -(\alpha + \eta_i )]/\sigma$. Indeed, the survival function might be obtained directly from $T_i$ or indirectly from $\epsilon_i$. One can write: \\
 \begin{equation}
   \begin{aligned}
 S(t \mid \eta_i) & = P(T_i > t ) \\
 &= P(\exp \{\alpha + \eta_i + \sigma \epsilon_i\} > t )\\
        &= P\left( \epsilon_i > \frac{\log t - (\alpha + \eta_i )}{\sigma} \right). 
  \end{aligned} \label{percaft} 
 \end{equation}\\
 where $\frac{1}{\sigma}[\log t - (\alpha + \eta_i)]$ represents the $t^{th}$ percentile of the  distribution assigned to $\epsilon_i$. For instance, the random variable $\epsilon_i$ might follow a standard Normal, Log-gamma or Logistic distribution. As a result, $T_i$ will have Log-normal, Gamma or Log-logistic distribution, respectively. In particular, consider that the AFT survival function can be rewritten in terms of the reference group survival as:\\
 \begin{equation}
    \begin{aligned}
    S(t \mid \eta_i) &= P(\exp\{\alpha +\eta_i+ \sigma \epsilon_i\} > t )  \\
    &= P(\exp\{\alpha +\eta_i -\eta_i + \sigma \epsilon_i \} > t \exp\{-\eta_i\})  \\
    &= P(\exp\{\alpha + \sigma \epsilon_i \} > t\exp\{-\eta_i\} ) \\
    &= S(t \exp\{-\eta_i\} \mid 0) \\
    &= S_0(t \exp\{-\eta_i\}).
    \end{aligned}\label{Formula:survaft}
\end{equation}\\
For an AFT model, the hazard function can be obtained with \ref{Formula:survaft} using the following relation:
\begin{equation}
\begin{aligned}
    h(t\mid \eta_i) &= -\frac{d~\log S(t \mid \eta_i)}{dt}\\
    &= -\frac{d~\log S_0(t \exp\{-\eta_i\})}{dt}\\
    &= -\exp\{-\eta_i\}\frac{-f_0(t \exp\{-\eta_i\})}{S_0(t \exp\{-\eta_i\})} \\
    &= \exp\{-\eta_i\} h_0(t \exp\{-\eta_i\}).\label{Formula:hazardAFT}
\end{aligned}
\end{equation}\\
Typically, the joint effect of covariates accelerates or decelerates the failure time, leading to the AFT family. Additionally, the AFT likelihood function is obtained with \ref{formula:likelihood} as well:\\
\begin{equation}
\begin{aligned}
L(\boldsymbol{\beta} \mid \boldsymbol{y}, \boldsymbol{\delta}, \boldsymbol{x}) &\propto \prod_{i = 1}^{n} P(T_i=t)~ P(T_i > t)\\ 
        &=  \prod \limits_{i=1}^n \left[ f(y_i, \boldsymbol \beta \mid \boldsymbol x)\right]^{\delta_i} \left[S(y_i, \boldsymbol \beta \mid \boldsymbol x)\right]^{1-\delta_i}\\ 
          &= \prod \limits_{i=1}^n \left[ h(y_i, \boldsymbol \beta \mid \boldsymbol{x})\right]^{\delta_i} S(y_i, \boldsymbol \beta \mid \boldsymbol{x}) \\
 &= \prod \limits_{i=1}^n \left[ h(y_i, \boldsymbol \beta \mid \boldsymbol x)\right]^{\delta_i} \exp\{-H(y_i, \boldsymbol \beta \mid \boldsymbol x)\}\\
        &=\prod \limits_{i=1}^n \left[ \exp\{-\eta_i\} h_0( y_i\exp\{-\eta_i\})\right]^{\delta_i} \exp\{-H_0(y_i\exp\{-\eta_i\})\}.
\end{aligned}
\label{loglikaft}
\end{equation}\\
The interpretation of the estimated AFT coefficients takes into account the logarithmic scale of the response. According to \cite{hosmer2008applied}, two patients can be compared in the absence (decrease) or presence (increase) of a certain characteristic through  the estimated time ratio (TR). This quantity is calculated with the median (or any percentile) of the survival time for a given group of patients divided by the median survival time for the another group. In order to obtain the TR for the median time point $t_{0.5}$ consider:\\
\begin{equation}
\begin{aligned}
 S(t_{0.5}\mid \eta_i) = S_0(t_{0.5}= \exp\{-\eta_i\})=0.5\\ 
\end{aligned}
\end{equation}\\
Accordingly,
\begin{equation}
\begin{aligned}
t_{0.5}(\eta_i,S_0) = S_0^{-1}(0.5)\exp\{\eta_i\}.\\ 
\end{aligned}
\end{equation}\\
Then, the TR is then given by:\\
\begin{equation}
\text{TR} =\frac{t_{0.5}(\eta_i,S_0)}{t_{0.5}(\eta_j,S_0)} =\frac{S_0^{-1}(0.5)\exp\{\eta_i\}}{S_0^{-1}(0.5)\exp\{\eta_j\}} =\exp\{\boldsymbol{\beta^\top}(\boldsymbol{x_i} - \boldsymbol{x_j})\}.\label{formula:tr} 
\end{equation}\\
This quantity is often referred to as
the acceleration factor \citep{hosmer2008applied}. \cite{Klein:1997} discuss about the  Exponential, Weibull, Log Normal, Log-logistic and Gamma, among other distributions, to describe time-to-event data using the AFT framework.  In fact, we can assume any distribution function for $T_i$ combined with any general class of survival model to assemble a parametric model. Some associations of that nature lead to parametric forms of known distributions, we will look at two  cases: the Weibull PH and AFT; and the Log-logistic PO and AFT.

The next section describes the relationship between some parametric forms for the three general families presented previously in this chapter. The models being discussed will assume parametric forms for the baseline functions. The patterns found in the next section will be explored in the simulation study and real applications. In particular, the parametric AFT models described have supported the generation of simulated data (Chapter 5).

\section{Relationship between the parametric AFT and other families}

$~~~~~$Although parametric models are not the focus of this work; it is essential to review some commonly used cases to study the advantages of the BP based models. In particular, we will focus on two of the most used probability distributions since they provide the three classes of models that are of interest to this dissertation. In this work, we will stick to the Weibull and the Log-logistic parametric formulations due to their specific relationship with the AFT class of models. The Weibull model is an option with both PH and AFT classes, while the Log-logistic formulation is considered for both PO and AFT. The artificial data sets investigated in Chapter 5 were obtained, assuming the parametric AFT versions for these two distributions. The main idea behind this strategy is to evaluate the BP based survival regression estimates obtained when fitting data sets originated from distinct data-generating models. 

\subsection{Weibull AFT and Weibull PH}
$~~~~~$The Weibull density function is often used to describe the lifetime of industrial products. Its popularity is since it can represent strictly increasing, strictly decreasing, or constant behaviors for the hazard function \citep{Klein:1997}. The Weibull distribution is parameterized in the present dissertation with the following configuration of density function:
\begin{equation*}
    f(t \mid \lambda, \kappa) = \lambda  \kappa \exp\{- \lambda t^\kappa\},~ y \ge 0.
\end{equation*}
where $\lambda > 0$ is the scale parameter and $\kappa > 0$ is the shape parameter. The survival and hazard functions for a Weibull distributed time-to-event random variable $T_i$ are respectively:\\
\begin{equation}
\begin{aligned}
S(t) & = \exp\{- \lambda t^\kappa\}, \\
h(t) & = \lambda  \kappa t^{ \kappa - 1}.  \label{Formula:hazweibull}
\end{aligned}
\end{equation}\\

According to \cite{Collett2015}, the Weibull distribution can be used to support a parametric AFT model, allowing $\lambda$ to differ between groups. For this, we need to keep the AFT structure in \ref{Formula:hazardAFT} and adopt the Weibull hazard function \ref{Formula:hazweibull} for the reference group. The Weibull AFT (WAFT) is defined as follows:
\begin{equation}
\begin{aligned}
    h(t\mid \eta_i)
    &= \exp\{-\eta_i\} \lambda  \kappa [t \exp\{-\eta_i\}]^{ \kappa - 1} = \lambda^*  \kappa t^{ \kappa - 1}. \label{Formula:waft}
\end{aligned}
\end{equation}\\
In  \ref{Formula:waft}, we can observe the Weibull hazard structure described in \ref{Formula:hazweibull}. As mentioned, the scale parameter differs between group and we can write $T_i \sim \mbox{Weibull}(\lambda^* = \lambda \exp\{-\eta_i \kappa\};  \kappa)$, with scale $\lambda^*$ and shape $ \kappa$. On the other hand, if a Weibull distribution is assumed for $T_i$ under the PH framework (WPH) in \ref{hazardPH}, it then follows that $T_i \sim \mbox{Weibull}(\lambda^* = \lambda \exp\{\eta_i\};  \kappa)$. The hazard function in this case is rewritten as:\\
\begin{equation*}
\begin{aligned}
h(t\mid \eta_i) &= \lambda  \kappa t^{ \kappa - 1}\exp\{\eta_i\}= \lambda^* \kappa t^{ \kappa - 1}.
\end{aligned}
\end{equation*}\\
Technically, it is possible to compare both models in terms of the resulting scale parameter $\lambda^*$. One of the goals of the simulation studies is to compare the BPPH and the BPAFT when fitted to the same data set originated from the WAFT model. Consider that $\eta^{\text{WAFT}}_i$ is the linear predictor for an individual under the WAFT framework and $\eta^{\text{WPH}}_i$ is the linear predictor for an individual under the WPH model. Explicitly, $\lambda^*= \lambda \exp\{-\eta^{\text{WAFT}}_i \kappa\}=\lambda \exp\{\eta^{\text{WPH}}_i\}$, that is,  $\eta^{\text{WPH}}_i = - \kappa\eta^{\text{WAFT}}_i$. Thus, the regression element $\eta_i$ of the WPH model is compatible with the regression of the WAFT model multiplied by the negative of the shape parameter.

In the case that a Weibull distribution is assumed in the PO model framework, or a Log-logistic distribution is assumed for the PH model, it is not feasible to find a parametric hazard function structure. The Weibull distribution model is the only parametric model that belongs to the AFT and the PH families at the same time. In the same way, the Log-logistic distribution model is the only parametric option belonging to the PO and AFT, simultaneously. The next topic describes how this relationship works within the parametric framework.

\subsection{Log-logistic AFT and Log-logistic PO}
$~~~~$The Log-logistic distribution can also provide a parametric model for analyzing right-censored survival data.  Unlike the Weibull distribution, which is more popular, its hazard function is not always monotonic. In turn, the hazard function can be unimodal,  monotonically decreasing and, also, it can have an increasing configuration at the beginning and a decreasing configuration for large time points. The Log-logistic distribution is parameterized in this dissertation as follows:\\
$$f(t \mid \nu, \zeta) = \frac{\zeta \nu t^{\zeta-1}}{[1+\nu t^\zeta]^2},~ t \ge 0, $$\\
where $\nu > 0$ is the scale and $\zeta > 0$ is the shape parameter.
Survival, hazard and odds functions are, respectively:\\
\begin{equation}
\begin{aligned}
S(t\mid \nu, \zeta) & = (1+\nu t^\zeta)^{-1},\\
h(t\mid \nu, \zeta) &= \frac{\zeta\nu t^{\zeta-1}}{1+\nu t^\zeta}, \\
R(t\mid \nu, \zeta) &=\frac{1-(1+\nu t^\zeta)^{-1}}{(1+\nu t^\zeta)^{-1}} = \nu t^\zeta.
\label{hazllogis}
\end{aligned}
\end{equation}\\
The Log-logistic distribution can be used as the basis for an parametric AFT model, allowing $\nu$ to differ between groups or, generally, introducing covariates that affect $\log(\zeta)$ as a linear function of the covariates \citep{Collett2015}. 

Similar to the WAFT and WPH models, the Log-logistic AFT (LLAFT) and the Log-logistic PO (LLPO) formulations are assembled with the general survival framework (either AFT or PO) and the Log-logistic parametric form of the baseline functions \ref{hazllogis}. Assuming that $T_i$ are independent Log-logistic random variables  in the AFT framework \ref{Formula:AFT}, we have:\\ 
\begin{equation}
  h(t \mid \eta_i) 
= \exp \{-\eta_i\} \frac{\zeta\nu(t\exp\{-\eta_i\})^{\zeta-1}}{1+\nu (t \exp\{-\eta_i\})^\zeta}
= \frac{\zeta\nu^* t^{\zeta-1}}{1+\nu^* t^\zeta}.  \label{Formula:LLAFT}
\end{equation}\\
As a result, we find the Log-logistic hazard structure described in \ref{hazllogis}. Thus, we write, $T_i \sim \mbox{L-Logis}(\nu^* =  \nu \exp\{-\zeta\eta_i\}; \zeta)$ with scale parameter $\nu^*$ and shape parameter $\zeta$. At the same time, the Log-logistic odds function \ref{hazllogis} can be rewritten in terms of the PO framework \ref{propodds2} as:\\
$$\begin{aligned}
R(t \mid \eta_i)= R_0(t) \exp\{\eta_i\}=\nu t^\zeta \exp\{\eta_i\}= (\nu \exp\{\eta_i\}) t^\zeta = \nu^* t^\zeta.
\end{aligned}$$\\
In this case, one can show that, $T_i$ follows a Log-logistic distribution with scale parameter $\nu^*$ and shape parameter $\zeta$, i.e, $T_i \sim \mbox{L-Logis}(\nu^*  = \nu \exp\{\eta_i\}, \zeta)$. The previous statements regarding the distributions of the $T_i$'s in the LLAFT and LLPO settings justify the strategy for comparing these
cases in terms of the scale parameter.
We write the following equality: $\nu^*= \nu \exp\{-\eta^{\text{LLAFT}}_i\zeta\}=\nu \exp\{\eta^{\text{LLPO}}_i\}$, \textit{i.e.} $\eta^{\text{LLPO}}_i = -\zeta \eta^{\text{LLAFT}}_i$, where $\eta^{\text{LLAFT}}_i$ is the linear predictor for an individual under the LLAFT framework and $\eta^{\text{LLPO}}_i$ linear predictor for a subject under the LLPO model. Other distributions do not provide similarities with respect to the relationship between the parametric Weibull and Log-logistic survival models and their respective AFT cases. 

Although parametric versions of the AFT model are often the first choice in many applications in the literature, the study
developed in this dissertation is focused on the semi-parametric configuration of the survival classes in this chapter. That is to say that we will not impose a probability distribution to describe the time response. In this sense, the BP based approach proposed in \cite{Osman:2012} arises as a flexible and appealing
semi-parametric alternative, allowing different shapes for the baseline function  without
need to pre-specify a distribution for the response variable. The approach explored in this dissertation has a non-parametric appeal concerning the baseline functions. In the sense that the parametric shape of the baseline functions is distribution-free. In other words, the parameters that determine the shape of the baseline functions are not related to the parameters of any probability distribution function. Despite the semi-parametric appeal of the BP based survival models, they indeed make use of fully parametric techniques.

The next chapter is dedicated to the BP presentation, along with facts, properties, and typical nuances resulting from its mathematical formulation. \cite{Lorentz:1953} has exhaustively exposed theoretical results and important facts about the BP in the context of mathematical analysis, such as proofs of theorems, generalizations, derivatives, asymptotic formulations, and approximation examples. In addition, \cite{farouki2012bernstein}
discussed relevant applications to the BP as a celebration for the centennial anniversary of this topic. The discussion accounts for aspects related to symmetry, non-negativity, and differentiation properties.  At the end of the next chapter, the BP is presented in the context of survival analysis.

% Chapter 3 = Chapter Title.
\chapter{Bernstein polynomial}

$~~~~$The BP were introduced by \cite{Bernstein:1912} for the approximation of any continuous function $c(x)$ whose domain is restricted to the interval $[0,1]$. According to \cite{Lorentz:1953}, the \textit{Bernstein Polynomial of degree m for the function c(x)} is given by:\\
\begin{equation}
B(x;m;c) = B_m^c(x) = \sum \limits_{k=0}^m c\left(\frac{k}{m}\right)b_{k,m}(x); ~~x \in [0,1] \label{formula:bp},
\end{equation}\\
where $b_{k,m}(x) = {{m}\choose{k}}x^k(1-x)^{m-k},~k \in \{ 1, 2, \cdots, m\}$ is called the BP basis polynomial (or just basis). \cite{Farouki:1987} discuss an alternative formulation to accommodate functions restricted to $[a,b]$; such that $a <b \in \mathbb{R}$. We write:

\begin{equation}
B_m^c(x) = \sum \limits_{k=0}^m c\left(a + \frac{k}{m}(b-a)\right)b_{k,m}((x-a)/b); ~~x \in [a,b] \label{formula:bp2}.
\end{equation}\\
In the present work, we will refer to the highest order of the BP basis polynomials (individual terms) as \textit{degree} and to the BP basis polynomial order simply as \textit{order}. \cite{Lorentz:1953} properly discusses on how to demonstrate the Weierstrass approximation theorem. The theorem say that every restricted continuous function can be approximate by a polynomial function, as close as desired, in the real domain. Using a finite BP, having a sufficiently high degree, it can be shown that  $\lim_{m \rightarrow \infty} B_m^c(x)  = c(x)$ uniformly. Briefly, consider a sequence of polynomials $\{B^c_m\}_{m\in\mathbb{N}}$ and the existence of some  polynomial of a high degree $M$. In such a way that, for every $m>M\in\mathbb{N}$ and $\epsilon >0$:\\
\begin{equation*}
  \mid B_m^c(x) - c(x) \mid < \epsilon, \text{ for all } x \in [0,1].  
\end{equation*}\\

The approximation ability of BP has been widely explored for the approximation of real-valued functions regarding several contexts in literature. BP is a crucial topic in the present dissertation,
since the semi-parametric survival analysis
via BP is the main structure to support the routines implemented
in the \texttt{R} package proposed here. In the next section, we highlight some useful BP properties and facts. At the end of this chapter, the BP is introduced in the context of survival regression.   

\section{BP basis properties}
$~~~~$The focus of this section is on the BP basis polynomial that will serve as the main structure for the approximation of the target function by the linear combination of lower (or equal) order polynomials. Consider the probability of $k$ successes in $m$ trials of a binomially distributed random variable $K$
with individual probability of success $x$ in each trial, that is $K \sim \mbox{Binomial}(m,x)$.  According to some authors \citep{Gzyl:1997,Kolarov:2007,cichon:2012,Alda:2017}, the BP formulation in \ref{formula:bp} can be interpreted as the expected value of the random variable $c(K/m)$, where $c(.)$ is the target function in \ref{formula:bp}:\\
$$E\left[c\left(\frac{K}{m}\right)\right] = \sum \limits_{k=0}^m c\left(\frac{k}{m}\right){{m}\choose{k}}x^k(1-x)^{m-k}.$$\\
The BP  basis polynomials should sum up to one, as well as the binomial probability mass function,  therefore, we write:\\
\begin{equation}
  \sum \limits_{k=0}^m b_{k,m}(x)=\sum \limits_{k=0}^m{{m}\choose{k}}x^k(1-x)^{m-k}=[x-(1-x)]^m = 1.  \label{oneness}
\end{equation}\\
Some properties of the BP basis are important to build the semi-parametric model described ahead in this dissertation. In particular, four key properties were described by \cite{farouki2012bernstein} as follows:\\
\begin{itemize}
    \item symmetry: \begin{equation}b_{k,m-k}(x) = b_{k,m}(1-x),    \label{symmetry}\end{equation}
    \item  recursion: \begin{equation}b_{k,m+1}(x) = x~b_{k-1,m}(x) + (1- x)~b_{k-1,m}(x),\label{recursion}\end{equation}
    \item non-negativity: \begin{equation}b_{k,m}(x) \ge 0, ~\forall~ x \in [0,1], \text{ if } 0 \le k \le m, \label{nonneg}\end{equation}
    \item derivatives: \begin{equation}\frac{d}{dx} b_{k,m}(x) = m [b_{k-1,m-1}(x) - b_{k,m-1}(x)] .\label{derivprop}\end{equation}
\end{itemize}
The BP properties are also broadly discussed in many applications of computer aided geometric design methodologies (computational mathematics methods to describe geometric objects). More details on basic properties of the BP can be found in  the references: \cite{Davis:1963}, \cite{Farouki:1987}, \cite{Farouki:1988} and \cite{Farouki:2008}.  

\begin{figure}[!htb]
 \centering
 $$
  \begin{array}{cc}
     \mbox{\textbf{(a)} $m = 4$} & \mbox{\textbf{(b) $m = 10$}}\\
     \includegraphics[width=0.45\textwidth]{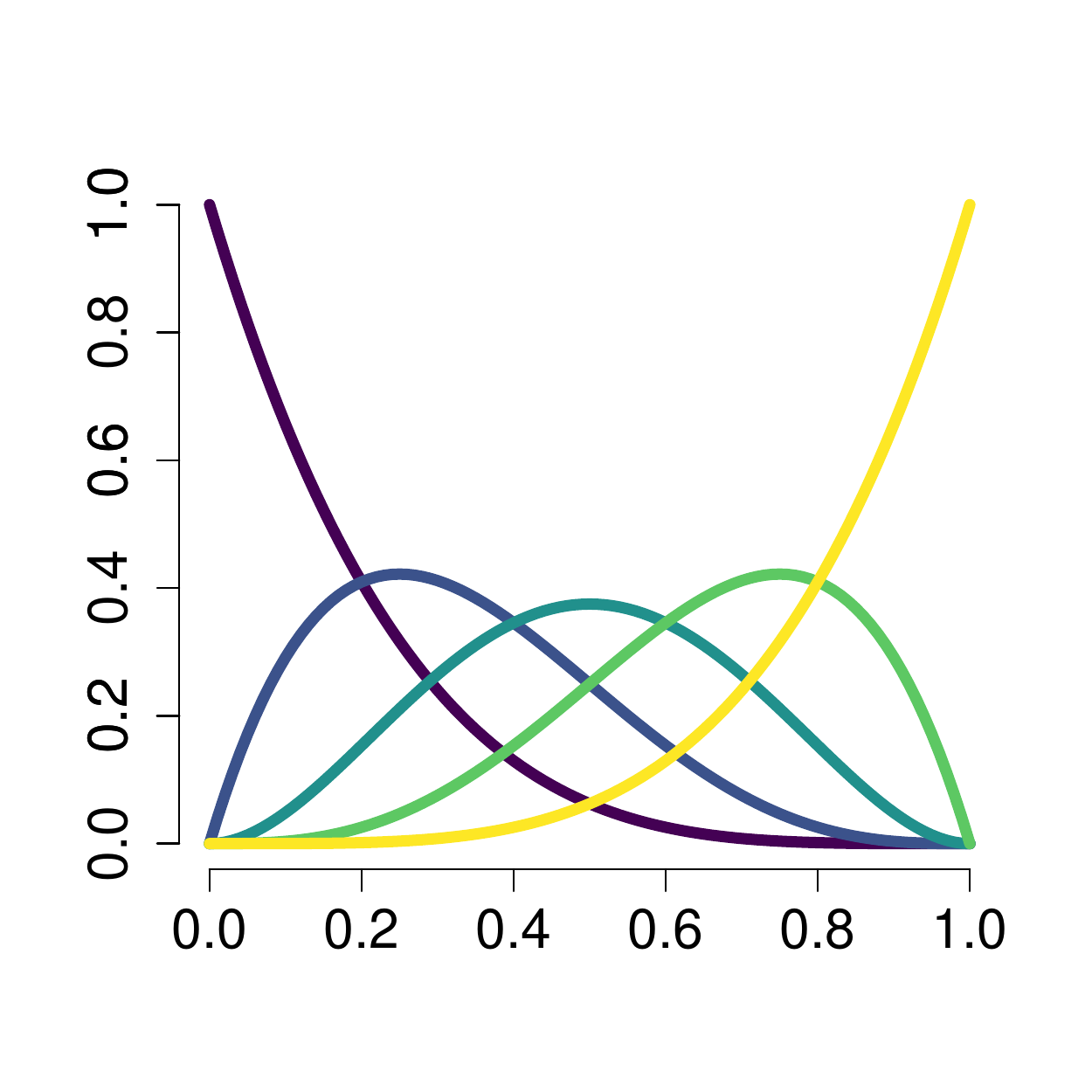} &
    \includegraphics[width=0.45\textwidth]{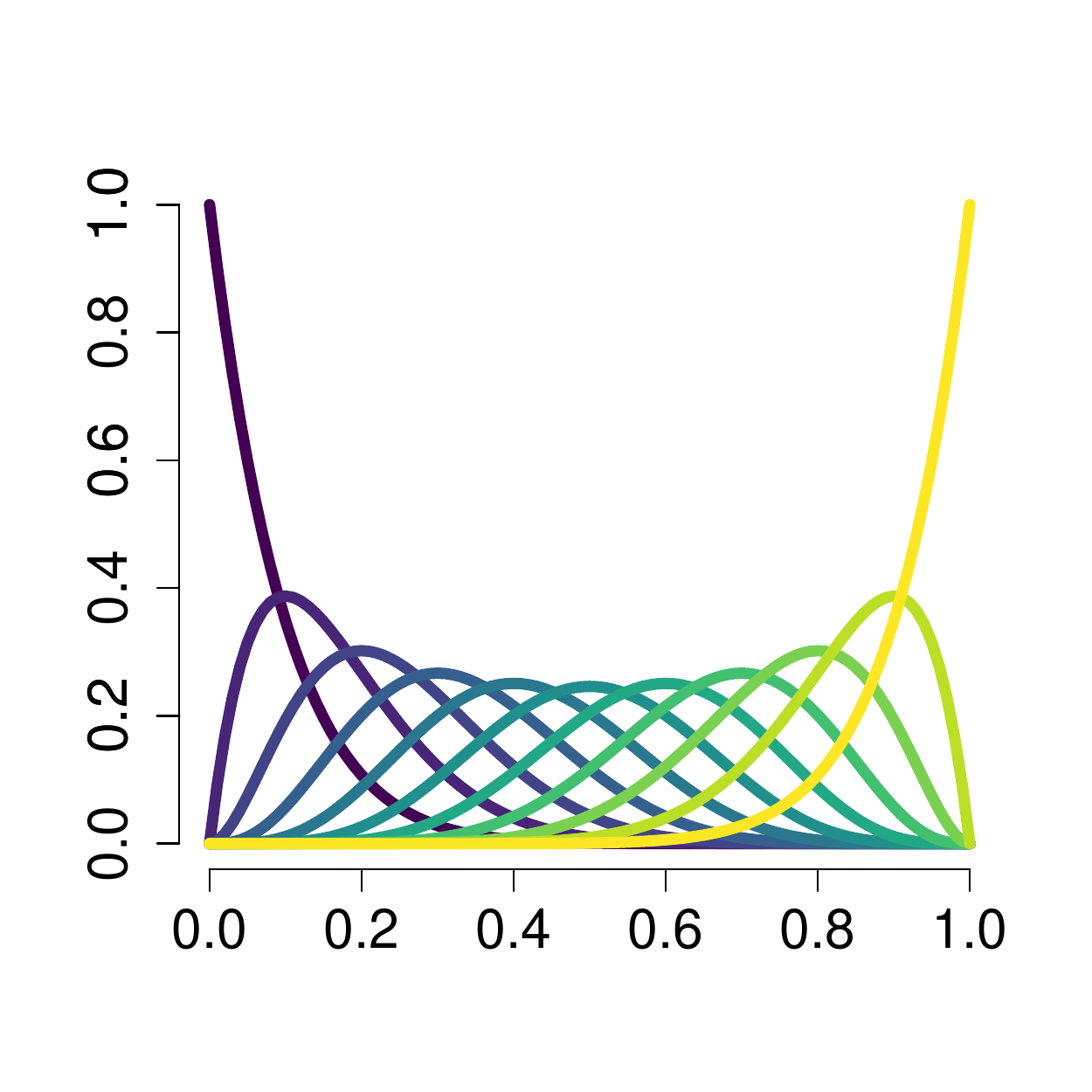}
    \end{array}
$$
    \label{Fig:bpbases}
 \caption{Bernstein polynomial basis . Panel \textbf{(a)}: polynomial basis  of order $m = 4$; Panel \textbf{(b)}: polynomial basis  of order $m = 10$. }
 \end{figure}

Figure \oldref{Fig:bpbases} illustrates the BP basis  symmetry \ref{symmetry}, recursion \ref{recursion},  and non-negativity \ref{nonneg} properties, considering $m = 4$ and $m = 10$ polynomial degrees. Panel (a) shows the Bernstein polynomial basis  of order $k\le m=4$, the left-hand side basis is $b_{0,m}(x)$ and the right-hand side basis is $b_{m,m}(x)$, corresponding to the black (darkest color) and the yellow (lightest color), respectively. All the basis polynomials, including these two, follow the properties presented in \ref{symmetry}, \ref{recursion}, \ref{nonneg} and \ref{derivprop}. As for instance, due to the the symmetry property we have: $b_{0,m-0}(x) = b_{0,m}(1-x)$. Also, the non-negative property can be observed at the image of the curves presented in the Panel (a).  In turn, Panel (b) illustrates the fact that the same properties are maintained for higher order polynomial basis. As discussed previous in this chapter, higher degree polynomials are expected to provide better approximations to the target function. It is also noteworthy that, for $m = 4$, the yellow curve begins
rising from 0.4 while, when $m = 10$, it starts rising from 0.7, that illustrates that the basis $b_{4,4}(x)$ order is lower than the basis $b_{10,10}(x)$ order. 
In particular, if the curves are evaluated at any fixed point in $x \in [0.1] $, the basis sum up to one \ref{oneness}, especially, $b_{0,m}(0) = b_{m,m}(1) = 1 $.\\

$~~~~$The BP differentiation can be obtained from the derivative property \ref{derivprop} of the basis.  This property provides the necessary results for the introduction of the BP in the survival analysis regression context, explored further in this dissertation. The last section of this chapter discusses how the BP differentiation expression can be used to estimate cumulative hazard, odds and related functions. The first partial derivative with respect to $x$ of the formulation in \ref{formula:bp} provides:\\
\begin{equation}
\small
\begin{aligned}
\frac{\partial B_m^c(x)}{\partial x} &= \sum \limits_{k = 0}^{m} c\left(\frac{k}{m}\right) {{m}\choose{k}}\{kx^{k-1}(1-x)^{m-k} - (m-k)x^k(1-x)^{m-k-1}\} \\ 
    &=m\sum\limits_{k=0}^{m}c\left(\frac{k}{m}\right)\left[{{m-1}\choose{k-1}} x^{k-1}(1-x)^{m-k}-{{m-1}\choose{k}}x^k(1-x)^{m-k-1}\right]\\
    &=m\sum \limits_{k=0}^{m}c\left(\frac{k}{m}\right)b_{k-1,m-1}(x)-m\sum \limits_{k=0}^{m}c\left(\frac{k}{m}\right)b_{k,m-1}(x)\\
    &= m\sum \limits_{i=-1}^{m-1}c\left(\frac{i+1}{m}\right)b_{i,m-1}(x)-m\sum \limits_{k=0}^{m}c\left(\frac{k}{m}\right)b_{k,m-1}(x).
\end{aligned}\label{formula:BPderiv}
\end{equation}\\
where  $i = k -1$. According to \cite{farouki2012bernstein}, consider by definition $b_{-1,m-1}(x) = b_{m,m-1}(x) =0$, thus we have:\\
\begin{equation}
\begin{aligned}
\frac{\partial B(x;m;c)}{\partial x} &= m \sum \limits_{i=0}^{m-1}\left\{c\left(\frac{i+1}{m}\right)-c\left(\frac{i}{m}\right)\right\}b_{i,m-1}(x)\\
&= m \sum \limits_{i=0}^{m-1} \Delta c^1_{i} b_{i,m-1}(x).
\label{formula:bpderiv}
\end{aligned}
\end{equation}\\
where $c\left(\frac{i+1}{m}\right)-c\left(\frac{i}{m}\right)$ is the difference of first order $\Delta c^1_{i}$ of the function $c(x)$ at $x= i/m$ \citep{Lorentz:1953}. It is important to notice that the basis derivatives are lower order basis written in terms of the $k^{th}$ difference $\Delta c^k_{i}$. The second difference, for example, is:\\
%\begin{itemize}
    %\item
%    if $k = 2$:
   $$
    \begin{aligned}
    \Delta c^2_{i} = \Delta(\Delta c^1_{i}) &=  \left\{c\left(\frac{i+2}{m}\right)-c\left(\frac{i+1}{m}\right)\right\} - \left\{c\left(\frac{i+1}{m}\right)-c\left(\frac{i}{m}\right)\right\} \\
   &=  c\left(\frac{i+2}{m}\right)-2~ c\left(\frac{i+1}{m}\right) + c\left(\frac{i}{m}\right). 
    \end{aligned}$$\\
%    \item if $k = r$:
%    $$\Delta c^r_{i} = c\left(\frac{i+r}{m}\right) - %{{r}\choose{1}} c\left(\frac{i+(r-1)}{m}\right) + \dots + (-1)^r %f\left(\frac{i}{m}\right).$$
%\end{itemize}\\
The BP derivative consists of a lower order basis  \ref{formula:bpderiv} due to the recursion property  \ref{recursion}. This property provides a recursive relationship for a sequential differentiation \citep{Lorentz:1953}, given by:\\
$$B_m^{(r)} = n(n-1) \dots (n-r+1) \sum \limits^{n-r}_{k = 0} \Delta c^k_{i}{{n-r}\choose{k}}x^k(1-x)^{m-k-r}, \forall k = \{1,2,\dots, m\}.$$\\

In this section, we have included some properties about the BP polynomial basis that will be important in building the BP  models for survival data. Primarily, the derivative properties are used in the last section of this chapter about BP in survival analysis.  Besides, the BP was presented from the perspective of a Binomial distributed random variable, 
this alternative perspective can facilitate the interpretation of the BP and make the presentation of its properties more intuitive.
 The next section shows how to use BP to approximate real-valued continuous functions. For example, the BP approximation should capture any continuous function, including non-negative cases such as the hazard rate functions.

\section{Finite BP approximation}
$~~~~$In the present section, we will illustrate how the BP is used to approximate known real valued functions, we have chosen a Weibull hazard function as an example of real-valued target function. The Weibull distribution is a very popular choice when it comes to parametric models, as mentioned in Chapter 2. Assuming that the target function has a non-negative domain restricted to $[a=0,b = \tau]$ , the equation \ref{formula:bp2} becomes:\\
\begin{equation}
 B_m^h(t) = \sum \limits_{k=0}^m h\left(\frac{k}{m}\tau \mid \lambda, \kappa \right) b_{k,m}(t/\tau).  \label{formula:bphaz}
\end{equation}\\
In order to obtain \ref{formula:bphaz}, we have assumed an upper bound limit $\tau$ for the hazard function in \ref{Formula:hazweibull}. Suppose that $m = 4$ and $\tau=5$, such that $x = t/5 \in [0,1]$. Then, we have:\\
\begin{equation}
\begin{aligned}
B_4^h(t \mid \lambda, \kappa) &=  h(0 \mid \lambda, \kappa)\;b_{0,4}(x) \\ 
&+ h\left(\frac{5}{4} \mid \lambda, \kappa \right)\;b_{1,4}(x) 
+h\left(\frac{10}{4} \mid \lambda, \kappa \right)\;b_{2,4}(x)
\\ &+ h\left(\frac{15}{4} \mid \lambda, \kappa \right)\;b_{3,4}(x) +h\left(\frac{20}{4} \mid \lambda, \kappa \right)\;b_{4,4}(x).
\end{aligned} \label{bpapprox}
\end{equation}\\
Accordingly, the Bernstein polynomial basis  in \ref{formula:bphaz} are:
\begin{equation}
\begin{aligned}
b_{0,4}(x) &= (1-x)^{4},\\
b_{1,4}(x) &= 4 x(1-x)^{3},   & \quad
b_{2,4}(x) &= 6 x^2(1-x)^{2}, \\
b_{3,4}(x) &= 4 x^3(1-x), & \quad
b_{4,4}(x) &= x^4.
\end{aligned}
\label{Formula:bpbasis}
\end{equation}\\
The finite BP approximation  $B_4^h(t \mid \lambda, \kappa)$ is obtained with the cross product between the BP basis in \ref{Formula:bpbasis} and the values of the target function $h(t \mid \lambda, \kappa)$ \ref{bpapprox} evaluated  at the equidistant points $\{0, \frac{5}{4}, \frac{10}{4}, \frac{15}{4}, \frac{20}{4}\}$. 

\begin{figure}[!htb]
 \centering
 $$
  \begin{array}{cc}
     \mbox{\textbf{(a)} $m=4$} & \mbox{\textbf{(b) $m=4$}}\\
     \includegraphics[width=0.44\textwidth]{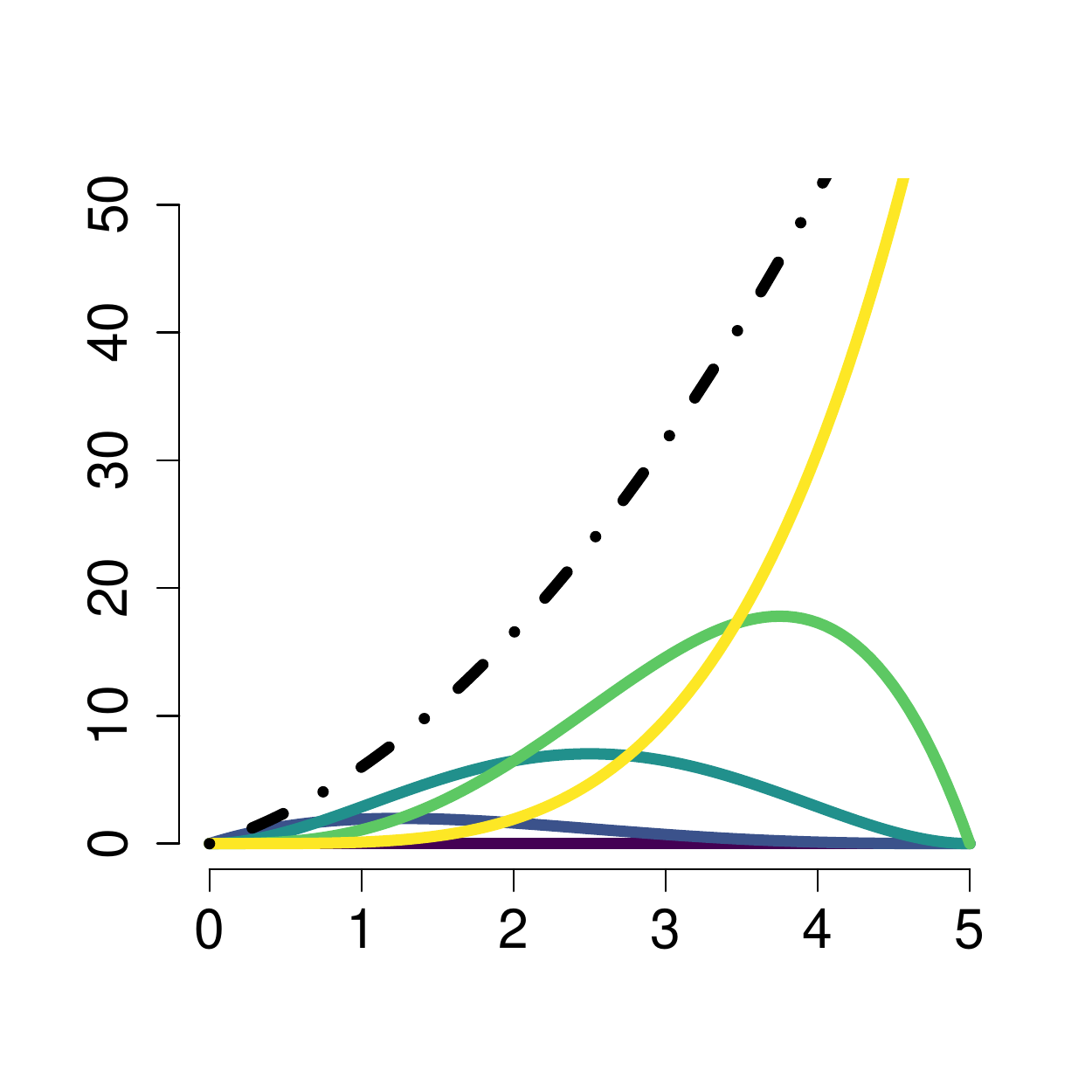} &
    \includegraphics[width=0.44\textwidth]{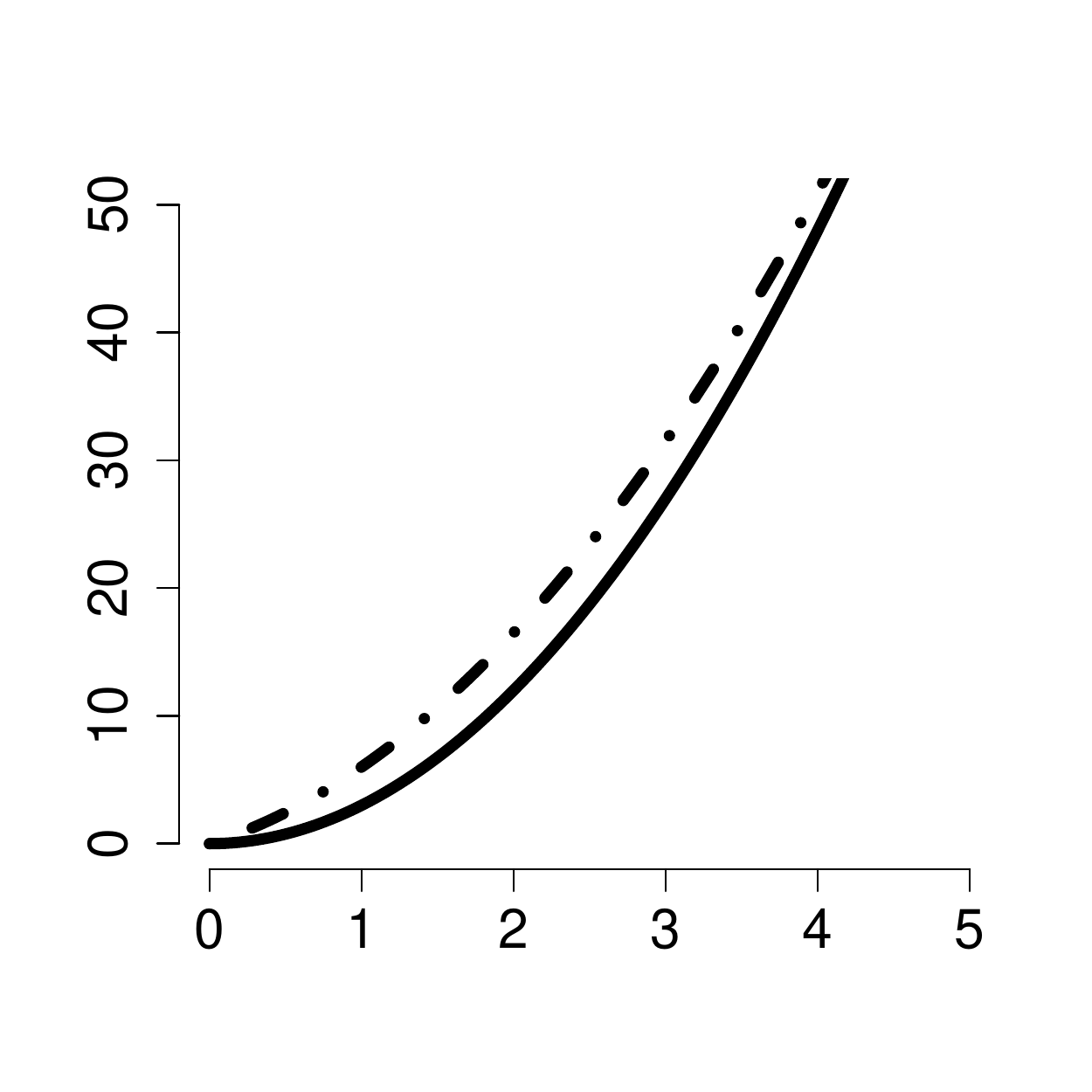}\\
         \mbox{\textbf{(c)} $m=10$} & \mbox{\textbf{(d) $m=10$}}\\
   \includegraphics[width=0.44\textwidth]{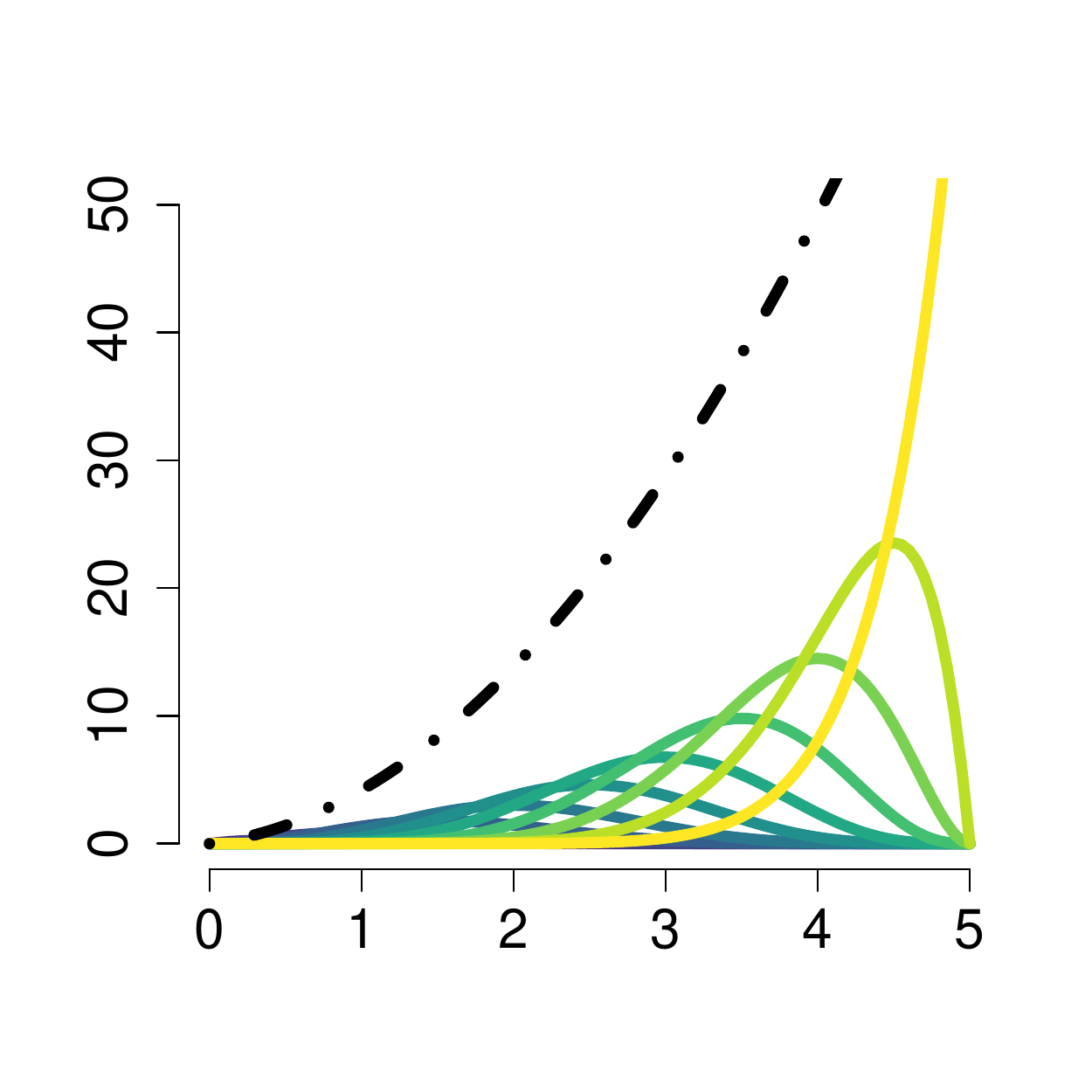} &
     \includegraphics[width=0.44\textwidth]{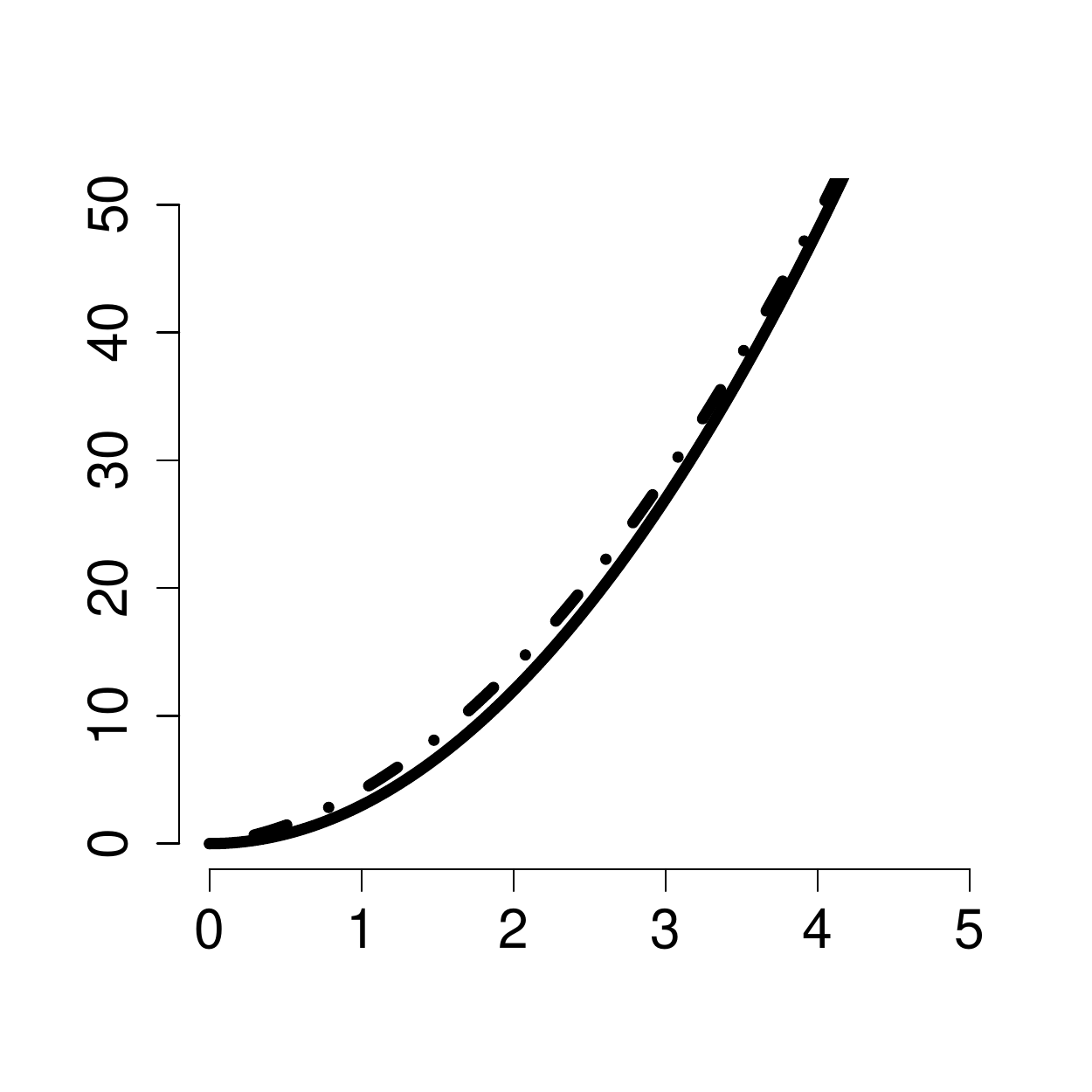} \\
  \end{array}
$$
\caption{Illustration of the BP approximation of degree $m$ for the Weibull hazard function (target). Panels \textbf{(a)}, \textbf{(b)}, \textbf{(c)} and \textbf{(d)}: the dot-dashed line represents the finite BP approximation $B_m(y \mid \lambda, \kappa)$; Panels \textbf{(a)} and \textbf{(c)}:  the solid colored lines (from dark blue to yellow) represent the Bernstein polynomial individual term $h\left(\frac{k}{m}\tau \mid \lambda, \kappa \right)b_{k,m}(y/\tau), ~\forall k = 1,2,\dots, m$, for $m = 4$ in Panel (a) and $m = 10$ in Panel (c); Panels \textbf{(b)} and \textbf{(d)}: the solid black line represents the target hazard function with scale $\lambda = 1$ and shape $\kappa = 3$.}  \label{Fig:approx1}
 \end{figure}

\begin{figure}[!htb]
 \centering
 $$
  \begin{array}{cc}
     \mbox{\textbf{(a) $m = 4$}} & \mbox{\textbf{(b) $m = 4$}}\\
     \includegraphics[width=0.43\textwidth]{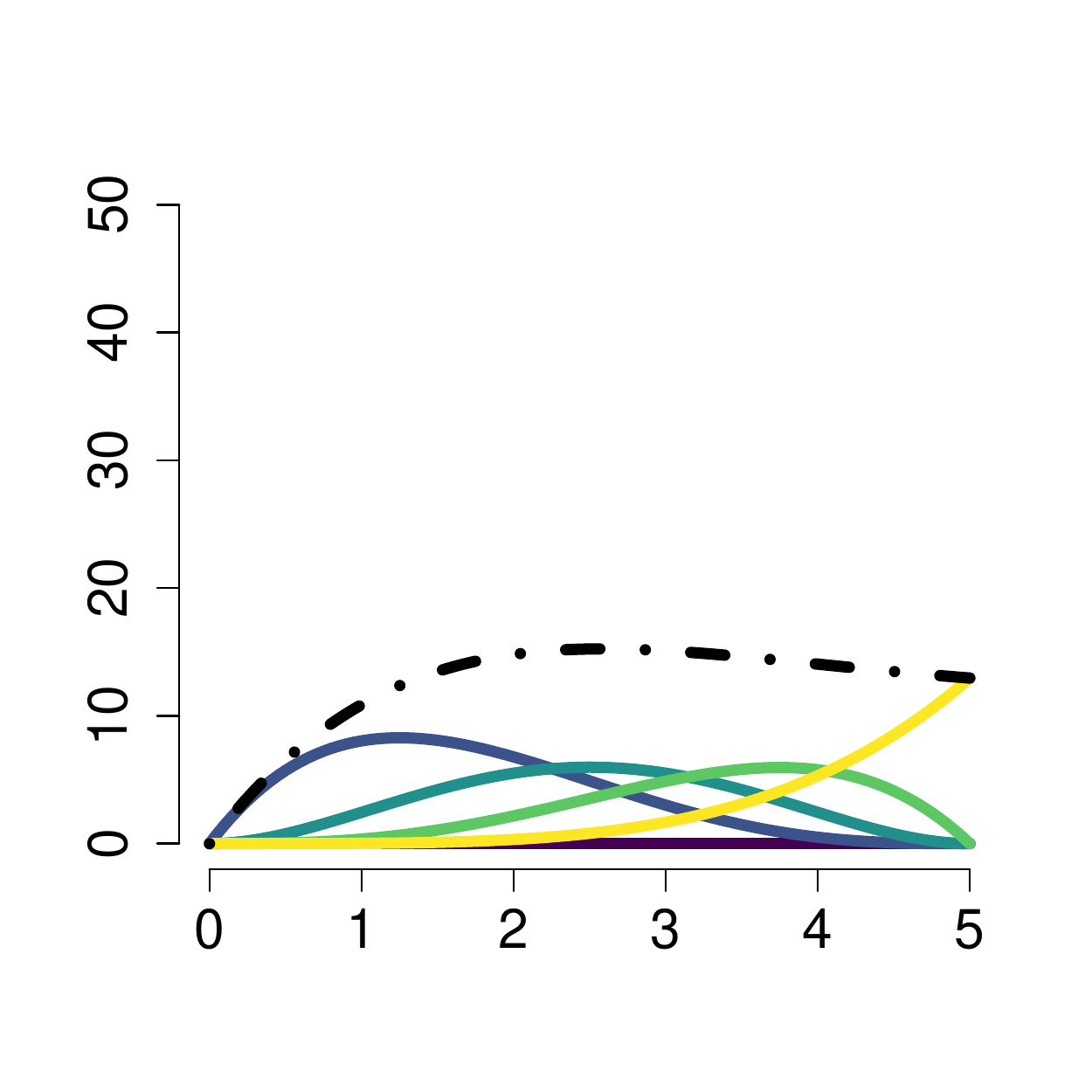} &
    \includegraphics[width=0.43\textwidth]{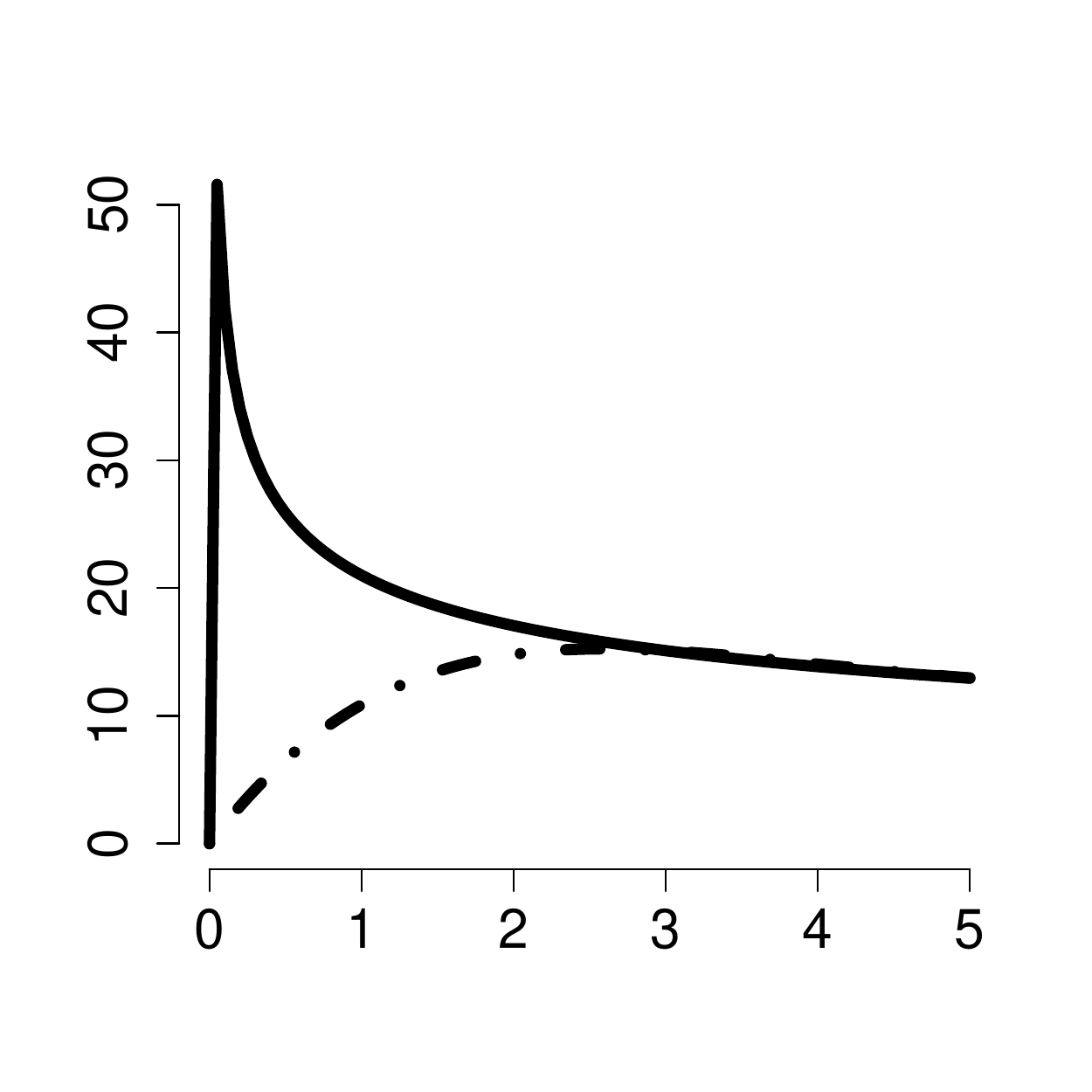}\\
         \mbox{\textbf{(c)  $m = 10$}} & \mbox{\textbf{(d)  $m = 10$}}\\
   \includegraphics[width=0.43\textwidth]{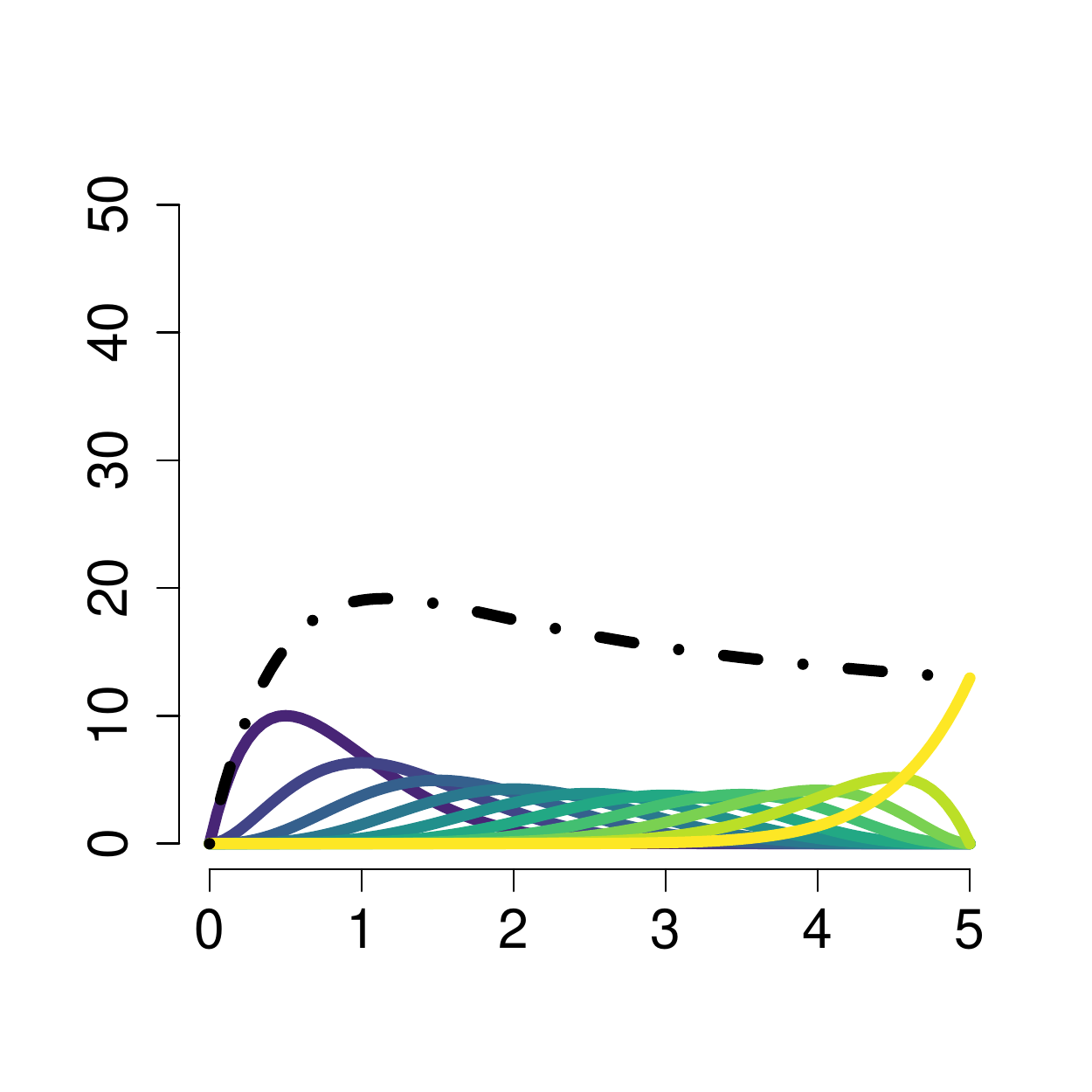} &
     \includegraphics[width=0.43\textwidth]{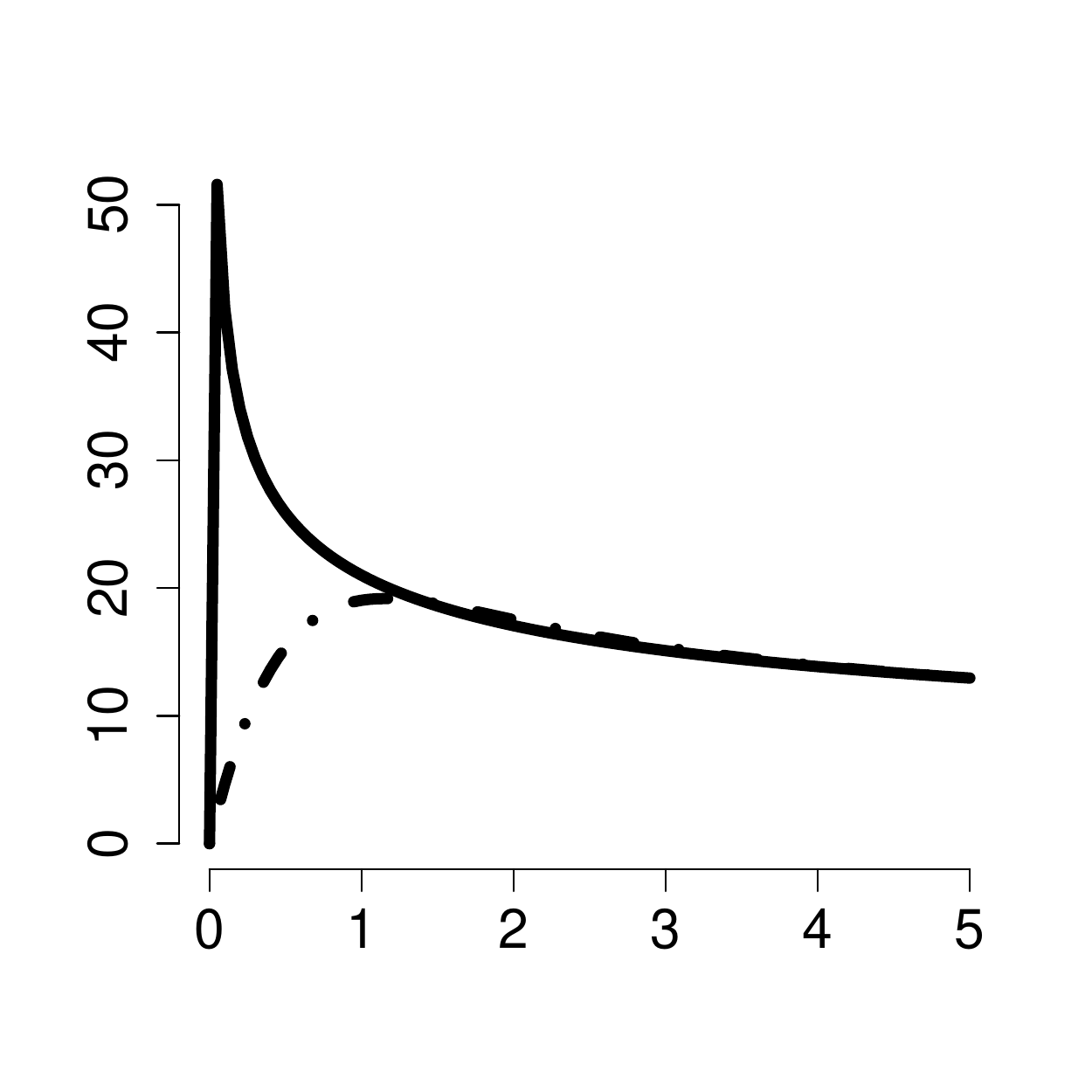} \\
  \end{array}
$$
\caption{Illustration of the BP approximation of order $m$ for the Weibull hazard function (target). Panels \textbf{(a)}, \textbf{(b)}, \textbf{(c)} and \textbf{(d)}: the dot-dashed line represents the finite BP approximation $B_m(y \mid \lambda, \kappa)$; Panels \textbf{(a)} and \textbf{(c)}:  the solid colored lines (from dark blue to yellow) represent the Bernstein polynomial  individual term $h\left(\frac{k}{m}\tau \mid \lambda, \kappa \right)b_{k,m}(y/\tau), ~\forall k = 1,2,\dots, m$, for $m = 4$ in Panel (a) and $m = 10$ in Panel (c); Panels \textbf{(b)} and \textbf{(d)}: the solid black line represents the target hazard function with scale $\lambda = 30$ and shape $\kappa = 0.7$.}\label{Fig:approx2}
\end{figure}

Figures \oldref{Fig:approx1} and \oldref{Fig:approx2} illustrate the behavior of every single product (individual term) of this finite BP \ref{bpapprox},  two configurations of the Weibull hazard function are explored. Note that, in this case, there is no symmetry. In Figure \oldref{Fig:approx1}, for example, the scaled (multiplied) basis polynomials are more concentrated to the right-hand corner compared to Figure \oldref{Fig:bpbases}. The resulting approximation consists of the sum of every (polynomial) curve in solid-colored lines; the finite BP curve is represented in dash-dotted lines. As expected, when $m$ increases,  from Panels (a) to (b) in Figures \oldref{Fig:approx1} and \oldref{Fig:approx2}, the BP approximation $B_m^h(y \mid \lambda, \kappa)$ gets closer to the actual target $h(y \mid \lambda, \kappa)$ curve. The higher degree in BP, Panel (b), indeed provides a closer approximation to the target function. The target function must be known so that it can play the role of modifying each basis form presented in Figure \oldref{Fig:bpbases}. 

Although the approximation of target hazard rate functions is often successful, the approach is not suitable for real data applications as the target function is unknown in practice. The researcher cannot know in advance the true distribution of the time until the occurrence of an event. Hence, the next section presents an alternative to estimate unknown cumulative hazard and odds functions for right-censored data, which uses the BP based survival regression modeling. \\
 \newpage
 \newpage
\newpage 

\section{BP in survival analysis}
$~~~~$The topic of this section is central to the proposal of the semi-parametric survival modeling developed in this dissertation. The methodology behind the elements that will be presented here was introduced in \cite{Osman:2012}. The authors took advantage of a finite BP in order to estimate positively bounded functions in survival regression modeling, such as the cumulative hazard function $H(t)$. For this, consider \ref{formula:bp2} to approximate $H(t) \text{ assuming } t \in [0, \tau]$:\\
\begin{equation}
B_m^H(t) = \sum \limits_{k=0}^m H\left(\frac{k}{m}\tau\right)b_{k,m}(t/\tau); ~~x \in [0, \tau].\label{bpsurv}
\end{equation}\\
The first BP differentiation \ref{formula:bpderiv} with respect to the time $t$ is:\\ 
\begin{equation}
\begin{aligned}
\footnotesize
\frac{\partial B_m^H(t)}{\partial t} 
& = \frac{m}{\tau} \sum \limits_{i=0}^{m-1} \left\{H\left(\frac{i+1}{m}\tau\right)-H\left(\frac{i}{m}\tau\right)\right\} b_{i,m-1}(t/\tau) \\
&=\frac{m}{\tau}\sum\limits_{k=1}^m \left\{{H\left(\frac{k}{m}\tau\right)-H\left(\frac{k-1}{m}\tau\right)}\color{black}{}\right\}{{m-1}\choose{k-1}}b_{k-1,m-1}(t/\tau) \\
&= \frac{m}{\tau}\sum\limits_{k=1}^m \left\{{H\left(\frac{k}{m}\tau\right)-H\left(\frac{k-1}{m}\tau\right)}\color{black}{}\right\}{{m-1}\choose{k-1}}\left(\frac{t}{\tau}\right)^{k-1}\left(1-\frac{t}{\tau}\right)^{(m-1)-(k+1)}\\
&=\frac{1}{\tau}\sum\limits_{k=1}^m \left\{{H\left(\frac{k}{m}\tau\right)-H\left(\frac{k-1}{m}\tau\right)}\color{black}{}\right\}\frac{\Gamma(m+1)}{\Gamma(m-k+1)\Gamma(k)}\left(\frac{t}{\tau}\right)^
{k-1}\left(1-\frac{t}{\tau}\right)^{m-k}\\
&= \sum\limits_{k=1}^m \left\{{H\left(\frac{k}{m}\tau\right)-H\left(\frac{k-1}{m}\tau\right)}\color{black}{}\right\} {\frac{f_\beta(t/\tau; k,m-k+1)}{\tau}}\color{black}{}.
\end{aligned}    
\label{formula:bpsurv}
\end{equation}\\
It is important to emphasize that the polynomial basis  were rewritten as $g_{k,m}(t)=f_\beta(t/\tau \mid k,m-k+1)\;\tau^{-1}$, in which $f_\beta$ corresponds to the beta density function. Moreover, note that the quantities $\gamma_k = \left\{H\left(\frac{k}{m}\tau\right)-H\left(\frac{k-1}{m}\tau\right)\right\}, ~\forall k = 1,2,\dots, m$ do not depend on the time. Thereby, these quantities should be estimated as we cannot know in advance the true cumulative hazard function value. Hereafter, these elements will be called BP parameters $~\boldsymbol{\gamma} = (\gamma_1, \gamma_2, \dots, \gamma_m)^\top,~\gamma_k \ge 0$. Some parametric models have been criticized because of the hazard decay for large time values \citep{Klein:1997}. Using the BP it is possible to estimate either increasing, decreasing, convex, concave, and other shapes of functions, such that the estimates for $~\boldsymbol{\gamma} = (\gamma_1, \gamma_2, .., \gamma_m)^\top$ dictates the shape that the functions related to the BP will assume, similar to what happened in Figures \oldref{Fig:approx1} and \oldref{Fig:approx2}. The hazard and cumulative hazard functions in the BP formulation are:
\begin{equation}
  h(t, \boldsymbol{\gamma}) = \boldsymbol{\gamma}^\top\boldsymbol{g}_m(t), ~ 0\le t < \infty
 ~~\text{ and }~~
 \begin{aligned}H(t, \boldsymbol{\gamma}) & = \int_0^t h(u, \boldsymbol{\gamma}) du = \boldsymbol{\gamma}^\top\boldsymbol{G}_m(t),\end{aligned} \label{formula:cumhaz}
\end{equation}\\
where $\boldsymbol{g}_m(t) = (g_{1,m}(t), g_{2,m}(t), \dots, g_{m,m}(t))^\top$, $\boldsymbol{G}_m(t) = (G_{1,m}(t), G_{2,m}(t), \dots, G_{m,m}(t))^\top$ and $G_{m,k}(t) = \int_0^t f_\beta(u/\tau; k,m-k+1) d(u/\tau)$. Since this model does not impose any functional form on the hazard function, it is said to have a non-parametric appeal. Some properties of the key functions for survival analysis, mentioned in Section 2, are preserved when the BP structure is assumed. For example, the cumulative hazard function is monotonic non-decreasing as a consequence of the restriction $\gamma_k \ge 0$.

As stated in Chapter 2, the inclusion of regression coefficients is required in order to achieve the statistical purpose of investigating the impact of covariates over the time response. Hence, the formulation \ref{formula:cumhaz} shall be included in the baseline hazard function of the PH likelihood function \ref{formula:loglikph}, as follows:\\ 
\begin{equation}
\begin{aligned}
L(\boldsymbol{\Theta} \mid \boldsymbol{y}, \boldsymbol{\delta}, \boldsymbol{x}) & = \prod \limits_{i=1}^n \left[ {h_0(y_i)}\color{black}\exp\{\eta_i\}\right]^{\delta_i} \exp\{-H_0(y_i) \exp\{\eta_i\}\} \\
&= \prod\limits_{i=1}^n  \left[\boldsymbol{\gamma}^\top\boldsymbol{g}_m(y_i) \exp\{\eta_i\} \right]^{\delta_i} -\exp\{\boldsymbol{\gamma}^\top\boldsymbol{G}_m(y_i) \exp\{\eta_i\}\}, 
\end{aligned}
\label{bpphlik}
\end{equation}\\ 
where $\boldsymbol{\Theta} = (\boldsymbol{\beta}, \boldsymbol{\gamma})$. \cite{Osman:2012} discuss proofs on the strictly concavity, the uniqueness of the maximum likelihood estimates and good differentiability properties that ease likelihood, gradient and Hessian matrix calculations.

The BP is also suitable to estimate the baseline odds on the event occurrence \citep{demarqui:2019}. This can be done due to the properties shared by the odds function and the cumulative hazard function. Similar to the cumulative hazard function, the odds function is a monotonic non decreasing function with $R_0(0) = 0$ and $\lim_{t \rightarrow \infty} R_0(t) = \infty$. The formulation \ref{formula:cumhaz} shall be included in the baseline odds function of the PO likelihood function \ref{loglikpo}, \textit{i.e.}, $R_0(t)= \boldsymbol{\xi}^\top \boldsymbol{G}_{m,k}(t)$ and  $r_0(t)= \boldsymbol{\xi}^\top \boldsymbol{g}_{m,k}(t)$, in order to define the next likelihood function:\\
\begin{equation}
\begin{aligned}
L(\boldsymbol{\Theta} \mid \boldsymbol{y}, \boldsymbol{\delta}, \boldsymbol{x}) &=\prod \limits_{i=1}^n \left[  \frac{\boldsymbol{\xi^\top g}_m(y_i) \exp\{\eta_i\}}{1 +\boldsymbol{\xi^\top G}_m(y_i) \exp\{\eta_i\}}\right]^{\delta_i} \left( \frac{1}{1 + \boldsymbol{\xi^\top G}_m(y_i) \exp\{\eta_i\}}\right),
\end{aligned}\label{bppolik}
\end{equation}\\
in this case $\boldsymbol{\Theta}=(\boldsymbol{\beta}, \boldsymbol{\xi})$. The BP parameters are denoted by $\boldsymbol{\xi} = (\xi_1, \xi_2, \dots, \xi_m)^\top$, to indicate the estimation of the odds instead of the cumulative hazard function. Finally, the likelihood function of the AFT model adapts the BP structure thought the hazard and cumulative hazard functions. The formulation \ref{formula:cumhaz} shall be included in the baseline odds function of the AFT likelihood \ref{loglikaft}. This provides:\\
\begin{equation}
\begin{aligned}
L(\boldsymbol{\Theta} \mid \boldsymbol{y}, \boldsymbol{\delta}, \boldsymbol{x}) &= \prod \limits_{i=1}^n \left[ \exp\{-\eta_i\} h_0( y_i\exp\{-\eta_i\})\right]^{\delta_i} \exp\{-H_0(y_i\exp\{-\eta_i\})\} \\
&= \prod \limits_{i=1}^n \left[ \exp\{-\eta_i\}\boldsymbol{\gamma}^\top\boldsymbol{g}_m(y_i \exp\{-\eta_i\})\right]^{\delta_i} \exp\{ -\boldsymbol{\gamma}^\top\boldsymbol{G}_m(y_i \exp\{-\eta_i\})\},
\end{aligned}
\label{bpaftlik}
\end{equation}\\ 
where $\boldsymbol{\Theta} = (\boldsymbol{\beta}, \boldsymbol{\gamma})$. The likelihood in \ref{bpphlik} refers to the Bernstein polynomial based proportional hazards (BPPH) model, the likelihood \ref{bppolik} refers to the Benstein polynomials based proportional odds (BPPO) model  and the likelihood in \ref{bpaftlik} refers to the Bernstein polynomial accelerated failure time (BPAFT) model under the right censoring mechanism. 

%Fully likelihood inference procedures are feasible based on these expressions to make inferences to time-to-event data in BP regression models. 

This chapter summarizes the usage of the BP in the finite approximation of positive continuous functions such as the hazard rate function. We emphasize that the target function must be known in advance to approximate it by the BP. Further, we also have comments on how to use BP under the statistical perspective of estimation. \cite{Osman:2012} proposed methods to estimate the hazard (target) function using the BP when this function is unknown. In this context, the next chapter presents some implementation details that were crucial to fit the BP based regression models.\\

\chapter{Inference procedures and technical issues}
 
$~~~~$Throughout the package implementation, some workarounds had to be done to guarantee a stable version of the \texttt{R} package implemented to fit the BP models. This chapter provides some perspective of what had to be done internally (without user interference) to achieve satisfactory performance in the simulation study results (Chapter 5). According to \cite{gjessing2010recurrent}, the presence of an exponential structure combined with other complex formulations might lead to an unstable model, with numerical problems. Our primary purpose is to avoid the explosive behavior of the survival regression models, also reported in \cite{aalen2008survival} and \cite{Kalbfleisch:2011}. 
 
Aware of the numerical stability problems, the survival package \citep{survival:2000}, for example, includes routines that internally standardize the covariates before evaluating the likelihood, in order to avoid an overflow in the argument of the exponential function. According to the \texttt{R} help documentation, the functions to fit models in the \texttt{survival} package, such as \texttt{survival::coxph}, internally scale and center data; see  the  ``\texttt{details}'' section of  \texttt{?coxph()} in the  \texttt{R} console. The adoption of this kind of feature brings some counterparts to the likelihood as a whole. Particularly,  these counterparts are related to the regression coefficients  and BP parameters scale. Consider the standardization of $p$ covariates as:\\
$$z_{ij} = \frac{x_{ij} - \bar{x}_j}{s_{x_j}};\quad j = 1, 2, \dots, p;$$\\
where $\bar{x}_j = \sum\limits_{i = 1}^n x_{ij}/n$ and $s_{x_j} = \sqrt{\sum\limits_{i = 1}^n (x_{ij} - \bar{x}_j)^2/(n-1)}$, such that  $\boldsymbol{\bar{x}} = (\bar{x}_1, \bar{x}_2, \dots, \bar{x}_p)^\top$ and $\boldsymbol{s_x} = (s_{x_1}, s_{x_2}, \dots,s_{x_p})^\top$ are the column vector of sample means and the column vector of sample standard deviations, respectively. For example, consider the inclusion of standardized covariates (explanatory variables) $\boldsymbol{z}_i = (z_1, z_2, \dots, z_p)^\top$ in the BPPH hazard function \ref{formula:loglikph}, as follows:\\
\begin{equation*}
\begin{aligned}
h(t \mid \boldsymbol{\Theta})
&= \boldsymbol{\gamma^*}^\top\boldsymbol{g}_m(t) \exp\{\boldsymbol{\beta^{*\top} z_i}\}\\
&= \boldsymbol{\gamma^*}^\top\boldsymbol{g}_m(t) \exp\{\boldsymbol{\beta^{*}}^\top [\boldsymbol{s}_{x}^{-1} \circ (\boldsymbol{x_i} - \bar{\boldsymbol{x}})]\}\\
&=  \boldsymbol{\gamma^*}^\top \exp\{-\boldsymbol{\beta^*}^\top (\boldsymbol{s}_{x}^{-1} \circ \bar{\boldsymbol{x}})\} ~\boldsymbol{g}_m(t) \exp\{\boldsymbol{\beta^*}^\top (\boldsymbol{s}_{x}^{-1} \circ \boldsymbol{x_i})\}\\
&=  [\boldsymbol{\gamma^*} \exp\{-\boldsymbol{\beta^*}^\top (\boldsymbol{s}_{x}^{-1} \circ \bar{\boldsymbol{x}})\}]^\top ~\boldsymbol{g}_m(t) \exp\{(\boldsymbol{\beta^*} \circ \boldsymbol{s}_{x}^{-1})^\top \boldsymbol{x_i})\}\\
&= \boldsymbol{\gamma}^\top\boldsymbol{g}_m(t) \exp\{\boldsymbol{\beta^\top x_i}\}\\
&= \boldsymbol{\gamma}^\top\boldsymbol{g}_m(t) \exp\{\eta_i\}.\\
\end{aligned}
\end{equation*}\\ 
where the symbol $\circ$ denotes the Hadamard (element-wise) product and the standard linear predictor is denoted by $\eta_i ^* = \boldsymbol{\beta^\top z_i}=\boldsymbol{\beta}^\top [\boldsymbol{s}_{x}^{-1} \circ (\boldsymbol{x_i} - \bar{\boldsymbol{x}})]$. The parametric space of interest is rewritten in terms of the new coefficients defined under the mentioned standardization $\boldsymbol{\Theta}^* =(\boldsymbol{\beta^*}; \boldsymbol{\gamma^*})$,   so that:\\
$$\boldsymbol{\Theta} = (\boldsymbol{\beta}; \boldsymbol{\gamma}) = \left(\boldsymbol{\beta^*} \circ \boldsymbol{s}_{x}^{-1};~ \boldsymbol{\gamma^*}\exp\{-\boldsymbol{\beta^*}^\top (\boldsymbol{s}_{x}^{-1} \circ \bar{\boldsymbol{x}})\}\right).$$ \\
Fully likelihood methods for ML estimation and Bayesian estimation were applied considering the transformed space. Thereafter, the quantities of interest could be recovered given the invariance property of the ML estimators or through the posterior mode (or any summary) of transformed chains,  following the relations:\\
\begin{equation}
\boldsymbol{\hat\beta} = \boldsymbol{\hat\beta^*} \circ \boldsymbol{s}_{x}^{-1} = \left(\frac{\hat\beta^*_1}{s_{x_1}}, \frac{\hat\beta^*_2}{s_{x_2}}, \dots, \frac{\hat\beta^*_p}{s_{x_p}}  \right)^\top,
\end{equation}\\
\begin{equation}\small
\boldsymbol{\widehat\Theta}=\begin{cases}
(\boldsymbol{\hat\beta}; \boldsymbol{\hat\gamma}) = \left(\boldsymbol{\hat\beta};~\boldsymbol{\hat\gamma^*}\exp\{-\boldsymbol{\hat\beta^*}^\top (\boldsymbol{s}_{x}^{-1} \circ \bar{\boldsymbol{x}})\}\right), \text{ if the model is BPPH,}\\
(\boldsymbol{\hat\beta}; \boldsymbol{\hat\xi})=\left(\boldsymbol{\hat\beta};~\boldsymbol{\hat\xi^*}\exp\{-\boldsymbol{\hat\beta^*}^\top (\boldsymbol{s}_{x}^{-1} \circ \bar{\boldsymbol{x}})\}\right), \text{ if the model is BPPO or}\\
(\boldsymbol{\hat\beta}; \boldsymbol{\hat\gamma})=\left(\boldsymbol{\hat\beta};~\boldsymbol{\hat\gamma^*}\exp\{\boldsymbol{~~\hat\beta^*}^\top (\boldsymbol{s}_{x}^{-1} \circ \bar{\boldsymbol{x}})\}\right), \text{ if the model is BPAFT.}
\end{cases}\label{gfunc}
\end{equation}\\

In this form, the BP survival regression estimates would be driven by the sample standard deviations $\boldsymbol{s}_{x} = (s_{x_1}, s_{x_2}, \dots,s_{x_p})^\top$ towards close to zero values, and the BP estimates would be inflated or deflated depending on the sample means $\bar{\boldsymbol{x}} = (\bar{x}_1, \bar{x}_2, \dots,\bar{x}_p)^\top$. It is expected that this technique brings more stability and accuracy to the proposed package. From the Bayesian perspective, the prior choice is made regarding the standardized coefficients. As mentioned,  we shall expect little deviations from zero and a great variability for the BP parameters as they shall depend on the arguments to the exponential function. 

From the Bayesian perspective, as a single effect can have a huge impact for the final result, one might assume, for example, the generic weakly informative and the weakly informative prior choices $\beta_j^* \sim N(0;\sigma^2_{\beta_j^*}=~16)$  and $\beta_j^* \sim N(0;\sigma^2_{\beta_j^*}=~100)$ for the regression coefficients, respectively. Also, generic or weakly informative Log-normal priors have been tested to express the lack of previous information about the BP parameters, the reader is reminded that, in this case, a Normal prior specification to the BP parameters in log-scale is equivalent to a Log-normal prior specification to the actual scale, that is: $\log(\gamma_k^*) \sim N(\mu_{\gamma^*}, \sigma^2_{\gamma^*})\equiv \gamma^*_k \sim LN(\mu_{\gamma^*}, \sigma^2_{\gamma^*}).$ Thus, the prior choice for the BP parameters is analogue, that is,  $\log (\gamma_k^*) \sim N(0;\sigma^2=~16)$ or $\log (\gamma_k^*) \sim N(0;\sigma^2=~100)$. According to the default configuration of the  \texttt{spsurv}, for each model fit, four chains of size 2000 are built for each quantity of interest, using the NUTS algorithm \citep{Hoffman:2014} provided in \texttt{Stan}. The first 1000 iterations are discarded as a burn-in (warm-up) period. After the sampling procedure, model comparison criteria are calculated using the \texttt{loo} package \citep{loo}. 

%The NUTS algorithm is the \texttt{Stan} software default MCMC, according to \cite{Carpenter:2017}. This sampling algorithm is an extension of the sampling dynamics proposed in the HMC algorithm. According to the Hamiltonian dynamics, the algorithm initializes with a random or arbitrary set of parameters. Then, a momentum vector is sampled, and the current value of the parameter  $\theta$ is updated using the leapfrog integrator; see \cite{rstan:2018}. After that, a decision is made whether to keep or update the existing state. The NUTS automatically selects an appropriate number of leapfrog steps at each step and avoids the random-walk behavior that arises when there is a correlation in the posterior. \\
Under the Frequentist perspective, the maximization algorithm is applied to the likelihood function to determine the ML estimates. The likelihood ratio (LR) test and Wald score tests are available for the BPPH, BPPO, or BPAFT models. The LR test statistic is not affected by the standardization as this statistic consists of twice the difference between two likelihood functions: $2\{L(\boldsymbol{\hat\Theta}\mid \boldsymbol{y, \delta, x}) - L(\boldsymbol{\hat\Theta_0}\mid \boldsymbol{y, \delta, x})\}$. Conversely, the Wald test statistic is affected by the standardization because the multivariate z-statistic is based on the Fisher information: $\boldsymbol{Z}_{\mbox{Wald}}=\sqrt{(\boldsymbol{\hat\Theta}-\boldsymbol{\hat\Theta_0})^{\top}  \displaystyle{\mathcal{\hat I}}(\boldsymbol{\hat\Theta})(\boldsymbol{\hat\Theta} - \boldsymbol{\hat\Theta_0})},$
where the observed Fisher information matrix is intended to be a sample-based version equivalent to the negative of the estimated Hessian matrix (log-likelihood second derivative). Nevertheless, the observed information matrix obtained refers to the information of standard configuration discussed at the beginning of this chapter. Hence, the Delta method is required to recover the observed information of interest \citep{oehlert1992note, Casella:2002, cooch2008program}. According to this point, we write:
\begin{equation}
  \sqrt{n}[\boldsymbol{\hat\Theta}^{*}-\boldsymbol{\Theta}^{*}]\,{\xrightarrow  {D}}\,{\mathcal  {N}}\left(0,\mathcal{\hat I}^{-1}(\boldsymbol{\hat\Theta^{*}})\right),
\end{equation}
\begin{equation}
\sqrt{n}[g(\boldsymbol{\hat\Theta}^{*})-g(\boldsymbol{\Theta}^{*})]\,{\xrightarrow {D}}\,{\mathcal {N}}\left(0,\nabla g(\boldsymbol{\hat\Theta}^{*})^\top \mathcal{\hat I}^{-1}(\boldsymbol{\hat\Theta^{*}})\nabla g(\boldsymbol{\hat\Theta}^{*})\right), \end{equation}\\
where $g$ represents the function described in \ref{gfunc} and $\nabla g(\boldsymbol{\hat\Theta}^{*})$ represents the gradient column vector  of partial derivatives  such that  $\mathcal{\hat I}^{-1}(\boldsymbol{\hat\Theta}) = \nabla g(\boldsymbol{\hat\Theta}^{*}) \mathcal{\hat I}^{-1}(\boldsymbol{\hat\Theta^{*}})\nabla g(\boldsymbol{\hat\Theta}^{*})^\top$.%The null hypothesis for both significance tests on BP based survival models is $H_0: \boldsymbol{\Theta} = \boldsymbol{\Theta_0} \text{  or  }  H_0: (\boldsymbol{\beta}, \boldsymbol{\gamma}) = (\boldsymbol{0},     \boldsymbol{\gamma})$. In this case, the LR test evaluates the semi-parametric model against the reference (only BP) model. Under the null hypothesis, the test statistic follows a Chi-square distribution with $p$-degrees of freedom. According to \cite{Therneau:2013}, the LR is more reliable than the Wald in the finite sample context.

The Fisher information is also needed to build confidence intervals. In this case, it is necessary to compute the estimated variance-covariance matrix, which is equivalent to the inverse of the observed Fisher information matrix  $\hat{V}(\boldsymbol{\hat\Theta}) = \mathcal{\hat I}(\boldsymbol{\hat\Theta}^{-1}$. In practice, some numerical problems may be experienced when dealing with the inversion of the observed information. The matrix $\mathcal{\hat I}(\boldsymbol{\hat\Theta})$ can be singular due to numerical approximations. To circumvent this issue, we consider a block-wise strategy for inversion that is defined for the following partition:

\begin{equation}
\mathcal{I}(\boldsymbol{\boldsymbol{\Theta})}=-\left[
\begin{array}{c|c}
  \frac{\partial^2 \ell (\boldsymbol{\Theta}\mid \boldsymbol{y, \delta, x})}{\partial^2\boldsymbol{\beta}} &   \frac{\partial^2 \ell (\boldsymbol{\Theta}\mid \boldsymbol{y, \delta, x})}{\partial\boldsymbol{\beta}\partial\boldsymbol{\gamma}}\\
   \hline 
    \frac{\partial^2\ell (\boldsymbol{\Theta}\mid \boldsymbol{y, \delta, x})}{\partial\boldsymbol{\beta}\partial\boldsymbol{\gamma}} &   \frac{\partial^2 \ell (\boldsymbol{\Theta}\mid \boldsymbol{y, \delta, x})}{\partial^2\boldsymbol{\gamma}}
\end{array} 
\right]= 
\left[
\begin{array}{c|c}
  \mathbf {A} & \mathbf {B} \\
  \hline
  \mathbf {C} & \mathbf {D} 
\end{array} \right].    
\end{equation}\\
Once the matrix is partitioned, it can be inverted block-wise as follows \citep{Bernstein:2009}:\\
\small
\begin{equation*}
{\displaystyle {\begin{bmatrix}\mathbf {A} &\mathbf {B} \\\mathbf {C} &\mathbf {D} \end{bmatrix}}^{-1}={\begin{bmatrix}\mathbf {A} ^{-1}+\mathbf {A} ^{-1}\mathbf {B} \left(\mathbf {D} -\mathbf {C A} ^{-1}\mathbf {B} \right)^{-1}\mathbf {C A} ^{-1}&-\mathbf {A} ^{-1}\mathbf {B} \left(\mathbf {D} -\mathbf {C A} ^{-1}\mathbf {B} \right)^{-1}\\-\left(\mathbf {D} -\mathbf {C A} ^{-1}\mathbf {B} \right)^{-1}\mathbf {C A} ^{-1}&\left(\mathbf {D} -\mathbf {C A} ^{-1}\mathbf {B} \right)^{-1}\end{bmatrix}},}  
\end{equation*}\\
where $\mathbf{A}$ refers to the regression block, $\mathbf {B}$ and $\mathbf {C}$ for the symmetric covariance blocks and $\mathbf {D}$ for the BP parameters block. Here we have that, $\mathbf {A}$ is $p\times p$ and $\mathbf {D}$ is a $m\times m$, so that both can be inverted. Also, $\mathbf {D} -\mathbf {C A} ^{-1}\mathbf {B}$ must be invertible.

Significance tests and confidence intervals were not proposed for the BP parameters since these parameters do not reflect any covariate effect or interpretative quantities. In turn, they dictate the shape of survival curves analogously to the scale parameters in parametric models. Beyond the necessary adjustments regarding the application of the model with transformed coefficients discussed earlier in this chapter, \cite{Osman:2012} comment on the choice of the polynomial degree and states that this is closely related to the true hazard function shape. There is a  bias-variance trade-off in which a small degree polynomial is likely to result in biased estimates, while a large degree polynomial might introduce excessive variation. The package structure will allow the users to choose the best polynomial degree for their applications. However, this choice must take into account the purpose of user-defined modeling. If a high-precise estimation of the baseline functions is not interesting to the study, the practitioner should choose low-degree polynomials. Otherwise, high-degree polynomials are preferable. The recommendation \citep{Osman:2012} is to the use of $m = \sqrt{n}$ as the choice of polynomial degree that is set as the default in the \texttt{spsurv} package.

The upper bound restriction, $\tau$ in \ref{bpsurv}, is not considered a parameter in the \texttt{spsurv}  package structure. According to \cite{Osman:2012}, in the context of survival analysis: $\tau<\infty \text{ such that } \tau = \inf\{t :
S(t \mid \boldsymbol{\Theta}, \boldsymbol{x}) = 0\}$ and $\hat\tau = \max\{t_1, t_2, \dots, t_n\} \rightarrow \tau = \inf\{t :
S(t \mid \boldsymbol{\Theta}, \boldsymbol{x}) = 0\}$ in probability. Thus, the main function of the package internally sets $\hat\tau=\max\{t_1, t_2, \dots, t_n\}$. Although this is applied to the implementation, one should know that any estimator for $\tau$ would generate an improper survival function, that is, we would have:\\
\begin{equation}
\begin{aligned}
  S(\hat\tau \mid \boldsymbol{\Theta}, \boldsymbol{x}) &= \exp\{-H(\hat\tau \mid \boldsymbol{\Theta}, \boldsymbol{x})\} \\
  &= \exp\{-\boldsymbol{\gamma}^\top\boldsymbol{G}_m(\hat\tau)\exp\{\eta_i\}\}\ge0,
\end{aligned}
\end{equation}\\
where we can not guarantee that $S(\hat\tau \mid \boldsymbol{\Theta}, \boldsymbol{x}) = 0$. In special,
\begin{equation}
\begin{aligned}
  S(\hat\tau =  \max\{t_1, t_2, \dots, t_n\} \mid \boldsymbol{\Theta}, \boldsymbol{x}) &= \exp\{-\boldsymbol{\gamma}^\top\boldsymbol{G}_m(\max\{t_1, t_2, \dots, t_n\})\exp\{\eta_i\}\}\\
   &= \exp\{-\boldsymbol{\gamma}^\top\boldsymbol{1}\exp\{\eta_i\}\} >0. \\
\end{aligned}
\end{equation}\\
As a solution to this inconsistency, a tail adjustment is necessary to satisfy $\lim_{t\rightarrow \infty}S(\tau\mid\boldsymbol{\Theta}, \boldsymbol{x})=0$; see \cite{Osman:2012}. Although there are some choices of corrections that provides a survival function that meets the requirement, these kind of corrections do not affect the estimates of the BP based survival models. The correction applied to the survival function formulation after the last observed time, will not change the estimates presented in this dissertation. Note that, the time region beyond $\hat{\tau} = \max\{t_1, t_2, \dots, t_n \}$ does not contain any time response.

We close here the third chapter of this dissertation. The next chapter is dedicated to a comprehensive simulation study that compares the overall performance of the BPPH, BPPO, and BPAFT. The main aim of the next chapter is to assess whether the BP estimates behave well in distinct scenarios.  After the analysis involving artificial data sets, the dissertation will be focused on applications of the BP models to real problems.

% Chapter 4 = Chapter Title.
\chapter{Monte Carlo simulation study}

$~~~~$In this chapter, we present a Monte Carlo (MC) simulation study to evaluate
the performance of the BP based models in terms of estimation. The analysis is divided in two scenarios: sample size $n = 100$ (Scenario I) and sample size $n = 200$ (Scenario II). For both cases, the same covariates  $\boldsymbol{x_i} = (x_{i1},x_{i2})$ were used to generate  1000 replications (data sets), from LLAFT and WAFT (parametric) models. Censoring times were produced based on the same distribution used to generate the failure times. Figure \oldref{Fig:censoringrate} (Appendix F) shows how the censoring rate was distributed for each simulated data set generated.

Ideally, the simulated  survival times can be generated from the relationship $T_i = S^{-1}(U_i)$, where $U_i$ is a uniformly distributed random variable and $T_i$ is the failure time random variable. In fact, it is feasible to generate a random variable observation from its own survival function $S$ (if the survival is invertible). One can introduce an observed value $u_i$ from $U_i \sim ~Uniform [0,1]$ and, therefore, obtain $y_i = S^{-1}(u_i), ~i \in \{1, \dots, n \}$. We can show that:\\ 
$$\begin{aligned} P(T_i > t_i) &= P(S^{-1}(U_i) > t_i)= P(S(S^{-1}(U_i)) > S(t_i))\\
       &= P(S(T_i) > S(t_i))= P(T_i > t_i).
\end{aligned}$$ \\
According to \cite{ross2012simulation}, this method is often called the Inverse Transform Sampling (ITS). In this case, $S(.)$ denotes the survival function adopted for $T_i$  either from WAFT \ref{Formula:waft} or LLAFT \ref{Formula:LLAFT}. Table \oldref{Table:settings} shows the settings of each data generator model:\\ 
\begin{table}[!htb]
\centering
\scriptsize
\begin{tabular}{ccccccccc}
\hline
Data set \# & Model & Covariate $x_1$ & Covariate $x_2$ & Shape & Censoring scale  & Average \% censoring \\ \hline
 1 & LLAFT & $x_1 \sim N(0,1)$                    & $x_2 \sim Hyper(500;250;250;1)$ &$\zeta = 2$ & $\nu_c=700$                              &$25.07 $       \\
 2 & WAFT      & $x_1 \sim N(0,1)$                    & $x_2 \sim Hyper(500;250;250;1)$ & $\gamma = 2$ & $\lambda_c=1.4$    & $25.68$      \\ \hline\\
\end{tabular}
\caption{Settings of the generator model assuming a sample size $n = 200$. The scale parameter $\nu_c$ (or $\lambda_c$) of the censoring time distributions was chosen to provide an approximately 25\% of censored observations in each data set. The total of 1000 MC replications were generated from each setting described in this table. $N$ and $Hyper$ refer to the  Normal and the Hypergeometric distributions, respectively.}
\label{Table:settings}
\end{table}\normalsize \\

As stated earlier, the censoring mechanism is non-informative. Hence, both event and censoring times were generated independently through the ITS method. The censoring times, for each data set, were also generated from the Weibull (WAFT) and the Logistic (LLAFT) survival functions. In order to achieve approximately 25\% average censoring (Table \oldref{Table:settings}), the scale parameter for the censoring times, were set to be $\nu^*=\nu_c=700$ in \ref{Formula:waft} and $\lambda^*=\lambda_c=1.4$ in  \ref{Formula:LLAFT}.  It should be mentioned that, the distribution of failure times is individual (not indentically distributed), that is, the value of the scale parameter $\lambda^*= \lambda \exp\{-\eta^{\text{WAFT}}_i \kappa\}$ (or $\nu^*=\nu\exp\{-\eta^{\text{LLAFT}}_i \alpha\}$) in the survival function differ between the artificial groups created (see Chapter 2). Summary statistics for these parameters  were presented in Table \oldref{Table:settings2}:\\
\begin{table}[!htb]
\centering
\footnotesize
\begin{tabular}{rrrrrrrr}
  \hline
 Model & Scale  & Min. & 1st Qu. & Median & Mean & 3rd Qu. & Max. \\ 
  \hline
LLAFT & $\nu^*$ & 3541.2876 & 16202.3240 & 28818.4750 & 79166.4723 & 57127.7961 & 4510736.9826 \\ 
  WAFT & $\lambda^*$ & 25.2166 & 40.4366 & 47.4655 & 50.9704 & 56.1337 & 189.7279 \\ 
   \hline \\ 
\end{tabular}
\caption{The individual scale parameter value differs for different groups artificially created in the generation of both data sets. Summary statistics for these quantities ($\nu^*$ or $\lambda^*$) are displayed in this table. The LLAFT and WAFT models were described previously in Section 2.4.}  
\label{Table:settings2}
\end{table}\normalsize

Note that the Table \oldref{Table:settings} shows only the settings for the Scenario II. However, Scenario I can be obtained by simply using the first 100 elements of the 200 sized data sets. We also emphasize that, the MC simulation study  were based on two distinct likelihood maximization methods and four distinct prior specifications for every replication in each of the two scenarios. A sensitivity analysis  for the prior specifications of the BP parameters was applied to evaluate the impact of the initial uncertainty about these unknown quantities. Table \oldref{Table:priors14} shows the four hyperparameter choices tested. In this context, independence between all parameters was assumed \textit{a priori}.\\
%including the independence between the BP parameters and the regression coefficients.
\begin{table}[!htb]
\centering
\scriptsize
\begin{tabular}{rrrrrrrr}
\hline
Prior \# & Prior for $\beta_j$ & $E(\beta_j)$ & $V(\beta_j)$ & Prior for $\gamma_k$ & $E(\gamma_k)$ & $V(\gamma_k)$ \\
\hline
 1  & $N(\mu = 0; \sigma = 4)$ & $0$ & $16$ & $LN(\mu = 0; \sigma = 4)$  & $2980.9580$ & $7.8963 \times 10^{13}$ \\
 2 & $N(\mu = 0; \sigma = 4)$ & $0$ & $16$ & $LN(\mu = 0; \sigma = 10)$  & $5.1847\times 10^{21}$ & $7.2260 \times 10^{86}$ \\
 3 & $N(\mu = 0; \sigma = 10)$ & $0$ & $100$ & $LN(\mu = 0; \sigma = 4)$ & $2980.9580$ & $7.8963 \times 10^{13}$ \\ 
 4    & $N(\mu = 0; \sigma = 10)$ & $0$ & $100$ & $LN(\mu = 0; \sigma = 10)$ & $ 5.1847\times 10^{21}$ & $7.2260 \times 10^{86}$ \\ 
\hline \\
\end{tabular}
\caption{Options of priors for the sensitivity analysis. $N$ refers to the Normal distribution, $LN$ refers to the Log-Normal probability density function. $E(.)$ and $V(.)$ refers to the expected value, respectively.}
\label{Table:priors14} \label{sensitivity}
\end{table}\normalsize\\
Under the Bayesian framework, Prior 1 and Prior 2 specifications provide more information to the regression coefficients (Table \oldref{Table:priors14}) due to the belief in slight deviations from zero. As mentioned in Chapter 4, the covariates were standardized before fitting the models, which means that atypical values will have less impact in the regression estimates once they have been internally standardized. In this context,  Priors 3 and 4, provided less information to the regression coefficients aiming to offer a contrasting choice. 

The primary goal of this analysis is to investigate how the regression estimates are affected by distinct levels of uncertainty attributed to the BP parameters. The next section shows the results of this simulation study for Scenario I. Results for Scenario II are explained later
in this dissertation. All the simulation study results are presented in Appendix B. The conclusions about the simulation studies are given in the last section of this chapter.

\section{Scenario I: sample size $n = 100$}
$~~~~$This section presents the results for the data sets with a sample size of $100$. In the investigations conducted in this dissertation, we choose to explore five statistics in the evaluation of the estimation under the MC scheme: the average estimate (est.), the average estimated standard error (se.), the standard error of the estimates (sde.), the relative bias (rb.) and the coverage probability. For this, consider $\phi$ a generic parameter, and assume that $\hat\phi$ is the ML (or the posterior) estimate. The analysis of the results from the MC replications takes into account the next elements:\\
\begin{itemize}
    \item the average estimate (est.):\begin{equation}
    \quad\bar\phi = \frac{1}{R}\sum\limits_{i=1}^R \hat\phi_i,\label{est.}
    \end{equation}
    \item the average estimated standard error (se.):\begin{equation}
     \quad\frac{1}{R}\sum\limits_{i=1}^R\widehat{se}(\hat\phi_i),   \label{se.}
    \end{equation}
    \item the standard error of the estimates (sde.):\begin{equation}
        \quad\frac{1}{R-1}\sum\limits_{i=1}^R (\hat\phi_i - \bar\phi)^2. \label{sde.}
    \end{equation}
\end{itemize}
where $\widehat{se}$ denotes the estimated standard error (or posterior standard
deviation) obtained for $\hat\phi$ and $\phi_{true}$ denotes the true value. In order to account for the distance between the reported estimate and the true estimate, the relative bias is considered with the following formulation:\\
\begin{equation}
  \text{rb($\phi$)} =100(\widehat{\phi} -\phi_{true})/ \mid \phi_{true} \mid.\label{rb.}
\end{equation}\\
The uncertainty related to the $R=1000$ replications is expressed based on interval estimates for the regression coefficients. Thus, another interesting element to be considered in the analysis is the coverage probability of the model. The coverage probability is the percentage of the MC replications that provide a 95\% interval that captures the true value of the parameter. In the frequentist approach, we consider 95\% confidence intervals. In the Bayesian framework, the investigation is based on 95\% HPD (Highest Probability Density) credibility intervals. 

The study presented here was developed with the following strategy. The BPPH and  BPAFT models were fitted to the data sets that originated from the WAFT model. Meanwhile, the BPPO and BPAFT models were fitted to the data sets that originated from the LLAFT model. These settings are comparable based on the relationships discussed in section 2.4. For comparative purposes, we have also fitted the generator models (parametric AFT) using the routine provided by \texttt{survreg::survival}.\\

\begin{table}[!htb]
\centering
\scriptsize
\begin{tabular}{rrrrrrrrrrrrrr}
\hline
ML     &      & \multicolumn{5}{l}{WAFT} &  & \multicolumn{5}{l}{LLAFT} \\ \cline{3-7} \cline{9-14} 
      & true & est.  & se. & sde. & rb. & cov. & & true & est.   & se.   & sde.  & rb.  & cov.  \\ \hline
$\beta_1$ & 2    &   2.0090   &  0.2282  &  0.2724  & 0.4253   &  0.8967   & 
          & 2 &   2.0030     &   0.2223    &  0.2325     &  0.1438    &   0.9434   \\
$\beta_2$ &  -1  &  -1.0080   &  0.3046 &  0.3729 & -0.7903    &  0.8947   &  
          &  -1  &   -1.0210    &   0.3065    &  0.3174     &  -2.1136    &  0.9434     \\ \hline\\
\end{tabular}
\caption{MC simulation study assuming the WAFT and LLAFT  generator model fits in the Scenario I ($n = 100$). The estimate of the regression coefficient (est.), average
standard error (se.), the standard deviation of the estimates (sde.), relative bias (rb in \%), and
coverage probability (nominal level 95\%). The routine used to supply data to this table was \texttt{survreg::survival}, the estimates reported here are obtained under the Maximum Likelihood (ML) estimation.} \label{ref100}
\end{table}\normalsize
The results in Table  \oldref{ref100} do not provide any indication that the procedure to generate data (ITS) contains errors. As expected, the relative bias is very close to zero (less than 3\%), which suggests that the generator model recovers well the true value of the coefficients. Furthermore, the coverage rates of the confidence intervals are close to the nominal value of 95\%, reflecting the good interval estimation provided by the model fit. In conclusion, Table \oldref{ref100} indicates that the generator model (either WAFT or LLAFT) can recover the true values of the coefficients. In this sense, we can consider that this table is a useful reference for the evaluation of the BP based survival regression model routines proposed here. The purpose of this section is to verify whether the BP based model's statistics provided are similar to the statistics provided by the model that generated the data (Table  \oldref{ref100}).

\begin{table}[!htb]
\centering
\scriptsize
\begin{tabular}{rrrrrrrrrrrrrr}
\hline
BFGS     &      & \multicolumn{5}{l}{BPPH} &  & \multicolumn{5}{l}{BPAFT$^{a}$} \\ \cline{3-7} \cline{9-14} 
      & true & est.  & se. & sde. & rb. & cov. & & true & est.   & se.   & sde.  & rb.  & cov.  \\ \hline
$\beta_1$ & -4    &  -4.4807 & 0.5814 & 0.7670 & -12.0172 & 0.8250 & 
          & 2 &   2.0398 & 0.1271 & 0.3209 & 1.9885 & 0.6497 \\
$\beta_2$ &  2   &    2.2373 & 0.5170 & 1.0465 & 11.8664 & 0.9080      &  
          &  -1  & -1.0169 & 540.4119 & 0.7781 & -1.6880 & 0.6562  \\ \hline
\end{tabular}
\begin{tabular}{rrrrrrrrrrrrrr}
\hline
LBFGS     &      & \multicolumn{5}{l}{BPPH} &  & \multicolumn{5}{l}{BPAFT$^{b}$} \\ \cline{3-7} \cline{9-14} 
      & true & est.  & se. & sde. & rb. & cov. & & true & est.   & se.   & sde.  & rb.  & cov.  \\ \hline
$\beta_1$ & -4    &  -4.4774 & 0.6129 & 0.7478 & -11.9343 & 0.8480 & 
          & 2 &   2.0368 & 0.1287 & 0.3209 & 1.8376 & 0.6596\\
$\beta_2$ &  2   &   2.2617 & 0.5275 & 1.8072 & 13.0832 & 0.9130     &  
          &  -1  & -1.0047 & 126.6637 & 0.5935 & -0.4669 & 0.6539   \\ \hline
\end{tabular}
\begin{tabular}{rrrrrrrrrrrrrr}
\hline
Prior 1    &      & \multicolumn{5}{l}{BPPH} &  & \multicolumn{5}{l}{BPAFT$^{c}$} \\ \cline{3-7} \cline{9-14} 
      & true & est.  & se. & sde. & rb. & cov. & & true & est.   & se.   & sde.  & rb.  & cov.  \\ \hline
$\beta_1$ & -4    & -3.4030      &  0.4419     & 0.3167      & 14.9300     & 0.7450     & 
          &  2    &  2.0950      &  0.2026     & 0.2158     &  4.7580  &  0.9248    \\
$\beta_2$ &  2   & 1.7060      & 0.4556      & 0.4021      & -14.6900     & 0.9200  & 
          &  -1  &  -1.0490 & 0.2569   & 0.3075  & -4.9360 & 0.9468  \\ \hline
\end{tabular}
\begin{tabular}{rrrrrrrrrrrrrr}
\hline
Prior 2    &      & \multicolumn{5}{l}{BPPH} &  & \multicolumn{5}{l}{BPAFT} \\ \cline{3-7} \cline{9-14} 
      & true & est.  & se. & sde. & rb. & cov. & & true & est.   & se.   & sde.  & rb.  & cov.  \\ \hline
$\beta_1$ & -4    & -3.9730       &  0.5486       & 0.4770  & 0.6630    &   0.9740  & 
          &  2    &  1.9890     &   0.5050    &  0.4935      & -0.5537     &   0.9660    \\
$\beta_2$ &  2   & 1.9540      & 0.5050      & 0.4854      & -2.2920     & 0.9660  & 
          &  -1  &  -1.0250  &  0.2421    & 0.3042  &  -2.4770  &  0.9250   \\ \hline
\end{tabular}
\begin{tabular}{rrrrrrrrrrrrrr}
\hline
Prior 3    &      & \multicolumn{5}{l}{BPPH} &  & \multicolumn{5}{l}{BPAFT} \\ \cline{3-7} \cline{9-14} 
      & true & est.  & se. & sde. & rb. & cov. & & true & est.   & se.   & sde.  & rb.  & cov.  \\ \hline
$\beta_1$ & -4    & -3.4800       &  0.4519     & 0.3335     & 13.0000    & 0.8250     & 
          &  2    &  2.1060      & 0.2058      &  0.2211     &   5.3111   &   0.9190    \\
$\beta_2$ &  2   & 1.7440      & 0.4590      & 0.4124     & -12.8100     & 0.9320  & 
          &  -1  &  -1.0620  &  0.2640    & 0.4569  &  6.2430  &  0.9420   \\ \hline
\end{tabular}
\begin{tabular}{rrrrrrrrrrrrrr}
\hline
Prior 4    &      & \multicolumn{5}{l}{BPPH} &  & \multicolumn{5}{l}{BPAFT} \\ \cline{3-7} \cline{9-14} 
      & true & est.  & se. & sde. & rb. & cov. & & true & est.   & se.   & sde.  & rb.  & cov.  \\ \hline
$\beta_1$ & -4    & -4.1230      & 0.5730      & 0.5326     & -3.0750    & 0.9670     & 
          &  2    &  2.0590      &   0.1957    &  0.2223     &  2.9660    &  0.9160     \\
$\beta_2$ &  2   & 2.0630      & 0.5144      & 0.5251      & 3.1550     & 0.9620  & 
          &  -1  & -1.0360   &   0.2484   & 0.4540  &  -3.5970  &   0.9170  \\ \hline \\ 
\end{tabular}
\caption{MC simulation study in scenario I ($n = 100$), models fitted to the WAFT data sets.  Estimate of the regression coefficient (est.), average
standard error (se.), standard deviation of the estimates (sde.), relative bias (rb in \%) and
coverage probability (nominal level 95\%). Symbols: ${a}$ indicates  $R = 925$ (6 non-finite Hessian matrices and 69 non-converging); ${b}$ indicates $R=890$ (1 non-finite hessian and 109 non-converging) and ${c}$ indicates $R=997$.} \label{sim1}
\end{table}\normalsize

The Table \oldref{sim1} displays the mentioned statistics for the estimates of the BP based models applied to the WAFT data set replications in Scenario I. This table shows twelve possible configurations for fitting the 1000 MC data sets using the \texttt{spsurv} package. The first two refer to the  Frequentist approaches either under the PH or AFT frameworks. The Broyden–Fletcher–Goldfarb–Shanno (BFGS) algorithm was first used to promote the likelihood maximization. Secondly, another option of optimization algorithm was tested to assess if there is any empirical evidence in preferring some of them; the Limited-memory BFGS (LBFGS) algorithm was used with that purpose, both algorithms are built-in \texttt{Stan}. Besides, the Bayesian model fit is also included assuming the prior specifications described in Table \oldref{sensitivity}.

It is noteworthy that we chose not to explore the Newton method, which was also made available to the user in the \texttt{spsurv} package, to save computational time. In our experience, this method tends to be slow to handle the BP models. We also highlight that the mentioned optimization methods can often find a local maximum rather than the global one.   The likelihood function was evaluated concerning four ML estimates that considered distinct random initial values for the optimizer to mitigate this kind of bias. Thus, the largest likelihood estimate obtained with these four possible fits were selected. Both BFGS and LBFGS are iterative methods for solving nonlinear optimization problems \citep{fletcher2000practical}. The LBFGS is a BFGS extension with a limited-memory; the BFGS algorithm accumulates all the gradient values, including the first ones. On the other hand, the LBFGS drops old gradients in favor of the new ones. The LBFGS is useful to avoid the bias of the initial gradient; even so, the estimated Hessian will still be biased by initial values until enough gradients are accumulated close to the solution. 

In short, according to Table \oldref{sim1}, we can consider that the BP based models provided good results when applied to Bayesian inference, especially looking for the  Prior 2 and 4 relative biases. Under the Frequentist approach, the BP based models did not produce results as good as the generator model fits (Table \oldref{ref100}); in particular, the BPPH model fits provided relatively biased estimates. In comparison, we might also suggest that the Bayesian approach is less error-prone, \textit{i.e.} we found here fewer problems regarding the computation of the estimates. For example,  an amount of 178 non-converging estimates were found in the BPAFT case (for both BFGS and LBFGS). At the same time, only the first prior specification provided fewer than expected replications ($R=997$).   As a result, the number of valid MC replications decreased due to issues in computing the estimates (see the caption of Table \oldref{sim1}). In the Bayesian case, the reason for detecting those missing values is related to the non-mixing chains that can occur due to the initial values that were randomly assigned by \texttt{Stan} internally.

Table \oldref{sim2} (Appendix B) shows the MC simulation study for the estimates of the BP models applied to LLAFT  data set. We found similar conclusions about interval estimates for the BPAFT model under the ML perspective applied to LLAFT data. Both average standard error and the coverage probability reflect the underestimated standard errors. Consequently, the BPAFT has presented narrower confidence intervals compared to the interval ranges from the generator model. Also, the optimization problems once more have caused a decrease in the number of valid MC replications. For the BPPO and BPAFT models, a total of 98 non-converging estimates and 17 non-finite Hessian matrices were found (see the caption of Table \oldref{sim2}). Regarding the Bayesian model fits sensitivity analysis, the same conclusion can be maintained: the prior specification that presented the best results was again Prior 2, which consists of attributing generic weakly informative priors to the regression coefficients and vague prior information to the BP parameters. Above all, we found that the Bayesian BP based survival regression models can provide accurate inferences (Prior 2) in using the three classes: PO, PH, and AFT.

\begin{figure}[!htb]
 \centering
 $$
  \begin{array}{cc}
   \mbox{\textbf{(a)} BPAFT}&
   \mbox{\textbf{(b)} BPPO}\\
     \includegraphics[width=0.5\textwidth]{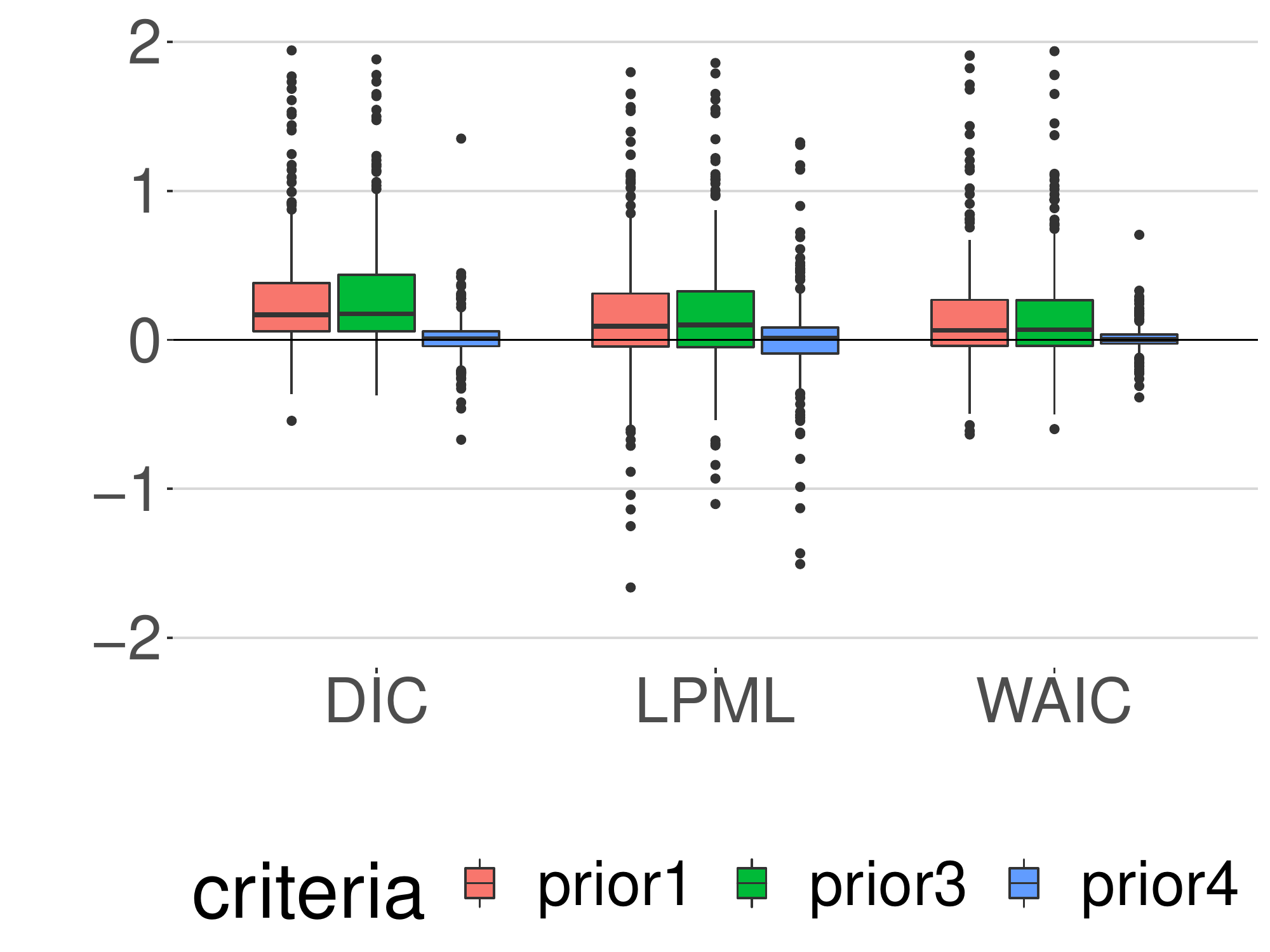}&
    \includegraphics[width=0.5\textwidth]{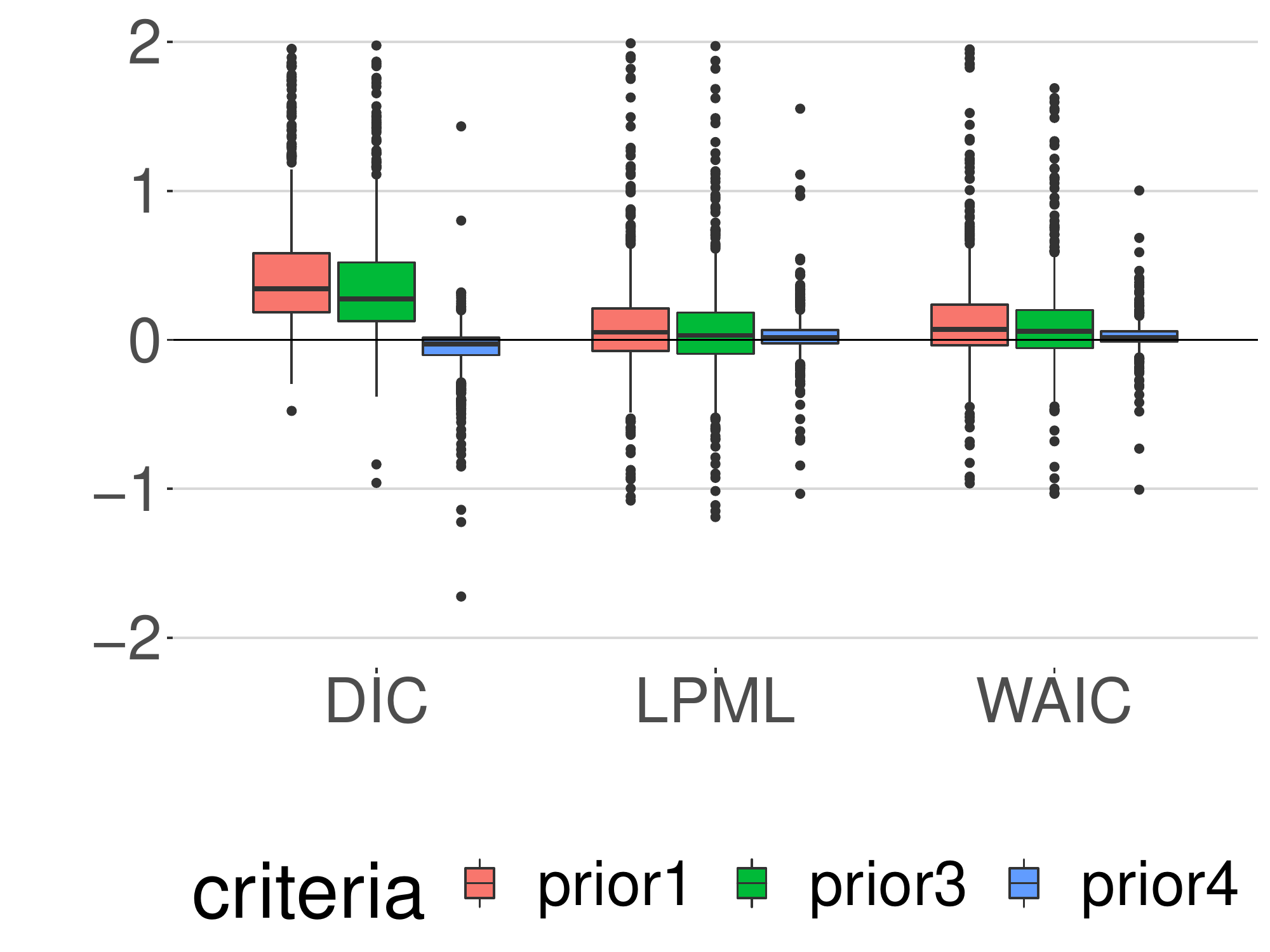} \\
       \mbox{\textbf{(c)} BPAFT}&
       \mbox{\textbf{(d)} BPPH}\\
     \includegraphics[width=0.5\textwidth]{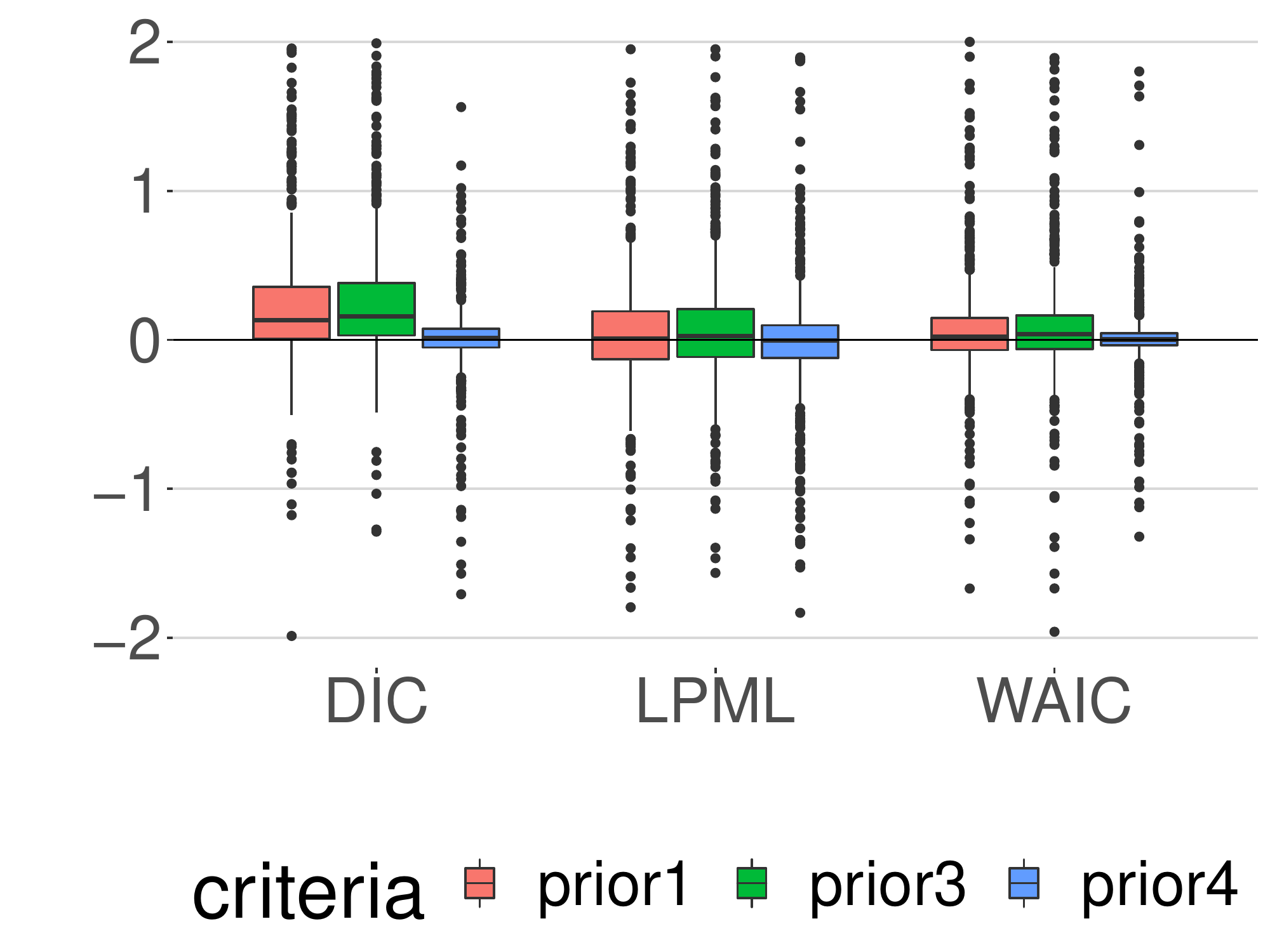}&
    \includegraphics[width=0.5\textwidth]{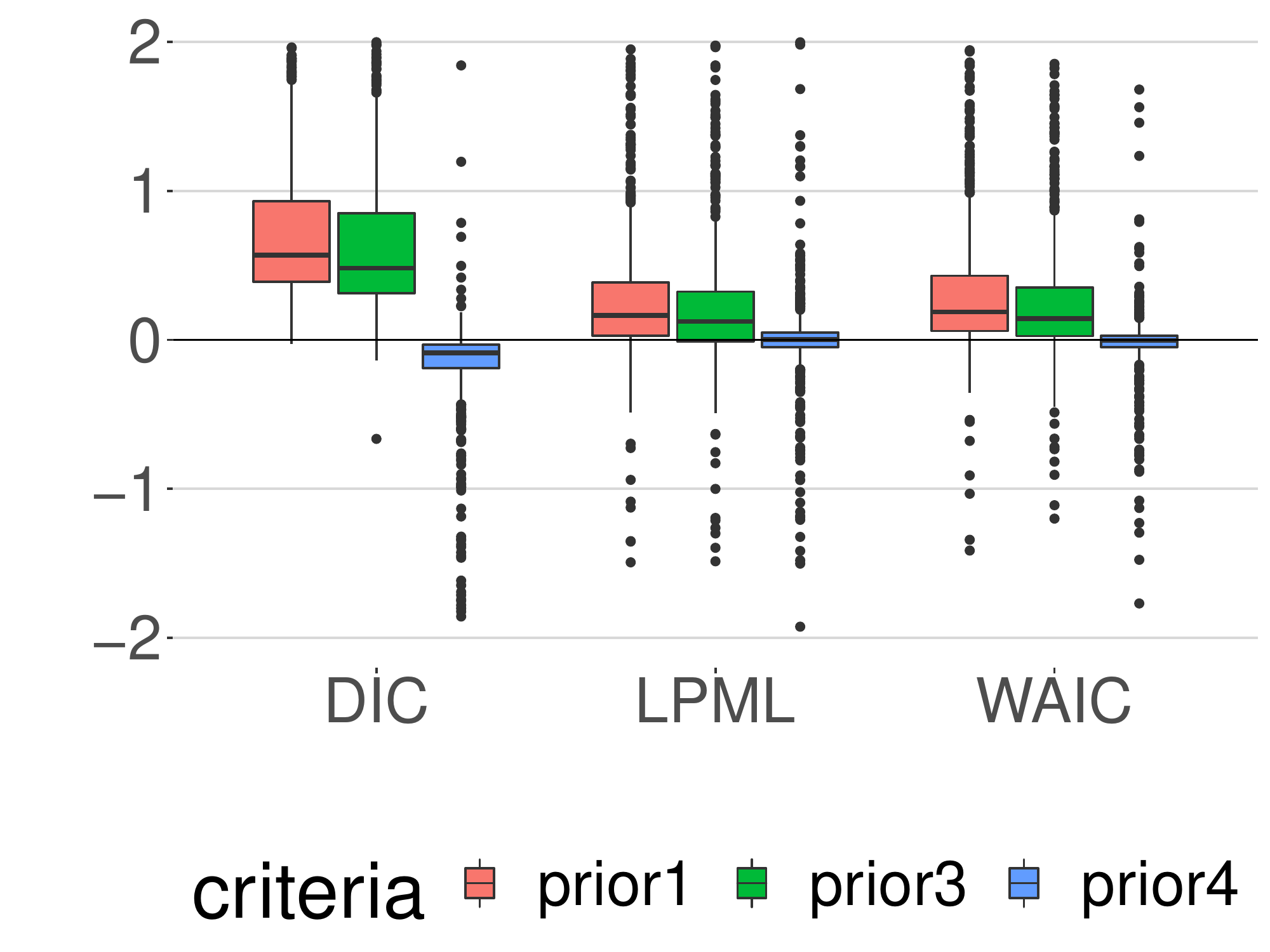} \\
  \end{array}
$$
 \caption{Box-plots of the relative differences between the values of the DIC, -2 WAIC and -2 LPML criteria from two models. The first model is the one identified by prior numbers 1, 2 and 3.
The reference (second) model is the one assuming Prior 2. Here, the models were fitted to data sets generated under Scenario I ($n = 100$). Panel \textbf{(a)}: Relative difference regarding the BPAFT model (LLAFT case). Panel \textbf{(b)}: Relative difference regarding the BPPO model. Panel \textbf{(c)}: Relative difference regarding the BPAFT model (WAFT case). Panel \textbf{(d)}: Relative difference regarding the BPPH model.}
  \label{criteria1}
 \end{figure}

In general, it is possible to state that the estimates provided by the LBFGS method are very similar when compared to the BFGS. Some model comparison criteria such as the LR statistic, the Akaike information criterion \citep{akalke1974new}, and the Bayesian information criterion  \citep{schwarz1978estimating}, for example, could be applied in this case to support the comparison between those algorithm estimates. However, we chose the one that indicated the highest coverage rate to BP based models applied to Frequentist inference. For this reason, the ML results of the LBFGS algorithm will be used in the comparison between the estimates of the two inferential approaches. From the Bayesian perspective, the deviance information criterion (DIC) \citep{spiegelhalter2002bayesian}, the Watanabe–Akaike information criterion (WAIC)  \citep{watanabe2013widely} and the logarithm of the pseudo-marginal likelihood (LPML) \citep{geisser1979predictive} was calculated to assist the evaluation of the prior choice. The LPML and WAIC criteria were multiplied by -2, so that the three criteria have a similar interpretation to facilitate the understanding, that is, the lower is the value of the criterion, the better the model fits the data. It is expected that the prior choice that presented consistently little relative bias, and near 95\% coverage rate will also outperform the other models concerning the above criteria. Accordingly, Figure \oldref{criteria1} shows the relative difference between some reported prior choice criterion and the value of the same criterion regarding Prior 2, for the identical MC replication.

The relative difference is somehow similar to relative bias \ref{rb.}. Instead, it accounts for the distance from another criterion (or estimate) rather than the distance from the true values. The relative difference accounts for the distance between the reported criterion (or estimate) and the value of some reference criterion:
\begin{equation}
  \text{rd($\phi$)} =100(\widehat{\phi} - \widehat{\phi}_r)/ \mid \widehat{\phi}_r \mid, \label{rb.}  
\end{equation}
where $\widehat{\phi}_r$ is the Prior 2 (reference) criterion (or estimate) and $\widehat{\phi}$ is the referred prior criterion (Prior 1, Prior 3 or Prior 4). The relative difference is a ratio with the numerator being the difference between estimates and the denominator being the magnitude of the reference value. Negative and positive results
indicate that the criterion is being evaluated above or below the reference estimate, respectively. The fraction is multiplied by 100, leading to a percentage representation of its magnitude. Some MC replicas provided non-finite values for the -2 WAIC and -2 LPML. Therefore the relative difference was only included if the three criteria were valid for both priors in the same MC replication comparison. 

 \newpage
  \newpage

Figure \oldref{criteria1} suggests that the model comparison criteria -2 LPML and -2 WAIC do not differ about the best model to be chosen. The results have indicated that Prior 2 fit is often preferable to other fits. Besides, we can state that the -2 WAIC and -2 LPML are suitable for comparing BP based survival regression models in Scenario I since most of them indicated the best fit concerning relative bias and coverage probability. Although \cite{christensen2011bayesian} states that LPML is preferable in many cases. We have also found that the DIC criterion is more sensitive than the -2 LPML and -2 WAIC when it comes to BP based models comparison. The relative difference presented by this one was higher than expected. Also, we discovered that the DIC criterion in BPPO and BPPH do not reflect the findings of this simulation study Scenario because it often indicates Prior 4 as the preferable model. From our analysis: we conclude that Prior 2 is slightly better than Prior 4 (making sense to be chosen). But we have learnt that the prior choice is essential for the best fit. 

Table \oldref{minisim1} shows the results of the Bayesian (Prior 2) and the Frequentist (LBFGS) approaches. Remarkably, the BPAFT coverage probabilities under the ML approach do not approximate to the nominal level of 95\%. Note that, the average standard error (se.) and the standard error of the estimates (sde.) under the ML perspective are not close. It is noteworthy that these statistics are tight in the Bayesian case. Therefore, we can suggest there exists a rough approximation error in the Delta Method applied.   The average estimated standard error of the BPAFT applied to the WAFT data set in (Table \oldref{minisim1}),  is, on average, underestimated, and exceedingly overestimated for the second coefficient. This conclusion is based on the comparison with the results from the generator model (Table \oldref{ref100}). In turn, this overly estimated average standard error is justified by the presence of discrepant values in the standard error estimation. Indeed, it can be seen that the below-expected coverage probability reflects the fact that most estimated confidence intervals for BPAFT models were exceptionally narrow. Despite the unsuccessful confidence interval estimation, the relative bias is very low for the BPAFT estimates (see Table \oldref{ref100}).\\

\begin{table}[!htb]
\centering
\scriptsize
%\begin{tabular}{rrrrrrrrrrrrrr}
%\hline
%BFGS     &      & \multicolumn{5}{l}{BPPH} &  & \multicolumn{5}{l}{BPAFT$^{a}$} \\ \cline{3-7} \cline{9-14} 
%      & true & est.  & se. & sde. & rb. & cov. & & true & est.   & se.   & sde.  & rb.  & cov.  \\ \hline
%$\beta_1$ & -4    &  -4.4807 & 0.5814 & 0.7670 & -12.0172 & 0.8250 & 
%          & 2 &   2.0398 & 0.1271 & 0.3209 & 1.9885 & 0.6497 \\
%$\beta_2$ &  2   &    2.2373 & 0.5170 & 1.0465 & 11.8664 & 0.9080      &  
 %         &  -1  & -1.0169 & 540.4119 & 0.7781 & -1.6880 & 0.6562  \\ \hline
%\end{tabular}

\begin{tabular}{rrrrrrrrrrrrrr}
\hline
LBFGS     &      & \multicolumn{5}{l}{BPPH} &  & \multicolumn{5}{l}{BPAFT$^{a}$} \\ \cline{3-7} \cline{9-14} 
(WAFT)      & true & est.  & se. & sde. & rb. & cov. & & true & est.   & se.   & sde.  & rb.  & cov.  \\ \hline
$\beta_1$ & -4    &  -4.4774 & 0.6129 & 0.7478 & -11.9343 & 0.8480 & 
          & 2 &   2.0368 & 0.1287 & 0.3209 & 1.8376 & 0.6596\\
$\beta_2$ &  2   &   2.2617 & 0.5275 & 1.8072 & 13.0832 & 0.9130     &  
          &  -1  & -1.0047 & 126.6637 & 0.5935 & -0.4669 & 0.6539   \\ \hline
\end{tabular}

%\begin{tabular}{rrrrrrrrrrrrrr}
%\hline
%Prior 1    &      & \multicolumn{5}{l}{BPPH} &  & \multicolumn{5}{l}{BPAFT$^{c}$} \\ \cline{3-7} \cline{9-14} 
%      & true & est.  & se. & sde. & rb. & cov. & & true & est.   & se.   & sde.  & rb.  & cov.  \\ \hline
%$\beta_1$ & -4    & -3.4030      &  0.4419     & 0.3167      & 14.9300     & 0.7450     & 
%          &  2    &  2.0950      &  0.2026     & 0.2158     &  4.7580  &  0.9248    \\
%$\beta_2$ &  2   & 1.7060      & 0.4556      & 0.4021      & -14.6900     & 0.9200  & 
%          &  -1  &  -1.0490 & 0.2569   & 0.3075  & -4.9360 & 0.9468  \\ \hline
%\end{tabular}

\begin{tabular}{rrrrrrrrrrrrrr}
\hline
Prior 2    &      & \multicolumn{5}{l}{BPPH} &  & \multicolumn{5}{l}{BPAFT} \\ \cline{3-7} \cline{9-14} 
(WAFT)      & true & est.  & se. & sde. & rb. & cov. & & true & est.   & se.   & sde.  & rb.  & cov.  \\ \hline
$\beta_1$ & -4    & -3.9730       &  0.5486       & 0.4770  & 0.6630    &   0.9740  & 
          &  2    &  1.9890     &   0.5050    &  0.4935      & -0.5537     &   0.9660    \\
$\beta_2$ &  2   & 1.9540      & 0.5050      & 0.4854      & -2.2920     & 0.9660  & 
          &  -1  &  -1.0250  &  0.2421    & 0.3042  &  -2.4770  &  0.9250   \\ \hline
\end{tabular}\\
%\begin{tabular}{rrrrrrrrrrrrrr}
%\hline
%Prior 3    &      & \multicolumn{5}{l}{BPPH} %&  & \multicolumn{5}{l}{BPAFT} \\ %\cline{3-7} \cline{9-14} 
%      & true & est.  & se. & sde. & rb. & cov. & & true & est.   & se.   & sde.  & rb.  & cov.  \\ \hline
%$\beta_1$ & -4    & -3.4800       &  0.4519     & 0.3335     & 13.0000    & 0.8250     & 
%          &  2    &  2.1060      & 0.2058      &  0.2211     &   5.3111   &   0.9190    \\
%$\beta_2$ &  2   & 1.7440      & 0.4590      & 0.4124     & -12.8100     & 0.9320  & 
%          &  -1  &  -1.0620  &  0.2640    & 0.4569  &  6.2430  &  0.9420   \\ \hline
%\end{tabular}

%\begin{tabular}{rrrrrrrrrrrrrr}
%\hline
%Prior 4    &      & \multicolumn{5}{l}{BPPH} &  & \multicolumn{5}{l}{BPAFT} \\ \cline{3-7} \cline{9-14} 
%      & true & est.  & se. & sde. & rb. & cov. & & true & est.   & se.   & sde.  & rb.  & cov.  \\ \hline
%$\beta_1$ & -4    & -4.1230      & 0.5730      & 0.5326     & -3.0750    & 0.9670     & 
 %         &  2    &  2.0590      &   0.1957    &  0.2223     &  2.9660    &  0.9160     \\
%$\beta_2$ &  2   & 2.0630      & 0.5144      & 0.5251      & 3.1550     & 0.9620  & 
 %         &  -1  & -1.0360   &   0.2484   & 0.4540  &  -3.5970  &   0.9170  \\ \hline \\ 
%\end{tabular}

\begin{tabular}{rrrrrrrrrrrrrr}
\hline
LBFGS     &      & \multicolumn{5}{l}{BPPO$^{b}$} &  & \multicolumn{5}{l}{BPAFT$^{c}$} \\ \cline{3-7} \cline{9-14} 
(LLAFT)      & true & est.  & se. & sde. & rb. & cov. & & true & est.   & se.   & sde.  & rb.  & cov.  \\ \hline
$\beta_1$ & -4    & -4.3523 & 0.6417 & 0.7813 & -8.8071 & 0.8548  & 
          & 2 & 2.0345 & 0.1586 & 0.3123 & 1.7269 & 0.6897 \\
$\beta_2$ &  2   & 2.2170 & 0.6935 & 0.7646 & 10.8513 & 0.9330 &  
          &  -1  &  -1.0222 & 0.2150 & 0.4025 & -2.2193 & 0.6886   \\ \hline
\end{tabular}

%\begin{tabular}{rrrrrrrrrrrrrr}
%\hline
%Prior 1    &      & \multicolumn{5}{l}{BPPO} &  & \multicolumn{5}{l}{BPAFT$^{e}$} \\ \cline{3-7} \cline{9-14} 
%      & true & est.  & se. & sde. & rb. & cov. & & true & est.   & se.   & sde.  & rb.  & cov.  \\ \hline
%$\beta_1$ & -4    & -3.6318 & 0.5333 & 0.4218 & 9.2040 & 0.9080     & 
%          &  2    &  2.0824 & 0.2432 & 0.2513 & 4.1185 & 0.9439    \\
%$\beta_2$ &  2   & 1.8568 & 0.6337 & 0.5762 & -7.1611 & 0.9570 & 
%          &  -1  &  -1.0680 & 0.3384 & 0.3506 & -6.8042 & 0.9429   \\ \hline
%\end{tabular}

\begin{tabular}{rrrrrrrrrrrrrr}
\hline
Prior 2    &      & \multicolumn{5}{l}{BPPO} &  & \multicolumn{5}{l}{BPAFT$^{d}$} \\ \cline{3-7} \cline{9-14} 
(LLAFT)      & true & est.  & se. & sde. & rb. & cov. & & true & est.   & se.   & sde.  & rb.  & cov.  \\ \hline
$\beta_1$ & -4    & -4.0437 & 0.6195 & 0.5562 & -1.0925 & 0.9600    & 
          &  2    &  2.0441 & 0.2321 & 0.2538 & 2.2041 & 0.9189      \\
$\beta_2$ &  2   & 2.0677 & 0.6770 & 0.6631 & 3.3839 & 0.9560  & 
          &  -1  &  -1.0426 & 0.3237 & 0.3453 & -4.2609 & 0.9319   \\ \hline \\
\end{tabular}
%\begin{tabular}{rrrrrrrrrrrrrr}
%\hline
%Prior 3    &      & \multicolumn{5}{l}{BPPO} &  & \multicolumn{5}{l}{BPAFT} \\ \cline{3-7} \cline{9-14} 
%      & true & est.  & se. & sde. & rb. & cov. & & true & est.   & se.   & sde.  & rb.  & cov.  \\ \hline
%$\beta_1$ & -4    & -3.7566 & 0.5533 & 0.4531 & 6.0854 & 0.9400     & 
%          &  2    &  2.0974 & 0.2469 & 0.2581 & 4.8714 & 0.9419     \\
%$\beta_2$ &  2   & 1.9204 & 0.6454 & 0.6005 & -3.9798 & 0.9600  & 
%          &  -1  &  -1.0752 & 0.3413 & 0.3550 & -7.5207 & 0.9449    \\ \hline
%\end{tabular}

%\begin{tabular}{rrrrrrrrrrrrrr}
% \hline
%Prior 4    &      & \multicolumn{5}{l}{BPPO} &  & \multicolumn{5}{l}{BPAFT} \\ \cline{3-7} \cline{9-14} 
%      & true & est.  & se. & sde. & rb. & cov. & & true & est.   & se.   & sde.  & rb.  & cov.  \\ \hline
%$\beta_1$ & -4    & -4.2490 & 0.6599 & 0.6346 & -6.2248 & 0.9509     & 
%          &  2    &  2.0576 & 0.2361 & 0.2583 & 2.8822 & 0.9200   \\
%$\beta_2$ &  2   & 2.1733 & 0.6980 & 0.7103 & 8.6649 & 0.9469 & 
%          &  -1  &  -1.0503 & 0.3268 & 0.3498 & -5.0311 & 0.9300    \\
%\end{tabular}
\caption{MC simulation study in scenario I ($n = 100$), models fitted to the WAFT and LLAFT data sets.  Estimate of the regression coefficient (est.), average
standard error (se.), standard deviation of the estimates (sde.), relative bias (rb in \%) and
coverage probability (nominal level 95\%). Symbols: ${a}$ indicates $R=890$ (1 non-finite hessian and 109 non-converging). ${b}$ indicates $R=985$ (15 non-finite Hessian matrices); ${c}$ indicates $R=941$ (59 non-converging), ${d}$ indicates  $R=999$ and  ${f}$ indicates $R=999$.} \label{minisim1} 
\end{table}\normalsize

Apart from the performance of the  ML interval estimation, the BP based models provided good estimates regarding the prior choices that attributed higher uncertainty to the BP parameters, such as  the Prior 2 and Prior 4 (Appendix Table \oldref{sim1}). As mentioned in Chapter 4, this should be justified by the fact that the prior specification and the maximization account for the transformed BP coefficients, for instance  $~\boldsymbol{\hat\gamma^*} = ~\boldsymbol{\hat\gamma}\exp\{\boldsymbol{\hat\beta^*}^\top (\boldsymbol{s}_{x}^{-1} \circ \bar{\boldsymbol{x}})\}$  in the PH class. This implies that the estimates in the standardized case are inflated or deflated according to the argument of the exponential function. For this reason, even a generic weakly informative prior can undermine the ability of the model to find the true value of the parameter. For instance, values near zero are considerably possible to the BP parameters and should often occur as the argument of the exponential function might be negative valued. In summary, when standardized covariates are passed to the BP based models, vague priors should be considered to express the uncertainty about the variability of the BP parameters. Compared to Prior 4, a more precise information about the regression coefficients in Prior 2 has lead, on  average, to lower posterior standard deviations and lower relative bias.\\

Considering Tables \oldref{sim1} and \oldref{sim2}, the average relative bias that accounted for the most considerable difference between estimated and actual values was reported in the application of the BPPH model under the ML approach; the reported average absolute bias was around 12\%. Simultaneously, the shortest distances captured by the relative bias were in the BPPH application under the Prior 2 approach, together with the BPAFT application under the ML approach, with approximate 2\% of average relative bias. Except for the BPPH model under ML, all modeling options provided an average absolute bias near or less than 10\%.

\begin{figure}[!htb]
 \centering
 $$
  \begin{array}{c}
         \mbox{\textbf{(a)} Relative bias} \\
   \includegraphics[width=0.65\textwidth]{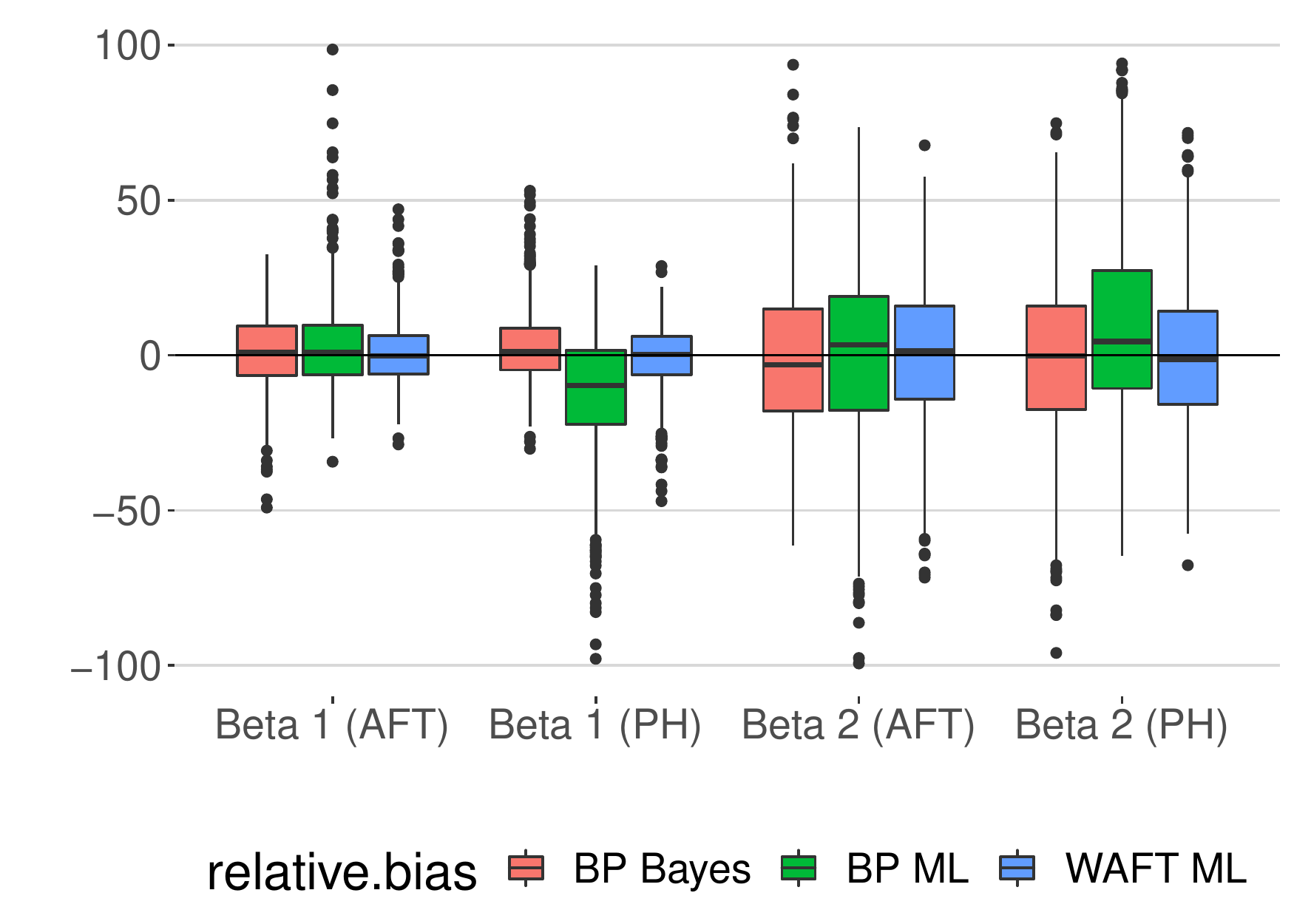} \\
   \mbox{\textbf{(b)} ML/Bayes ratio}\\
     \includegraphics[width=0.65\textwidth]{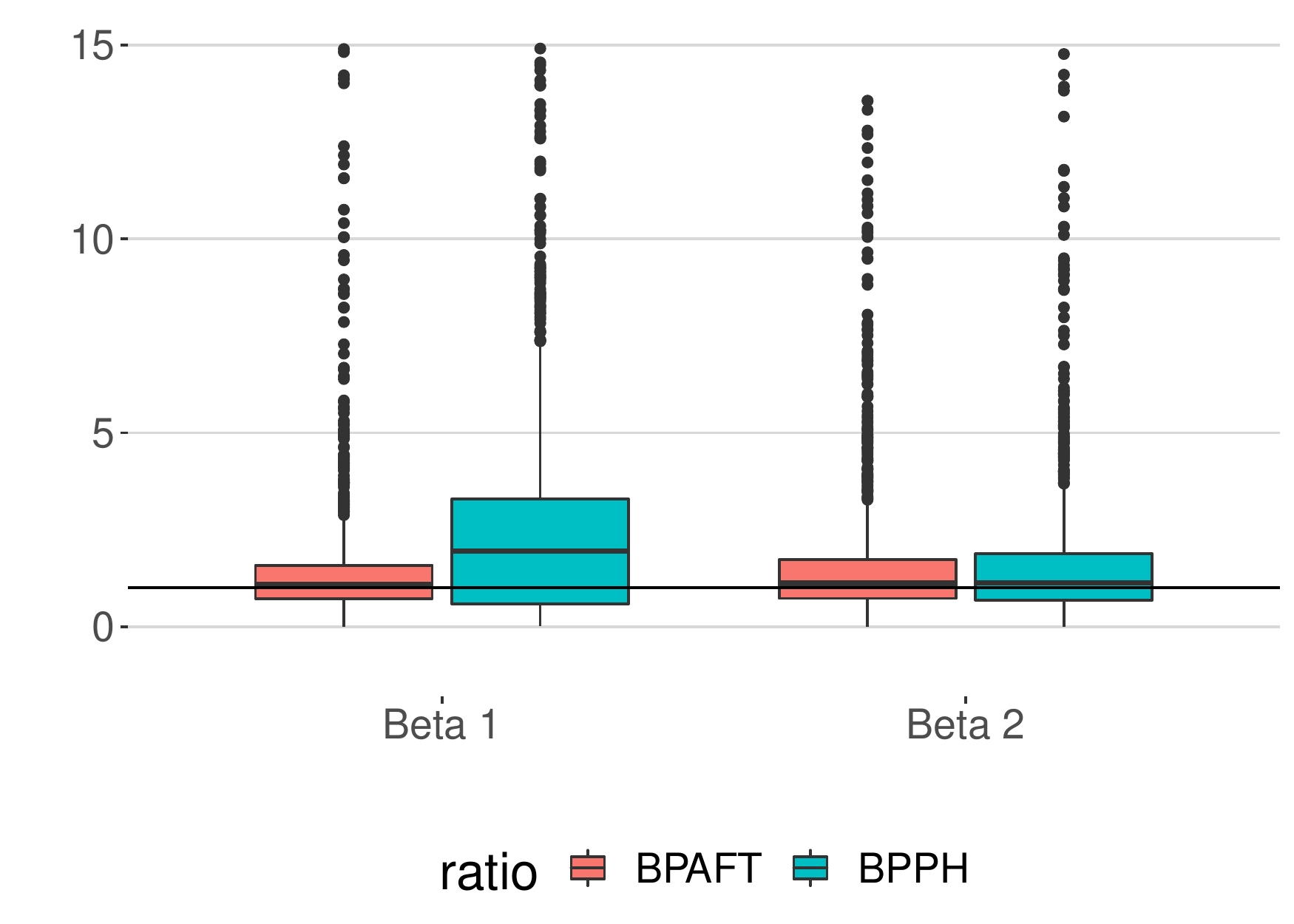} \\
  \end{array}
$$
 \caption{Box-plots of the relative bias and absolute relative bias ratio in Scenario I ($n=100$). Panel \textbf{(a)}: Relative bias of the BP based Bayesian estimates in red, the BP based ML estimates in green and the parametric  ML estimates obtained with the WAFT model in  blue.  Panel \textbf{(b)}: Ratio between the BP based ML absolute relative bias over the  Bayesian relative bias (Prior 2) for the same MC replication; BPAFT in red and BPPH in cyan.} \label{rbias1}
 \end{figure}

\begin{figure}[!htb]
 \centering
 $$
  \begin{array}{c}
     \mbox{\textbf{(a)} Relative bias} \\
     \includegraphics[width=.65\textwidth]{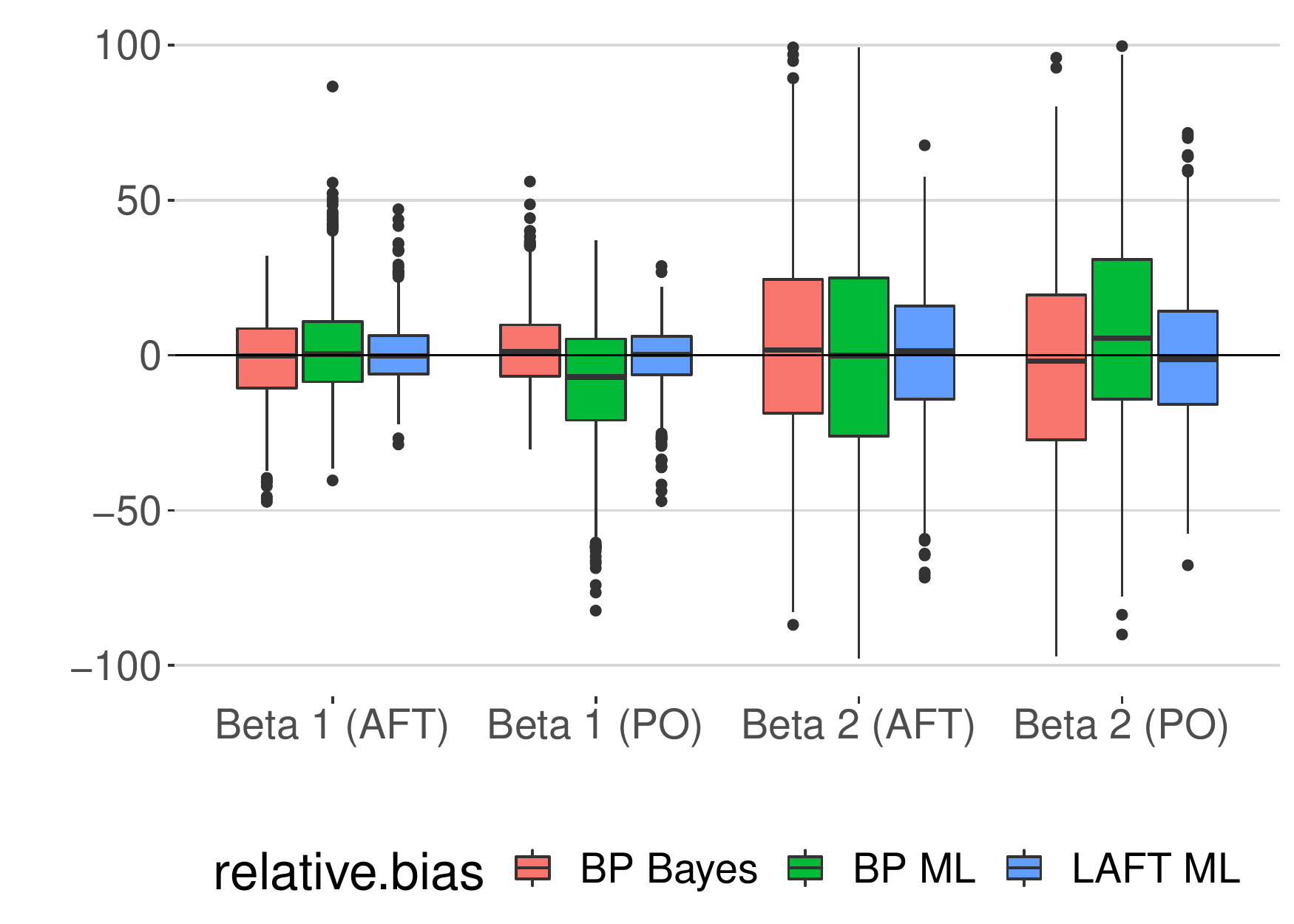} \\
     \mbox{\textbf{(b)} ML/Bayes ratio} \\
    \includegraphics[width=.65\textwidth]{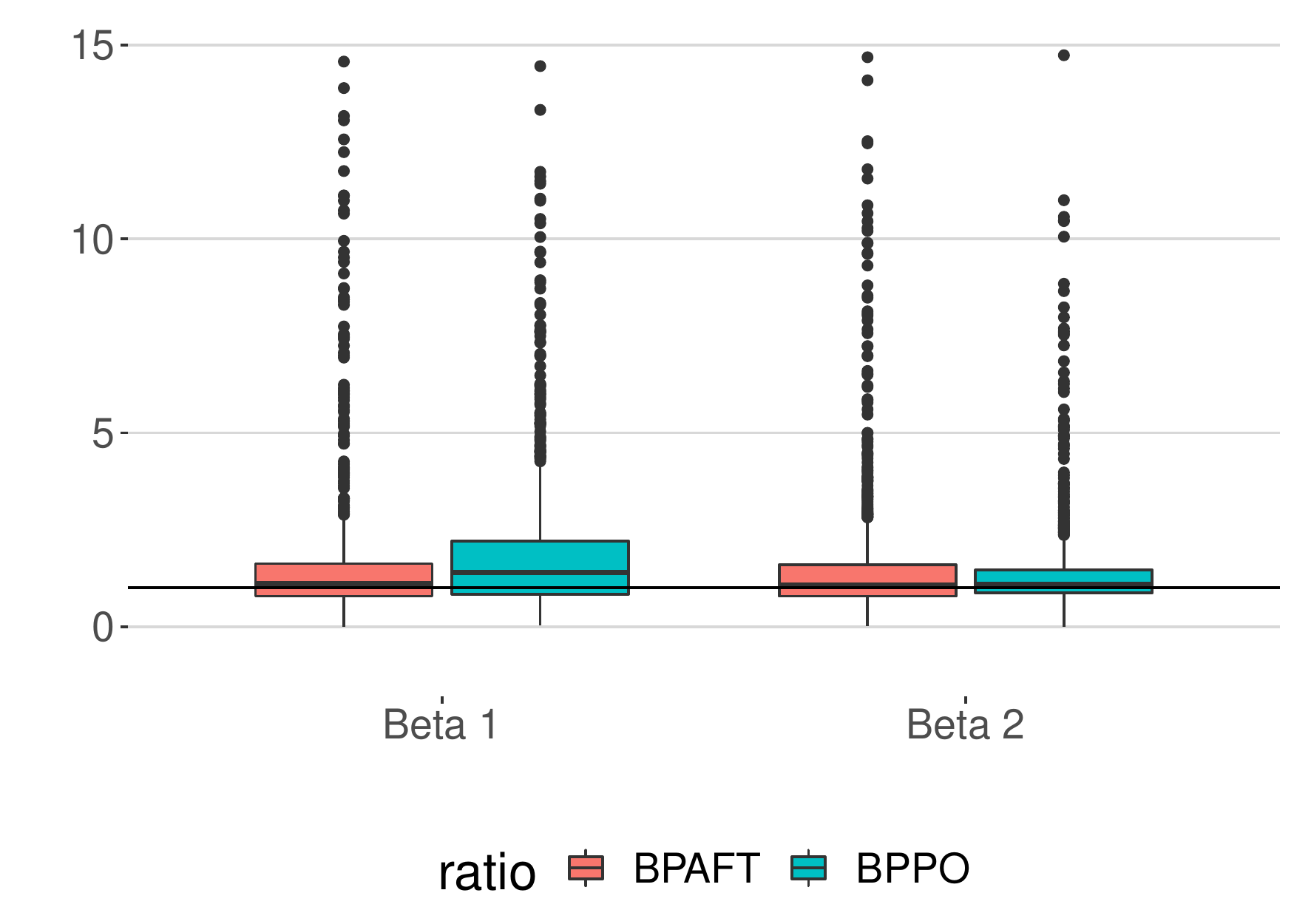}\\
  \end{array}
$$
 \caption{Box-plots of the relative bias and absolute relative bias ratio in Scenario I ($n=100$). Panel \textbf{(a)}: Relative bias of the BP based Bayesian estimates in red, the BP based ML estimates in green and the parametric  ML estimates obtained with the LLAFT model in  blue.  Panel \textbf{(b)}: Ratio between the BP based ML absolute relative bias over the  Bayesian relative bias (Prior 2) for the same MC replication; BPAFT in red and BPPO in cyan.} \label{rbias2}
 \end{figure}

Also, the Figures \oldref{rbias1} and \oldref{rbias2} show the relative bias comparison between the models that provided, in most cases, the adequate results concerning the relative bias. From a Frequentist perspective, we can say that there is no difference in using the BFGS or the LBFGS. Therefore, we chose to stick with the second, as it consists of an extension of the BFGS. On the other hand, the applications under the Prior 2 model presented a lower bias. Thus, those results were chosen to represent Bayesian applications. The panels refer to the illustration of the pairwise estimates that were previously summarized separately (Tables \oldref{sim1} and \oldref{sim2}),  an estimate was only displayed in these graphs if both estimation approaches were valid for the same MC data set. The Panel (a) shows the dispersion of the relative bias for the BP based Bayesian estimates in red, the BP based ML estimates in green, and the parametric  ML estimates obtained with the generator model in blue (either WAFT or LLAFT). The relationship between the parametric models that are under comparison in these figures was previously described in Section 2.4.  It is possible to compare the PH, or the PO,  with the AFT, only by multiplying the negative value of the scale parameter to the regression coefficients estimates provided by the AFT model. In this case, the shape parameter chosen for the two generator models was 2 (see Table \oldref{Table:settings}). The Panel (b) illustrates the dispersion of the ratio between the absolute relative bias of the ML estimates over the Bayesian estimates for the same MC replication. 

It is noteworthy that,  under both approaches, the relative biases for the effect of the categorical variable $\hat \beta_2$ have presented more considerable deviations from the median value when compared to the deviation from the median for the continuous covariate estimated effect $\hat \beta_1$; see the covariates settings in Table \oldref{Table:settings}. This fact was reported previously in the mentioned tables through summary statistics such as the standard deviation of the estimates. Also, it should be noted that, accordingly, the approach that presented the most considerable relative difference from the real values of 12\%, was the BPPH under the ML  approach in Figure 5.2 (a). Moreover, the estimates obtained with the Bayesian approach are quite good in comparison to the ML case. Not least, the absolute relative bias ratio is useful to compare the estimates obtained from the distinct approaches for the MC data set. In this case, we found in Figures 5.2 (b) and 5.3 (b) that the BP Bayesian estimates are, in general, closer to the real value of the parameter than the corresponding ML outcome.  However, yet, on average, they exceed the standard deviations observed in the generator model in Figures 5.2 (a) and 5.3 (a).

 \newpage
 \newpage 
\newpage
 \newpage

In this section, we have found problems regarding BPAFT model confidence interval estimation for both WAFT and LLAFT data sets. Nevertheless, this problem can be solved by implementing specific routines that rely on optimization algorithms that are not provided in the \texttt{rstan} integration, which is part of the proposed \texttt{spsurv} package. Such methods may be considered for future work. However, they are not part of the scope of the present work. Conversely, the optimization of other BP based survival regression models, such as the BPPH and BPPO, proved to be more efficient in terms of interval estimation (coverage probability).

In summary, considering all the tables, figures, and analyses that have been included in this section, we can conclude that for Scenario I, the Bayesian model with Prior 2 specification provided the closest results to the outcomes of the generator model and, consequently, the best results concerning the estimation of the actual values of the parameters. Still, it is expected that, in addition to other eventual prior choices that were not tested in this dissertation, the estimates of this model might be even closer to those from the generator model when large sample sizes are considered. Also, we firmly believe that the prior choice we are proposing in this section (Prior 2) is reasonable weakly informative rather than uninformative because the model covariates have been standardized. The next part of this simulation study presents the results in Scenario II. The average estimate \ref{est.}, the average standard error of the estimates (or posterior deviation) \ref{se.}, the average standard deviation of the estimates \ref{sde.}, the relative bias \ref{rb.}, and the coverage probability, described at the beginning of this section, were also used to evaluate the next section scenario (Scenario II).

\newpage
\section{Scenario II: sample size $n = 200$}

$~~~~$This section presents the results for the Scenario with sample size 200 in the MC scheme. The main aim is to investigate how the survival regression estimates of the BP based models are affected by larger sample size. It is expected that, with twice as many samples, we will find a relevant improvement concerning the BP estimates obtained in the previous Scenario.  The increase from $n = 100$ to $n = 200$ has resulted in a higher computation cost to fit the Bayesian BPAFT model. For this reason, a smaller part of the MC replications was explored here. A total of 100  artificial data sets were evaluated in this case.  Even though these models had already demonstrated an outstanding ability to recover the true values of the parameters in Scenario I (with Prior 2), we expect an improved performance even with fewer replications and longer computing time. 

As a first step of the analysis, an investigation was done to find the causes of the higher computational cost for fitting the BPAFT. Table \oldref{time1}  presents summary statistics about timing on ten posterior samplings of four chains with distinct prior specifications, in Scenario I. The evaluation accounted for the computational time, given in seconds, for each of the twelve Bayesian configurations analyzed earlier. The results were obtained through
the function \texttt{spsurv::spbp} in \texttt{R} and using the same computer (Intel Core i7-4700HQ CPU @ 2.40GHz, 12 Gb of memory RAM and 8 CPU threads). The simulations were not performed in parallel with other activities
on the computer.

\begin{table}[ht]
\centering
\scriptsize
\begin{tabular}{rlrrrrrrr}
  \hline
 & Model & Prior \# & Min. & 1st Qu. & Mean & Median & 3rd Qu. & Max.  \\ 
  \hline
1 & BPPH &Prior 1 & 7.15 & 8.30 & 8.58 & 8.52 & 8.87 & 10.16  \\ 
  2 & BPPH &Prior 2 & 11.59 & 12.15 & 12.56 & 12.35 & 12.88 & 13.78  \\ 
  3 & BPPH &Prior 3 & 6.83 & 8.28 & 9.00 & 8.56 & 9.19 & 11.92  \\ 
  4 & BPPH &Prior 4 & 9.32 & 11.75 & 12.40 & 11.98 & 14.14 & 14.84  \\ \hline  \hline
  5 & BPPO &Prior 1 & 8.08 & 8.18 & 8.50 & 8.37 & 8.54 & 9.45  \\ 
  6 & BPPO &Prior 2 & 10.96 & 11.21 & 12.16 & 11.87 & 12.75 & 13.82  \\ 
  7 & BPPO &Prior 3 & 7.06 & 8.37 & 8.83 & 9.01 & 9.30 & 9.92  \\ 
  8 & BPPO &Prior 4 & 11.22 & 11.47 & 12.54 & 12.34 & 13.25 & 14.84  \\ \hline \hline
  9 & BPAFT &Prior 1 & 168.73& 173.19 & 175.69 & 174.47 & 179.40 & 185.29  \\ 
  10 & BPAFT &Prior 2 & 324.64 & 326.68 & 333.98 & 329.29 & 344.48 & 349.55  \\ 
  11 & BPAFT &Prior 3 & 151.89 & 172.20 & 174.80 & 177.81 & 179.25 & 184.21  \\ 
  12 & BPAFT &Prior 4 & 317.88 & 328.29 & 331.57 & 330.83 & 334.93 & 349.20  \\ 
   \hline \\ 
\end{tabular}
     \caption{Summary statistics for the computational time under distinct prior specifications and survival regression classes in Scenario I. The response summarized here is the computational time (in seconds) to fit ten models in each configuration.}
     \label{time1}
\end{table}\normalsize

According to the reported outcomes, we can conclude that the prior specification directly impacts on the computational time that the NUTS algorithm requires to complete the routine. It should be noted that the prior choices that attributed greater variability to the BP parameters, \textit{i.e.}, Prior 2, and Prior 4, have presented approximately 1.3 to 2 times longer computational times compared to others. Especially, the minimum reported time to run a BPAFT model fit is approximately more than 20 times greater than the BPPH and the BPPO average times; the BPAFT model fit takes about 5 minutes to be completed. The difference between the computation time from the AFT model to the other cases is because the AFT survival regression must compute the new BP basis at each MCMC iteration. In this case, the observed survival time points need to be rescaled in terms of the regression coefficients. In other words, the quantities, $g_{k,m}(y_i \exp\{-\eta_i\})$ and $G_{k,m}(y_i \exp\{-\eta_i\})$, where $\eta_i=\boldsymbol{\beta^\top x_i}$; see the likelihood \oldref{loglikaft}. Table \oldref{time2}  presents summary statistics for the computational time under the Prior 2 specification for the distinct configurations of models and sample sizes.\\
\begin{table}[ht]
\centering
\scriptsize
\begin{tabular}{rrlrrrrrrr}
  \hline
 & & Model & Sample size & Min. & 1st Qu. & Mean & Median & 3rd Qu. & Max.  \\ 
  \hline
\multirow{6}{*}{\texttt{Prior 2}}&1 & BPPH &$n=100$ & 10.84 & 11.47 & 12.23 & 11.66 & 13.35 & 14.53 \\ 
&  2 & BPPH &$n=200$ & 25.10 & 25.28 & 26.30 & 25.97 & 27.09 & 28.50\\ \cline{2-10}  
&  3 & BPPO &$n=100$ & 9.58 & 11.16 & 11.48 & 11.42 & 11.68 & 14.09  \\ 
&  4 & BPPO &$n=200$ & 21.49 & 23.66 & 24.06 & 24.14 & 24.61 & 26.57  \\ \cline{2-10}    
&  5 & BPAFT &$n=100$ & 312.13 & 325.88 & 328.41 & 327.80 & 330.13 & 346.49 \\ 
&  6 & BPAFT &$n=200$ & 1258.25 & 1271.54 & 1318.17 & 1290.30 & 1381.70 & 1396.06 \\ 
  % \hline  \hline
%\multirow{6}{*}{\texttt{LBFGS}}&1 & BPPH &$n=100$ & 0.02 & 0.02 & 0.03 & 0.03 & 0.03 & 0.11 \\ 
%&  2 & BPPH &$n=200$ & 0.06 & 0.06 & 0.08 & 0.07 & 0.08 & 0.20  \\  \cline{2-10}  
%&  3 &BPPO &$n=100$ & 0.02 & 0.03 & 0.03 & 0.03 & 0.03 & 0.05  \\ 
%&  4 & BPPH &$n=200$ & 0.05 & 0.05 & 0.06 & 0.05 & 0.07 & 0.07 \\ \cline{2-10}
%&  5 & BPAFT &$n=100$ & 0.34 & 0.35 & 0.35 & 0.35 & 0.36 & 0.37  \\ 
%&  6 & BPAFT &$n=200$ & 1.79 & 1.79 & 2.03 & 1.80 & 1.84 & 3.01  \\ 
   \hline \\ 
\end{tabular}
     \caption{Summary statistics for the computational times obtained under distinct model classes and both sample size scenarios (Prior 2). The results are reported in seconds, and they correspond to fitting ten models for each configuration.}
     \label{time2}
\end{table}\normalsize

In Table  \oldref{time2}, note that the computational time to fit the BPAFT with $n = 200$ is four times bigger when compared to the $n=100$ case. The inclusion of more data has also doubled the fitting time of the BPPH and BPPO. In Scenario II, the BPPH and BPPO take about 30 seconds to run the routine while a BPAFT spends approximately 23 minutes in the same task. Therefore, an amount of 2000 BPAFT fits, under the MC scheme, would take approximately 46000 minutes, which is equivalent to approximately 32 days to handle the calculations related to the Prior 2 case. In short, we have epirical evidence to conclude that the BPAFT computational cost is directly affected by the sample size.

In Table 5.8, MC simulation study comparison purposes, we have fitted the WAFT and the LLAFT models in Scenario II. As expected, the generator models have provided a good estimation. Furthermore, our findings indicate that they recover the values of the actual regression parameters accurately. Therefore, we can conclude that the ITS was applied correctly in generating the survival times and that the results reported here should be considered as a useful reference to evaluate the BP based survival regression models in Scenario II.\\

\begin{table}[!htb]
\centering
\scriptsize
\begin{tabular}{rrrrrrrrrrrrrr}
\hline
ML     &      & \multicolumn{5}{l}{WAFT} &  & \multicolumn{5}{l}{LLAFT} \\ \cline{3-7} \cline{9-14} 
      & true & est.  & se. & sde. & rb. & cov. & & true & est.   & se.   & sde.  & rb.  & cov.  \\ \hline
$\beta_1$ & 2 & 2.0376 & 0.1285 & 0.1375 & 1.8816 & 0.9400  & 
          & 2 &  2.0251 & 0.1399 & 0.1454 & 1.2575 & 0.9600    \\
$\beta_2$ &  -1  & -1.0097 & 0.1560 & 0.1532 & -0.9711 & 0.9800  &  
          &   -1  & -0.9948  &  0.1514 & 0.1576 &  0.5194   & 0.9379    \\ \hline\\
\end{tabular}
\caption{MC simulation study assuming the WAFT and LLAFT generator models in Scenario II ($n = 200$). Estimate of the regression coefficient (est.), average
standard error (se.), standard deviation of the estimates (sde.), relative bias (rb in \%) and
coverage probability (nominal level 95\%).  The routine used to supply data to this table was \texttt{survreg::survival}, the estimates reported here are obtained under the Maximum Likelihood (ML) estimation.}
\label{ref200}
\end{table}\normalsize

In Table \oldref{sim3}, emphasis should be given to the accuracy of  the point estimates of the BPAFT model. The average of the BPPH estimates under the ML approach have reduced the distance in 0.3\% towards the true value of the coefficients compared to Table \oldref{sim1}. This improvement on the estimation is also reflected in terms of reduction on the average relative bias, of approximately 6\% less bias when compared to Table \oldref{sim1}. In general, the  HPD interval probability coverage improves. On the other hand, the BPAFT standard errors of the estimates were still underestimated, providing unsatisfactory coverage probabilities below the  nominal level (95\%), in Scenario II. From the Bayesian fits, highlight should be given to the Prior 2 and Prior 4 estimates.\\
 \begin{table}[!htb]
 \centering
 \scriptsize
\begin{tabular}{rrrrrrrrrrrrrr}
\hline
BFGS     &      & \multicolumn{5}{l}{BPPO$^{a}$} &  & \multicolumn{5}{l}{BPAFT$^{b}$} \\ \cline{3-7} \cline{9-14} 
      & true & est.  & se. & sde. & rb. & cov. & & true & est.   & se.   & sde.  & rb.  & cov.  \\ \hline
$\beta_1$ & -4 & -4.1910 & 0.4547 & 0.5305 & -4.7745 & 0.9053 & 
          & 2 &  2.0299 & 0.1140 & 0.2251 & 1.4932 & 0.7193     \\
$\beta_2$ &  2   & 2.1112 & 0.4780 & 0.5180 & 5.5586 & 0.9287  &  
          &  -1  &  -1.0119 & 0.1529 & 0.2798 & -1.1884 & 0.6821    \\ \hline
\end{tabular}
 \begin{tabular}{rrrrrrrrrrrrrr}
 \hline
 LBFGS     &      & \multicolumn{5}{l}{BPPO$^{c}$} &  & \multicolumn{5}{l}{BPAFT$^{d}$} \\ \cline{3-7} \cline{9-14} 
       & true & est.  & se. & sde. & rb. & cov. & & true & est.   & se.   & sde.  & rb.  & cov.  \\ \hline
 $\beta_1$ & -4    & -4.1970 & 0.4653 & 0.5236 & -4.9257 & 0.9116   & 
           & 2 &   2.0190     &   0.1081    &    0.2416   &  0.9466    &  0.6978    \\
 $\beta_2$ &  2   & 2.1175 & 0.4810 & 0.5145 & 5.8726 & 0.9311   &  
           &  -1  & -1.0175 & 0.1611 & 0.2848 & -1.7524 & 0.6986   \\ \hline
 \end{tabular}

\begin{tabular}{rrrrrrrrrrrrrr}
\hline
Prior 1    &      & \multicolumn{5}{l}{BPPO} &  & \multicolumn{5}{l}{BPAFT$^{\dagger}$} \\ \cline{3-7} \cline{9-14} 
      & true & est.  & se. & sde. & rb. & cov. & & true & est.   & se.   & sde.  & rb.  & cov.  \\ \hline
$\beta_1$ & -4    & -3.7550       &  0.4047     & 0.3514      & 6.136    & 0.9190     & 
          &  2  & 2.0834 & 0.1404 & 0.1462 & 4.1680 & 0.8900        \\
$\beta_2$ &  2   & 1.9040       & 0.4524      & 0.4346      & -4.7910     & 0.9510  & 
          &  -1  & -1.0450 & 0.1722 & 0.1676 & -4.5027 & 0.9400  \\ \hline
\end{tabular}

 \begin{tabular}{cccccccccccccc}
 \hline
 Prior 2    &      & \multicolumn{5}{l}{BPPO} &  & \multicolumn{5}{l}{BPAFT$^{\dagger}$} \\ \cline{3-7} \cline{9-14} 
       & true & est.  & se. & sde. & rb. & cov. & & true & est.   & se.   & sde.  & rb.  & cov.  \\ \hline
 $\beta_1$ & -4    & -4.0230       & 0.4522      & 0.4726     & -0.5808   & 0.9420     & 
           &  2  & 2.0438 & 0.1315 & 0.1322 & 2.1915 & 0.9254      \\
 $\beta_2$ &  2   & 2.0380      & 0.4745      & 0.4869      & 1.9125     & 0.9420  & 
           &  -1  &  -1.0218 & 0.1566 & 0.1629 & -2.1792 & 0.9254   \\ \hline
 \end{tabular}

\begin{tabular}{rrrrrrrrrrrrrr}
\hline
Prior 3    &      & \multicolumn{5}{l}{BPPO} &  & \multicolumn{5}{l}{BPAFT$^{\dagger}$} \\ \cline{3-7} \cline{9-14} 
      & true & est.  & se. & sde. & rb. & cov. & & true & est.   & se.   & sde.  & rb.  & cov.  \\ \hline
$\beta_1$ & -4    & -3.8260      & 0.4149      & 0.3674     & 4.3410    & 0.9450     & 
          &  2 & 2.0810 & 0.1392 & 0.1354 & 4.0481 & 0.8955    \\
$\beta_2$ &  2   & 1.9410      & 0.4571      & 0.4409      & -2.9430     & 0.9530  & 
          &  -1 & -1.0410 & 0.1688 & 0.1598 & -4.1005 & 0.9552     \\ \hline
\end{tabular}

\begin{tabular}{rrrrrrrrrrrrrr}
\hline
Prior 4    &      & \multicolumn{5}{l}{BPPO} &  & \multicolumn{5}{l}{BPAFT$^{\dagger}$} \\ \cline{3-7} \cline{9-14} 
      & true & est.  & se. & sde. & rb. & cov. & & true & est.   & se.   & sde.  & rb.  & cov.  \\ \hline
$\beta_1$ & -4    & -4.1270       & 0.4676      & 0.5123     & -3.1690    & 0.9350     & 
          &  2  & 2.0486 & 0.1327 & 0.1334 & 2.4297 & 0.8955     \\
$\beta_2$ &  2   & 2.0840     & 0.4811      & 0.5054      & 4.1880
& 0.9350
& 
          &  -1 & -1.0240 & 0.1577 & 0.1634 & -2.3975 & 0.9254  \\ \hline
\end{tabular}
 \caption{MC simulation study in Scenario II ($n = 200$), models fitted to the WAFT data set.  Estimate of the regression coefficient (est.), average
 standard error (se.), standard deviation of the estimates (sde.), relative bias (rb in \%) and
 coverage probability (nominal level 95\%). Symbols: $a$ indicates $R = 996$ (4 non-finite Hessian matrices); $b$ indicates $R = 933$ (60 non-converging and 7 non-finite Hessian matrices); $c$ indicates $R= 999$ (1 non-finite Hessian matrix); $d$ indicates $R = 973$ (2 non-finite Hessian matrices and 124 non-converging); $\dagger$ indicates $R=100$. }\label{sim3}
 \end{table}\normalsize\\

In addition to  the results reported in Table \oldref{sim3}, the  Table \oldref{sim4} provided results related to the BPAFT and BPPO to fit the artificial LLAFT data in Scenario II. Under the ML perspective, the average estimate of the BPPO model fits have reduced in 0.2\% the distance to the true value of the coefficients, this improvement can also be noted through a decrease of approximatily more than 4\% in the absolute relative bias. In few cases, such as the BBPO, we can find that the BFGS performs slightly better than the LBFGS in relation to the relative bias and coverage probability. A total of  166 non-finite Hessian matrices and 286 non-converging estimates were found in the procedure to reach the results in Tables  \oldref{sim3}  and  \oldref{sim4}. In general, the analysis here confirm that some improvement with respect to the relative bias of the BP based survival regression, is obtained when compared to the case $n = 100$ (Scenario I). In the Scenario II, both models presented posterior standard deviations close to the generator model estimated standard error. Table \oldref{minisim2} displays MC simulation study descriptive statistics under the Frequentist and the Bayesian perspectives for both datasets generated.\\
\begin{table}[!htb]
\centering
\scriptsize
%\begin{tabular}{cccccccccccccc}
%\hline
%BFGS     &      & \multicolumn{5}{l}{BPPH $^{a}$} &  & \multicolumn{5}{l}{BPAFT $^{b}$} \\ \cline{3-7} \cline{9-14} 
%      & true & est.  & se. & sde. & rb. & cov. & & true & est.   & se.   & sde.  & rb.  & cov.  \\ \hline
%$\beta_1$ & 4 &--4.2421 & 0.3997 & 0.4858 & -6.0527 & 0.8825    & 
%          & 2 & 1.9864 & 0.0831 & 0.4730 & -0.6778 & 0.6259   \\
%$\beta_2$ &  2  & 2.1033 & 0.3511 & 0.3887 & 5.1650 & 0.9187   &  
%          &  -1  &  -0.9683 & 0.1039 & 0.3079 & 3.1656 & 0.5959 \\ \hline
%\end{tabular}
\begin{tabular}{rrrrrrrrrrrrrr}
\hline
LBFGS     &      & \multicolumn{5}{l}{BPPH $^{a}$} &  & \multicolumn{5}{l}{BPAFT $^{b}$} \\ \cline{3-7} \cline{9-14} 
 (WAFT)     & true & est.  & se. & sde. & rb. & cov. & & true & est.   & se.   & sde.  & rb.  & cov.  \\ \hline
$\beta_1$ & 4 &-4.2426 & 0.4233 & 0.4682 & -6.0661 & 0.9159    & 
          & 2 & 1.9795 & 0.0845 & 0.3934 & -1.0237 & 0.6510   \\
$\beta_2$ &  2  & 2.1033 & 0.3585 & 0.3830 & 5.1656 & 0.9249  &  
          &  -1  &  -0.9736 & 0.1037 & 0.2950 & 2.6371 & 0.6178  \\ \hline
\end{tabular} \\
%\begin{tabular}{rrrrrrrrrrrrrr}
%\hline
%Prior 1    &      & \multicolumn{5}{l}{BPPH} &  & \multicolumn{5}{l}{BPAFT^{\dagger}} \\ \cline{3-7} \cline{9-14} 
%      & true & est.  & se. & sde. & rb. & cov. & & true & est.   & se.   & sde.  & rb.  & cov.  \\ \hline
%$\beta_1$ & -4    & -3.5421 & 0.3284 & 0.2608 & 11.4477 & 0.7200    & 
%          &  2    &  2.0834 & 0.1404 & 0.1462 & 4.1680 & 0.8900    \\
%$\beta_2$ &  2   & 1.7670 & 0.3201 & 0.2892 & -11.6493 & 0.8970  & 
%          &  -1  &  -1.0450 & 0.1722 & 0.1676 & -4.5027 & 0.9400  \\ \hline
%\end{tabular}

\begin{tabular}{rrrrrrrrrrrrrr}
\hline
Prior 2    &      & \multicolumn{5}{l}{BPPH} &  & \multicolumn{5}{l}{BPAFT$^{\dagger}$} \\ \cline{3-7} \cline{9-14} 
(WAFT)      & true & est.  & se. & sde. & rb. & cov. & & true & est.   & se.   & sde.  & rb.  & cov.  \\ \hline
$\beta_1$ & -4    & -3.9692 & 0.3884 & 0.3648 & 0.7699 & 0.9520    & 
          &  2    &  2.044      &   0.1566    &   0.1322    &  2.1920    &    0.9254   \\
$\beta_2$ &  2   & 1.9788 & 0.3459 & 0.3385 & -1.0576 & 0.9480 & 
          &  -1  &  -1.022  &  0.1566    & 0.1629  &  -2.1790  &  0.9254   \\ \hline
\end{tabular}
\begin{tabular}{rrrrrrrrrrrrrr}
\hline
LBFGS     &      & \multicolumn{5}{l}{BPPO$^{c}$} &  & \multicolumn{5}{l}{BPAFT$^{d}$} \\ \cline{3-7} \cline{9-14} 
(LLAFT)      & true & est.  & se. & sde. & rb. & cov. & & true & est.   & se.   & sde.  & rb.  & cov.  \\ \hline
$\beta_1$ & -4    & -4.1970 & 0.4653 & 0.5236 & -4.9257 & 0.9116   & 
          & 2 &   2.0190     &   0.1081    &    0.2416   &  0.9466    &  0.6978    \\
$\beta_2$ &  2   & 2.1175 & 0.4810 & 0.5145 & 5.8726 & 0.9311   &  
          &  -1  & -1.0175 & 0.1611 & 0.2848 & -1.7524 & 0.6986   \\ \hline
\end{tabular}
%\begin{tabular}{rrrrrrrrrrrrrr}
%\hline
%Prior 1    &      & \multicolumn{5}{l}{BPPO} &  & \multicolumn{5}{l}{BPAFT^{\dagger}} \\ \cline{3-7} \cline{9-14} 
%      & true & est.  & se. & sde. & rb. & cov. & & true & est.   & se.   & sde.  & rb.  & cov.  \\ \hline
%$\beta_1$ & -4    & -3.7550       &  0.4047     & 0.3514      & 6.136    & 0.9190     & 
%          &  2  & 2.0834 & 0.1404 & 0.1462 & 4.1680 & 0.8900        \\
%$\beta_2$ &  2   & 1.9040       & 0.4524      & 0.4346      & -4.7910     & 0.9510  & 
%          &  -1  & -1.0450 & 0.1722 & 0.1676 & -4.5027 & 0.9400  \\ \hline
%\end{tabular}

\begin{tabular}{cccccccccccccc}
\hline
Prior 2    &      & \multicolumn{5}{l}{BPPO} &  & \multicolumn{5}{l}{BPAFT$^{\dagger}$} \\ \cline{3-7} \cline{9-14} 
(LLAFT)      & true & est.  & se. & sde. & rb. & cov. & & true & est.   & se.   & sde.  & rb.  & cov.  \\ \hline
$\beta_1$ & -4    & -4.0230       & 0.4522      & 0.4726     & -0.5808   & 0.9420     & 
          &  2  & 2.0438 & 0.1315 & 0.1322 & 2.1915 & 0.9254      \\
$\beta_2$ &  2   & 2.0380      & 0.4745      & 0.4869      & 1.9125     & 0.9420  & 
          &  -1  &  -1.0218 & 0.1566 & 0.1629 & -2.1792 & 0.9254   \\ \hline \\ 
\end{tabular}
%\begin{tabular}{rrrrrrrrrrrrrr}
%\hline
%Prior 3    &      & \multicolumn{5}{l}{BPPO} %&  & \multicolumn{5}{l}{BPAFT^{\dagger}} \\ \cline{3-7} \cline{9-14} 
%      & true & est.  & se. & sde. & rb. & %cov. & & true & est.   & se.   & sde.  & rb.  & cov.  \\ \hline
%$\beta_1$ & -4    & -3.8260      & 0.4149      & 0.3674     & 4.3410    & 0.9450     & 
%          &  2 & 2.0810 & 0.1392 & 0.1354 & 4.0481 & 0.8955    \\
%$\beta_2$ &  2   & 1.9410      & 0.4571      & 0.4409      & -2.9430     & 0.9530  & 
%          &  -1 & -1.0410 & 0.1688 & 0.1598 & -4.1005 & 0.9552     \\ \hline
%\end{tabular}

%\begin{tabular}{rrrrrrrrrrrrrr}
%\hline
%Prior 4    &      & \multicolumn{5}{l}{BPPO} &  & \multicolumn{5}{l}{BPAFT^{\dagger}} \\ \cline{3-7} \cline{9-14} 
%      & true & est.  & se. & sde. & rb. & cov. & & true & est.   & se.   & sde.  & rb.  & cov.  \\ \hline
%$\beta_1$ & -4    & -4.1270       & 0.4676      & 0.5123     & -3.1690    & 0.9350     & 
%          &  2  & 2.0486 & 0.1327 & 0.1334 & 2.4297 & 0.8955     \\
%$\beta_2$ &  2   & 2.0840     & 0.4811      & 0.5054      & 4.1880
%& 0.9350
%& 
%          &  -1 & -1.0240 & 0.1577 & 0.1634 & -2.3975 & 0.9254  \\ \hline
%\end{tabular}
\caption{MC simulation study in Scenario II ($n = 200$), models fitted to the WAFT data sets.  Estimate of the regression coefficient (est.), average
standard error (se.), standard deviation of the estimates (sde.), relative bias (rb in \%) and
coverage probability (nominal level 95\%). Symbols: ${a}$ indicates $R = 996$ (4 non-finite Hessian matrices); ${b}$ indicates $R = 933$ (7 non-finite Hessian matrices and 60 non-converging); ${a}$ indicates $R = 999$ (1 non-finite hessian); ${b}$ indicates  $R= 973$ (2 non-finite Hessian matrices and 124 non-converging); $c$ indicates $R= 973$ (27 non-converging); $d$ indicates $R = 969$ (51 non-converging); $\dagger$ indicates $R=100$.}\label{minisim2}
\end{table}\normalsize\\

Considering Tables \oldref{sim3} and \oldref{sim4}, the average relative bias associated with the most considerable difference to the actual values of the parameters was reported for the BPPH model under the ML approach. The reported average absolute bias was approximately  $6\%$. Simultaneously, the shortest distance concerning the actual values was obtained in the BPPH under the Prior 2 approach, with less than 2\% of average relative bias. All the modeling options provided an average absolute bias around or less than 6\%. As mentioned in the previous section, the quality of the interval estimates might have been influenced by optimization biases (BFGS or LBFGS algorithms) or due to the rough approximation provided by the Delta method. We have found problems in the interval estimation of the BPAFT model applied to both WAFT and LLAFT data sets. Again, possible solutions to this problem are not part of the scope of this work. The estimates of other BP based survival regression models, such as the BPPH and BPPO, indicated good results in terms of coverage probability. These BP cases proved to be more reliable in terms of interval estimation for both scenarios. 

\begin{figure}[!htb]
 \centering
 $$
  \begin{array}{cc}
   \mbox{\textbf{(a)} BPAFT}&
   \mbox{\textbf{(b)} BPPO}\\
     \includegraphics[width=0.5\textwidth]{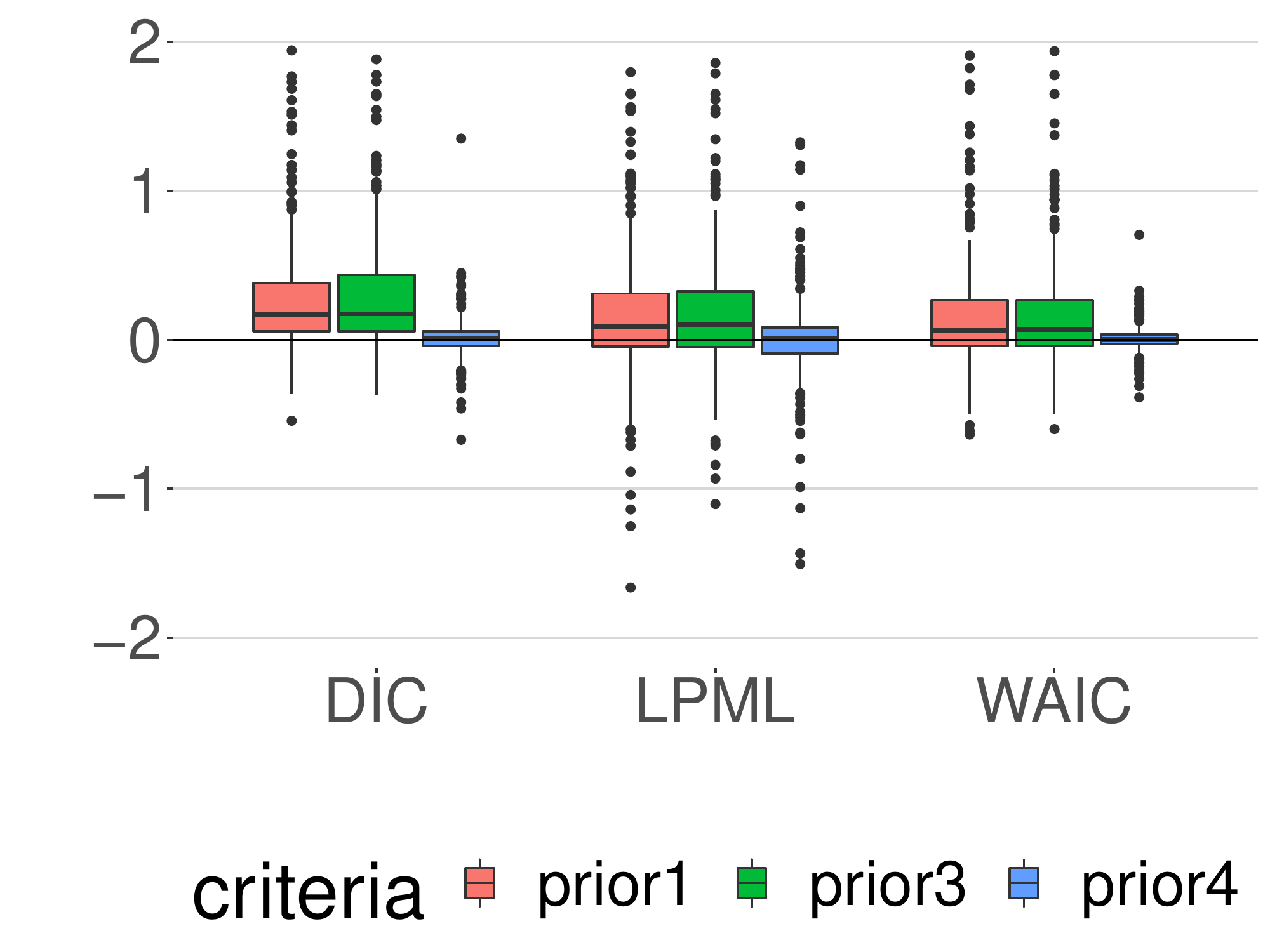}&
    \includegraphics[width=0.5\textwidth]{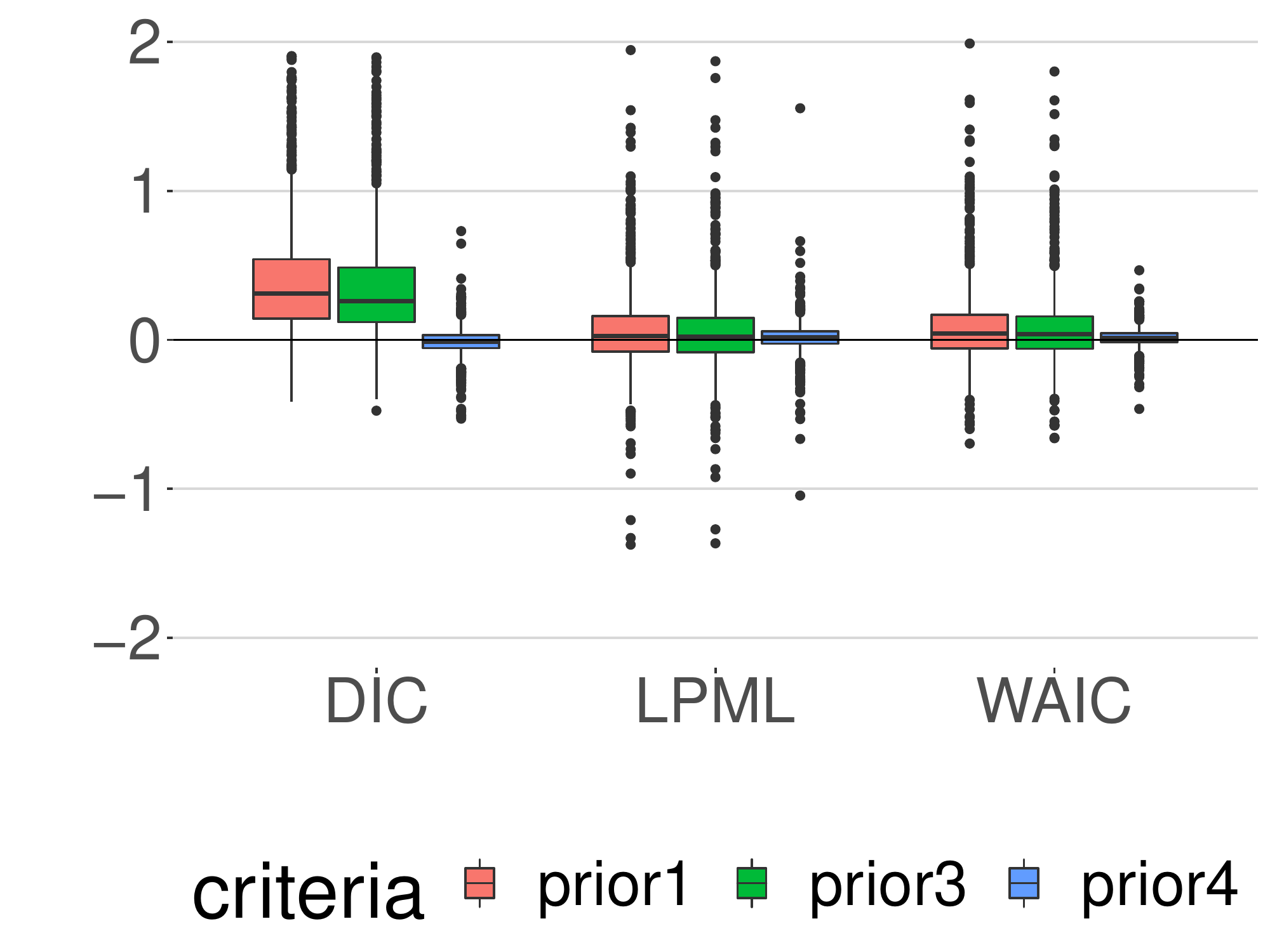} \\
       \mbox{\textbf{(c)} BPAFT}&
       \mbox{\textbf{(d)} BPPO}\\
     \includegraphics[width=0.5\textwidth]{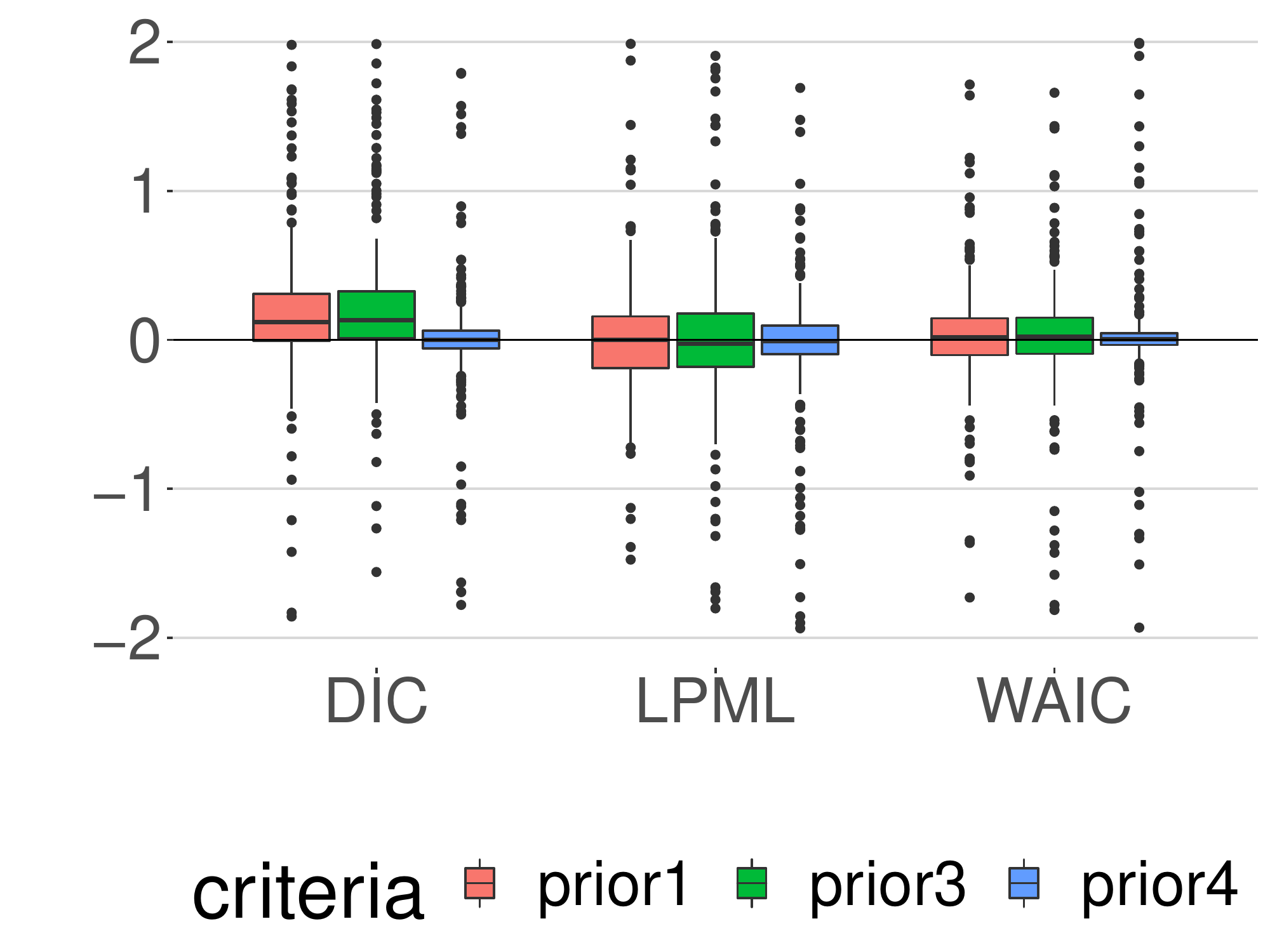}&
    \includegraphics[width=0.5\textwidth]{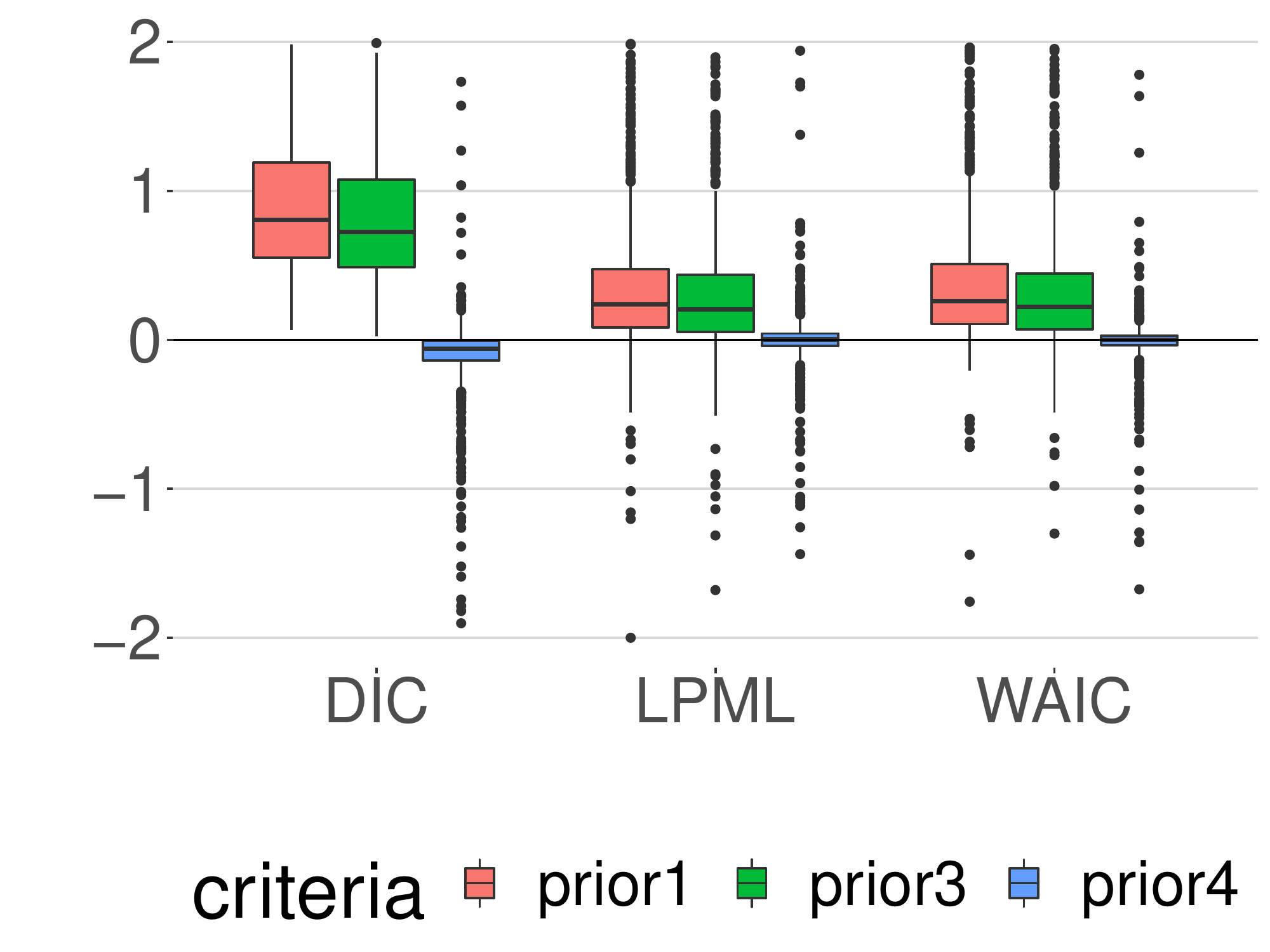} \\
  \end{array}
$$
 \caption{Box-plots of the relative differences between the values of the DIC, -2 WAIC and -2 LPML criteria from two models. The first model is the one identified by prior number 1, 2 or 3.
The reference (second) model is the one assuming Prior 2. Here, the models were fitted to data sets generated through Scenario II ($n = 200$). Panel \textbf{(a)}: Relative difference regarding the BPAFT model (LLAFT case). Panel \textbf{(b)}: Relative difference regarding the BPPO model. Panel \textbf{(c)}: Relative difference regarding the BPAFT model (WAFT case). Panel \textbf{(d)}: Relative difference regarding the BPPH model.} \label{criteria2}
 \end{figure}

\begin{figure}[!htb]
 \centering
 $$
  \begin{array}{c}
         \mbox{\textbf{(a)} Relative bias} \\
   \includegraphics[width=0.65\textwidth]{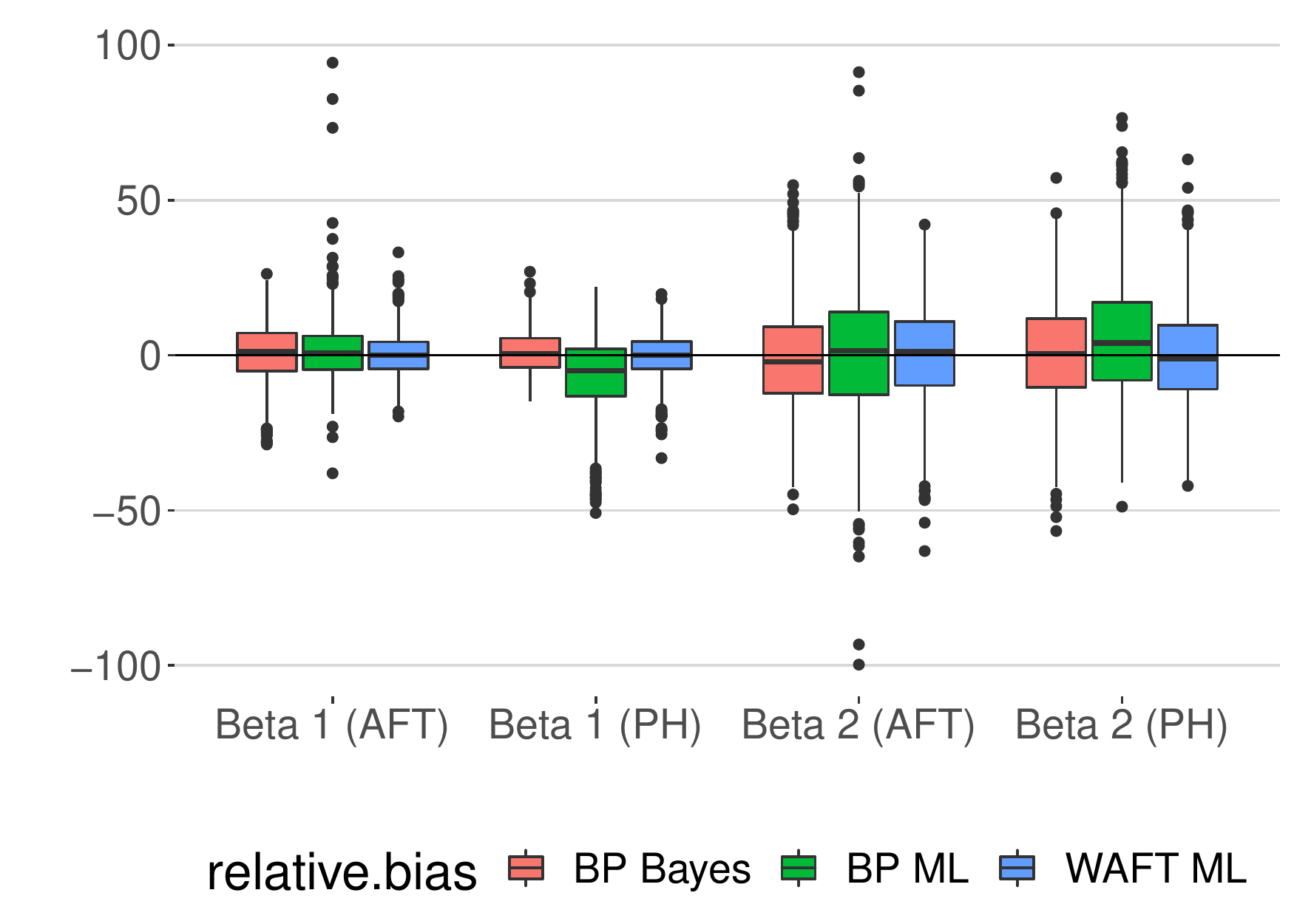} \\
   \mbox{\textbf{(b)} ML/Bayes ratio}\\
     \includegraphics[width=0.65\textwidth]{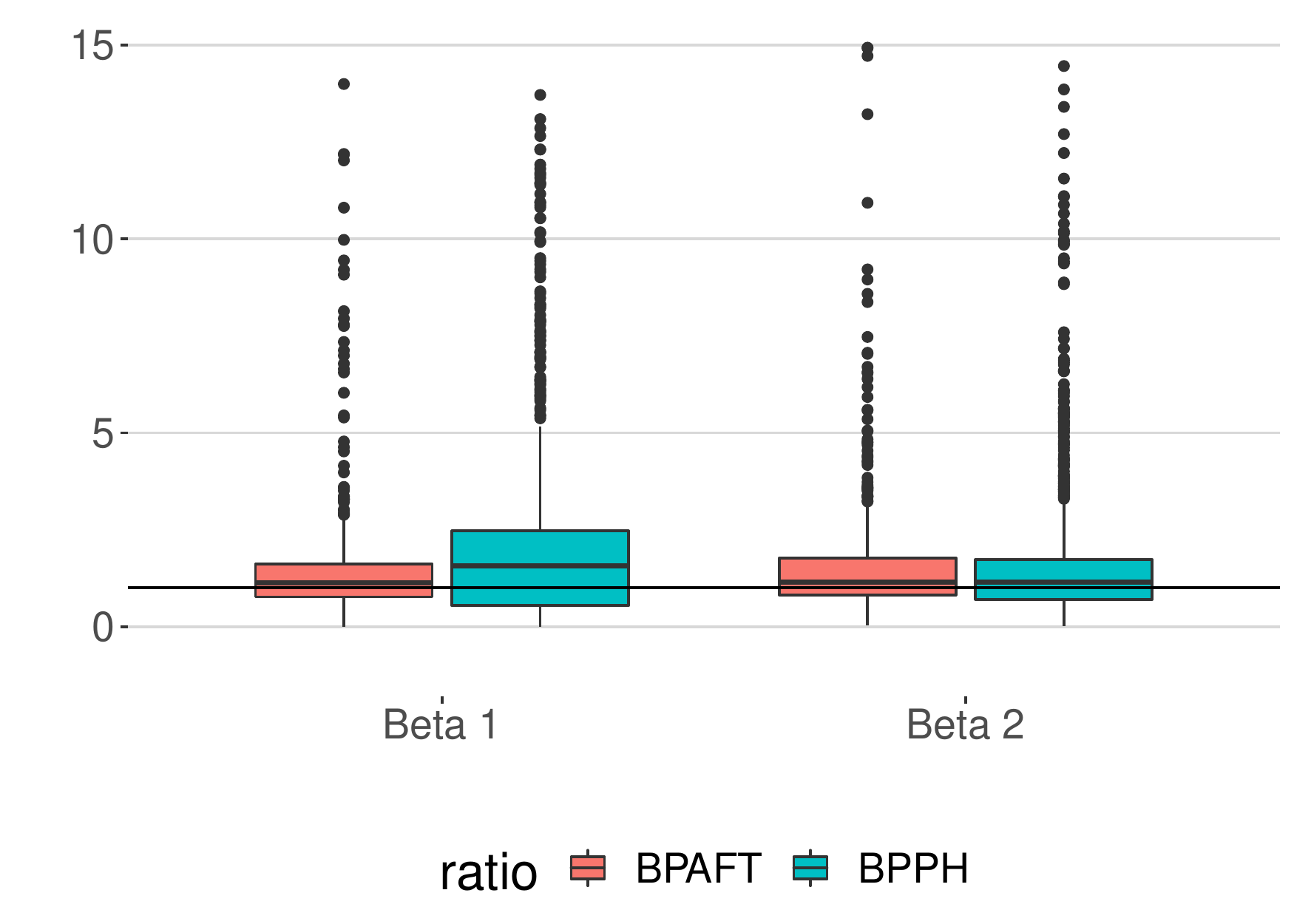} \\
  \end{array}
$$
 \caption{Box-plots of the relative bias and absolute relative bias ratio in Scenario II ($n=200$). Panel \textbf{(a)}: Relative bias of the BP based Bayesian estimates in red, the BP based ML estimates in green and the parametric  ML estimates obtained with the WAFT model in  blue.  Panel \textbf{(b)}: Ratio between the BP based ML absolute relative bias over the  Bayesian relative bias (Prior 2) for the same MC replication; BPAFT in red and BPPH in cyan.} \label{rbias3}
 \end{figure}

\begin{figure}[!htb]
 \centering
 $$
  \begin{array}{c}
     \mbox{\textbf{(a)} Relative bias} \\
     \includegraphics[width=.65\textwidth]{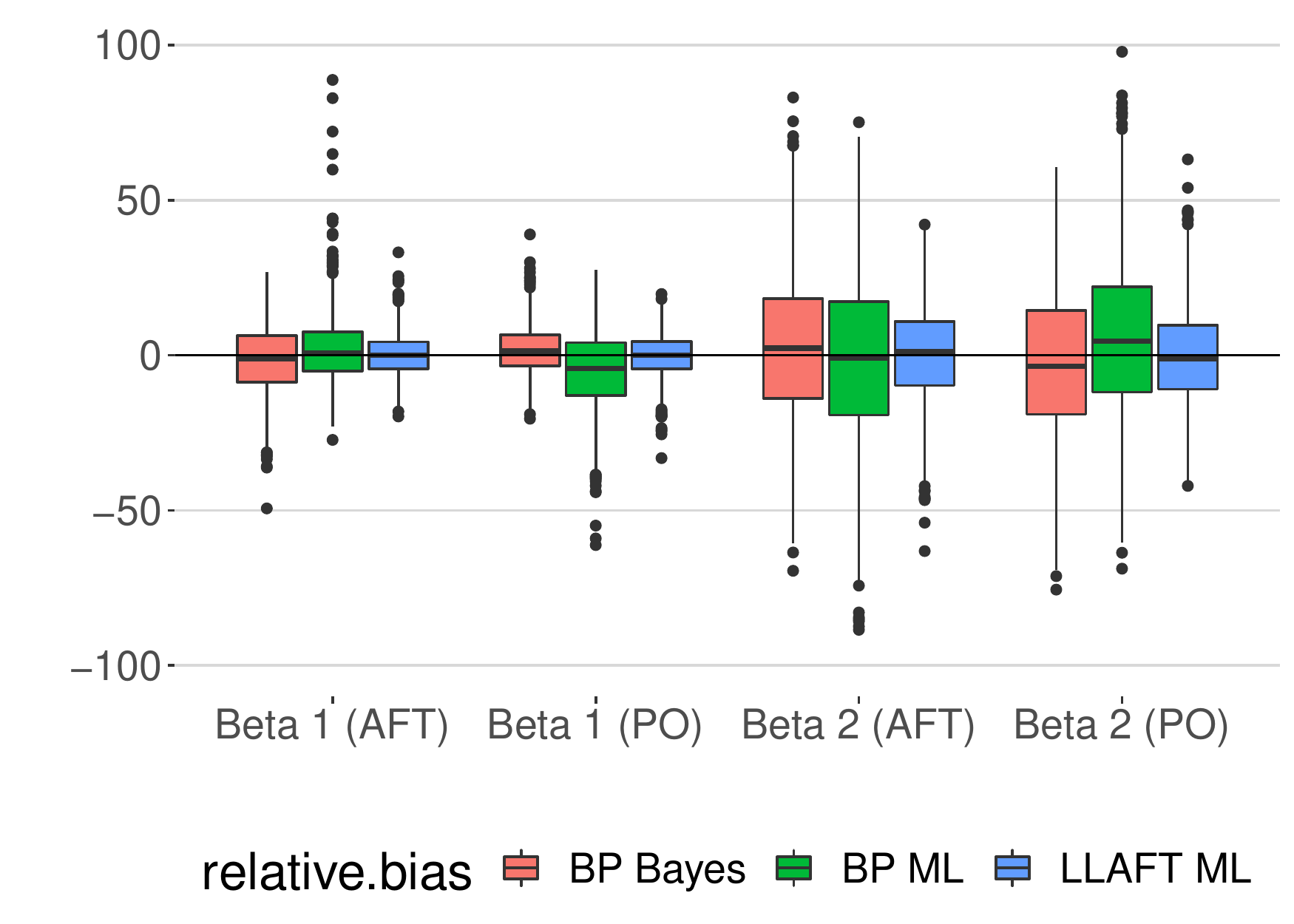} \\
     \mbox{\textbf{(b)} ML/Bayes ratio} \\
    \includegraphics[width=.65\textwidth]{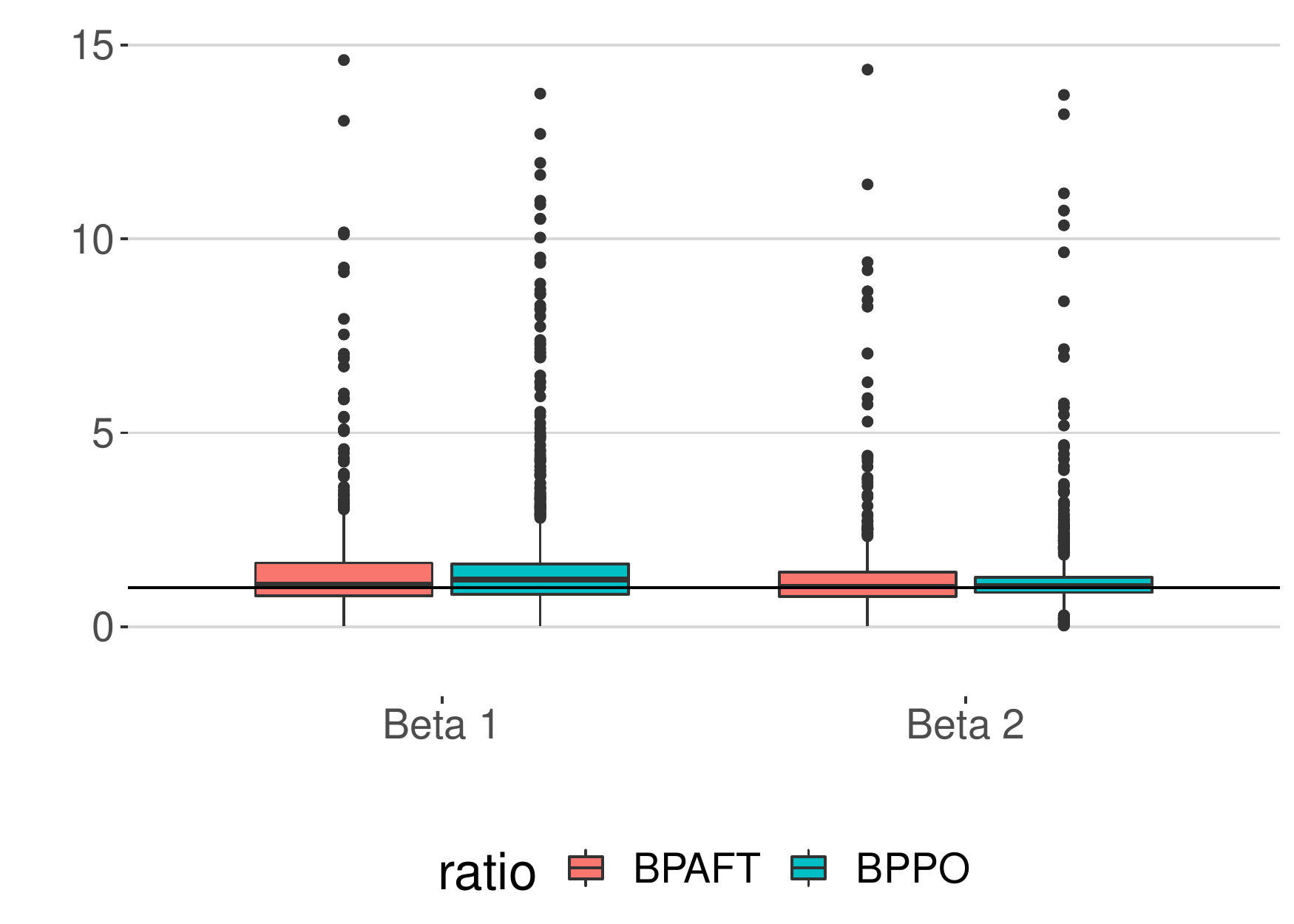}\\
  \end{array}
$$
 \caption{Box-plots of the relative bias and absolute relative bias ratio in Scenario II ($n=200$). Panel \textbf{(a)}: Relative bias of the BP based Bayesian estimates in red, the BP based ML estimates in green and the parametric  ML estimates obtained with the LLAFT model in  blue.  Panel \textbf{(b)}: Ratio between the BP based ML absolute relative bias over the  Bayesian relative bias (Prior 2) for the same MC replication; BPAFT in red and BPPH in cyan.} \label{rbias4}
 \end{figure}

Model comparison criteria (DIC, WAIC, and LPML) were also calculated to assist in the evaluation of the prior choice in Scenario II. It is also expected that the Prior 2 outperforms the other models to the above criteria because it was the model that presented the smallest relative bias. In this context, the Figures \oldref{criteria2} illustrates the relative difference between a reported prior choice criterion and the value of the same criterion regarding Prior 2, for the same MC replication. The analysis here confirms the conclusions based on the same model comparison criteria developed for the Scenario I. We have also found that the DIC often points to the Prior 4 case as the most appropriate model in BPPH, Figure \oldref{criteria2} (d), which is not true. \\
 \\
 \newpage
Figures \oldref{rbias3} and \oldref{rbias4} show the relative bias comparison between the estimates obtained with BP based survival regression models under the ML and Bayesian approaches, \textit{i.e.} LBFGS and Prior 2, respectively. The panels refer to the illustration of the pairwise estimates. The Panel (a) shows the dispersion of the relative bias to the: BP based Bayesian estimates in red, the BP based ML estimates in green, and the parametric  ML estimates obtained with the generator model in blue. The Panel (b) illustrates the dispersion of the ratio between the absolute relative bias of the ML estimates over the Bayesian estimates for the same MC replication. Under both approaches, the corresponding bias for all estimates (Figures \oldref{rbias3} and \oldref{rbias4}) have presented, in general, smaller deviations from the median relative difference compared to Scenario I relative biases. This fact was reported in Tables \oldref{sim3} and \oldref{sim4}. 

Also, it should be noted that, accordingly, the approach having the largest (6\%)  average relative bias from the real values in Tables \oldref{sim3} and \oldref{sim4},  is the  BPPH under the ML perspective (green boxplot) represented in Figure \oldref{rbias1} (a). Moreover, the estimates obtained with the Bayesian approach are quite good in comparison to the ML case. We can also observe that the standard deviations of the estimates are quite similar to the standard deviation observed in the generator model. Besides, even though the gap between the Bayesian and ML estimates have narrowed, we have also found that the BP Bayesian estimates are often closer to the real value of the parameter than the ML outcomes in Figures \oldref{rbias3} (b) and \oldref{rbias4} (b).

We finally reach the end of Chapter 5, which has investigated several statistics summarizing the results of more than 40000 BP based regression model fits, using exclusively the unpublished routines proposed as part of the contribution of this dissertation. The main conclusions from this simulation study can be listed as follows:
\begin{itemize}
    \item the BPAFT presented confidence interval estimation issues, as a consequence, we have observed coverage probabilities below the nominal level,
    
    \item there was no empirical evidence to prefer the LBFGS or BFGS. According to the results related to  relative bias and coverage probability, both algorithms provided very similar outcomes,
    
     \item  the best prior choice in the sensitivity analysis promoted within the simulation study was the Prior 2 (Table \oldref{sensitivity}). This choice indicated the best estimates in all explored scenarios. Also, the model comparison criterion that worse reflected the lowest relative bias was the DIC. 
     
     \item the BPAFT model under the Bayesian approach presented the longest computational time of approximately five minutes for the Scenario I and around 23 minutes for Scenario II. These cases have a costly computational time. The other frameworks take less than 30 seconds to build 4 chains with 1000 samples each.
\end{itemize}
In conclusion, the BP based regression models tend to provide low bias estimates and reliable confidence (except BPAFT) and HPD intervals.
 The newly \texttt{R}  package \texttt{spsurv} serves as an alternative tool to fit survival regression models being freely available for the \texttt{R} the community users. The main objective of the next chapter is to illustrate the use of BP based models to fit well known real data sets in the literature.
  
 \newpage
 \newpage 
\newpage

 \newpage 
\chapter{Real applications}

$~~~~$In this Chapter, we present the analysis of two real case studies. For the first application, consider the reports on male larynx-cancer patients diagnosed during the 1970s \citep{Kardaun:1983}. In the second illustration, consider the 137 randomized observations of two treatments for lung cancer \citep{Prentice:1973}. The first application consists of a comparison between estimates obtained from the WAFT,  CoxPH,  BPPH, and  BPPAFT  models; the last two are available from the \texttt{spsurv} package. In the second illustration,  the assumption of proportional hazards is violated (Appendix F). Thus, the AFT and PO models are alternatives in this case. Following the evidence found in the application performed in \cite{Bennett:1983}, we are also considering that the LLAFT may be adequate here. The relative difference defined in \ref{rb.} account for the distance between the reported \texttt{spsurv}  package estimate and the estimate of some reference routine in this Chapter. The reference estimate $\widehat{\phi}_r$ is obtained with the freely available \texttt{R} packages \texttt{survival} \citep{survival:2000} and   \texttt{timereg} \citep{timereg}, and $\widehat{\phi}$ is the ML or posterior estimate from \texttt{spsurv}. Negative and positive results
indicate that the \texttt{spsurv} is estimating, in percentage, above or below the reference estimate, respectively.  

\section{Application I: laryngeal cancer data}

$~~~~$This study involves 90 patients with laryngeal cancer. The idea here is to compare the survival times of cancer patients hospitalized during the 1970s, in the Netherlands, see \cite{Kardaun:1983}. Following the settings assumed by \cite{Klein:1997}, the two covariates in this study are the patient age (in years) and the cancer stage (I, II, III, or IV). The study aims to investigate whether the patient's lifetime is affected by aging or disease development in more critical stages, such as III and IV. Hypothetically, zero aged patients in phase I refer to the baseline group such that $h(t \mid 0)=h_0(t)$. Tables \oldref{table:mlelarynx} and \oldref{table:belarynx} show the results according to the ML and Bayesian inferential approaches, respectively, for the PH class. \\
 
\begin{table}[!htb]
\centering
\scriptsize
\begin{tabular}{rlrrrcrrr}
  \hline
&\multicolumn{1}{l}{}   & coef & exp(coef) & se(coef) & CI$_{\text{coef}}$ & p$_{\text{coef}}$ & rd$_{\text{coef}}$ & rd$_{\text{exp(coef)}}$\\ 
  \hline
\multirow{4}{*}{\texttt{coxph}} 
 & $X_1:$age & 0.0190 & 1.0192 & 0.0143 & [-0.0089, 0.0470] & 0.1820 & -& -\\ 
 & $X_2:$stage II & 0.1400 & 1.1503 & 0.4625 & [-0.7664, 1.0465] & 0.7620 & -& -\\
 & $X_3:$stage III & 0.6424 & 1.9010 & 0.3561 & [-0.0556, 1.3404] & 0.0712 & -& -\\
 & $X_3:$stage IV & 1.7060 & 5.5068 & 0.4219 & [0.8790, 2.5330] & 0.0001 & -& -\\ 
   \hline
   \hline
\multirow{4}{*}{\shortstack{\texttt{spbp}\\\texttt{(model="ph")}}}  
& $X_1:$age & 0.0193 & 1.0195 & 0.0144 & [-0.0089,0.0474] & 0.18008 & 1.2525 & 0.0294\\ 
& $X_2:$stage II & 0.1720 & 1.1876 & 0.4626 & [-0.7347, 1.0786] & 0.7100 & 22.7894 & 3.1407\\
& $X_3:$stage III & 0.6585 & 1.9318 & 0.3556 & [-0,0386, 1.3555] & 0.0640 & 2.5018 & 1.5944\\ 
& $X_4:$stage IV & 1.7991 & 6.0442 & 0.4288 & [0.9586, 2.6396] & $<$0.0001 & 5.4582 & 8.8912 \\
\hline\\
\end{tabular}
\caption{ML outcome for the CoxPH and the BPPH models applied to the  Laryngeal cancer data set (Application I). Consider $\equiv$ coef: ML point estimate; exp(coef): the estimated HR; se(coef): estimated standard errors of the ML point estimate; CI$_{\text{coef}}$ : estimated $95\%$ confidence intervals; p$_{\text{coef}}$ p-value of the Wald test; rd: percentage relative difference  for the \texttt{survival::coxph} estimates.}
\label{table:mlelarynx}
\end{table}\normalsize

Table \oldref{table:mlelarynx} provides the ML point estimates for the regression effects, estimated HRs,  estimated standard errors, the $95\%$ confidence intervals and the univariate Wald test p-value for the CoxPH and BPPH fits. As expected, based on Table \oldref{table:mlelarynx}, the risk of laryngeal cancer death for stage IV patients is significantly greater in comparison to the Stage I patients in the same age. The estimated HR for the stage IV patients is around $6$ (\texttt{spsurv::spbp} function), which means that the risk of death is about six times higher for patients in stage IV when compared to the same age patients in stage I. Similar conclusions (inferences) about the risk of laryngeal cancer death in other groups of patients (heterogeneous population) regarding the patients treated in the period 1970-1978  at a peripheral hospital in Netherlands \citep{Kardaun:1983}, can be obtained with the \texttt{spsurv} estimates. For example, the estimated  HR \ref{formula:hr} between a 86 years old patient diagnosed with Stage II cancer and a 41 years old patient diagnosed with Stage IV cancer is $\widehat{\text{HR}} =\exp\{(0.0193)(77-41)+(0.1720)(1-0)+(1.7060)(0-1)\} = 35.1675$, which means that the risk of death is  35 times higher for the 86 years old patient. The 95\% interval for $\log \text{HR} = 36\times\beta_1 + \beta_2-\beta_4$ can be calculated to confirm that this result is statistically significant as $\text{CI}_{\log \text{HR}} = [\log \text{HR} \pm 1.96 \times \sqrt{V(36\times\beta_1 + \beta_2-\beta_4)}] = [2.0186;5.1016]$. Despite the remarkable difference of $22\%$ between the Stage II estimated effect provided by \texttt{spsurv::spbp} relative to the \texttt{survival::coxph} estimate, both confidence intervals contains zero, meaning that there is not a significant difference in the risk of death for patients in Stages I and II, according to the outcomes of both packages.

Table \oldref{table:mlelarynxaft} provides the point and interval estimates (similar to Table \oldref{table:mlelarynx}) regarding the estimates of the WAFT and BPAFT models under the ML approach. It should be noted that the relative difference in Table  \oldref{table:mlelarynxaft} reflects the big distance between the two models estimate. The AFT and PH families are different, and the fits, except for the particular case of the Weibull parametric survival model (see Section 2.4), allow different interpretations. In this sense, we found that the BPPH and BPAFT models applied to the Application II allowed different interpretations, favoring the suspicion that the parametric Weibull model is not suitable. According to the simulation study (Chapter 5) findings, the BPAFT models provided very low biased estimates either under Frequentist or Bayesian perspective in Scenario I ($n = 100$). Therefore, the BPAFT might also offer a very low biased estimate for Application I data set ($n=90$). Indeed, the discrepancy between the estimates indicates that the Weibull parametric model is not well fitted to the laryngeal cancer data.  \\

\begin{table}[!htb]
\centering
\scriptsize
\begin{tabular}{rlrrrcrrr}
  \hline
&\multicolumn{1}{l}{}   & coef & exp(coef) & se(coef) & CI$_{\text{coef}}$ & p$_{\text{coef}}$ & rd$_{\text{coef}}$ & rd$_{\text{exp(coef)}}$\\ 
  \hline
\multirow{4}{*}{\shortstack{\texttt{survreg}\\\texttt{("weibull")}}} 
 & Intercept & 3.5288 & 34.0817 & 0.9041 & [1.7567, 5.3008] & $<$0.0001 &  -&- \\
  & Log(scale) & -0.1223 & 0.8849  & 0.1225 & [0.6447,1.1249] & 0.3180 & - & -\\
 & $Z_1:$age & -0.0175 & 0.98267 & 0.0128 & [-0.0425, 0.0076] & 0.1820 & -& -\\ 
 & $Z_2:$stage II & -0.1477 & 0.8627 & 0.4076 & [-0.9465, 0.6511] & 0.7170 & -& -\\
 & $Z_3:$stage III & -0.5866 & 0.5563 & 0.3199 & [-1.2136, 0.0405] & 0.0670 & -& -\\
 & $Z_3:$stage IV & -1.5441 & 0.2135 & 0.3633 & [-2.2561, -0.8320] & 0.0001 & -& -\\ 
   \hline
   \hline
\multirow{4}{*}{\shortstack{\texttt{spbp}\\\texttt{(model="aft")}}}  
& $Z_1:$age & -0.0073 & 0.9927 & 0.0050 & [-0.0172,0.0025] & 0.1445 & 58.0596 & 1.0191\\ 
& $Z_2:$stage II & -0.0905 & 0.9134 & 0.3954 & [-0.8655, 0.6844] & 0.8189 & 38.7039 & 5.8833\\
& $Z_3:$stage III & -1.1058 & 0.3310 & 0.3047 & [-1.7030,-0.5085] & 0.0003 & -88.5155 & -40.5001\\ 
& $Z_4:$stage IV & -2.1106 & 0.1212 & 0.2738 & [-2.6472, -1.5740] & $<$0.0001 & -36.6898 & -43.2502 \\
\hline \\
\end{tabular}\\
\caption{ML outcome for the WAFT and the BPAFT models applied to the  Laryngeal cancer data set (Application I). Consider coef: ML Point Estimate; exp(coef): the estimated TR (or acceleration factor); se(coef): estimated standard errors of the ML point estimate; CI$_{\text{coef}}$ = Estimate $95\%$ confidence intervals; p$_{\text{coef}}$: p-value of the Wald test; rd: percentage relative difference  to \texttt{survival::survreg} estimates.}
\label{table:mlelarynxaft}
\end{table}\normalsize

Before proceeding with the WAFT model estimates interpretation, it is desirable to confirm the suitability of the Weibull distribution to this data set. As discussed in \cite{Colosismo:2001}, graphical methods are widely used for this purpose. A parametric model is well fitted to the data if the distribution of the Cox-Snell residuals follow the a Exponential distribution with mean 1. As a result,  the Cox-Snell
residuals against the cumulative hazard of Cox-Snell residuals is presented should present a straight
line with; see the logarithmic relation in \ref{Formula:AFT}. Considering the censoring times, we use the nonparametric Nelson-Aalen estimator to compare the estimated cumulative hazard of the residuals \citep{Colosismo:2001, lawless2011statistical, qi2009comparison}. 

Figure \oldref{residwei} shows the scatter-plot  of the Cox-Snell
residuals against the cumulative hazard of Cox-Snell residual in Panel (a) and the respective Kaplan-Meier curve compared to the Exponential curve in Panel (b). We can observe that the WAFT model is not well fitted in this case as the Kaplan-Meier estimates do not approximate to the Exponential (reference) survival curve. Despite that, the adequacy of the PH model class was confirmed by the analysis of the Schoenfeld residuals, included in Appendix C. It is important to note that the PH class is adequate. However, the functional form imposed by the Weibull form is not suitable. In turn, distribution-free models are more flexible in this regard. It is expected that, once the assumption is verified, both CoxPH and BPPH are suitable for several baseline formats, especially non-monotonic ones.

%In many cases, AIC and BIC criteria are used to compare non-nested models. However,   AIC and BIC are both criteria likelihood-penalized by the number of parameters in the model. As the number of parameters increases according to the sample size, we know that the BP models are not as parsimonious as other semi-parametric models. In that case, we suggest not to compare other models AIC or BIC.
\begin{figure}[!htb]
\centering
$$
  \begin{array}{cc}
     \mbox{\textbf{(a)} Residuals scatter-plot} & \mbox{\textbf{(b)} Survival curves} \\
     \includegraphics[width=.45\textwidth]{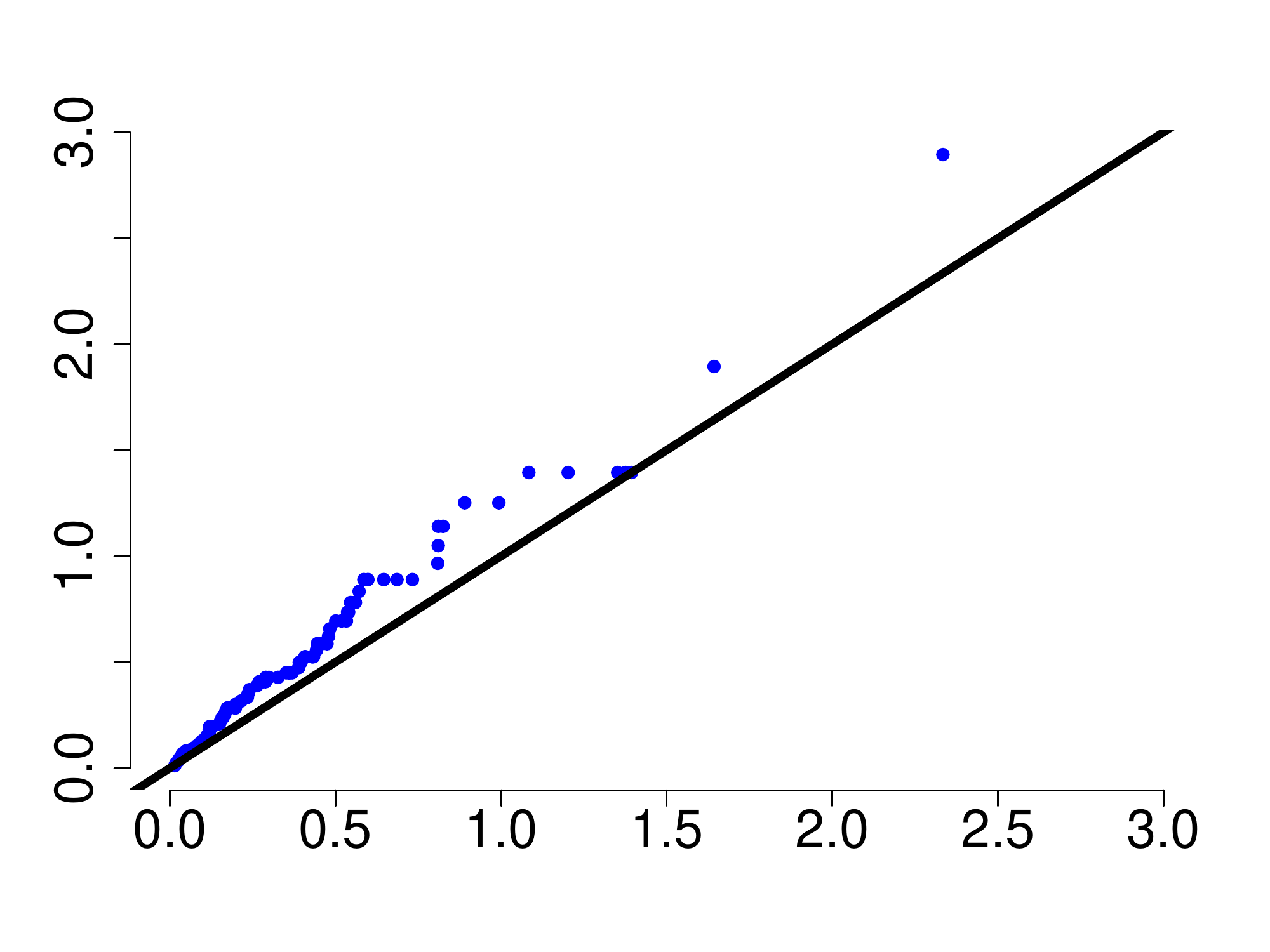} &
    \includegraphics[width=.45\textwidth]{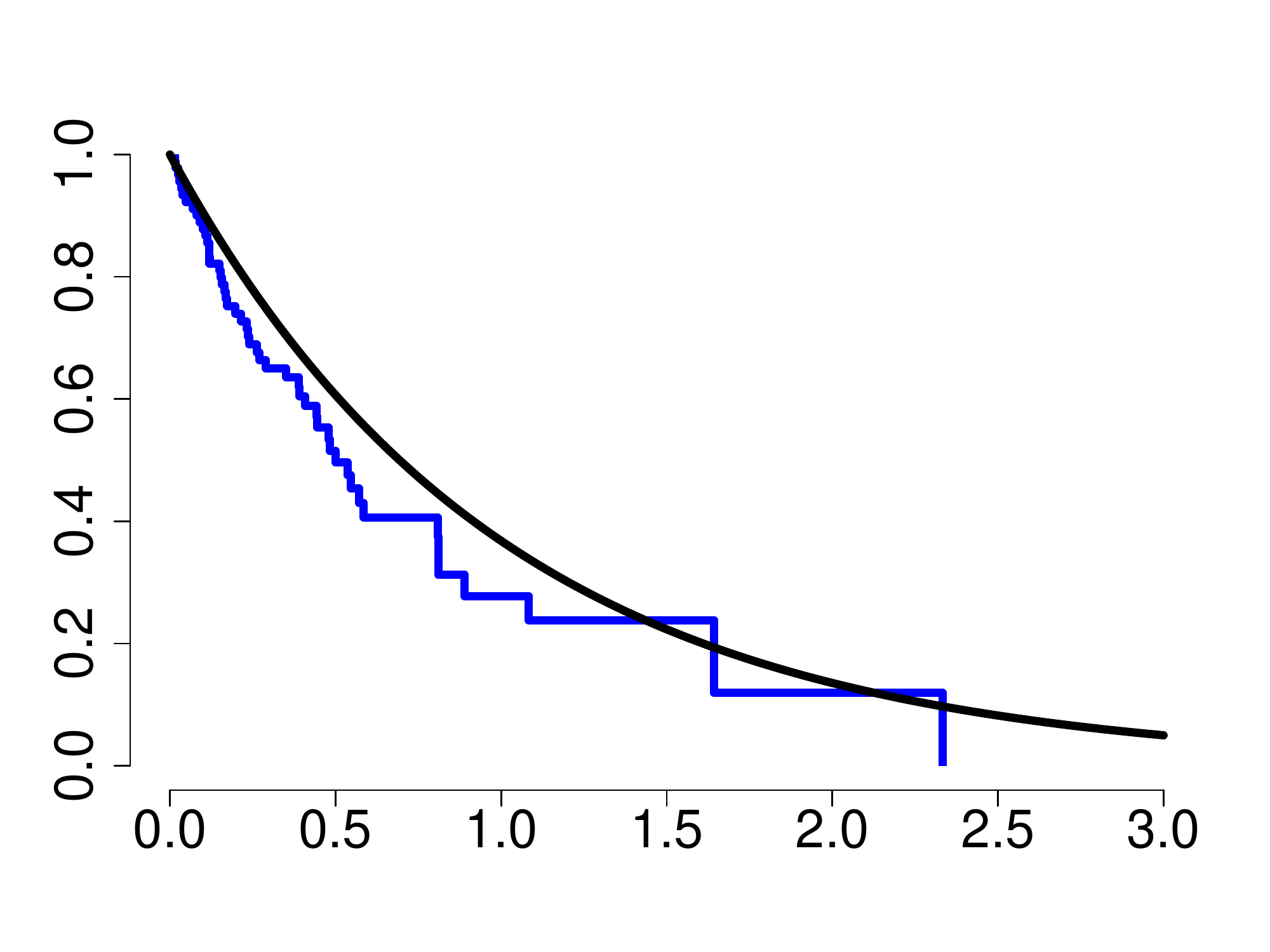}\\
  \end{array}
$$
\caption{\label{residwei} Cox-Snell residuals analysis for the WAFT model in Application I. Panel \textbf{(a)} Scatter-plot of the Cox-Snell residuals against the cumulative hazard (Nelson Aalen) of Cox-Snell residuals; Panel \textbf{(b)} survival curves of the Exponential survival and the Kaplan-Meier estimated step curve for the residuals. In Panel \textbf{(a)}, the (black) solid line refers to the straight identity and the blue points to the paired survival values. In Panel  \textbf{(b)}, the dashed (black) line refers to the Exponential (mean 1) survival curve, and the step (blue) curve to the Kaplan-Meier curve of the residuals.}
\end{figure}
\newpage
The \texttt{spsurv::spbp} package also interfaces to  \texttt{Stan} built-in MCMC algorithms to allow the Bayesian analysis. The prior choice here corresponds to the Prior 2 setting (Table \oldref{sensitivity}) used in the simulation study. The MCMC setup is configured with 4 chains containing 1000 observations (warm-up = 1000 iterations). The summary statistics were calculated based on the 4000 posterior samples for each quantity of interest. The posterior density plots, presented in Appendix E, are dedicated to the diagnostic of convergence of the Bayesian models applied to Application I and II real data sets. In most of the situations described, the posterior densities are unimodal, and the trace plots show chains with good mixing, which suggests a good behavior of the MCMC algorithm.

The prior information for the regression coefficients is chosen based on the idea of small deviations (from zero) given that the covariates are standardized, as explained in Chapters 4 and 5. Table \oldref{table:belarynx} shows the results of the Bayesian BPPH, and the Bayesian BPAFT model fits. Compared to Table \oldref{table:mlelarynx}, the posterior mean for the regression coefficients of the BPPH are relatively close to the Frequentist estimates from the CoxPH model.\\

\begin{table}[!htb]
\centering
\scriptsize
\begin{tabular}{rlrrrcrrr}
  \hline
 & & mean(coef) & exp(mean) & sd(coef)  & HPD & rd$_{\text{mean}}$ & rd$_{\text{exp(coef)}}$ \\ 
\hline
\multirow{4}{*}{\shortstack{\texttt{spsurv::spbp}\\(\texttt{model = "ph"})}} 
& $X_1:$age & 0.0197 & 1.0199 & 0.0141 & [-0.0102,~~0.0457] &3.6618&0.0697\\ 
& $X_2:$stage II & 0.1330  & 1.1422 & 0.4627 & [-0.7877,~~1.0059] &-5.0108&-0.6993\\ 
& $X_3:$stage III & 0.6501 &  1.9157 & 0.3587 & [-0.0566,~~1.3561] &1.1954&0.7709\\ 
& $X_4:$stage IV & 1.7742 & 5.8956 & 0.4358 & [0.9832,~~2.6519] &4.0018&7.0654\\  \hline \hline
\multirow{4}{*}{\shortstack{\texttt{spsurv::spbp}\\(\texttt{model = "aft"})}} 
& $X_1:$age & -0.0142 & 0.9860 & 0.0146  & [-0.0445,~~0.0126] &18.6883&0.3269\\ 
& $X_2:$stage II & -0.1341  & 0.9660 & 0.4420 & [-0.9908,~~0.7450] &9.2104&1.3697\\ 
& $X_3:$stage III & -0.8919 &  0.4500 & 0.4269 & [-1.6594,~~-0.0907] &-52.0565&-26.3128 \\ 
& $X_4:$stage IV & -1.7361 & 0.1980 & 0.4679 & [-2.5804,~~-0.7174] &-12.4362&-17.4713\\  \hline \\
\end{tabular}
\caption{ \label{table:belarynx} Bayesian outcome for the BPPH and the BPAFT models applied to the Laryngeal cancer data set (Application I). Consider $\equiv$ mean(coef): posterior mean ; exp(mean): the estimated HR or the estimated TR (respectively); se(coef): posterior standard errors; HPD: Highest posterior density interval with 95\% probability; rd: percentage of relative difference with respect to \texttt{survival::coxph} estimates or \texttt{survival::survreg} estimates.}
\end{table}\normalsize

The relative differences between the BPPH estimates compared to the CoxPH model are lower than 10\% for every coefficient. Conversely, the Bayesian BPAFT did not provide estimates close to the WAFT model. Here, we also find empirical evidence to state that the difference in the results obtained may be due to an inadequate specification of the WAFT model confirmed by the graphical analysis of the residuals in  Figure \oldref{residwei}. According to Chapter 5 simulation studies,  we have found that the relative bias are very low when the Bayesian model is applied to artificial data sets of sample size $n=100$; for a data set of size $n=90$, it might also provide reasonable estimates.

\begin{figure}[!htb]
\centering
$$
  \begin{array}{c}
    \mbox{\textbf{(a)} Posterior survival} \\
      \includegraphics[width=0.6\textwidth]{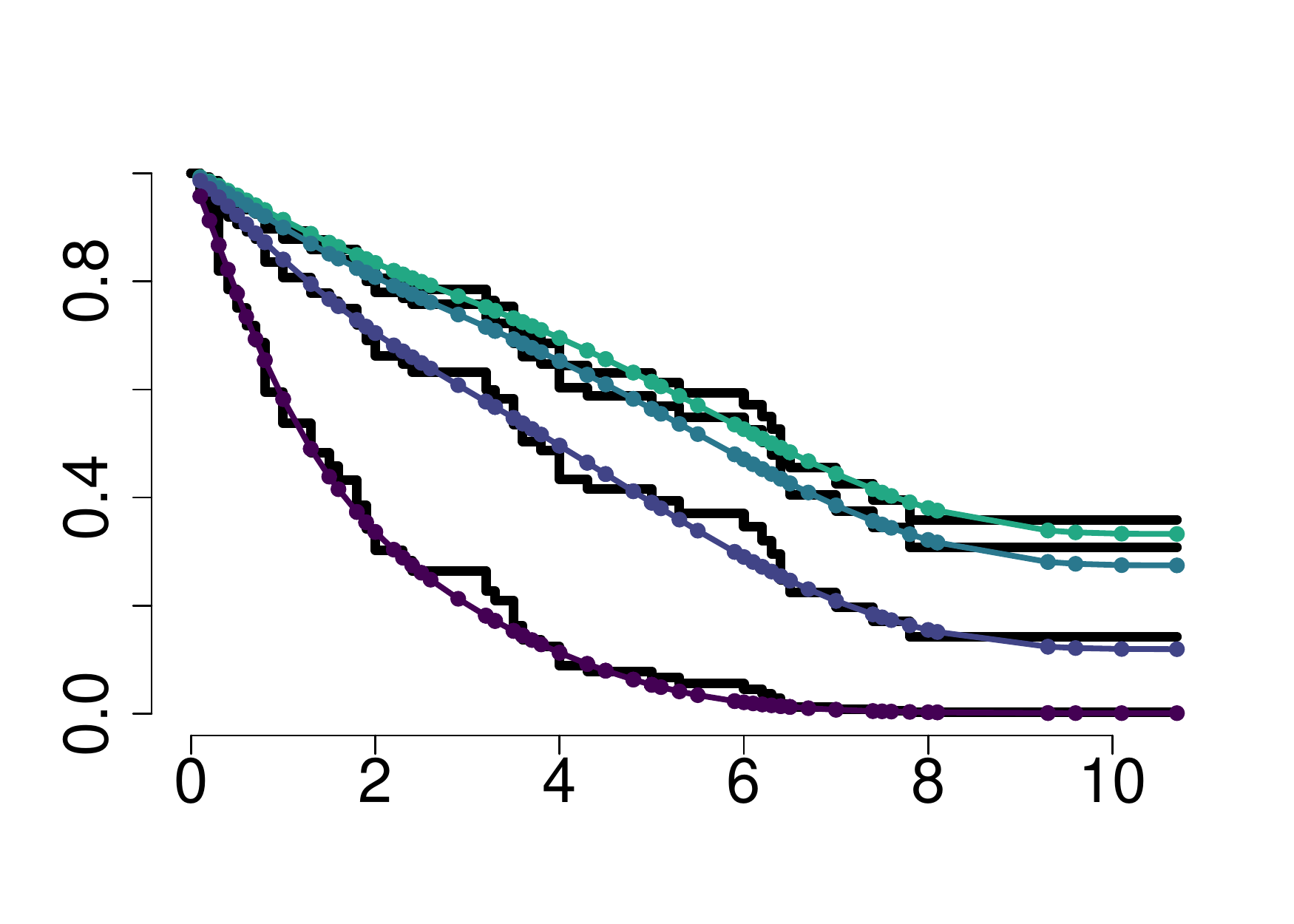}   \\ 
      \mbox{\textbf{(b)} ML estimated survival}  \\ \includegraphics[width=0.6\textwidth]{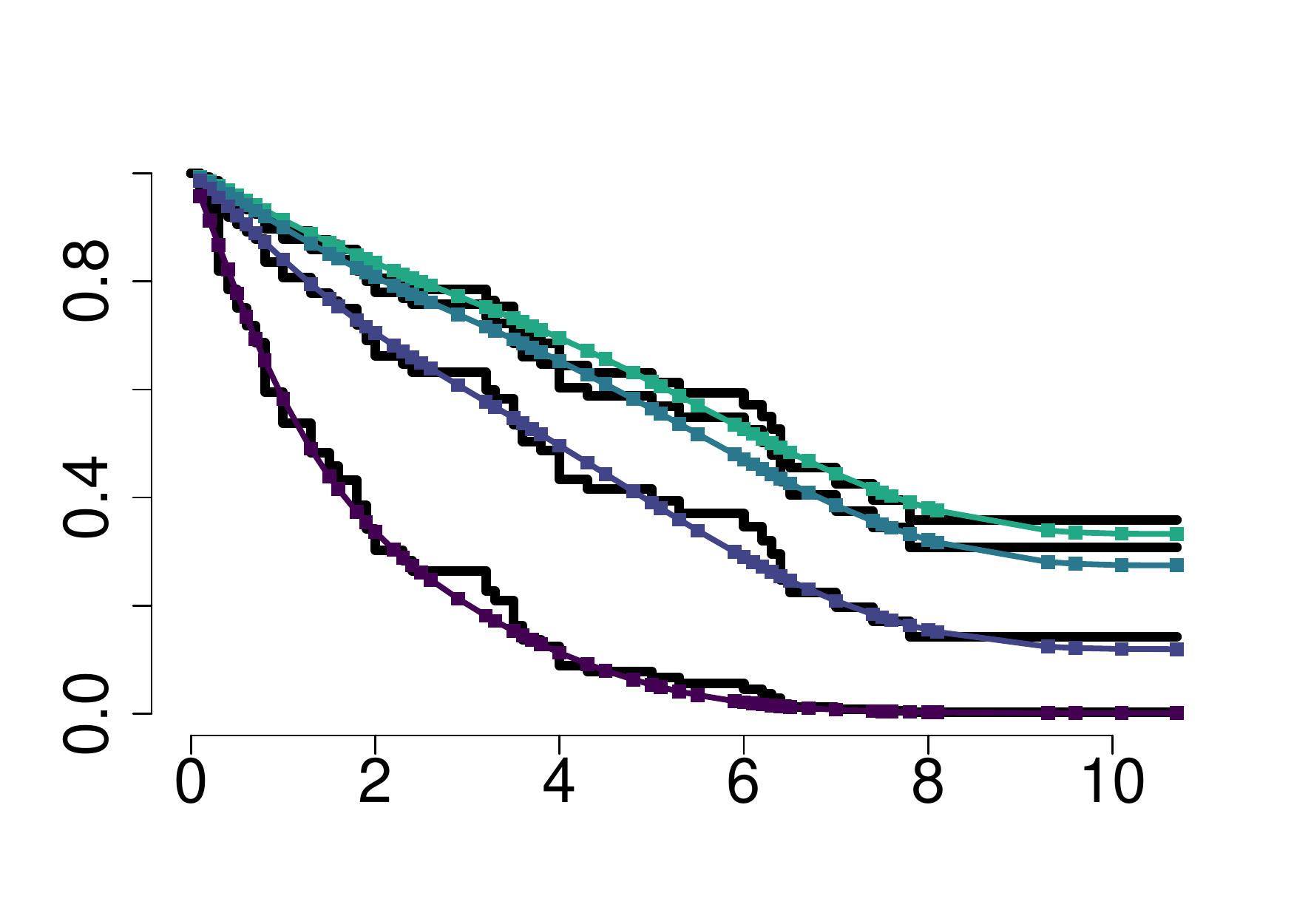} \\
  \end{array}
$$
\caption{\label{Fig:surv}Survival curves for a 77 years old patient. Consider: \textbf{(a)} Posterior survival mean; \textbf{(b)} ML estimated survival. Breslow non-parametric estimate is given in solid lines. The rounded black points represent the posterior mean at the observed time points while the squared refer to the ML estimates of the survival function at the observed time points. The lightest (green) to the darkest (purple) colored lines  refer to the patient cancer stage from Stage I at the top of the graph to Stage IV at the bottom.}\end{figure}

Table \oldref{table:belarynx} might be used for the same type of analysis developed based on Table \oldref{table:mlelarynx}, however
the results here are from the Bayesian approach. Based on the Bayesian AFT model, the median (or any percentile) lifetime for stage IV patients is significantly lower in comparison to Stage I subjects in the same age (HPD does not contain zero). The estimated TR for the Stage IV patients is approximately $0.20$, which means that the lifetime is $80\%$ lower for patients in Stage IV when compared to same-aged patients (in Stage I). The same occurs with the Stage III patients that have lifetimes $55\%$ lower than Stage I patients (equally aged). In addition, there is no statistical evidence to reject that the lifetime is not affected by the age of the patient. 
Figure \oldref{Fig:surv} shows survival curves for a 77 years old patient provided by the Breslow estimator and the BPPH model. 
Figure \oldref{Fig:surv} (a) compares survival curves obtained from the Bayesian approach and  Breslow estimator. Note that, the result is similar to the one achieved in Figure Panel (b). At the end of the  MCMC algorithm, the mean of the 4000 survival values at each failure time is taken and then used to build Panel (a), the plot shows how close the posterior mean of the survival curve is to the Breslow step curve estimate, indicating that the BPPH model is also performing as expected under the Bayesian framework.
The survival estimates presented in Figure \oldref{Fig:surv} are calculated in terms of the BP based cumulative hazard function given in \ref{formula:cumhaz}. It is worth noting that each estimated survival curve is close to the non-parametric Breslow estimator step function, indicating that the BPPH model works as expected with no implementation issues. 

In this section, the  BPPH model applicability was confirmed. However, the conclusions obtained with the BPPH and BPAFT models applied in Application I are not the same. The main explanation is that the WAFT is not the appropriate parametric model for this real data set. Since the BP based models are distribution-free methods, this suggests that other survival parametric models such as Log-logistic, Log-Normal, and others,  might provide more accurate fits to this concrete case. The similarity of the conclusions only occurs in cases where the actual baseline functions come from the Weibull distribution.  In this situation, the PH and AFT relationship described in Section 2.4 is valid. One of the advantages of estimating survival models based on BP is that distributions to handle the baseline functions are not required. As a result, the model is expected to deal with several shapes, according to the properties discussed in Chapter 3. 

In the next section, another popular real data set that is suitable for survival modeling will be analyzed to evaluate the applicability of the BPPO model. The study is related to the status of lung cancer patients. In this example, there are 137 randomized observations of two treatments for lung cancer. In this case, only 97 patients who did not have previous therapy were analyzed.

\newpage

\section{Application II: veterans administration data}
 
$~~~~$ This study was conducted by the US Department of Veterans Affairs \citep{Prentice:1973}, patients with inoperable lung cancer were given either standard therapy or test chemotherapy. The goal of this application is to investigate the prognosis regarding the 97 patients that did not have previous therapy. In medicine, the Karnofsky's performance status  \citep{karnofsky1949clinical} is a measure that quantifies the general welfare of a patient. The performance score (PS) can be used to support professionals' decisions regarding the possibility of receiving chemotherapy, the need to adjust medications, among other purposes. Although PS lower than 50 is considered low by specialists, we chose not to categorize this covariate, following one of the applications in \cite{Bennett:1983}. The continuous covariate PS measures a status ranging from 0 (bad) to 100 (good). Besides, there are four kinds of cells in the study: squamous, small cell, adeno, and large. Patients whose PS is zero and the cell type is ``large'' refer to the baseline group such that $R(t \mid \mathbf{0}) = R_0(t)$. 

\begin{figure}[!htb]
\centering
      \includegraphics[width=\textwidth]{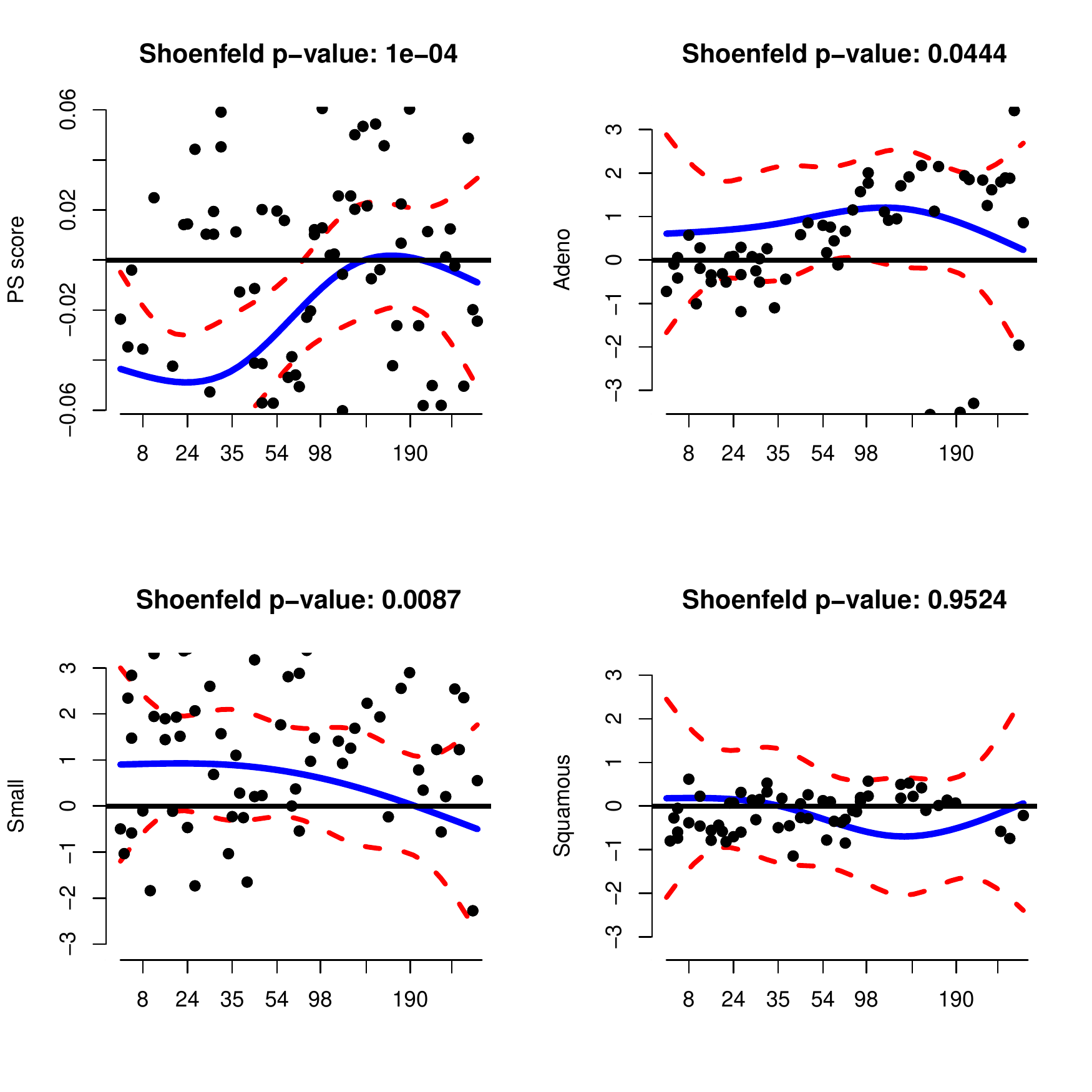} 
\caption{Schoenfeld residuals analysis in Application I. The panels shows the p-value of the Schoenfeld test for each explanatory variable whose null hypothesis is that the residuals are independent of the time. The (blue) line represents the spline interpolation of the residuals, and the dashed (red) lines illustrate its respective confidence intervals to assist the graphical analysis.}\label{notokhaz}
\end{figure}

The CoxPH model is flexible regarding the distribution-free characteristic shared with the BP based models. Yet, those are sometimes not suitable for real cases. In particular, the PH class models are not well fitted to the data if the assumption of PH is violated. To assess the validity of this adequacy, we have analyzed the Schoenfeld residuals of the CoxPH model applied to Application II data; the residuals were calculated using the routines available in the \texttt{survival} package. Figure  \oldref{notokhaz} shows the scatter-plot of the Schoenfeld scaled residuals versus time. In principle, the Shoenfeld residuals are independent of time. Thus, non-random patterns against time suggest that the PH assumption is violated. However, decision-based on the splines showed in Figure \oldref{notokhaz} are subjective. For this reason, the Shoenfeld test p-values are then very useful to support the model diagnosis. Appendix Figure \oldref{scho1} shows the adequacy of the PH model to the Application I, it can be compared to the scatter-plots obtained for the Application II.\\

\newpage
As the PH assumption is violated, the LLPO, partial ML PO, BPPO, and BPAFT become alternatives to be considered to fit the Application II data. In this sense, Table \oldref{table:mlelung} provides estimates for the Karnofsky's PS and the cell type effects, regarding the BPPO model. For comparison purposes, these estimates are showed together with the outcomes of the PO model under the partial ML, proposed in the \texttt{timereg} package \citep{timereg} that is based on transformation models \citep{martinussen2007dynamic}. Similar to what happens with the BPPH model (Table \oldref{table:mlelarynx}), the BPPO model also presented results close to the results provided by the partial ML distribution-free method available in the literature. The outcomes suggest that the implemented routines (\texttt{spsurv}) to fit the BPPO showed similar estimates  to the \texttt{timereg} outcomes when it comes to the percentage of relative difference using the partial ML as a reference, all of them are below $5\%$. For example, one can use the estimates provided by the  \texttt{spsurv} package to state that, the odds on occurring death is almost four times greater for a patient with small cell type compared to the large cell type patients case (assuming the same PS status).\\

\begin{table}[!htb]
\centering
\scriptsize
\begin{tabular}{rlrrrcrrr}
  \hline
&\multicolumn{1}{l}{}   & coef & exp(coef) & se(coef) & CI$_{\text{coef}}$ & p$_{\text{coef}}$ & rd$_{\text{coef}}$ & rd$_{\text{exp(coef)}}$\\ 
  \hline
\multirow{4}{*}{\texttt{timereg}} 
 & $X_1:$PS & -0.0597 & 0.9420 &0.0079 & [-0.0751, -0.0443] & $<$0.0001 & -& -\\ 
 & $X_2:$adeno & 1.3200 & 3.7514 & 0.4170 & [0.5030, 2.1400] & 0.0002 & -& -\\
 & $X_3:$small & 1.1400 & 3.1327 & 0.4090 & [0.3380, 1.9400] & 0.0027 & -& -\\
 & $X_3:$squamous & -0.1130 & 0.8935 & 0.4820 & [-1.0600, 0.8320] & 0.8190 & -& -\\ 
   \hline
   \hline
\multirow{4}{*}{\shortstack{\texttt{spbp}\\\texttt{(model="po")}}}  
& $X_1:$PS & -0.0613 & 0.9406 & 0.0088 & [-0.0784,-0.0441] & $<$0.0001  & -2.5948 & -0.1548\\ 
& $X_2:$adeno & 1.3302 & 3.7818 & 0.4633 & [0.4222, 2.2382] & 0.0041 & 0.6108 & 0.8108\\
& $X_3:$small & 1.1807 & 3.2567 & 0.4301 & [0.3377, 2.0237] & 0.0060 & 3.4000 & 3.9587\\ 
& $X_4:$squamous & -0.1077 & 0.8979 & 0.4666 & [-1.0221, 0.8068] & 0.8174 & 4.3608 & 0.4923 \\
\hline\\
\end{tabular}
\caption{ML outcome for the partial PO and BPPO models applied to the  Veteran administration data set (Application II). Consider coef: ML point estimate; exp(coef): the estimated OR; se(coef): standard errors of the ML point estimate; CI$_{\text{coef}}$ = estimated $95\%$ confidence intervals; p$_{\text{coef}}$: p-value of the Wald test; rd: percentage of relative difference  for the \texttt{timereg::prop.odds} estimates.}
\label{table:mlelung}
\end{table}\normalsize
Also, according to the results, there is statistical evidence to state that the only estimate that presented a big discrepancy with respect to the reference model is not significant in both packages outcomes. That means, there is no statistical evidence to confirm that the odds on occurring death is affected by the cell type when small cell patients of the same status are compared to patients with  large cell (equal PS). Remarkably, when compared to the semi-parametric PO model (distribution-free) and the LLAFT in the Application II, the BP based models, BPPO and BPAFT, has provided similar conclusions about the Application II, respectively. As a consequence, the Log-logistic parametric form might describe well the data in Application II as the Log-logistic is the only parametric model that comprises both PO and AFT classes. Figure \oldref{distfit1} confirm this empirical evidence, as we found that the Kaplan-Meier estimate of the residuals approximate to the Log-logistic curve.\\
\begin{table}[!htb]
\centering
\scriptsize
\begin{tabular}{rlrrrcrrr}
  \hline
&\multicolumn{1}{l}{}   & coef & exp(coef) & se(coef) & CI$_{\text{coef}}$ & p$_{\text{coef}}$ & rd$_{\text{coef}}$ & rd$_{\text{exp(coef)}}$\\ 
  \hline
\multirow{4}{*}{\shortstack{\texttt{survreg}\\\texttt{("loglogistic")}}} 
 & Intercept & 2.4512 & 11.6022 & 0.3435 & [1.7779, 3.1245] & $<$0.0001 &  & \\
  & Log(scale) & -0.5430 & 1.7878  & 0.0742 & [ 0.4356,0.7264] & $<$0.0001 & - & -\\
 & $X_1:$PS & 0.0361 & 1.0367 & 0.0044 & [0.0274, 0.0447] & $<$0.0001 & -& -\\ 
 & $X_2:$adeno & -0.7494 & 0.4727 & 0.2614 & [-1.2618, -0.2370] & 0.0042 & -& -\\
 & $X_3:$small & -0.6608 & 0.5164 & 0.2403 & [-1.1317, -0.1899] & 0.0060 & -& -\\
 & $X_3:$squamous & 0.0290 & 1.0294 & 0.2635 & [-0.4876, 0.5455] & 0.9125 & -& -\\ 
   \hline
   \hline
\multirow{4}{*}{\shortstack{\texttt{spbp}\\\texttt{(model="aft")}}}  
& $X_1:$PS & 0.0342 & 1.0348 & 0.0038 & [0.0268,0.0415] & $<$0.0001 & -5.2163 & -0.1879\\ 
& $X_2:$adeno & -0.7443 & 0.4751 & 0.2596 & [-1.2532, -0.2355] & 0.0041 & 0.6738 & 0.5062\\
& $X_3:$small & -0.5678 & 0.5668 & 0.2574 & [-1.0724,-0.0633] & 0.0274 & 14.0704 & 9.7438\\ 
& $X_4:$squamous & 0.1248 & 1.1330 & 0.2635 & [-0.3916, 0.6413] & 0.6356 & 330.8831 & 10.0614 \\
\hline\\
\end{tabular}
\caption{ML outcome for the LLAFT and BPAFT models applied to the  Veteran administration data set (Application II). Consider $\equiv$ coef: ML point estimate; exp(coef): the estimated TR (or acceleration factor); se(coef): standard errors of the ML point estimate; CI$_{\text{coef}}$ = estimated $95\%$ confidence intervals; p$_{\text{coef}}$: p-value of the Wald test; rd: percentage of relative difference  for the \texttt{survival::survreg} estimates.}
\label{table:mlelungaft}
\end{table}\normalsize

Likewise, the \texttt{spsurv} package also allows Bayesian analysis for the BPPO and the BPAFT models. The prior information given to the parameters is the same used in the Application I, which corresponds to the Prior 2 (Table \oldref{sensitivity}). In Table \oldref{table:belung}, the Bayesian outcome for the veteran lung cancer trials are displayed. As expected, similar conclusions are obtained with BPAFT and LLAFT models. For example, we have found that the median lifetime for small-cell patients is significantly lower in comparison to large type cell patients of the same PS status. The estimated TR for the small cell patients were around $0.55$ (\texttt{spbp}), which means the median lifetime is 1.8 times greater for the large cell type patients when compared to the adeno cell type subjects of the same status. Besides, there is no statistical evidence to reject that the median lifetime is not affected adeno cell type when compared to the large cell type.

In addition, conclusions (inferences) about the lifetime of heterogeneous groups (population) of the US veterans diagnosed with laryngeal cancer can be obtained with the Bayesian BPAFT model fit outcomes provided by the proposed \texttt{spsurv} package, showed in Table \oldref{table:belung}. For example, the TR \ref{formula:tr} posterior mean can express the effect of the covariates in the time (in days) to lung cancer death. For example, the time to lung cancer death for a squamous cell type patient with PS score 70  is approximately 6 times greater when compared to the lifetime of a adeno cell type patient with PS score 40. From the Bayesian outcomes,  the posterior mean of the TR denoted as $\exp\{\beta_1 (70-40) + \beta_2(0-1) +\beta_3(0-0) + \beta_4(1-0)\}$ is $5.9166$. Also, the HPD interval of the logarithm of the TR posterior: $\log\text{TR} = 30 \times \beta_1  -\beta_2 + \beta_4$,  confirms that this result is statistically significant as $\text{HPD}_{\log\text{TR}}=[1.2601;2.3029]$. \\

\begin{table}[!htb]
\centering
\scriptsize
\begin{tabular}{rlrrrcrrr}
  \hline
 & & mean(coef) & exp(mean) & sd(coef)  & HPD & rd$_{\text{mean}}$ & rd$_{\text{exp(coef)}}$ \\ 
\hline
\multirow{4}{*}{\shortstack{\texttt{spsurv::spbp}\\(\texttt{model = "po"})}} 
& $X_1:$PS & -0.0626 & 0.9394 & 0.0090 & [-0.0794,~~-0.0443] & -4.3505 &-0.2594\\ 
& $X_2:$adeno & 1.3260  & 4.1939 & 0.4640 & [0.4359,~~2.2189] & 0.7282 &0.9674\\ 
& $X_3:$small & 1.1616 &  3.5157 & 0.4345 & [0.3451,~~2.0196] & 3.4147 &3.9762\\ 
& $X_4:$squamous & -0.1261 & 0.9851 & 0.4700 & [-1.0685,~~0.7338] & -2.6131 &-0.2938\\  \hline \hline
\multirow{4}{*}{\shortstack{\texttt{spsurv::spbp}\\(\texttt{model = "aft"})}} 
& $X_1:$PS & 0.0335 & 1.0341 & 0.0042  & [0.0254,~~0.0419] & -7.1250 & -0.2557\\ 
& $X_2:$adeno & -0.7564 & 0.4844 & 0.2514 & [-1.2562,~~-0.2788] & -0.9437 & 2.4924 \\ 
& $X_3:$small & -0.6188 &  0.5554 & 0.2476 & [-1.0813,~~-0.1185] & 6.3631 & 7.5513 \\ 
& $X_4:$squamous & 0.1478 & 1.2052 & 0.2805 & [-0.4005,~~0.6940] & 410.0063 & 17.0815\\  \hline\\
\end{tabular}
\caption{ \label{table:belung} Bayesian outcome for the BPPO and the BPAFT models applied to the Veteran administration data set (Application II). Consider: mean(coef): posterior mean; exp(mean): the estimated HR or the estimated TR (respectively); se(coef): posterior standard errors; HPD: Highest posterior density intervals; rd: percentage relative difference to  \texttt{survival::coxph} estimates or \texttt{survival::survreg} estimates (respectively).}
\end{table}\normalsize

In summary, this Chapter outlined how to interpret the results from the BP based regression models in the context of survival analysis. In both real applications, it was clear that the BPPH and BPPO can be applied for the same purposes considered for the distribution-free models in the literature (CoxPH and partial PO). However, in the application of BAFT, it is possible to see the importance of distribution-free methods with respect to the choice of an appropriate model family (AFT, PH, or PO). This difference was apparent between the first and the second applications. In short, in the Application I, the results differed from the WAFT parametric model, and, in the Application II, the results between BPAFT and LLAFT were closer. Suggesting that BP modeling is flexible in the sense that it does not impose functional forms to base distributions and, thus, comprises several cases. Note that, the conclusions obtained with the BPPO and BPAFT are similar because both models are applicable when the Log-Logistic parametric model is well fitted. The main advantage of the BPAFT model is that it does not require any probability distribution to handle baseline functions. As a consequence, this a distribution-free method that encompasses a wide variety of situations, such as those situations illustrated in this Chapter.

The guidelines on how to fit these data using the \texttt{spsurv} package in \texttt{R} are provided in Appendix G along with details on how to use the main fitter function \texttt{spsurv::spbp} to meet all the modeling options presented. The next chapter presents the main remarks,
conclusions, and future work-related.

\newpage

% Chapter 5 = Conclusion.
\chapter{Conclusions}
$~~~~$This dissertation aimed to introduce a set of new routines that initially compose in the proposed package \texttt{spsurv}. This package was implemented using the \texttt{R} language and is intended to allow fully likelihood-based procedures for survival regression models. Based on a semi-parametric appealing structure, the \texttt{spsurv} refers to the acronym ``semi-parametric survival analysis''. Throughout this dissertation, we have explored fundamental concepts of survival analysis, along with the approximation structure inherent from the Bernstein polynomial. This methodology was referred to as Bernstein polynomial based survival regression models. This innovative proposal for semi-parametric survival modeling was previously introduced by \cite{Osman:2012}. The methods provided in the package do not rely on probability distributions to handle baseline terms; however, they depend on the parameter based structure of the polynomials. Unlike the Cox model, under a Bernstein polynomial based survival regression model, baseline functions are specified with polynomial structures, which we refer to as the Bernstein basis polynomials.

As part of the theoretical framework, we have presented the three classes of survival models provided by the \texttt{spsurv} package: the proportional hazards model, the proportional odds model, and the accelerated failure time model. Also, this work discusses some of the relationships between parametric survival model families. To develop simulation studies to
explore the performance of the leading models, we have considered parametric versions based on the Weibull and Log-Logistic distribution to generate artificial data sets. Since we can make simulated data, the actual values of all unknown parameters are available for the study focused on
investigating how well the model can handle the data information (compared to the generator model). Although in the context of survival analysis, the approximation of the baseline functions is not feasible, instructions on how to apply Bernstein polynomial in approximating continuous functions in the real domain were also explored. Afterward, we have described both Frequentist and Bayesian full likelihood inferential procedures for the Bernstein based survival regression models to provide the necessary estimation to the baseline functions. 

The simulation study developed here was based on two scenarios differing in terms of sample size. In these cases, we have considered twelve variations of the proposed routines under a Monte Carlo simulation scheme. In general, the results indicate good approximations
of the estimates to the actual values. For example, the average relative bias in the second scenario was below 6 \%, regarding the estimates of all the tested models. We have also discussed the coverage probability of confidence intervals and credible intervals ranges. The resulting achievements, real data applications, and differences between approaches were also considered in this study. We found that BP based models can be widely applied to studies in the literature that earlier made use of models that correspond to the model classes used here. The results of the Bayesian model (Prior 2) showed low bias and coverage rate, very close to the nominal level of 95\%, in the simulation Scenarios I and II, analyzed in Chapter 5.
On the other hand, BP based models should be applied with caution under the ML approach, we found estimates with very low relative bias, but, the confidence intervals tend to be narrow as a consequence of the underestimation of the standard errors. In the simulation study, we have observed that the Delta method combined with the Hessian estimates provided by the LBFGS (or  BFGS) algorithm did not provide good interval estimation. At the end of this dissertation, we have presented a summary of some new routine.

$~~~~$The current version of the package can be improved or extended in many directions. First, the routines provided here explore just right-censored data modeling options. Therefore, one possible extension is to include models for other censoring configurations. Also, the methodologies do not consider non-crossing survival curves or clustered data, as seen in other models in the literature \citep{lipsitz1994jackknife, glidden2004modelling, logan2008comparing, li2015statistical}. Another interesting aspect to be considered, as an extension, is the inclusion of frailty and cure fraction models \citep{lambert2007modeling, gutierrez2002parametric, mcgilchrist1991regression}. Indeed, many improvements in the general aspects of the \texttt{spsurv} package should still be considered, new routines, more S3 methods extensions, specific graphical presentations, and the implementation of new models are critical to the next upcoming works.
Along with that, some alternatives might reduce the computational time of posterior samplings, such as the relationship between the power and Bernstein polynomials described in \cite{Farouki:1987}. Although several topics have to be considered, we highlight the main positive aspects: integration with the \texttt{Stan}, the ability to estimate well (point estimates) either in Bayesian or Frequentist inferential approaches, the availability of three survival regression classes and the possibility to choose among six distinct prior specifications in a Bayesian analysis. We believe that the \texttt{spsurv} has a great potential to become a comprehensive package including
different tools and models to deal with a wide range of applications in survival analysis.\\

% Appendix
\newpage
\appendix

\addcontentsline{toc}{chapter}{Appendix}
\appendixpage
\renewcommand{\thetable}{A.\arabic{table}}
\renewcommand{\thefigure}{A.\arabic{figure}} \setcounter{figure}{0}
%%%%%%%%%%%%%%%%%%%%%%%%%%%%%%%%%%%%%%%%%%%%%%
%%%%%%%%%%%%%  Appendix A     %%%%%%%%%%%%%%%%
%%%%%%%%%%%%%%%%%%%%%%%%%%%%%%%%%%%%%%%%%%%%%%

{\flushleft \Large \textbf{Appendix A: Real applications cancer data sets}}\vspace{15pt}\\

$~~~~$In Appendix A, Tables \oldref{data:larynx} and \oldref{data:lung} show the real data sets used in Applications I and II (Chapter 6). The first data set refers to the study that allowed records of patients with inoperable laryngeal cancer, carried out between the years 1970-1978 in a dutch hospital  \citep{Kardaun:1983}. Table \oldref{data:larynx} shows the patient number,  cancer stage classification, the lifetime (in months), and the failure indicator (delta). In turn, the veteran's administration data set is part of the public health data found in the National Center for Veterans Analysis and Statistics by the US Department of Veteran Affairs \citep{Prentice:1973}. Table  \oldref{data:lung} shows the patient number, lifetime (in days) of each patient, the Karnofsky performance score (see Chapter 6), and the kind of cell. In both tables, the header is repeated to accommodate all observations side by side. The censoring times are indicated with the $^+$ symbol.

In Table  \oldref{data:larynx}, we have 90 larynx cancer patients. The patient number 1, for example, has experienced the death approximately 18 days (0.6 months) after this patient was included in the study. The referred patient was 77 years old and was diagnosed with Stage I laryngeal cancer. In Table \oldref{data:lung}, we have 97 patients without prior chemotherapy treatment. The patient number 1, for example, has experienced the death after 72 days counted from the day that this patient was included in the study. The cell type of this patient was squamous, and the performance score of 60 was considered good (above 50).
\newpage
\begin{table}[!htb]
\centering
\tiny
\begin{tabular}{cccccccccccccccccc}
  \hline
patient & stage & time & age & delta & patient & stage & time & age & delta  & patient & stage & time & age & delta \\ 
  \hline
  1 & 1 & 0.60 & 77 & 1 & 40 & 2 & 6.20 &  74 & 1 & 79 &   4 & $4.30^+$ &  48 & 0\\ 
  2 &   1 & 1.30 &  53  &   1 & 41 &   2 & 7.00 &  62 &   1 & 80 &   4 & 3.80 &  84 & 1\\ 
  3 &   1 & 2.40 &  45  &   1 & 42 &   2 & $7.50^+$ &  50 &   0 & 81 &   4 & 3.60 &  71 & 1\\ 
  4 &   1 & $2.50^+$ &  57  &   0 & 43 &   2 & $7.60^+$ &  53 &   0 & 82 &   4 & $2.90^+$ &  74 &   0 \\ 
  5 &   1 & 3.20 &  58 & 1 & 44 &   2 & $9.30^+$ &  61 &  0 & 83 &   4 & 2.30 &  62 & 1\\ 
  6 &   1 & $3.20^+$ &  51 &   0 & 45 &   3 & 0.30 &  49 & 1 & 84 &   4 & 2.00 &  69 & 1\\ 
  7 &   1 & 3.30 &  76 & 1 & 46 &   3 & 0.30 &  71 & 1&   85 &   2 & 4.00 &  81 &  1\\ 
  8 &   1 & $3.30^+$ &  63 &   0 & 47 &   3 & 0.50 &  57 & 1 &   86 &   2 & $4.30^+$ &  47 & 0\\ 
  9 &   1 & 3.50 &  43 & 1 & 48 &   3 & 0.70 &  79 &  1 & 87 &   2 & $4.30^+$ &  64 & 0\\ 
  10 &   1 & 3.50 &  60 & 1 & 49 &   3 & 0.80 &  82 & 1 & 88 &   2 & $5.00^+$ &  66 & 0\\ 
  11 &   1 & 4.00 &  52 & 1 & 50 &   3 & 1.00 &  49 & 1 &  89 &   2 & 3.60 &  70 & 1 \\ 
  12 &   1 & 4.00 &  63 & 1 & 51 &   3 & 1.30 &  60 & 1 &   90 &   2 & $3.60^+$ &  72 & 0 \\ 
  13 &   1 & 4.30 &  86 & 1 & 52 &   3 & 1.60 &  64 &  1\\ 
  14 &   1 & $4.50^+$ &  48 &  0 & 53 &   3 & 1.80 &  74 & 1  \\ 
  15 &   1 & $4.50^+$ &  68 &  0 & 54 &   3 & 1.90 &  72 & 1 \\ 
  16 &   1 & 5.30 &  81 & 1 & 55 &   3 & 1.90 &  53 &    1 \\ 
  17 &   1 & $5.50^+$ &  70 &   0 & 56 &   3 & 3.20 &  54 &  1 \\ 
  18 &   1 & $5.90^+$ &  58 &   0 & 57 &   3 & 3.50 &  81 &   1 \\ 
  19 &   1 & $5.90^+$ &  47 &   0 & 58 &   3 & $3.70^+$ &  52  &   0\\ 
  20 &   1 & 6.00 &  75 &   1 & 59 &   3 & $4.50^+$ &  66 &    0\\ 
  21 &   1 & $6.10^+$ &  77 &  0 & 60 &   3 & $4.80^+$ &  54 &    0 \\ 
  22 &   1 & $6.20^+$ &  64 & 0 & 61 &   3 & $4.80^+$ &  63 &    0\\ 
  23 &   1 & 6.40 &  77 & 1 & 62 &   3 & 5.00 &  59 &  1\\ 
  24 &   1 & 6.50 &  67 & 1 & 63 &   3 & $5.00^+$ &  49 &  0\\ 
  25 &   1 & $6.50^+$ &  79 & 0 & 64 &   3 & $5.10^+$ &  69 & 0\\ 
  26 &   1 & $6.70^+$ &  61 & 0 & 65 &   3 & 6.30 &  70 & 1\\ 
  27 &   1 & $7.00^+$ &  66 & 0 & 66 &   3 & 6.40 &  65 &  1\\ 
  28 &   1 & 7.40 &  68 & 1 & 67 &   3 & $6.50^+$ &  65 &  0\\ 
  29 &   1 & $7.40^+$ &  73 & 0 & 68 &   3 & 7.80 &  68 &  1\\ 
  30 &   1 & $8.10^+$ &  56 & 0 & 69 &   3 & $8.00^+$ &  78 & 0\\ 
  31 &   1 & $8.10^+$ &  73 & 0 & 70 &   3 & $9.30^+$ &  69 & 0\\ 
  32 &   1 & $9.60^+$ &  58 & 0 & 71 &   3 & $10.10^+$ &  51 & 0\\ 
  33 &   1 & $10.70^+$ &  68 & 0 & 72 &   4 & 0.10 &  65 &  1 \\ 
  34 &   2 & 0.20 &  86 & 1 & 73 &   4 & 0.30 &  71 & 1\\ 
  35 &   2 & 1.80 &  64 & 1 & 74 &   4 & 0.40 &  76 & 1 \\ 
  36 &   2 & 2.00 &  63 & 1 & 75 &   4 & 0.80 &  65 & 1 \\ 
  37 &   2 & $2.20^+$ &  71 & 0 & 76 &   4 & 0.80 &  78 &   1\\ 
  38 &   2 & $2.60^+$ &  67 & 0 & 77 &   4 & 1.00 &  41 &  1 \\ 
  39 &   2 & $3.30^+$ &  51 & 0 & 78 &   4 & 1.50 &  68 & 1\\
  \hline
\end{tabular}
\caption{\label{data:larynx} Laryngeal cancer stage study; source: \cite{Kardaun:1983}. Consider $\equiv$  patient: unique identifier; stage: stage of disease 1 (stage 1), 2 (stage 2), 3 (stage 3), 4 (stage 4); time: time to death in months; age: age at diagnosis of larynx cancer; symbol $\mathbf{^+}$: censoring indicator assuming 0 (alive) and 1 (dead).}
\end{table}

\newpage
\begin{table}[!htb]
\centering
\tiny
\begin{tabular}{cccccccccccccc}
  \hline
patient & time & karno & celltype  & patient & time & karno & celltype & patient & time & karno & celltype\\ 
  \hline
  1 & 72  & 60 & 1 &  34 & 117  & 80 & 3 & 67 & 24  & 60 & 2 \\
  2 & 228  & 60 & 1 & 35 & 132  & 80 & 3 & 68 & 99  & 70 & 2 \\ 
  3 & 10  & 20 & 1 & 36 & 162  & 80 & 3 & 69 & 8  & 80 & 2 \\  
  4 & 110  & 80 & 1 & 37 & 3  & 30 & 3 & 70 & 99  & 85 & 2 \\
  5 & 314  & 50 & 1 & 38 & 95  & 80 & 3 & 71 & 61  & 70 & 2\\  
  6 & $100^+$ & 70 & 1 & 39 & 162  & 80 & 4 & 72 & 25  & 70 & 2\\
  7 & 42  & 60 & 1 & 40 & 216  & 50 & 4   & 73 & 95  & 70 & 2\\ 
  8 & 144  & 30 & 1 & 41 & 553  & 70 & 4  & 74 & 80  & 50 & 2 \\ 
  9 & 30  & 60 & 2 & 42 & 278  & 60 & 4   & 75 & 29  & 40 & 2\\ 
  10 & 384  & 60 & 2 & 43 & 260  & 80 & 4 & 76 & 24  & 40 & 3 \\ 
  11 & 4  & 40 & 2 & 44 & 156  & 70 & 4   & 77 & $83^+$ & 99 & 3\\ 
  12 & 13  & 60 & 2 &  45 & $182^+$  & 90 & 4 & 78 & 31  & 80 & 3\\  
  13 & $123^+$  & 40 & 2 &  46 & 143  & 90 & 4 & 79 & 51  & 60 & 3\\ 
  14 & $97^+$  & 60 & 2 & 47 & 105  & 80 & 4  & 80 & 52  & 60 & 3\\ 
  15 & 59  & 30 & 2 & 48 & 103  & 80 & 4  & 81 & 73  & 60 & 3\\ 
  16 & 117  & 80 & 2 & 49 & 112  & 80 & 1 & 82 & 8  & 50 & 3 \\ 
  17 & 151  & 50 & 2 &  50 & $87^+$  & 80 & 1 & 83 & 36  & 70 & 3\\ 
  18 & 22  & 60 & 2 & 51 & 242  & 50 & 1 & 84 & 48  & 10 & 3\\  
  19 & 18  & 20 & 2 & 52 & 111  & 70 & 1 &   85 & 7  & 40 & 3 \\ 
  20 & 139  & 80 & 2 & 53 & 587  & 60 & 1 &   86 & 140  & 70 & 3  \\  
  21 & 20  & 30 & 2 & 54 & 389  & 90 & 1 &   87 & 186  & 90 & 3 \\  
  22 & 31  & 75 & 2 & 55 & 33  & 30 & 1 &   88 & 19  & 50 & 3 \\ 
  23 & 52  & 70 & 2 &  56 & 25  & 20 & 1 &   89 & 45  & 40 & 3 \\ 
  24 & 18  & 30 & 2 & 57 & 357  & 70 & 1 &   90 & 80  & 40 & 3 \\  
  25 & 51  & 60 & 2 & 58 & 467  & 90 & 1 &   91 & 52  & 60 & 4 \\  
  26 & 122  & 80 & 2 & 59 & 1  & 50 & 1  &  92 & 53  & 60 & 4 \\  
  27 & 27  & 60 & 2 & 60 & 30 & 70 & 1   &   93 & 15  & 30 & 4 \\ 
  28 & 54  & 70 & 2 & 61 & 283 & 90 & 1    &  94 & 133  & 75 & 4 \\  
  29 & 7  & 50 & 2 &  62 & 25  & 30 & 2    &   95 & 111  & 60 & 4 \\  
  30 & 63  & 50 & 2 &  63 & 21 & 20 & 2    &   96 & 378  & 80 & 4 \\  
  31 & 392  & 40 & 2 & 64 & 13 & 30 & 2    &   97 & 49  & 30 & 4 \\ 
  32 & 92  & 70 & 3 &  65 & 87  & 60 & 2\\ 
  33 & 35  & 40 & 3 &  66 & 7  & 20 & 2 \\ 
\hline
\end{tabular}
\caption{\label{data:lung}Veteran administration lung cancer study; source: \cite{Prentice:1973}. Consider $\equiv$ patient: unique identifier; time: survival time; karno: Karnofsky performance score; celltype: kind of cell 1 (squamous), 2 (small), 3 (adeno) and 4 (large); symbol $\mathbf{^+}$: censoring indicator assuming 0 (alive) and 1 (dead).}
\end{table}
%%%%%%%%%%%%%%%%%%%%%%%%%%%%%%%%%%%%%%%%%%%%%%
%%%%%%%%%%%%%  Appendix B     %%%%%%%%%%%%%%%%
%%%%%%%%%%%%%%%%%%%%%%%%%%%%%%%%%%%%%%%%%%%%%%

\newpage
{\flushleft \Large \textbf{Appendix B: Monte Carlo simulation study outcomes}} \vspace{15pt}

In Appendix B, Tables \oldref{sim1}, \oldref{sim2}, \oldref{sim3} and \oldref{sim4} show the Monte Carlo Simulation study results. In Chapter 5, the five statistics used to evaluate the MC replications fits were described. Consider the average estimate of the regression coefficient (est.), average standard error (se.) standard derivation of the estimation, relative bias (rb.), and coverage probability(cov.). The MC simulation results were distributed in four tables, Tables A.3 and A.4 refer to the fits of the survival models applied to the Scenario  I WAFT and LLAFT data sets. Table A.4 and A.5 to the Scenario II WAFT and LLAFT data sets, respectively. The tables that contain the MC simulation results are divided into twelve sub-tables that describe the two values of the parameter and the five statistics mentioned. The BPPH and BPAFT are applied to the WAFT data sets, and the BPPO is applied to the LLAFT data sets. Note that the BPAFT model is applied in all tables.

The model is well fitted if relative bias to the true value of the parameters is close to zero; this indicates that low biased estimates are obtained with the model. Also, the coverage probability must be close to the nominal value of 95\%; this indicates that good interval estimation is provided. Along with that, the standard error values must be close to the standard deviation of the estimates; this indicates that the standard error (or posterior deviation) is close to the standard deviation of the 1000 MC replica. In general, low or high average standard error (compared to sde.) might indicate that those values are being underestimated or overestimated, respectively. 

\newpage

\begin{table}[!htb]
\centering
\scriptsize
\begin{tabular}{rrrrrrrrrrrrrr}
\hline
BFGS     &      & \multicolumn{5}{l}{BPPO$^{a}$} &  & \multicolumn{5}{l}{BPAFT$^{b}$} \\ \cline{3-7} \cline{9-14} 
      & true & est.  & se. & sde. & rb. & cov. & & true & est.   & se.   & sde.  & rb.  & cov.  \\ \hline
$\beta_1$ & -4    &  -4.3582 & 0.6328 & 0.7973 & -8.9558 & 0.8534   & 
          & 2 & 2.0463 & 0.1543 & 0.3150 & 2.3157 & 0.6603 \\
$\beta_2$ &  2   & 2.2116 & 0.6919 & 0.7695 & 10.5789 & 0.9314      &  
          &  -1  &  -1.0251 & 0.2123 & 0.4086 & -2.5076 & 0.6713   \\ \hline
\end{tabular}
\begin{tabular}{rrrrrrrrrrrrrr}
\hline
LBFGS     &      & \multicolumn{5}{l}{BPPO$^{c}$} &  & \multicolumn{5}{l}{BPAFT$^{d}$} \\ \cline{3-7} \cline{9-14} 
      & true & est.  & se. & sde. & rb. & cov. & & true & est.   & se.   & sde.  & rb.  & cov.  \\ \hline
$\beta_1$ & -4    & -4.3523 & 0.6417 & 0.7813 & -8.8071 & 0.8548  & 
          & 2 & 2.0345 & 0.1586 & 0.3123 & 1.7269 & 0.6897 \\
$\beta_2$ &  2   & 2.2170 & 0.6935 & 0.7646 & 10.8513 & 0.9330 &  
          &  -1  &  -1.0222 & 0.2150 & 0.4025 & -2.2193 & 0.6886   \\ \hline
\end{tabular}
\begin{tabular}{rrrrrrrrrrrrrr}
\hline
Prior 1    &      & \multicolumn{5}{l}{BPPO} &  & \multicolumn{5}{l}{BPAFT$^{e}$} \\ \cline{3-7} \cline{9-14} 
      & true & est.  & se. & sde. & rb. & cov. & & true & est.   & se.   & sde.  & rb.  & cov.  \\ \hline
$\beta_1$ & -4    & -3.6318 & 0.5333 & 0.4218 & 9.2040 & 0.9080     & 
          &  2    &  2.0824 & 0.2432 & 0.2513 & 4.1185 & 0.9439    \\
$\beta_2$ &  2   & 1.8568 & 0.6337 & 0.5762 & -7.1611 & 0.9570 & 
          &  -1  &  -1.0680 & 0.3384 & 0.3506 & -6.8042 & 0.9429   \\ \hline
\end{tabular}
\begin{tabular}{rrrrrrrrrrrrrr}
\hline
Prior 2    &      & \multicolumn{5}{l}{BPPO} &  & \multicolumn{5}{l}{BPAFT$^{f}$} \\ \cline{3-7} \cline{9-14} 
      & true & est.  & se. & sde. & rb. & cov. & & true & est.   & se.   & sde.  & rb.  & cov.  \\ \hline
$\beta_1$ & -4    & -4.0437 & 0.6195 & 0.5562 & -1.0925 & 0.9600    & 
          &  2    &  2.0441 & 0.2321 & 0.2538 & 2.2041 & 0.9189      \\
$\beta_2$ &  2   & 2.0677 & 0.6770 & 0.6631 & 3.3839 & 0.9560  & 
          &  -1  &  -1.0426 & 0.3237 & 0.3453 & -4.2609 & 0.9319   \\ \hline
\end{tabular}
\begin{tabular}{rrrrrrrrrrrrrr}
\hline
Prior 3    &      & \multicolumn{5}{l}{BPPO} &  & \multicolumn{5}{l}{BPAFT} \\ \cline{3-7} \cline{9-14} 
      & true & est.  & se. & sde. & rb. & cov. & & true & est.   & se.   & sde.  & rb.  & cov.  \\ \hline
$\beta_1$ & -4    & -3.7566 & 0.5533 & 0.4531 & 6.0854 & 0.9400     & 
          &  2    &  2.0974 & 0.2469 & 0.2581 & 4.8714 & 0.9419     \\
$\beta_2$ &  2   & 1.9204 & 0.6454 & 0.6005 & -3.9798 & 0.9600  & 
          &  -1  &  -1.0752 & 0.3413 & 0.3550 & -7.5207 & 0.9449    \\ \hline
\end{tabular}
\begin{tabular}{rrrrrrrrrrrrrr}
\hline
Prior 4    &      & \multicolumn{5}{l}{BPPO} &  & \multicolumn{5}{l}{BPAFT} \\ \cline{3-7} \cline{9-14} 
      & true & est.  & se. & sde. & rb. & cov. & & true & est.   & se.   & sde.  & rb.  & cov.  \\ \hline
$\beta_1$ & -4    & -4.2490 & 0.6599 & 0.6346 & -6.2248 & 0.9509     & 
          &  2    &  2.0576 & 0.2361 & 0.2583 & 2.8822 & 0.9200   \\
$\beta_2$ &  2   & 2.1733 & 0.6980 & 0.7103 & 8.6649 & 0.9469 & 
          &  -1  &  -1.0503 & 0.3268 & 0.3498 & -5.0311 & 0.9300    \\ \hline \\
\end{tabular}
\caption{MC simulation study in Scenario I ($n = 100$), models fitted to the LLAFT data sets.  Estimate of the regression coefficient (est.), average
standard error (se.), standard deviation of the estimates (sde.), relative bias (rb in \%) and
coverage probability (nominal level 95\%). Symbols: ${a}$ indicates $R=962$ (38 non-converging), ${b}$ indicates $R=998$ (2 non-finite Hessian matrices); ${c}$ indicates $R=985$ (15 non-finite Hessian matrices); ${d}$ indicates $R=941$ (59 non-converging), ${e}$ indicates  $R=999$ and  ${f}$ indicates $R=999$.}
 \label{sim2}
\end{table}\normalsize

\newpage

\begin{table}[!htb]
\centering
\scriptsize
\begin{tabular}{rrrrrrrrrrrrrr}
\hline
BFGS     &      & \multicolumn{5}{l}{BPPO$^{a}$} &  & \multicolumn{5}{l}{BPAFT$^{b}$} \\ \cline{3-7} \cline{9-14} 
      & true & est.  & se. & sde. & rb. & cov. & & true & est.   & se.   & sde.  & rb.  & cov.  \\ \hline
$\beta_1$ & -4    & -4.1910 & 0.4547 & 0.5305 & -4.7745 & 0.9053   & 
          & 2 &  2.0299 & 0.1140 & 0.2251 & 1.4932 & 0.7193     \\
$\beta_2$ &  2   & 2.1112 & 0.4780 & 0.5180 & 5.5586 & 0.9287  &  
          &  -1  &  -1.0119 & 0.1529 & 0.2798 & -1.1884 & 0.6821    \\ \hline
\end{tabular}
\begin{tabular}{rrrrrrrrrrrrrr}
\hline
LBFGS     &      & \multicolumn{5}{l}{BPPO$^{c}$} &  & \multicolumn{5}{l}{BPAFT$^{d}$} \\ \cline{3-7} \cline{9-14} 
      & true & est.  & se. & sde. & rb. & cov. & & true & est.   & se.   & sde.  & rb.  & cov.  \\ \hline
$\beta_1$ & -4    & -4.1970 & 0.4653 & 0.5236 & -4.9257 & 0.9116   & 
          & 2 &   2.0190     &   0.1081    &    0.2416   &  0.9466    &  0.6978    \\
$\beta_2$ &  2   & 2.1175 & 0.4810 & 0.5145 & 5.8726 & 0.9311   &  
          &  -1  & -1.0175 & 0.1611 & 0.2848 & -1.7524 & 0.6986   \\ \hline
\end{tabular}
\begin{tabular}{rrrrrrrrrrrrrr}
\hline
Prior 1    &      & \multicolumn{5}{l}{BPPO} &  & \multicolumn{5}{l}{BPAFT$^{\dagger}$} \\ \cline{3-7} \cline{9-14} 
      & true & est.  & se. & sde. & rb. & cov. & & true & est.   & se.   & sde.  & rb.  & cov.  \\ \hline
$\beta_1$ & -4    & -3.7550       &  0.4047     & 0.3514      & 6.136    & 0.9190     & 
          &  2  & 2.0834 & 0.1404 & 0.1462 & 4.1680 & 0.8900        \\
$\beta_2$ &  2   & 1.9040       & 0.4524      & 0.4346      & -4.7910     & 0.9510  & 
          &  -1  & -1.0450 & 0.1722 & 0.1676 & -4.5027 & 0.9400  \\ \hline
\end{tabular}
\begin{tabular}{cccccccccccccc}
\hline
Prior 2    &      & \multicolumn{5}{l}{BPPO} &  & \multicolumn{5}{l}{BPAFT$^{\dagger}$} \\ \cline{3-7} \cline{9-14} 
      & true & est.  & se. & sde. & rb. & cov. & & true & est.   & se.   & sde.  & rb.  & cov.  \\ \hline
$\beta_1$ & -4    & -4.0230       & 0.4522      & 0.4726     & -0.5808   & 0.9420     & 
          &  2  & 2.0438 & 0.1315 & 0.1322 & 2.1915 & 0.9254      \\
$\beta_2$ &  2   & 2.0380      & 0.4745      & 0.4869      & 1.9125     & 0.9420  & 
          &  -1  &  -1.0218 & 0.1566 & 0.1629 & -2.1792 & 0.9254   \\ \hline
\end{tabular}
\begin{tabular}{rrrrrrrrrrrrrr}
\hline
Prior 3    &      & \multicolumn{5}{l}{BPPO} &  & \multicolumn{5}{l}{BPAFT$^{\dagger}$} \\ \cline{3-7} \cline{9-14} 
      & true & est.  & se. & sde. & rb. & cov. & & true & est.   & se.   & sde.  & rb.  & cov.  \\ \hline
$\beta_1$ & -4    & -3.8260      & 0.4149      & 0.3674     & 4.3410    & 0.9450     & 
          &  2 & 2.0810 & 0.1392 & 0.1354 & 4.0481 & 0.8955    \\
$\beta_2$ &  2   & 1.9410      & 0.4571      & 0.4409      & -2.9430     & 0.9530  & 
          &  -1 & -1.0410 & 0.1688 & 0.1598 & -4.1005 & 0.9552     \\ \hline
\end{tabular}
\begin{tabular}{rrrrrrrrrrrrrr}
\hline
Prior 4    &      & \multicolumn{5}{l}{BPPO} &  & \multicolumn{5}{l}{BPAFT$^{\dagger}$} \\ \cline{3-7} \cline{9-14} 
      & true & est.  & se. & sde. & rb. & cov. & & true & est.   & se.   & sde.  & rb.  & cov.  \\ \hline
$\beta_1$ & -4    & -4.1270       & 0.4676      & 0.5123     & -3.1690    & 0.9350     & 
          &  2  & 2.0486 & 0.1327 & 0.1334 & 2.4297 & 0.8955     \\
$\beta_2$ &  2   & 2.0840     & 0.4811      & 0.5054      & 4.1880
& 0.9350
& 
          &  -1 & -1.0240 & 0.1577 & 0.1634 & -2.3975 & 0.9254  \\ \hline
\end{tabular}
\caption{MC simulation study in Scenario II ($n = 200$), models fitted to the LLAFT data set.  Estimate of the regression coefficient (est.), average
standard error (se.), standard deviation of the estimates (sde.), relative bias (rb in \%) and
coverage probability (nominal level 95\%). Symbols: $a$ indicates $R = 855$ (145 non-finite Hessian matrices matrices); $b$ indicates $R = 983$ (24 non-converging and 7 non-finite Hessian matrices matrices); $c$ indicates $R= 973$ (27 non-converging); $d$ indicates $R = 969$ (51 non-converging); $\dagger$ indicates $R=100$. }\label{sim4}
\end{table}\normalsize

\newpage

{\flushleft \Large \textbf{Appendix C: Shoenfeld residuals analysis}} \vspace{15pt}

The assumption of proportional hazards (PH) is verified using the statistical tests and graphical diagnostics based on the Schoenfeld residuals. Ideally, if the model is well fitted, the Schoenfeld residuals should be random over time. For each covariate, the scatter-plot of the Schoenfeld residuals versus time can be used to evaluate the violation of the PH assumption. To test the independence between residuals and time, the scaled Schoenfeld statistic provided by the \texttt{survival::cox.zph} routine  is defined in \cite{therneau2000modeling}. Besides, an individual version of the test, for each covariate, can be also be derived from the multivariate version \citep{hosmer2008applied, Colosismo:2001, Collett2015}.

\begin{figure}[!htb]
\centering
      \includegraphics[width=\textwidth]{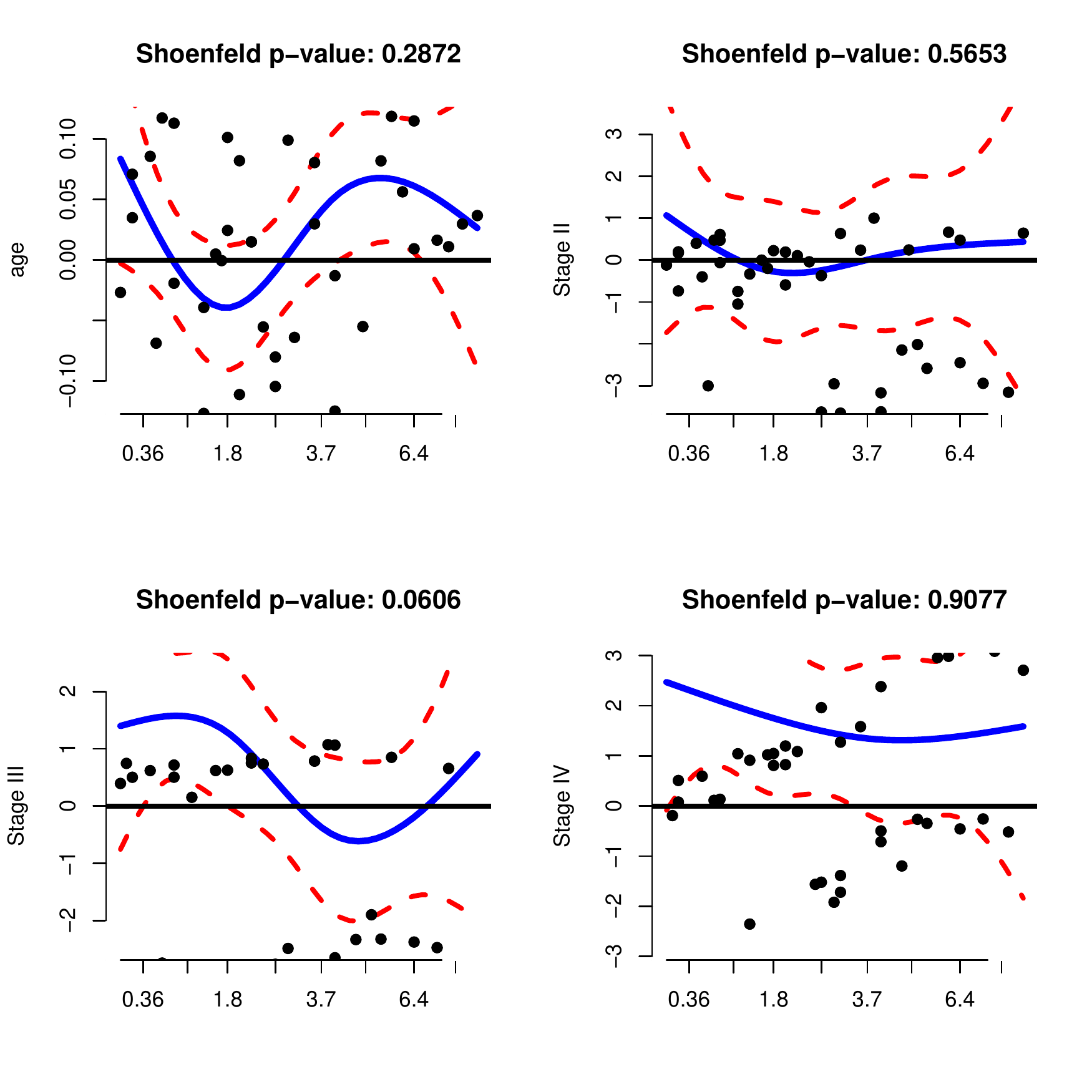} 
\caption{Schoenfeld residuals analysis in Application I. The panels show the p-value of the Schoenfeld test for each explanatory variable whose null hypothesis is that the residuals are independent of the time. The (blue) line represents the spline interpolation of the residuals and the dashed (red) lines illustrate its respective confidence intervals to assist the graphical analysis.}\label{scho1}
\end{figure}

\begin{figure}[!htb]
\centering
      \includegraphics[width=\textwidth]{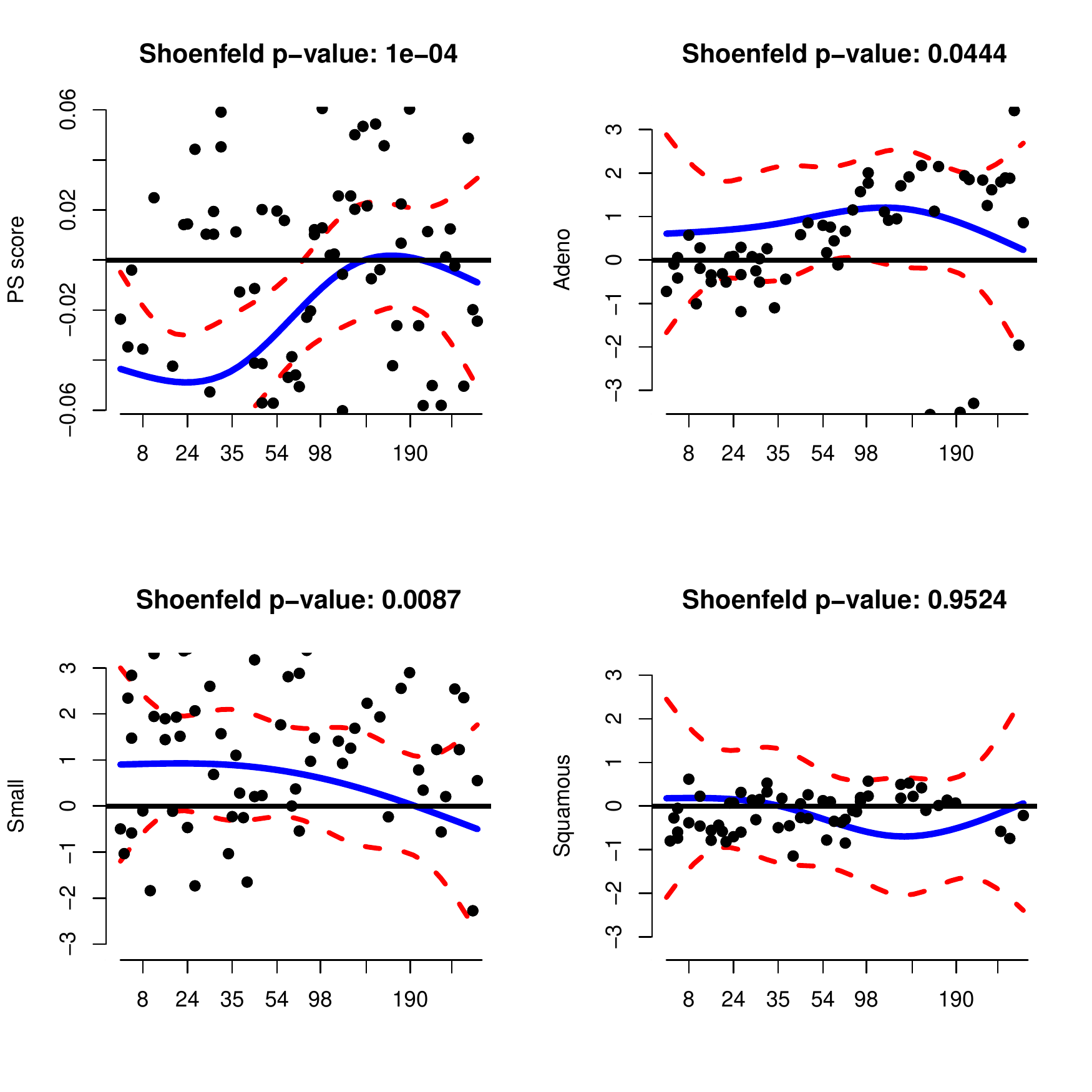} 
\caption{Schoenfeld residuals analysis in Application II. The panels show the p-value of the Schoenfeld test for each explanatory variable whose null hypothesis is that the residuals are independent of the time. The (blue) line represents the spline interpolation of the residuals and the dashed (red) lines illustrate its respective confidence intervals to assist the graphical analysis.}
\label{scho2}
\end{figure}

The p-value showed in Figures \oldref{scho1}, and \oldref{scho2} refer to the individual Shoenfeld test, whose null hypothesis is that the residuals and time are independent. Thus, if the p-value is lower than the significance value of $0.5$, adopted in this dissertation, this hypothesis is rejected, and the PH assumption is violated. The solid line (in blue) refers to the spline interpolation of the observed residuals, and the dashed (red) lines represent its respective confidence interval to assist the graphical analysis, that is, to identify an increasing (or decreasing) trending splines can be used to as the graphical evidence in favour of the PH assumption violation.

\newpage

\newpage

\newpage
{\flushleft \Large \textbf{Appendix D: Standard residuals analysis}} \vspace{15pt}

In Appendix D, the graphical model diagnosis is based on the residual analysis for the parametric AFT models applied in Chapter 6. Here we can identify if a model is well fitted to the data using graphical diagnosis methods. The parametric AFT model assumption is that the exponential to the standard residual follow the same standard distribution chosen to describe the time-to-event data. As a result, if the Kaplan- Meier estimate of the residual does not approximate to the standard distribution survival curve, thus, the model is not well fitted to data. In figure A.3, we can see the residual analysis for the application I. Figure A.4 shows analogous plots to the application II parametric AFT fit.   According to the Panels (a) and (b) plots showed it is possible to conclude that the parametric WAFT model is not well fitted to Application I data set. Conversely, according to Panels (c) and (d) we can concluded that the LLAFT model is well fitted to the data in application II. 
\newpage
\begin{figure}[!htb]
\centering
$$
  \begin{array}{cc}
     \mbox{\textbf{(a)} WAFT residuals scatter-plot} & \mbox{\textbf{(b)} WAFT survival curves} \\
     \includegraphics[width=.45\textwidth]{Figuras/distfit1.pdf} &
    \includegraphics[width=.45\textwidth]{Figuras/kmres1.pdf}\\
         \mbox{\textbf{(c)} LLAFT residuals scatter-plot} & \mbox{\textbf{(d)} LLAFT survival curves} \\
     \includegraphics[width=.45\textwidth]{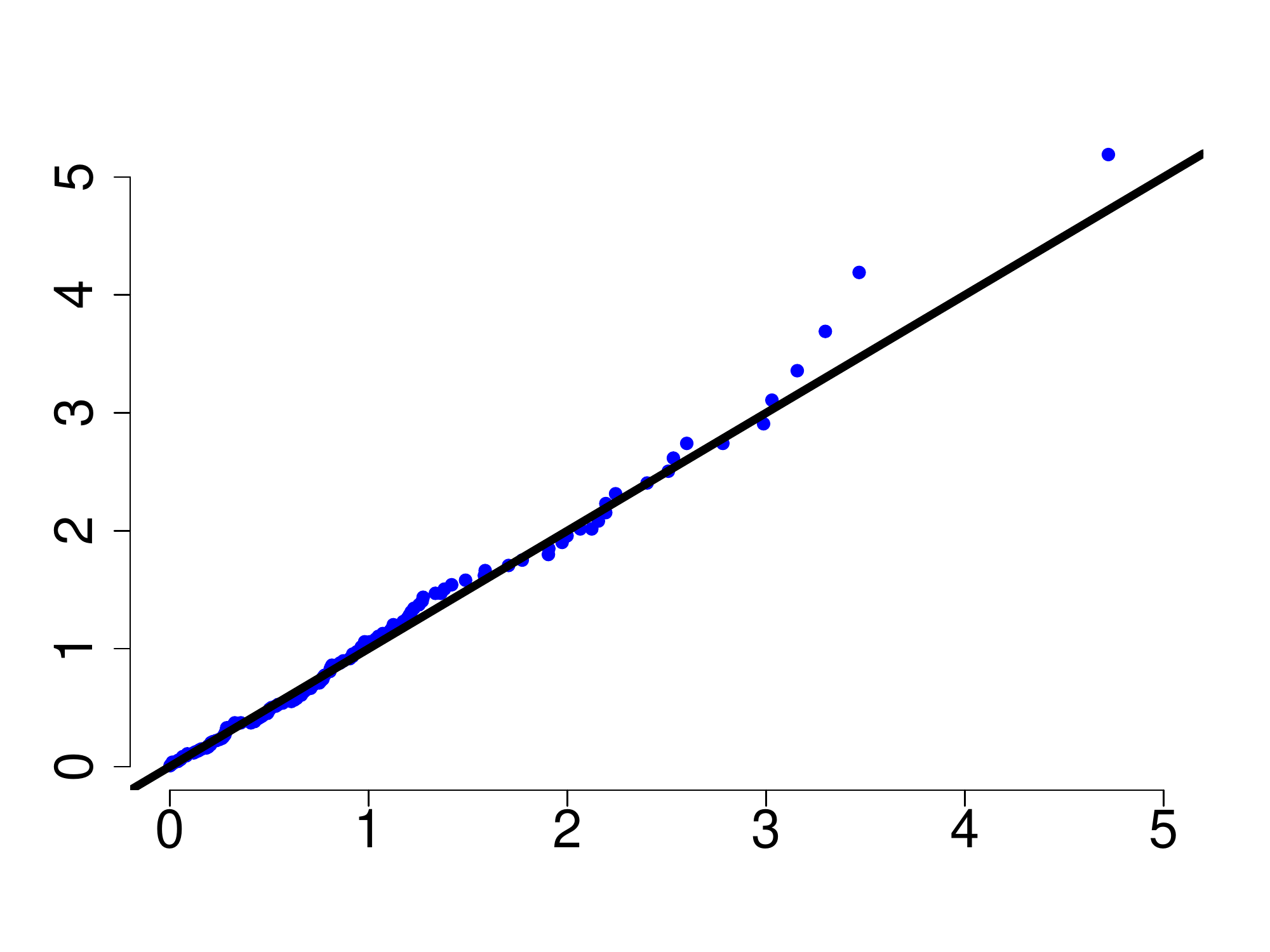} &
    \includegraphics[width=.45\textwidth]{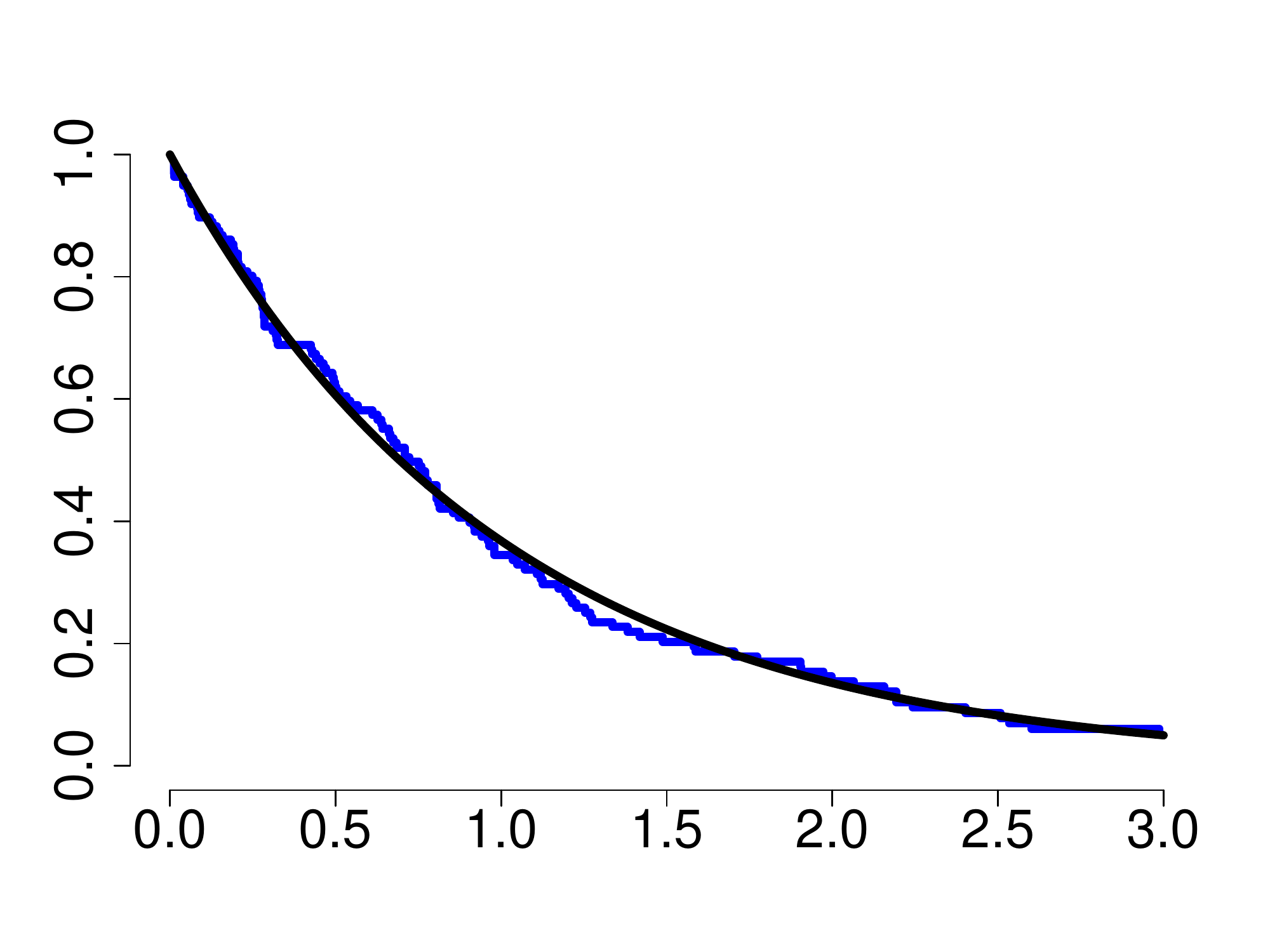}\\
  \end{array}
$$
\caption{\label{residwei} Cox-Snell residuals analysis for the WAFT model in Application I and for the LLAFT model in Application II. Panel \textbf{(a)} Scatter-plot of the Cox-Snell residuals against the cumulative hazard (Nelson Aalen) of Cox-Snell residuals; Panel \textbf{(b)} survival curves of the Exponential survival and the Kaplan-Meier estimated step curve for the residuals. In Panel \textbf{(a)}, the (black) solid line refers to the straight identity and the blue points to the paired survival values. In Panel  \textbf{(b)}, the dashed (black) line refers to the Exponential (mean 1) survival curve, and the step (blue) curve to the Kaplan-Meier curve of the residuals.}
\end{figure}\label{distfit1}
\newpage

{\flushleft \Large \textbf{Appendix E: Posterior plots}} \vspace{15pt}
\begin{figure}[!htb]
    \centering
    \includegraphics[width=.85\textwidth]{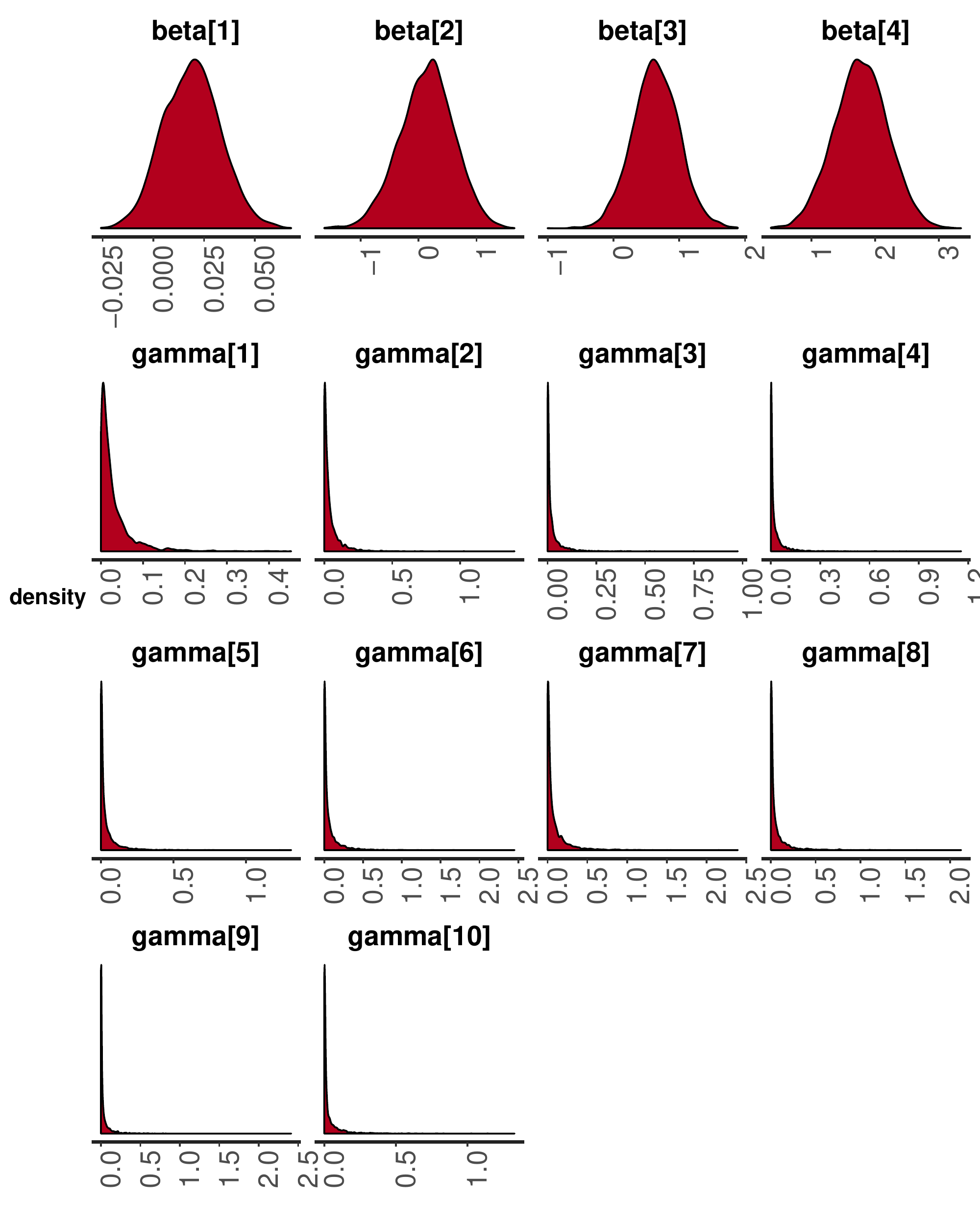}
    \caption{Posterior density plots for the posterior samples obtained through the MCMC applied to the BPPH model (Application I). The name of the parameter is identified at the top of the graphs. Recall that $\beta_1$ is related to age. In addition, $\beta_2, \beta_3$ and $\beta_4$
are connected with Stage II, III, and IV, respectively. The remaining chains are associated with the BP coefficients.}
     \label{spsurvtrace}
\end{figure} 

\newpage
\begin{figure}[!htb]
    \centering
    \includegraphics[width=.85\textwidth]{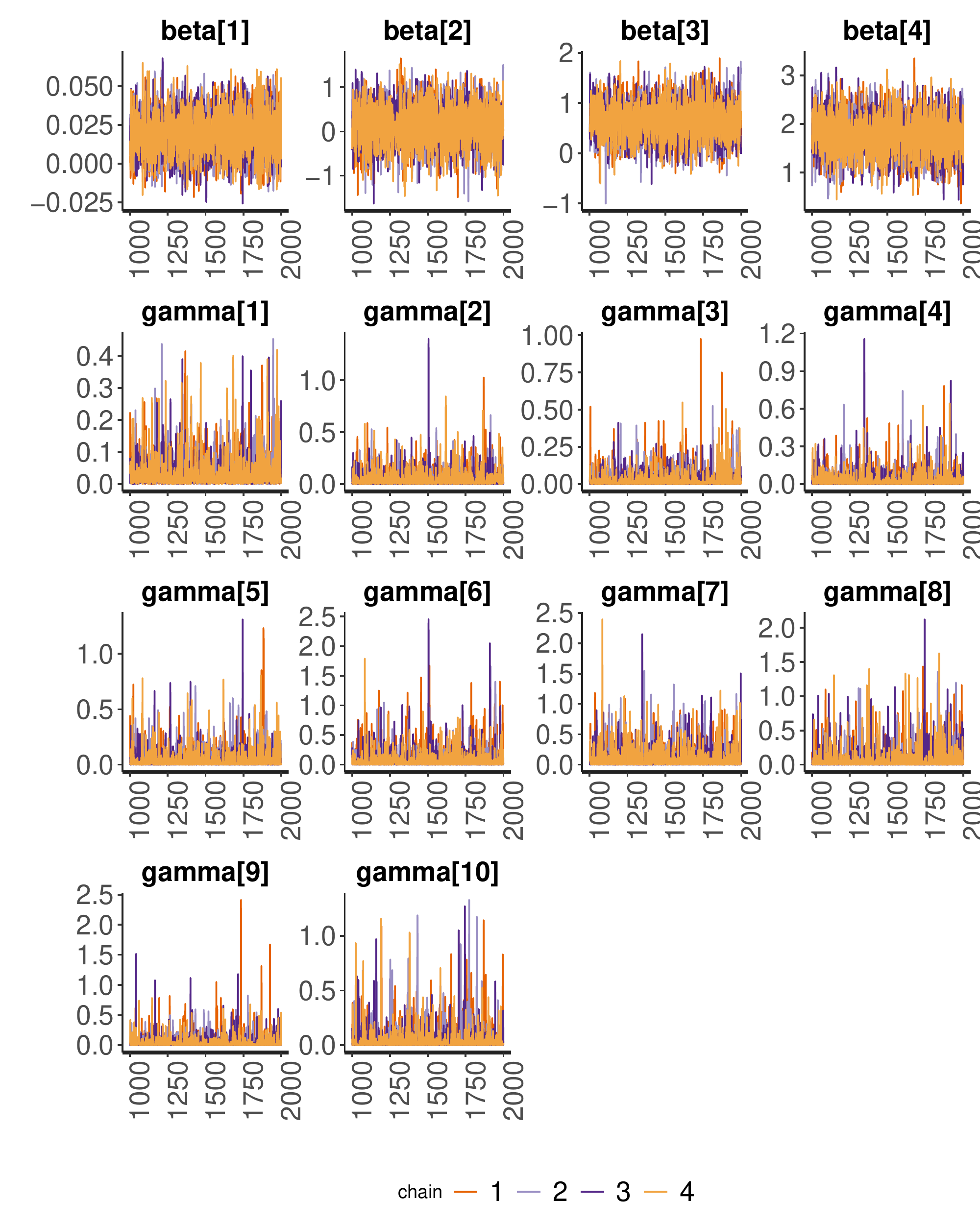} 
    \caption{Posterior trace plots for the posterior samples obtained through the MCMC applied to the BPPH model (Application I). The name of the parameter is identified at the top of the graphs. Recall that $\beta_1$ is related to age. In addition, $\beta_2, \beta_3$ and $\beta_4$
are connected with Stage II, III, and IV, respectively. The remaining chains are associated with the BP coefficients.}
    \label{spsurvchain}
\end{figure} 
\newpage

\begin{figure}[!htb]
    \centering
    \includegraphics[width=.85\textwidth]{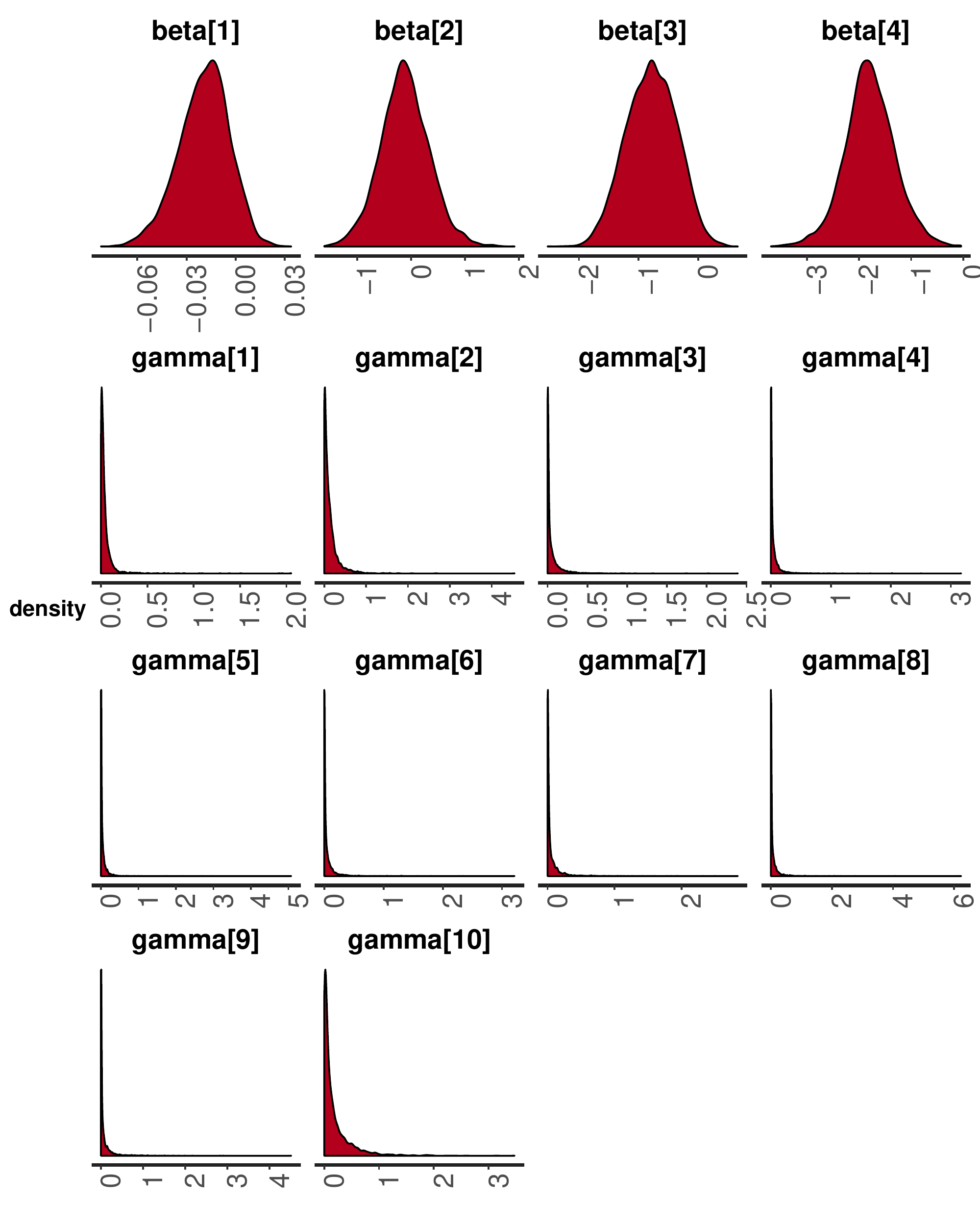}
    \caption{Posterior density plots for the posterior samples obtained through the MCMC applied to the BPAFT model (Application I). The name of the parameter is identified at the top of the graphs. Recall that $\beta_1$ is related to age. In addition, $\beta_2, \beta_3$ and $\beta_4$
are connected with Stage II, III, and IV, respectively. The remaining chains are associated with the BP coefficients.}
     \label{spsurvtrace}
\end{figure} 

\newpage
\begin{figure}[!htb]
    \centering
    \includegraphics[width=.85\textwidth]{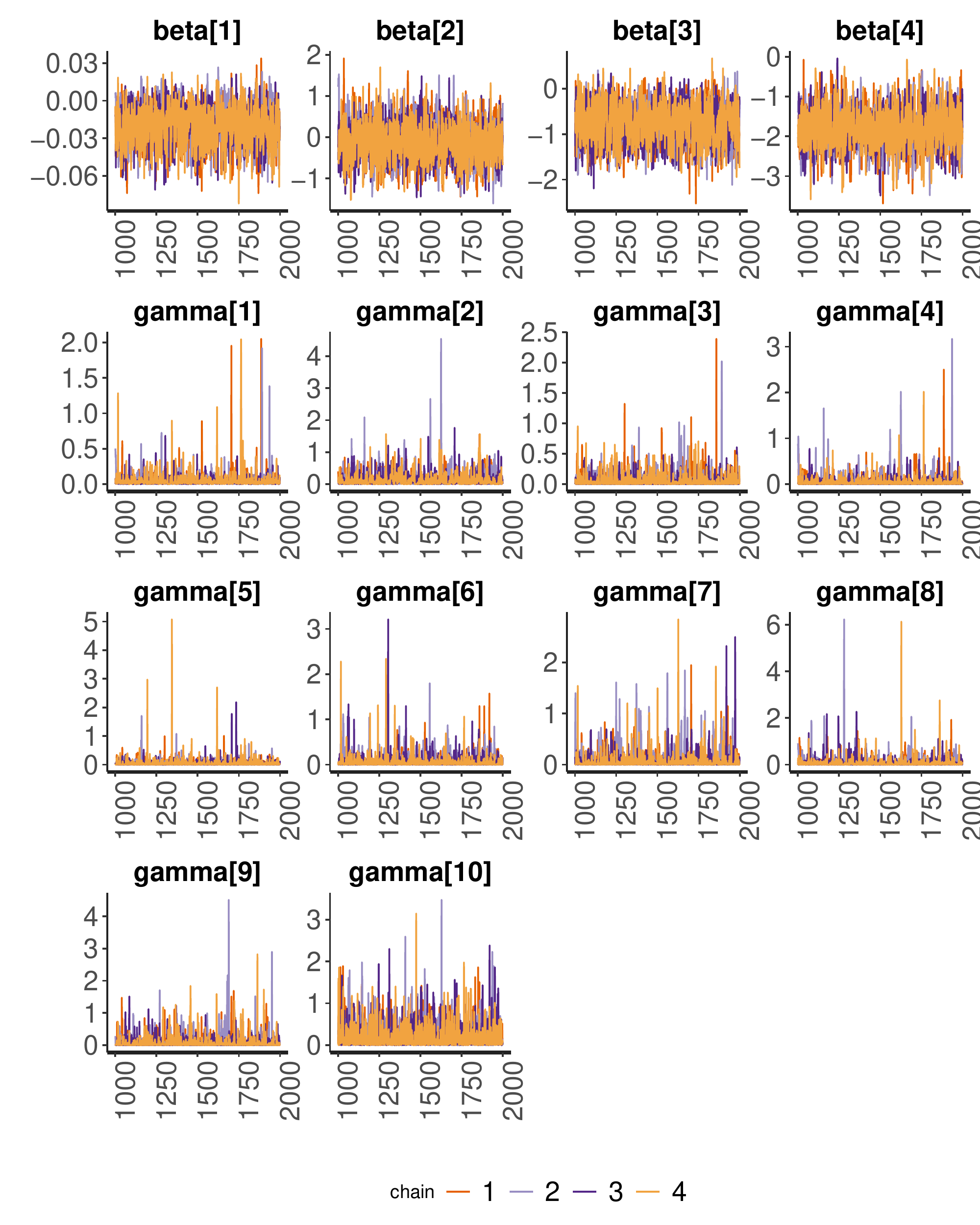} 
    \caption{Posterior trace plots for the posterior samples obtained through the MCMC applied to the BPAFT model (Application I). The name of the parameter is identified at the top of the graphs. Recall that $\beta_1$ is related to age. In addition, $\beta_2, \beta_3$ and $\beta_4$
are connected with Stage II, III, and IV, respectively. The remaining chains are associated with the BP coefficients.}
    \label{spsurvchain}
\end{figure} 
\newpage

\begin{figure}[!htb]
    \centering
    \includegraphics[width=.85\textwidth]{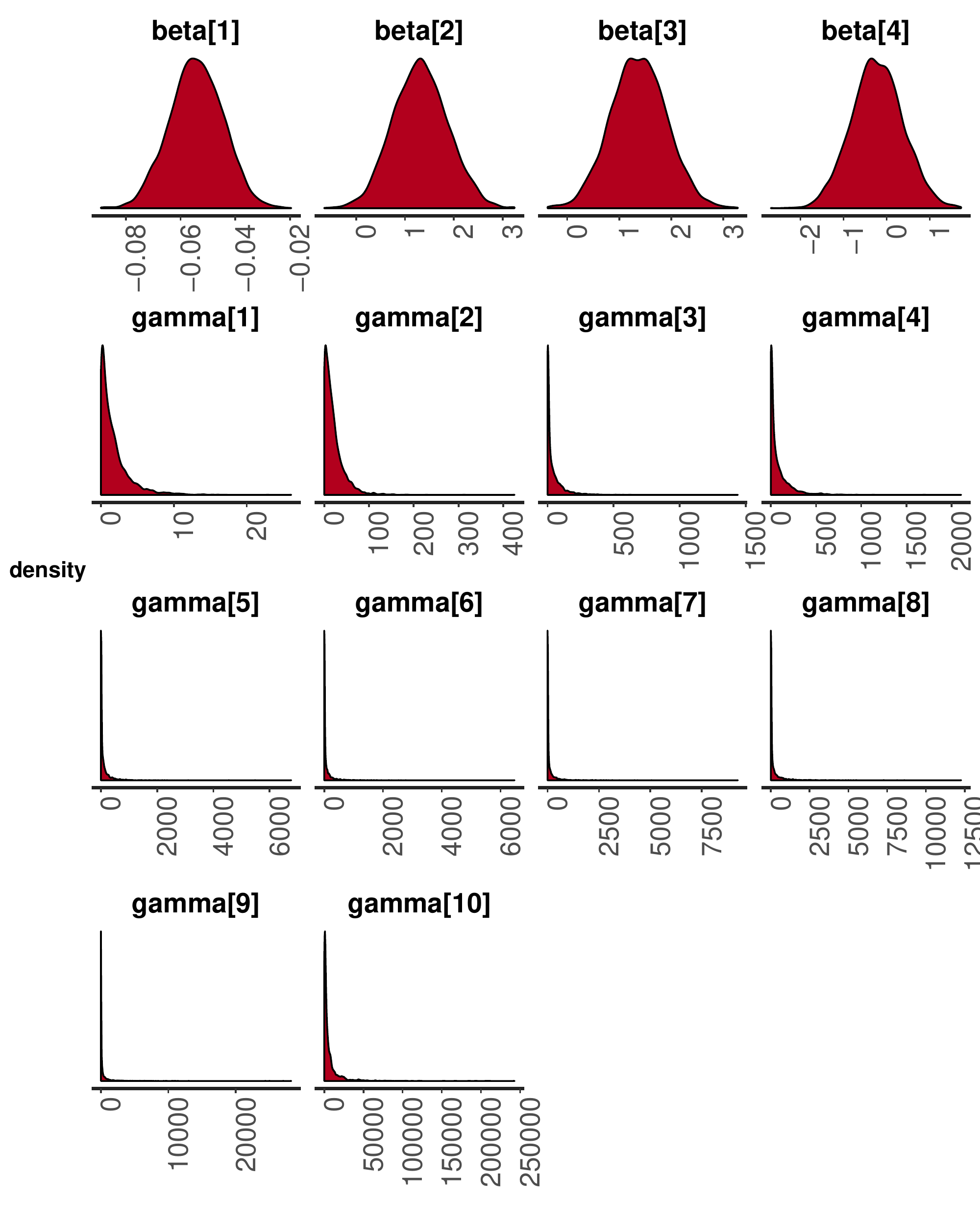}
    \caption{Posterior density plots for the posterior samples obtained through the MCMC applied to the BPPO model (Application II). The name of the parameter is identified at the top of the graphs. Recall that $\beta_1$ is related to PS score. In addition, $\beta_2, \beta_3$ and $\beta_4$
are connected with adeno cell, small cell, and squamous cell types, respectively. The remaining chains are associated with the BP coefficients.}
     \label{spsurvtrace}
\end{figure} 

\newpage
\begin{figure}[!htb]
    \centering
    \includegraphics[width=.85\textwidth]{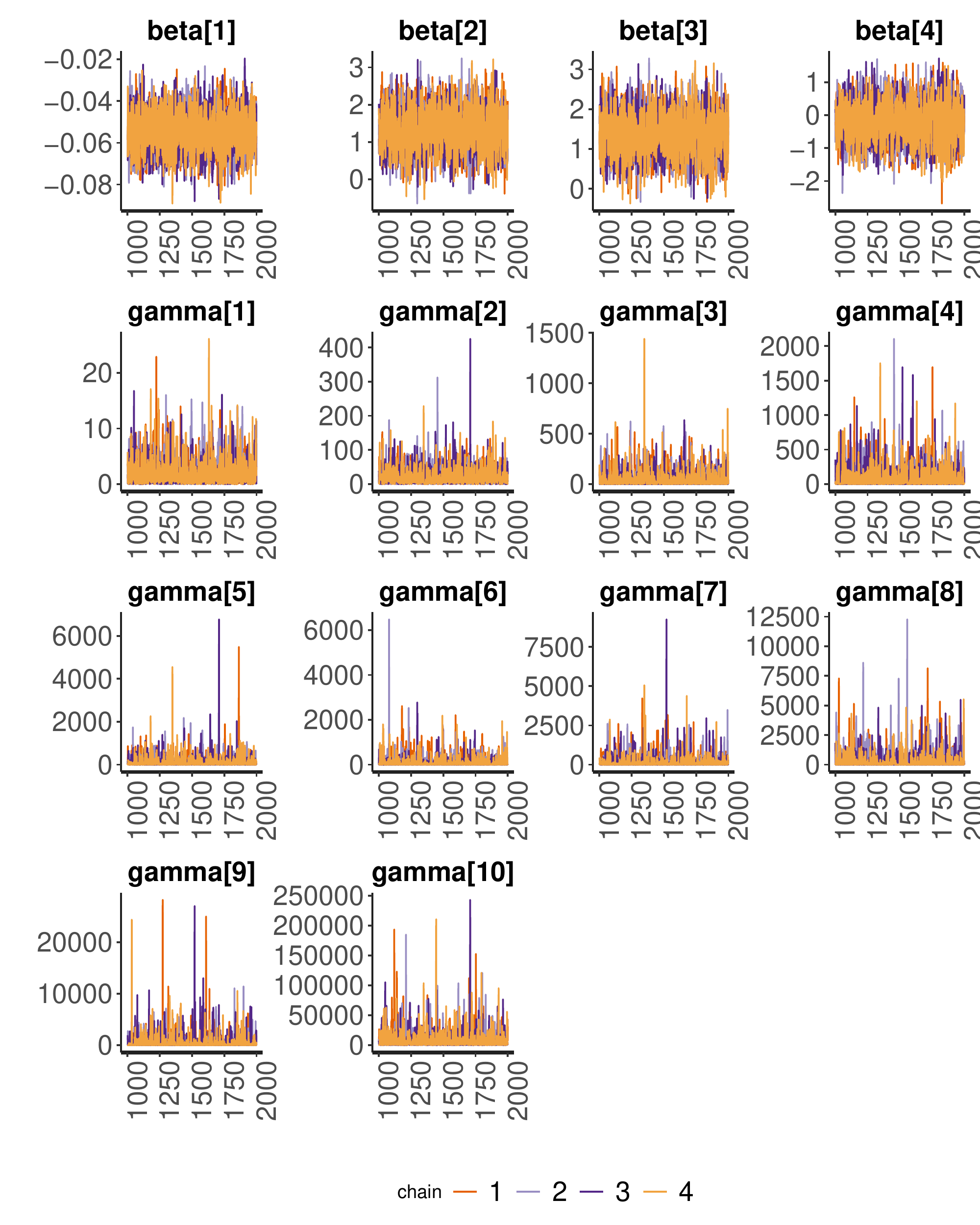} 
    \caption{Posterior trace plots for the posterior samples obtained through the MCMC applied to the BPPO model (Application II). The name of the parameter is identified at the top of the graphs. Recall that $\beta_1$ is related to PS score. In addition, $\beta_2, \beta_3$ and $\beta_4$
are connected with adeno cell, small cell, and squamous cell types, respectively. The remaining chains are associated with the BP coefficients.}
    \label{spsurvchain}
\end{figure} 
\newpage

\begin{figure}[!htb]
    \centering
    \includegraphics[width=.85\textwidth]{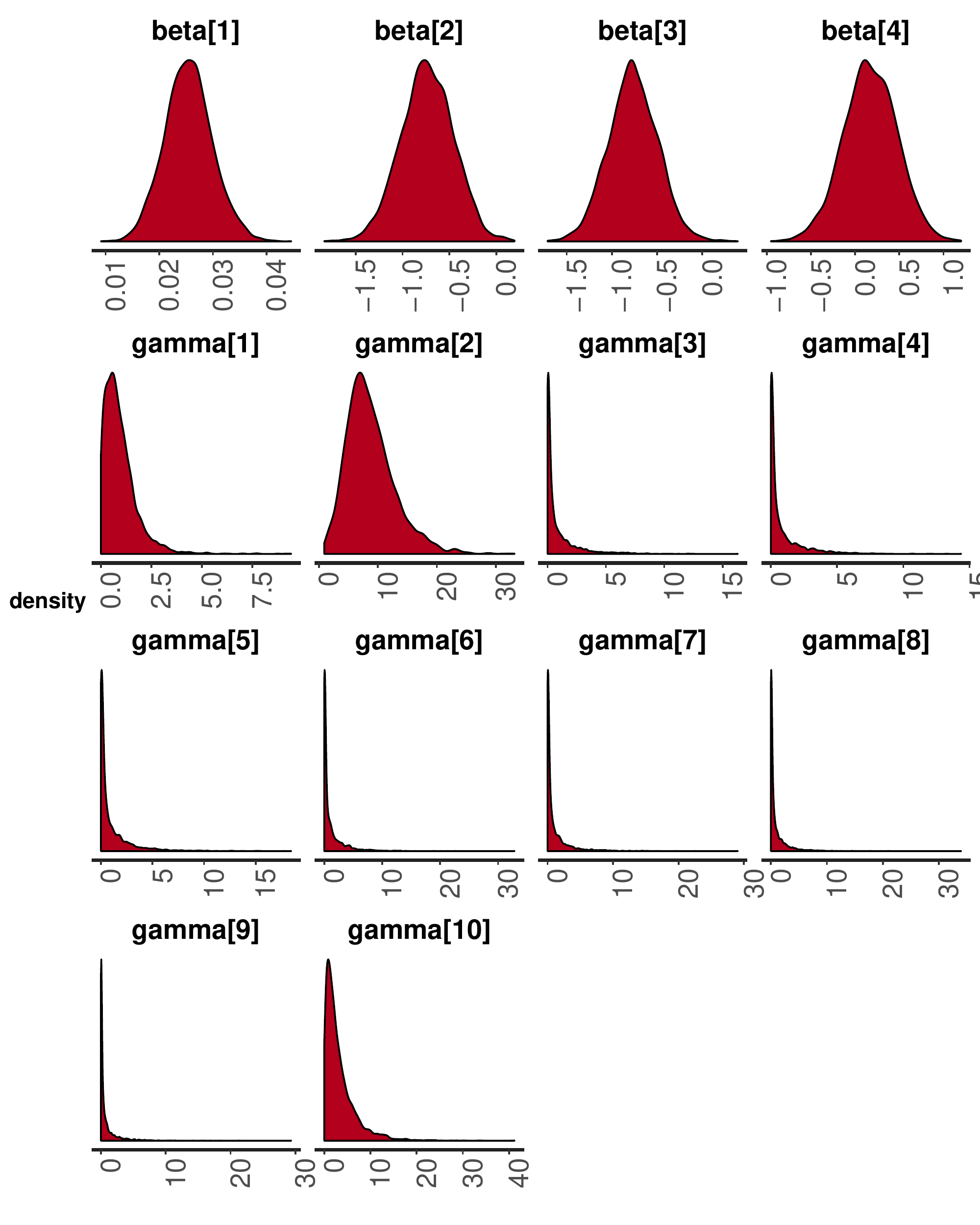}
    \caption{Posterior density plots for the posterior samples obtained through the MCMC applied to the BPAFT model (Application II). The name of the parameter is identified at the top of the graphs. Recall that $\beta_1$ is related to PS score. In addition, $\beta_2, \beta_3$ and $\beta_4$
are connected with adeno cell, small cell, and squamous cell types, respectively. The remaining chains are associated with the BP coefficients.}
     \label{spsurvtrace}
\end{figure} 

\newpage
\begin{figure}[!htb]
    \centering
    \includegraphics[width=.85\textwidth]{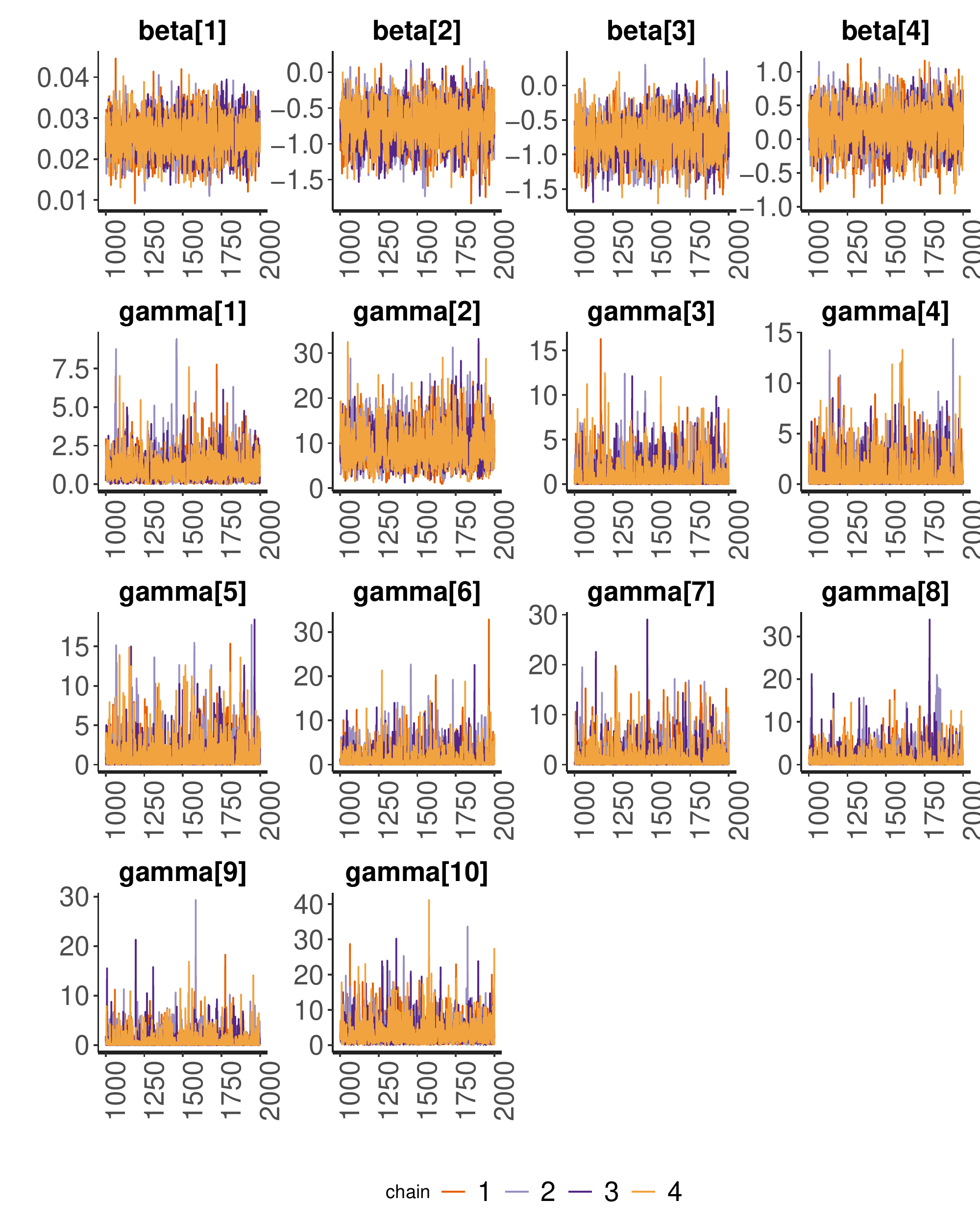} 
    \caption{Posterior trace plots for the posterior samples obtained through the MCMC applied to the BPAFT model (Application II). The name of the parameter is identified at the top of the graphs. Recall that $\beta_1$ is related to PS score. In addition, $\beta_2, \beta_3$ and $\beta_4$
are connected with adeno cell, small cell, and squamous cell types, respectively. The remaining chains are associated with the BP coefficients.}
    \label{spsurvchain}
\end{figure} 
\newpage
{\flushleft \Large \textbf{Appendix F: Box-plots of the percentage censoring}} \vspace{15pt}

\begin{figure}[!htb]
\centering
\includegraphics[width=\textwidth]{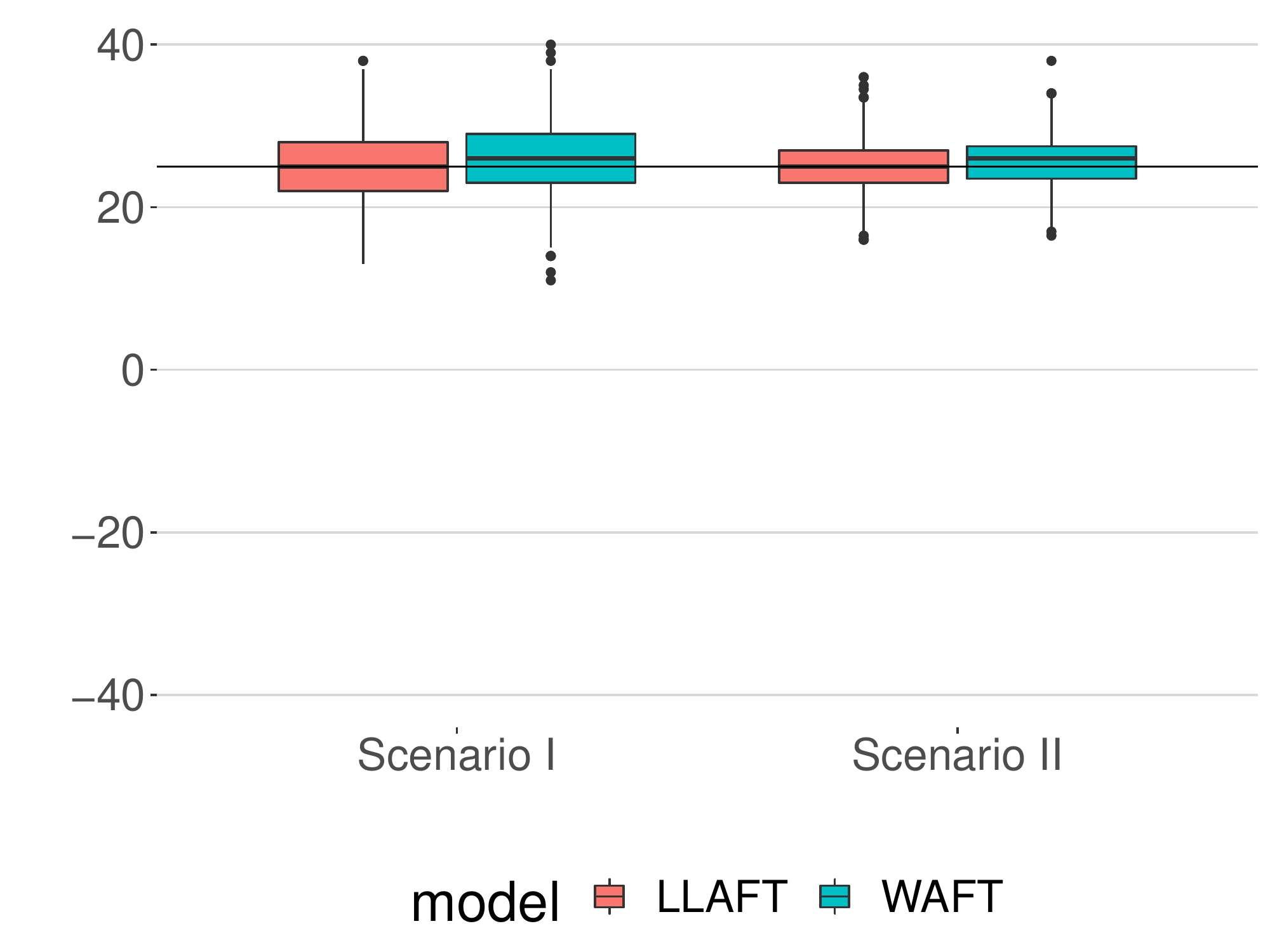}
\caption{\label{Fig:censoringrate}Percentage of censored observations in the simulated data. LLAFT refers to the Log Logistic accelerated failure time data generator model. WAFT refers to the  Weibull accelerated failure time data generator model.}
\end{figure} 

\newpage
{\flushleft \Large \textbf{Appendix G: Semi-parametric survival analysis with the \texttt{spsurv} package}} \vspace{15pt}

$~~~~$This Chapter gives the instructions on how to fit the BP based survival regression models using the new routines implemented in the \texttt{spsurv} package. The \texttt{spsurv::spbp}  function is the main routine of this package, as it allows to fit all BP related survival regression approaches presented in the previous chapters. The acronym \texttt{spbp} refers to ``\textit{semi-parametric Bernstein polynomial based regression}''. The \texttt{formula} argument in \texttt{spsurv::spbp} makes use of the same structure available at the \texttt{survival} package in order to provide a familiar environment to the public.
Indeed, the \texttt{spsurv} package imports  specific routines that provide the necessary support for internal calculations. During the installation, other dependencies are required, such as the libraries \texttt{survival}, \texttt{loo} \citep{loo}, \texttt{coda} \citep{coda},  \texttt{rstan} and \texttt{MASS} \citep{MASS}.

The target data set is passed to the \texttt{spsurv::spbp} function through a mandatory \texttt{data.frame} object class. Also, it is possible to switch between  ``\texttt{bayesian}'' and ``\texttt{mle}'' approaches through the ``\texttt{approach}'' argument and between the ``\texttt{ph}'',  ``\texttt{po}'' and ``\texttt{aft}'' frameworks using the \texttt{model} argument. Naturally, prior choices are ignored if the \texttt{approach} argument is set to ``\texttt{mle}'' (ML estimation); a warning is displayed in this case. In addition, consider extra arguments that may be passed directly to \texttt{Stan} software to apply \texttt{rstan::optimizing} (if ML method), or \texttt{rstan::sampling} (if MCMC method), function control options. As mentioned in Chapter 4, the polynomial degree (highest basis order) can be chosen arbitrarily. In particular, the polynomial degree must be greater than zero,  the default value of the polynomial degree  $\sqrt{n}$ is rounded to the greatest integer. Note that, the domain restriction for the BP, referred to as $\tau$ in this dissertation, is defined internally, see the discussion in Chapter 5. The reason for not allowing a user-defined $\tau$ is to avoid an improper definition that  would cause computing problems. 

Considering the variety of settings that \texttt{Stan} can provide and the modeling options above, we believe that the package is flexible regarding user-defined applications of the BP based models. Beyond that, a class, namely ``\texttt{spbp}'' was created to extend some S3 methods to meet the \texttt{R} community need for printing, summarizing, and plotting functions. Accordingly, we had developed S3 methods extensions to accomplish these tasks such as the \texttt{spsurv::print.spbp}, \texttt{spsurv::summary.spbp}, \texttt{spsurv::model.matrix.spbp} and other summary printing extensions such as \texttt{print.summary.bpph.bayes}. Further, there are some coding instructions on how to fit the semi-parametric models: BPPH, BPPO, and BPPAFT, under the Bayesian or Frequentist approach. In the Bayesian perspective, Normal prior distributions are attributed to the regression coefficients while Log-Normal, Gamma, or Inverse Gamma can be attributed to the BP parameters. The default arguments for the \texttt{spbp} functions were set as follows:
\begin{shaded}
\scriptsize
\begin{lstlisting}[language=R]
spbp.default <-
  spbp(formula, degree, data,
            approach = c("mle", "bayes"),
            model = c("ph", "po", "aft"),
            priors = list(beta = c("normal(0,4)"),
                         gamma = "lognormal(0,10)"),
           scale = TRUE,
           ...)
\end{lstlisting}
\end{shaded}
Consider \texttt{formula} an object of class formula, with the \texttt{Surv} object (\texttt{survival} package) for right censored time-to-event data on the left side of ``$\sim$'' and the explanatory terms on the right; \texttt{degree} for the integer value of the BP degree, non-integer values are rounded to the greatest valid degree; \texttt{data} for a mandatory  \texttt{data.frame} object with variables named in the formula; \texttt{approach} for either Bayes or ML estimation methods, default is ``\texttt{bayes}''; \texttt{model}
PH, PO or AFT for the modeling classes discussed in Chapter 2, default is ``\texttt{ph}''; \texttt{priors} list of prior settings, which is ignored when  ``\texttt{mle}'', and \texttt{"scale"} for a logical value that indicates whether to apply the standardization discussed in Chapter 5. Recall that extra arguments can be passed to \texttt{rstan::sampling} (e.g. \texttt{iter}, \texttt{chains}, \texttt{init}), more details in  \url{https://mc-stan.org/users/documentation/}.

Following, \cite{Klein:1997}, most statistical packages about survival regression returns an ANOVA table. In this sense, the object of class \texttt{spbp} follows the design provided in the \texttt{survival} package. The output corresponding to the ANOVA table can be obtained with: 
\begin{shaded}
\scriptsize
\begin{lstlisting}[language=R]
library("KMsurv")
data("larynx")

library(spsurv)
str(larynx)
fit <- spsurv::spbp(Surv(time, delta)~age+factor(stage),
                    approach = "mle",  data = larynx)
summary(fit)                    
\end{lstlisting}
\end{shaded}

One can reproduce this example by copying and pasting the indicated code in the \texttt{R} console. The output is as follows:
\begin{shaded}
\scriptsize
\begin{lstlisting}[language=R]
> library("KMsurv")
> data("larynx")
>   
> library(spsurv)
> fit <- spsurv::spbp(Surv(time, delta)~age+factor(stage),
+                     approach = "mle",  data = larynx)
<@\begin{verbatim}
Priors are ignored due to mle approach.
\end{verbatim}@>
> summary(fit)
<@\begin{verbatim}
Bernstein Polynomial based Proportional Hazards model
Call:
spbp.default(formula = Surv(time, delta) ~ age + factor(stage), 
    data = larynx, approach = "mle", model = "ph")

  n= 90, number of events= 50 

                 coef exp(coef) se(coef)    z Pr(>|z|)    
age            0.0193    1.0195   0.0144 1.34    0.180    
factor(stage)2 0.1720    1.1876   0.4626 0.37    0.710    
factor(stage)3 0.6585    1.9318   0.3556 1.85    0.064 .  
factor(stage)4 1.7991    6.0442   0.4288 4.20  2.7e-05 ***
---
Signif. codes:  0 ‘***’ 0.001 ‘**’ 0.01 ‘*’ 0.05 ‘.’ 0.1 ‘ ’ 1

Likelihood ratio test= 19.6  on 4 df,   p=6e-04
Wald test            = 22.6  on 4 df,   p=2e-04
\end{verbatim}@>
\end{lstlisting}
\end{shaded}
Consider that \texttt{coef} refers to the ML point estimates; \texttt{exp(coef)} is the point estimate for the hazard ratio; \texttt{se(coef)} represents the standard errors; \texttt{z}  is the test statistic for the Wald test and \texttt{p} is the p-value of the  Wald test. The estimated BP parameters, the value of the evaluated log-likelihood of the null (reference) model and the  \texttt{stan}  object can be obtained having access to the \texttt{spbp} class object elements.
Moreover, apart from the \texttt{fit} object, it is also possible to obtain the matrix of covariates, the covariance matrix and the likelihood value. This can be done using the following code:

\begin{shaded}
\scriptsize
\begin{lstlisting}[language=R]
> fit$coefficients
<@\begin{verbatim}
           age factor(stage)2 factor(stage)3 factor(stage)4         gamma1 
  1.926946e-02   1.719545e-01   6.584531e-01   1.799095e+00   1.687151e-02 
        gamma2         gamma3         gamma4         gamma5         gamma6 
  4.164110e-02   8.170170e-50   1.401060e-02   4.677284e-02   9.401934e-34 
        gamma7         gamma8         gamma9        gamma10 
  1.331014e-01   1.189283e-62  4.517613e-112  3.128413e-131  
\end{verbatim}@>
> head(model.matrix(fit))
<@\begin{verbatim}
  age factor(stage)2 factor(stage)3 factor(stage)4
1  77              0              0              0
2  53              0              0              0
3  45              0              0              0
4  57              0              0              0
5  58              0              0              0
6  51              0              0              0
\end{verbatim}@>
> diag(fit$var)
<@\begin{verbatim}
 [1]  2.064355e-04  2.139849e-01  1.264739e-01  1.838831e-01  3.846090e+56
 [6]  3.622219e+56  8.390800e+29  3.976866e+58  6.059285e+57  3.873257e-06
[11]  2.271415e+56  3.337039e-64 2.987286e-163 7.737132e-203
\end{verbatim}@>
> fit$loglik
<@\begin{verbatim}
[1] -149.8360 -140.0512
\end{verbatim}@>
\end{lstlisting}
\end{shaded}

From the Bayesian point of view, the \texttt{spbp} class contains posterior summary statistics such as the mode, median, mean and standard deviation, along with 95\% HPD interval based on the posterior density. Note that the arguments passed after \texttt{data} are considered \texttt{Stan} specific control parameters. For instance, the argument \texttt{chain} allows to choose the number of chains in the MCMC . Other settings such as \texttt{iter} and \texttt{warmup} are also flexible and might be set at convenience. The user can simply type in the \texttt{R}
console the code  
$``??\texttt{rstan::sampling}"$ for help. The following \texttt{R} console outcome refers to the Bayesian estimation for the \texttt{larynx} data set:

\begin{shaded}
\scriptsize
\begin{lstlisting}[language=R]
> fit <- spsurv::spbp(Surv(time, delta)~age+factor(stage),
+                     approach = "bayes",  data = larynx,
+                     iter = 2000, chains = 1, warmup = 1000)
<@\begin{verbatim}
SAMPLING FOR MODEL 'spbp' NOW (CHAIN 1).
Chain 1: 
Chain 1: Gradient evaluation took 9.2e-05 seconds
Chain 1: 1000 transitions using ten leapfrog steps per transition would take 0.92
seconds.
Chain 1: Adjust your expectations accordingly!
Chain 1: 
Chain 1: 
Chain 1: Iteration:    1 / 2000 [  0%]  (Warmup)
Chain 1: Iteration:  200 / 2000 [ 10%]  (Warmup)
Chain 1: Iteration:  400 / 2000 [ 20%]  (Warmup)
Chain 1: Iteration:  600 / 2000 [ 30%]  (Warmup)
Chain 1: Iteration:  800 / 2000 [ 40%]  (Warmup)
Chain 1: Iteration: 1000 / 2000 [ 50%]  (Warmup)
Chain 1: Iteration: 1001 / 2000 [ 50%]  (Sampling)
Chain 1: Iteration: 1200 / 2000 [ 60%]  (Sampling)
Chain 1: Iteration: 1400 / 2000 [ 70%]  (Sampling)
Chain 1: Iteration: 1600 / 2000 [ 80%]  (Sampling)
Chain 1: Iteration: 1800 / 2000 [ 90%]  (Sampling)
Chain 1: Iteration: 2000 / 2000 [100%]  (Sampling)
Chain 1: 
Chain 1:  Elapsed Time: 3.63369 seconds (Warm-up)
Chain 1:                2.53932 seconds (Sampling)
Chain 1:                6.17301 seconds (Total)
Chain 1: 
Warning messages:
1: Relative effective sample sizes ('r_eff' argument) not specified.
For models fit with MCMC, the reported PSIS effective sample sizes and 
MCSE estimates will be over-optimistic. 
2: Some Pareto k diagnostic values are slightly high.
See help('pareto-k-diagnostic') for details.
3: 2 (2.2%) p_waic estimates greater than 0.4. We recommend trying loo instead.
\end{verbatim}@>
\end{lstlisting}
\end{shaded}

As with the ML estimation, the summary method is extended to the \texttt{spsurv::spbp} class when applying to a Bayesian fit. Along with the regression estimates, this output also contains descriptive statistics for the posterior hazard ratio denoted by \texttt{\_exp} (in the console output) and the diagnosis statistics from the \texttt{loo} package. The effective sample size  \texttt{n\_eff} gives an estimate of the independent draws from the posterior distribution, and \texttt{Rhat} referred to as the potential scale reduction statistic,  is one of the useful ways to monitor whether a chain has converged to the equilibrium distribution. This statistic measures the ratio between the average variation of the samples within each chain and the variation of the combined samples in the chains; if the chains have not converged to a common distribution, this statistic will be greater than one \citep{Manual:2016}. It is worth noting that all, the information provided by the \texttt{Stan} output, including warnings, is passed to the final user. One can have access to the \texttt{stanfit} object with the \texttt{fit\$stanfit} command. In particular, one can have access to built-in plot functions and even to a \texttt{shiny} app (details in \url{https://shiny.rstudio.com/}) developed by \texttt{Stan} developer's team. The summary outcome is as follows:

\begin{shaded}
\scriptsize
\begin{lstlisting}
> summary(fit)
<@\begin{verbatim}
Bayesian Bernstein Polynomial based Proportional Hazards model

Call:
spbp.default(formula = Surv(time, delta) ~ age + factor(stage), 
    data = larynx, approach = "bayes", iter = 2000, chains = 1, 
    warmup = 1000, model = "ph")

  n= 0, number of events= 0 

                 mode  mean  se_mean     sd    50% n_eff  Rhat lowerHPD upperHPD
age            0.0172 0.019 0.000415 0.0148 0.0189  1276 1.000 -0.00821   0.0488
factor(stage)2 0.2135 0.126 0.015843 0.4634 0.1527   856 0.999 -0.81920   0.9446
factor(stage)3 0.6030 0.627 0.012488 0.3331 0.6197   712 0.999 -0.09428   1.2192
factor(stage)4 1.8345 1.765 0.013916 0.3984 1.7758   819 1.000  1.01502   2.5609
---
               mean_exp median_exp sd_exp lowerHPD_exp upperHPD_exp
age                1.02       1.02 0.0151        0.992         1.05
factor(stage)2     1.26       1.16 0.5825        0.326         2.31
factor(stage)3     1.98       1.86 0.6900        0.910         3.38
factor(stage)4     6.33       5.91 2.6719        2.118        11.49
---

Computed from 1000 by 90 log-likelihood matrix

          Estimate   SE
elpd_waic   -149.2  9.3
p_waic         8.6  0.8
waic         298.4 18.7

Computed from 1000 by 90 log-likelihood matrix

         Estimate   SE
elpd_loo   -149.3  9.3
p_loo         8.7  0.9
looic       298.7 18.7
------
Monte Carlo SE of elpd_loo is 0.1.

Pareto k diagnostic values:
                         Count Pct.    Min. n_eff
(-Inf, 0.5]   (good)     88    97.8%   507       
 (0.5, 0.7]   (ok)        2     2.2%   370       
   (0.7, 1]   (bad)       0     0.0%   <NA>      
   (1, Inf)   (very bad)  0     0.0%   <NA>      

All Pareto k estimates are ok (k < 0.7).
See help('pareto-k-diagnostic') for details.
Warning message:
2 (2.2%) p_waic estimates greater than 0.4. We recommend trying loo instead. 
\end{verbatim}@>
\end{lstlisting}
\end{shaded}
The next code chunk shows the code for trace and density plotting and to give access to the \texttt{shiny} app from the \texttt{shinystan} package  \citep{shinystan}.
Figures \oldref{spsurvtrace} and \oldref{spsurvchain} illustrate the trace plot and the density plot of the BPPH for the \texttt{larynx} data set. The graphs show unimodal posterior densities and well behaved chains with good mixing, this is a good behavior indication. 
 \begin{shaded}
 \scriptsize
\begin{lstlisting}
rstan::traceplot(fit$stanfit, pars = c("beta", "gamma"))
rstan::stan_dens(fit$stanfit, pars = c("beta", "gamma"))
shinystan::launch_shinystan(fit$stanfit)
\end{lstlisting}
\end{shaded}

Not least, a S3 method had to be created rather than extended. The \texttt{survivor} method was created to accomplish the calculation of the survival function evaluated in each time point. The goal is similar to the \texttt{survival::survfit} S3 method, that could be extended instead. The difference is that \texttt{spbp} classes allows both Bayesian and Frequentist approaches. The following code was used to generate Figure \oldref{Fig:surv}: 

\begin{shaded}
 \scriptsize
\begin{lstlisting}
## CoxPH model
fitcoxph <- survival::coxph(Surv(time , delta)~age+factor(stage),
data = larynx)

## Determine the groups of patients
newdata <-  data.frame(age =c(77,77,77,77), stage = c(1,2,3,4))

## survfit Breslow estimator
breslowsurv <- survival::survfit(fitcoxph, newdata = newdata)

## spbp point-wise estimate
spbpsurv <- spbp::survivor(fit, newdata = newdata)

plot(breslowsurv)
points(spbpsurv)
\end{lstlisting}
\end{shaded}

The content of this dissertation is now complete. Here, the analysis was dedicated to illustrating, in practice, the commands
implemented in the proposed package \texttt{spsurv}. We still have work to do to improve and update this tool, however,
the present version is ready for the main statistical study in the field of survival analysis. The routines presented in this dissertation are unprecedented. Therefore, they have not yet been published in The Comprehensive R Archive Network (CRAN). Many efforts with regard to the instruction manuals and routines documentation were carried out concurrently with the \texttt{spsurv} package implementation. The package is already in public use and is available at the  \texttt{github} development platform, the link is: \url{https: //github.com/rvpanaro/spsurv}. The submission to CRAN was be made after the comments and suggestions  concerning this dissertation.

{\bibliographystyle{jasa}
\renewcommand{\refname}{\normalsize References}
\bibliography{references}}

\begin{thebibliography}{}
\newcommand{\enquote}[1]{``#1''}

\bibitem[Aalen et~al.(2008)Aalen, Borgan, and Gjessing]{aalen2008survival}
Aalen, O., Borgan, O., and Gjessing, H. (2008), \emph{Survival and event
  history analysis: a process point of view}, Springer, New York.

\bibitem[Akalke(1974)Akalke]{akalke1974new}
Akalke, H. (1974), \enquote{A new look at the statistical model
  identification,} \emph{IEEE transactions on automatic control}, 19, 716723.

\bibitem[Ald{\`a} and Rubinstein(2017)Ald{\`a} and Rubinstein]{Alda:2017}
Ald{\`a}, F. and Rubinstein, B.~I. (2017), \enquote{The Bernstein mechanism:
  Function release under differential privacy,} \emph{Thirty-First AAAI
  Conference on Artificial Intelligence,
  http://www.aaai.org/Conferences/AAAI/2017/PreliminaryPapers/12-Alda-14735.pdf}.

\bibitem[Babu et~al.(2002)Babu, Canty, and Chaubey]{Babu:2002}
Babu, G.~J., Canty, A.~J., and Chaubey, Y.~P. (2002), \enquote{Application of
  Bernstein polynomials for smooth estimation of a distribution and density
  function,} \emph{Journal of Statistical Planning and Inference}, 105,
  377--392.

\bibitem[Bennett(1983)Bennett]{Bennett:1983}
Bennett, S. (1983), \enquote{Analysis of survival data by the proportional odds
  model,} \emph{Statistics in Medicine}, 2, 273--277.

\bibitem[Bernstein(2009)Bernstein]{Bernstein:2009}
Bernstein, D.~S. (2009), \emph{Matrix mathematics: theory, facts, and
  formulas}, Princeton University Press, Princeton.

\bibitem[Bernstein(1912)Bernstein]{Bernstein:1912}
Bernstein, S.~N. (1912), \enquote{(in Russian) On the best approximation of
  continuous functions by polynomials of a given degree,} \emph{Communications
  of the Mathematical Society}, Series 2, 49--194.

\bibitem[Breslow(1972)Breslow]{Breslow:1972}
Breslow, N.~E. (1972), \enquote{Contribution to discussion of paper by DR Cox,}
  \emph{Journal of the Royal Statistical Society, Series B}, 34, 216--217.

\bibitem[Carpenter et~al.(2017)Carpenter, Gelman, Hoffman, Lee, Goodrich,
  Betancourt, Brubaker, Guo, Li, and Riddell]{Carpenter:2017}
Carpenter, B., Gelman, A., Hoffman, M.~D., Lee, D., Goodrich, B., Betancourt,
  M., Brubaker, M., Guo, J., Li, P., and Riddell, A. (2017), \enquote{Stan: A
  probabilistic programming language,} \emph{Journal of Statistical Software},
  76.

\bibitem[Casella and Berger(2002)Casella and Berger]{Casella:2002}
Casella, G. and Berger, R.~L. (2002), \emph{Statistical inference}, vol.~2,
  Duxbury, Pacific Grove.

\bibitem[Chang et~al.(2005)Chang, Hsiung, Wu, and Yang]{Chang:2005}
Chang, I.~S., Hsiung, C.~A., Wu, Y.~J., and Yang, C.~C. (2005),
  \enquote{Bayesian survival analysis using Bernstein polynomials,}
  \emph{Scandinavian Journal of Statistics}, 32, 447--466.

\bibitem[Chang et~al.(2007)Chang, Chien, Hsiung, Wen, Wu, et~al.]{Chang:2007}
Chang, I.-S., Chien, L.-C., Hsiung, C.~A., Wen, C.-C., Wu, Y.-J., et~al.
  (2007), \enquote{Shape restricted regression with random Bernstein
  polynomials,} in \emph{Complex datasets and inverse problems}, pp. 187--202,
  Institute of Mathematical Statistics, Beachwood.

\bibitem[Chen et~al.(2014)Chen, Hanson, and Zhang]{Chen2:2014}
Chen, Y., Hanson, T., and Zhang, J. (2014), \enquote{Accelerated hazards model
  based on parametric families generalized with Bernstein polynomials,}
  \emph{Biometrics}, 70, 192--201.

\bibitem[Choudhuri et~al.(2004)Choudhuri, Ghosal, and Roy]{Choudhuri:2004}
Choudhuri, N., Ghosal, S., and Roy, A. (2004), \enquote{Bayesian estimation of
  the spectral density of a time series,} \emph{Journal of the American
  Statistical Association}, 99, 1050--1059.

\bibitem[Christensen et~al.(2011)Christensen, Johnson, Branscum, and
  Hanson]{christensen2011bayesian}
Christensen, R., Johnson, W., Branscum, A., and Hanson, T.~E. (2011),
  \emph{Bayesian ideas and data analysis: an introduction for scientists and
  statisticians}, CRC Press, Boca Raton.

\bibitem[Cicho{\'n} and Go{\l}{\k{e}}biewski(2012)Cicho{\'n} and
  Go{\l}{\k{e}}biewski]{cichon:2012}
Cicho{\'n}, J. and Go{\l}{\k{e}}biewski, Z. (2012), \enquote{On Bernoulli sums
  and Bernstein polynomials,} p. 179—190, Nancy, France, Discrete Mathematics
  and Theoretical Computer Science (DMTCS).

\bibitem[Collett(2015)Collett]{Collett2015}
Collett, D. (2015), \emph{Modelling survival data in medical research}, Chapman
  and Hall/CRC, Boca Raton.

\bibitem[Colosimo and Giolo(2006)Colosimo and Giolo]{Colosismo:2001}
Colosimo, E.~A. and Giolo, S.~R. (2006), \emph{Análise de sobrevivência
  aplicada}, Edgard Blucher, São Paulo.

\bibitem[Cooch(2008)Cooch]{cooch2008program}
Cooch, E. (2008), \enquote{Program MARK," A gentle introduction",}
  \emph{http://www. phidot. org/software/mark/docs/book/}.

\bibitem[Cox(1972)Cox]{Cox:1972}
Cox, D.~R. (1972), \enquote{Regression models and life-tables,} \emph{Journal
  of the Royal Statistical Society, Series B}, 34, 187--202.

\bibitem[Davis(1963)Davis]{Davis:1963}
Davis, P.~J. (1963), \emph{Interpolation and approximation}, Blaisdell, New
  York.

\bibitem[De~Iorio et~al.(2009)De~Iorio, Johnson, M{\"u}ller, and
  Rosner]{DeIorio:2009}
De~Iorio, M., Johnson, W.~O., M{\"u}ller, P., and Rosner, G.~L. (2009),
  \enquote{Bayesian nonparametric nonproportional hazards survival modeling,}
  \emph{Biometrics}, 65, 762--771.

\bibitem[Demarqui et~al.(2019)Demarqui, Mayrink, and Ghosh]{demarqui:2019}
Demarqui, F.~N., Mayrink, V.~D., and Ghosh, S.~K. (2019), \enquote{An Unified
  Semiparametric Approach to Model Lifetime Data with Crossing Survival
  Curves,} \emph{arXiv preprint arXiv:1910.04475}.

\bibitem[Duane et~al.(1987)Duane, Kennedy, Pendleton, and Roweth]{Duane:1987}
Duane, S., Kennedy, A.~D., Pendleton, B.~J., and Roweth, D. (1987),
  \enquote{Hybrid Monte Carlo,} \emph{Physics Letters B}, 195, 216--222.

\bibitem[Farouki(2008)Farouki]{Farouki:2008}
Farouki, R. (2008), \emph{Pythagorean-hodograph Curves: Algebra and Geometry
  Inseparable (9783540733973)}, Springer, New York.

\bibitem[Farouki(2012)Farouki]{farouki2012bernstein}
Farouki, R.~T. (2012), \enquote{The Bernstein polynomial basis: A centennial
  retrospective,} \emph{Computer Aided Geometric Design}, 29, 379--419.

\bibitem[Farouki and Rajan(1987)Farouki and Rajan]{Farouki:1987}
Farouki, R.~T. and Rajan, V. (1987), \enquote{On the numerical condition of
  polynomials in Bernstein form,} \emph{Computer Aided Geometric Design}, 4,
  191--216.

\bibitem[Farouki and Rajan(1988)Farouki and Rajan]{Farouki:1988}
Farouki, R.~T. and Rajan, V. (1988), \enquote{Algorithms for polynomials in
  Bernstein form,} \emph{Computer Aided Geometric Design}, 5, 1--26.

\bibitem[Fletcher(2000)Fletcher]{fletcher2000practical}
Fletcher, R. (2000), \emph{Practical methods of optimization}, John Wiley and
  Sons, 2 edition, Chichester.

\bibitem[Gabry(2018)Gabry]{shinystan}
Gabry, J. (2018), \emph{shinystan: Interactive Visual and Numerical Diagnostics
  and Posterior Analysis for Bayesian Models}, R package version 2.5.0.

\bibitem[Geisser and Eddy(1979)Geisser and Eddy]{geisser1979predictive}
Geisser, S. and Eddy, W.~F. (1979), \enquote{A predictive approach to model
  selection,} \emph{Journal of the American Statistical Association}, 74,
  153--160.

\bibitem[Gjessing et~al.(2010)Gjessing, R{\o}ysland, Pena, and
  Aalen]{gjessing2010recurrent}
Gjessing, H.~K., R{\o}ysland, K., Pena, E.~A., and Aalen, O.~O. (2010),
  \enquote{Recurrent events and the exploding Cox model,} \emph{Lifetime Data
  Analysis}, 16, 525--546.

\bibitem[Glidden and Vittinghoff(2004)Glidden and
  Vittinghoff]{glidden2004modelling}
Glidden, D.~V. and Vittinghoff, E. (2004), \enquote{Modelling clustered
  survival data from multicentre clinical trials,} \emph{Statistics in
  medicine}, 23, 369--388.

\bibitem[Gutierrez(2002)Gutierrez]{gutierrez2002parametric}
Gutierrez, R.~G. (2002), \enquote{Parametric frailty and shared frailty
  survival models,} \emph{The Stata Journal}, 2, 22--44.

\bibitem[Gzyl and Palacios(1997)Gzyl and Palacios]{Gzyl:1997}
Gzyl, H. and Palacios, J.~L. (1997), \enquote{The Weierstrass approximation
  theorem and large deviations,} \emph{The American Mathematical Monthly}, 104,
  650--653.

\bibitem[Hacking(2006)Hacking]{Hacking:2006}
Hacking, I. (2006), \emph{The emergence of probability: A philosophical study
  of early ideas about probability, induction and statistical inference},
  Cambridge University Press, Cambridge.

\bibitem[Hoffman and Gelman(2014)Hoffman and Gelman]{Hoffman:2014}
Hoffman, M.~D. and Gelman, A. (2014), \enquote{The No-U-Turn sampler:
  adaptively setting path lengths in Hamiltonian Monte Carlo.} \emph{Journal of
  Machine Learning Research}, 15, 1593--1623.

\bibitem[Hosmer~Jr et~al.(2008)Hosmer~Jr, Lemeshow, and May]{hosmer2008applied}
Hosmer~Jr, D.~W., Lemeshow, S., and May, S. (2008), \emph{Applied survival
  analysis: regression modeling of time-to-event data}, vol. 618,
  Wiley-Interscience.

\bibitem[Ibrahim et~al.(2001)Ibrahim, Chen, and Sinha]{Ibrahim:2014}
Ibrahim, J.~G., Chen, M.-H., and Sinha, D. (2001), \emph{Bayesian Survival
  Analysis}, Springer, New York.

\bibitem[Jackson(2016)Jackson]{flexsurv:2016}
Jackson, C. (2016), \enquote{{flexsurv}: A Platform for Parametric Survival
  Modeling in {R},} \emph{Journal of Statistical Software}, 70, 1--33.

\bibitem[Kalbfleisch and Prentice(2011)Kalbfleisch and
  Prentice]{Kalbfleisch:2011}
Kalbfleisch, J.~D. and Prentice, R.~L. (2011), \emph{The statistical analysis
  of failure time data}, vol. 360, John Wiley and Sons, Hoboken.

\bibitem[Kardaun(1983)Kardaun]{Kardaun:1983}
Kardaun, O. (1983), \enquote{Statistical survival analysis of male
  larynx-cancer patients-a case study,} \emph{Statistica Neerlandica}, 37,
  103--125.

\bibitem[Karnofsky(1949)Karnofsky]{karnofsky1949clinical}
Karnofsky, D.~A. (1949), \enquote{The clinical evaluation of chemotherapeutic
  agents in cancer,} \emph{Evaluation of chemotherapeutic agents}, pp.
  191--205.

\bibitem[Kassambara and Kosinski(2018)Kassambara and Kosinski]{survminer:2018}
Kassambara, A. and Kosinski, M. (2018), \emph{survminer: drawing survival
  curves using 'ggplot2'}, R package version 0.4.3.

\bibitem[Klein and Moeschberger(1997)Klein and Moeschberger]{Klein:1997}
Klein, J.~P. and Moeschberger, M.~L. (1997), \emph{Survival analysis:
  techniques for censored and truncated data}, Springer Science and Business
  Media, Berlin.

\bibitem[Koralov and Sinai(2007)Koralov and Sinai]{Kolarov:2007}
Koralov, L. and Sinai, Y.~G. (2007), \emph{Theory of probability and random
  processes}, Springer Science and Business Media, Berlin.

\bibitem[Lambert(2007)Lambert]{lambert2007modeling}
Lambert, P.~C. (2007), \enquote{Modeling of the cure fraction in survival
  studies,} \emph{The Stata Journal}, 7, 351--375.

\bibitem[Lawless(2011)Lawless]{lawless2011statistical}
Lawless, J.~F. (2011), \emph{Statistical models and methods for lifetime data},
  vol. 362, John Wiley \& Sons, Hoboken.

\bibitem[Li et~al.(2015)Li, Han, Hou, Chen, and Chen]{li2015statistical}
Li, H., Han, D., Hou, Y., Chen, H., and Chen, Z. (2015), \enquote{Statistical
  inference methods for two crossing survival curves: a comparison of methods,}
  \emph{PLoS One}, 10.

\bibitem[Lipsitz et~al.(1994)Lipsitz, Dear, and Zhao]{lipsitz1994jackknife}
Lipsitz, S.~R., Dear, K.~B., and Zhao, L. (1994), \enquote{Jackknife estimators
  of variance for parameter estimates from estimating equations with
  applications to clustered survival data,} \emph{Biometrics}, pp. 842--846.

\bibitem[Logan et~al.(2008)Logan, Klein, and Zhang]{logan2008comparing}
Logan, B.~R., Klein, J.~P., and Zhang, M.-J. (2008), \enquote{Comparing
  treatments in the presence of crossing survival curves: an application to
  bone marrow transplantation,} \emph{Biometrics}, 64, 733--740.

\bibitem[Lorentz(1953)Lorentz]{Lorentz:1953}
Lorentz, G.~G. (1953), \emph{Bernstein polynomials}, University of Toronto
  Press, Toronto.

\bibitem[Martinussen and Scheike(2007)Martinussen and
  Scheike]{martinussen2007dynamic}
Martinussen, T. and Scheike, T.~H. (2007), \emph{Dynamic regression models for
  survival data}, Springer Science and Business Media, New York.

\bibitem[McGilchrist and Aisbett(1991)McGilchrist and
  Aisbett]{mcgilchrist1991regression}
McGilchrist, C. and Aisbett, C. (1991), \enquote{Regression with frailty in
  survival analysis,} \emph{Biometrics}, pp. 461--466.

\bibitem[McLain and Ghosh(2013)McLain and Ghosh]{Mclain:2013}
McLain, A.~C. and Ghosh, S.~K. (2013), \enquote{Efficient sieve maximum
  likelihood estimation of time-transformation models,} \emph{Journal of
  Statistical Theory and Practice}, 7, 285--303.

\bibitem[Oehlert(1992)Oehlert]{oehlert1992note}
Oehlert, G.~W. (1992), \enquote{A note on the delta method,} \emph{The American
  Statistician}, 46, 27--29.

\bibitem[Osman and Ghosh(2012)Osman and Ghosh]{Osman:2012}
Osman, M. and Ghosh, S.~K. (2012), \enquote{Nonparametric regression models for
  right-censored data using Bernstein polynomials,} \emph{Computational
  Statistics and Data Analysis}, 56, 559--573.

\bibitem[Petrone(1999)Petrone]{Petrone:1999}
Petrone, S. (1999), \enquote{Bayesian density estimation using Bernstein
  polynomials,} \emph{Canadian Journal of Statistics}, 27, 105--126.

\bibitem[Pettitt(1984)Pettitt]{Pettitt:1984}
Pettitt, A. (1984), \enquote{Proportional odds models for survival data and
  estimates using ranks,} \emph{Journal of the Royal Statistical Society,
  Series C}, 33, 169--175.

\bibitem[Plummer et~al.(2006)Plummer, Best, Cowles, and Vines]{coda}
Plummer, M., Best, N., Cowles, K., and Vines, K. (2006), \enquote{CODA:
  convergence diagnosis and output analysis for MCMC,} \emph{R News}, 6, 7--11.

\bibitem[Prentice(1973)Prentice]{Prentice:1973}
Prentice, R.~L. (1973), \enquote{Exponential survivals with censoring and
  explanatory variables,} \emph{Biometrika}, 60, 279--288.

\bibitem[Qi(2009)Qi]{qi2009comparison}
Qi, J. (2009), \enquote{Comparison of proportional hazards and accelerated
  failure time models,} Ph.D. thesis.

\bibitem[{R Core Team}(2019){R Core Team}]{R:2018}
{R Core Team} (2019), \emph{R: A Language and Environment for Statistical
  Computing}, R Foundation for Statistical Computing, Vienna, Austria.

\bibitem[Rickert(2017)Rickert]{Rviews:2017}
Rickert, J. (2017), \enquote{{Survival Analysis with R,} R Views Blog,}
  \emph{Online:
  \url{https://rviews.rstudio.com/2017/09/25/survival-analysis-with-r/}}.

\bibitem[Ross(2012)Ross]{ross2012simulation}
Ross, S. (2012), \emph{Simulation}, Knovel Library, Elsevier Science.

\bibitem[Scheike and Zhang(2011)Scheike and Zhang]{timereg}
Scheike, T.~H. and Zhang, M.-J. (2011), \enquote{Analyzing Competing Risk Data
  Using the {R} {timereg} Package,} \emph{Journal of Statistical Software}, 38,
  1--15.

\bibitem[Schwarz et~al.(1978)Schwarz et~al.]{schwarz1978estimating}
Schwarz, G. et~al. (1978), \enquote{Estimating the dimension of a model,}
  \emph{The Annals of Statistics}, 6, 461--464.

\bibitem[Spiegelhalter et~al.(2002)Spiegelhalter, Best, Carlin, and Van
  Der~Linde]{spiegelhalter2002bayesian}
Spiegelhalter, D.~J., Best, N.~G., Carlin, B.~P., and Van Der~Linde, A. (2002),
  \enquote{Bayesian measures of model complexity and fit,} \emph{Journal of the
  Royal Statistical Society, Series B}, 64, 583--639.

\bibitem[{Stan Development Team}(2016){Stan Development Team}]{Manual:2016}
{Stan Development Team} (2016), \enquote{Stan modeling language users guide and
  reference manual,} \emph{Online:
  \url{https://mc-stan.org/docs/2_19/reference-manual/index.html}}.

\bibitem[{Stan Development Team}(2018){Stan Development Team}]{rstan:2018}
{Stan Development Team} (2018), \enquote{{RStan}: the {R} interface to {Stan},}
  R package version 2.18.2.

\bibitem[Tenbusch(1997)Tenbusch]{Tenbusch:1997}
Tenbusch, A. (1997), \enquote{Nonparametric curve estimation with Bernstein
  estimates,} \emph{Metrika}, 45, 1--30.

\bibitem[{Terry M. Therneau} and {Patricia M. Grambsch}(2000){Terry M.
  Therneau} and {Patricia M. Grambsch}]{survival:2000}
{Terry M. Therneau} and {Patricia M. Grambsch} (2000), \emph{Modeling Survival
  Data: Extending the {C}ox Model}, Springer, New York.

\bibitem[Therneau and Grambsch(2000)Therneau and
  Grambsch]{therneau2000modeling}
Therneau, T.~M. and Grambsch, P.~M. (2000), \emph{Modeling survival data:
  extending the Cox model}, Springer Science and Business Media, Berlin.

\bibitem[Vehtari et~al.(2019)Vehtari, Gabry, Magnusson, Yao, and Gelman]{loo}
Vehtari, A., Gabry, J., Magnusson, M., Yao, Y., and Gelman, A. (2019),
  \enquote{loo: Efficient leave-one-out cross-validation and WAIC for Bayesian
  models,} R package version 2.2.0.

\bibitem[Venables and Ripley(2002)Venables and Ripley]{MASS}
Venables, W.~N. and Ripley, B.~D. (2002), \emph{Modern Applied Statistics with
  S}, Springer, New York, 4 edn., ISBN 0-387-95457-0.

\bibitem[Vitale(1975)Vitale]{Vitale:1975}
Vitale, R.~A. (1975), \emph{A Bernstein polynomial approach to density function
  estimation}, Elsevier, Amsterdam.

\bibitem[Watanabe(2013)Watanabe]{watanabe2013widely}
Watanabe, S. (2013), \enquote{A widely applicable Bayesian information
  criterion,} \emph{Journal of Machine Learning Research}, 14, 867--897.

\bibitem[Wu et~al.(2018)Wu, Fang, Cheng, Chu, Shih, and Chien]{Wu:2018}
Wu, Y.-J., Fang, W.-Q., Cheng, L.-H., Chu, K.-C., Shih, Y.-T., and Chien, L.-C.
  (2018), \enquote{A flexible Bayesian non-parametric approach for fitting the
  odds to case II interval-censored data,} \emph{Journal of Statistical
  Computation and Simulation}, 88, 3132--3150.

\bibitem[Zhou and Hanson(2018)Zhou and Hanson]{Zhou:2018}
Zhou, H. and Hanson, T. (2018), \enquote{A unified framework for fitting
  Bayesian semiparametric models to arbitrarily censored survival data,
  including spatially referenced data,} \emph{Journal of the American
  Statistical Association}, 113, 571--581.

\bibitem[Zhou et~al.(2017)Zhou, Hu, and Sun]{Zhou:2017}
Zhou, Q., Hu, T., and Sun, J. (2017), \enquote{A sieve semiparametric maximum
  likelihood approach for regression analysis of bivariate interval-censored
  failure time data,} \emph{Journal of the American Statistical Association},
  112, 664--672.

\end{thebibliography}

\end{document}